\documentclass[journal]{IEEEtran}

\ifCLASSINFOpdf
\usepackage[pdftex]{graphicx}
\else
\usepackage[dvips]{graphicx}
\fi

\usepackage{amsmath,amsthm,amssymb,amsfonts,bm}
\usepackage[table]{xcolor}
\usepackage{enumitem}
\usepackage{dsfont}
\usepackage{soul}
\usepackage{mathtools}
\usepackage{graphicx}
\usepackage[utf8]{inputenc}
\usepackage{multirow} 
\usepackage[belowskip=0pt,aboveskip=0pt]{caption}
\usepackage[belowskip=0pt,aboveskip=-0.5pt]{subcaption}
\usepackage[tablename=TABLE]{caption}
\usepackage{bbm}
\usepackage{algorithm,algorithmic}
\usepackage[normalem]{ulem}
\usepackage{xr}
\usepackage{xr-hyper}
\usepackage[colorlinks]{hyperref}
\usepackage{cleveref}
\usepackage{jhldef}

\newcommand{\beq}{\begin{equation}}
\newcommand{\eeq}{\end{equation}}

\long\def\/*#1*/{}

\begin{document}
%
\title{A Worker-Task Specialization Model for Crowdsourcing: Efficient Inference and Fundamental Limits}

\author{Doyeon Kim, Jeonghwan Lee and Hye Won Chung
\thanks{{Doyeon Kim (\href{mailto:highlowzz@alumni.kaist.ac.kr}{\texttt{highlowzz@alumni.kaist.ac.kr}}) and Hye Won Chung (\href{mailto:hwchung@kaist.ac.kr}{\texttt{hwchung@kaist.ac.kr}}) are with the School of Electrical Engineering at KAIST.
Jeonghwan Lee (\href{mailto:jhlee97@uchicago.edu}{\texttt{jhlee97@uchicago.edu}}) is a Ph.D. student at the Department of Statistics at the University of Chicago.
This research was supported by the National Research Foundation of Korea under Grant 2021R1C1C11008539, and by the Ministry of Science and ICT, South Korea, under the ITRC support program under Grant IITP-2023-2018-0-01402. This research was presented in part at 2021 and 2022 IEEE International Symposium on Information Theory \cite{kim2021crowdsourced,kim2022generalized}.}


}}

\maketitle


\begin{abstract}
Crowdsourcing system has emerged as an effective platform for labeling data with relatively low cost by using non-expert workers. Inferring correct labels from multiple noisy answers on data, however, has been a challenging problem, since the quality of the answers varies widely across tasks and workers. Many existing works have assumed that there is a fixed ordering of workers in terms of their skill levels, and focused on estimating worker skills to aggregate the answers from workers with different weights. In practice, however, the worker skill changes widely across tasks, especially when the tasks are heterogeneous. In this paper, we consider a new model, called $d$-type specialization model, in which each task and worker has its own (unknown) type and the reliability of each worker can vary in the type of a given task and that of a worker. We allow that the number $d$ of types can scale in the number of tasks. In this model, we characterize the optimal sample complexity to correctly infer the labels within any given accuracy, and propose label inference algorithms achieving the order-wise optimal limit even when the types of tasks or those of workers are unknown. We conduct experiments both on synthetic and real datasets, and show that our algorithm outperforms the existing algorithms developed based on more strict model assumptions.
\end{abstract}


\section{Introduction}
\label{sec:introduction}

\indent In recent years, crowdsourcing system has emerged as an effective platform to collect a large amount of useful data from low-cost human workers, by assigning tasks to workers, requesting them to respond to these tasks, and offering them compensations in monetary terms. One of the main goals of the crowdsourcing platform is the reliable estimation of the unknown ground-truth labels of a large volume of tasks, so-called the \emph{crowdsourced labeling}. Throughout this paper, we focus on the crowdsourced labeling problem for binary tasks. In this problem, one wants to guarantee the reliable recovery of the binary labels with the minimum number of queries to save the budget for crowdsourcing. The main challenges arise from the fact that human workers are often non-experts and their responses can be quite noisy. Moreover, the quality of the answers may change widely across workers and tasks. Thus, it is significant to have a realistic model that can describe the erroneous responses collected from the crowd worker, and to develop an efficient inference algorithm that can infer the correct labels from the noisy responses at the minimum number of queries.

\indent Many existing works on the crowdsourced labeling problem have assumed simple error model on worker responses to estimate the model parameters from the observed data. To explain the relation between the worker responses and the observation model, we consider the following crowdsourcing model for the binary labeling problems. Let $m$ and $n$ denote the numbers of tasks and workers, respectively, and let $a_i \in \left\{ \pm 1 \right\}$, $i \in [m]$, denote the ground-truth label associated with the $i$-th task. If the $j$-th worker is requested to provide a response for the $i$-th task, the answer $M_{ij}$ equals to $a_i$ with probability $F_{ij}$, while it equals to $- a_i$ with probability $1 - F_{ij}$, where $F_{ij} \in [0,1]$ represents the \emph{fidelity} of the $j$-th worker on the $i$-th task. Note that it holds that $\mathbb{E} \left[ M_{ij} \right] = a_i \left( 2 F_{ij} - 1 \right)$. Most existing works have assumed a rank-1 model for $\mathbb{E} \left[ \mathbf{M} \right]$ to make the recovery of the \emph{fidelity matrix} $\mathbf{F} := \left( F_{ij} : (i, j) \in [m] \times [n] \right)$ feasible from the the binary observation matrix $\mathbf{M} := \left( M_{ij} : (i, j) \in [m] \times [n] \right)$. For example, the single-coin Dawid-Skene (D\&S) model \cite{dawid1979maximum}, the most widely studied crowdsourcing model in the literature, assumes that the workers make errors independently and the probability that each worker makes an error is associated with a fixed unknown skill parameter $r_j \in [0,1]$, \emph{i.e.}, $F_{ij} = r_j$ for every task $i \in [m]$. Thus, it holds that $\mathbb{E} \left[ M_{ij} \right] = a_i \left( 2 F_{ij} - 1 \right) = a_i \left( 2 r_j - 1 \right)$ for every $(i, j) \in [m] \times [n]$, which implies $\textnormal{rank} \left( \mathbb{E} \left[ \mathbf{M} \right] \right) = 1$. 

\indent Under the low-rank assumption on the expected observation matrix $\mathbb{E} \left[ \mathbf{M} \right]$, a variety of inference algorithms have been proposed to estimate the worker skills and to utilize this information in inferring the ground-truth labels of the items. As an example, the existing works proposed inference algorithms to estimate the worker skills based on the spectral method \cite{jeong2023recovering, zhang2016spectral,dalvi2013aggregating, ghosh2011moderates, karger2013efficient}, belief propagation or iterative algorithms \cite{karger2014budget, karger2011iterative,li2014error, liu2012variational,ok2016optimality}, the expectation-maximization (EM) algorithms \cite{dawid1979maximum, gao2013minimax} or non-convex optimization algorithms for the rank-1 matrix completion problem \cite{pmlr-v80-ma18b,ma2020adversarial,ibrahim2019crowdsourcing}. The estimated skill parameters are then used to infer the ground-truth labels by approximating the maximum likelihood (ML) estimator \cite{jeong2023recovering, gao2013minimax,gao2016exact,zhang2016spectral,karger2013efficient,li2014error,raykar2010learning,smyth1994inferring,ipeirotis2010quality,berend2014consistency}, where the responses are aggregated by the weighted majority voting with the weights equal to the log-odds of the worker's fidelity. For the single-coin D\&S model, each worker possesses a weight fixed for all tasks.

\indent In practice, however, the D\&S model often falls short of describing the real-world datasets collected via crowdsourcing platforms, especially when the tasks are heterogeneous and the workers' skills change widely across tasks. For instance, when the tasks are to classify the images of athletes depending on certain criteria, \emph{e.g.}, whether the athlete in an image is an Olympic medalist, and the images include athletes of different sports types, \emph{e.g.}, baseball, soccer, volleyball, and basketball, then the quality of the responses may vary widely depending on the type of the task and the worker's interest or expertise. For example, a worker who knows little about the baseball may not be able to provide high-quality answers for the baseball players, but still can provide high-quality answers for the athletes of other sport types related to his/her expertise or interest. Since the single-coin DS model assumes a fixed skill level for each worker regardless of the task types, it cannot properly model the above scenarios.

\indent Accordingly, in this paper, we introduce a model that better describes the real-world datasets collected through crowdsourcing when the tasks are heterogeneous and the worker reliability can vary with the type of a given task. Specifically, we assume that each task and worker has its own (unknown) type among the set $[d] := \left\{ 1, 2, \cdots, d \right\}$, and the reliability of a worker on a certain task is determined by the worker-task type pair. Here, the type of a task may mean the category of the content associated with the task, and the type of a worker may indicate the speciality or the domain of interest of that worker. We call this model the \emph{$d$-type specialization model}. In this general model, the reliability of workers or the subset of high-fidelity workers can be completely changed depending on each task type. Contrast to the existing models assuming a low-rank structure, the maximum rank of the fidelity matrix $\mathbf{F}$ is now equal to $d$, which we allow to scale in the number of tasks $m$, and it has a block structure.  Since the types of workers or those of tasks are unknown in practical scenarios, the exact block structure of the fidelity matrix $\mathbf{F}$ is not revealed to the inference algorithms.

\indent In the $d$-type specialization model, we first fully characterize the optimal sample complexity required to correctly estimate the labels within a desired accuracy with the known  $\mathbf{F}$, and then design inference algorithms achieving the order-wise optimal sample complexity under mild assumptions even when $\mathbf{F}$ is unknown. We devise two types of inference algorithms depending on whether or not the task types are approximately known, while we assume that the worker types are unknown to both algorithms. Since the worker crowd is often anonymous and transient in crowdsourcing systems, revealing their type (speciality) in advance of collecting answers is impossible in general. Our algorithms are composed of two stages, where the first stage clusters workers into $d$ different groups based on their estimated types, and the second stage identifies \emph{reliable} vs. \emph{unreliable} worker clusters for each task (type) and then aggregates the responses from each worker cluster by upweighting the answers from the reliable worker clusters and downweighting or ignoring the answers from the unreliable worker clusters. 

\indent When the task types are revealed in advance, by grouping the tasks of the same estimated type and measuring the average bias on the responses of each worker cluster for each task type, one can identify \emph{reliable} vs. \emph{unreliable} worker clusters for each task type.  By aggregating the answers only from the reliable worker clusters, our algorithm achieves the order-wise optimal sample complexity of the ML estimator, even when the worker types and the exact fidelity matrix $\mathbf{F}$ are unknown. When the task types are unknown, on the other hand, our second algorithm identifies a unique worker cluster whose type matches the type of a given task, and it estimates the ground-truth labels by upweighting the responses from the matched worker cluster and downweighting the responses from the remaining worker clusters with properly designed weights. For this case, our algorithm achieves the order-wise optimal sample complexity for a special case of the $d$-type specialization model where the reliability of the matched worker cluster is equal to $p$ and those of the all unmatched worker clusters are equal to $q \in \left[ 1/2, p \right)$ for each task type. We prove that our second algorithm achieves the order-wise optimality for any $(p,q)$ range satisfying $1/2 \leq q < p \leq 1$.


\indent We also demonstrate the benefits of our proposed crowdsourcing model and the two inference algorithms in real applications by conducting experiments both on synthetic and real-world datasets. In particular, we collect the real-world datasets through a widely used crowdsourcing platform, Amazon Mechanical Turk, to classify the images of athletes and those of movies depending on a certain criteria. The athlete images possess some inherent types that can be categorized into four sports including football, baseball, soccer and basketball, and the inherent types of movie images can be categorized based on the related genres including action, romance, thriller and science-fiction (SF). By analyzing the responses provided by human workers, we first demonstrate that our $d$-type specialization model describes well the collected datasets, and the fidelity of each worker indeed widely varies depending on the type of a given task. We then show that our two inference algorithms, each of which assumes the presence/absence of an approximately known task-type vector, achieve better performances in inferring the ground-truth labels of the items, compared to the state-of-the-art baselines, which are mainly developed for the models assuming a fixed order of the worker skills for all tasks. Our code is publicly available at \href{https://github.com/iids88/dtype/}{our code}.

\subsection{Related works}
\label{subsec:related_works}

\subsubsection{Prior works on crowdsourcing systems}

One of the most extensively studied crowdsourcing models in the literature is the Dawid-Skene (D\&S) model \cite{dawid1979maximum}. Under this model, each worker is associated with a confusion matrix, which models the probability of giving a label $b \in [K]$ for the true label $a \in [K]$ for the $K$-ary classification tasks. The original D\&S model thus allows non-symmetric error probabilities over the labels. In the single-coin D\&S model, the model is further simplified such that each worker possesses a fixed skill level regardless of the true labels, and the error probabilities are symmetric over the labels. For the binary case (\emph{i.e.}, $K=2$), when the skill of the $j$-th worker is parameterized by $r_j \in [0,1]$, the worker provides a correct label (regardless of the true label) with probability $r_j$ and flips the label with probability $1 - r_j$ for every task. In the single-coin D\&S model, the key challenge is the absence of the exact knowledge of the skill parameters, and many existing works have focused on designing algorithms for estimating the worker skill parameters based on the ML estimation, moment matching or Bayesian methods \cite{dawid1979maximum, gao2013minimax, gao2016exact, zhang2016spectral, karger2011iterative, raykar2010learning, raykar2010learning, dalvi2013aggregating, pmlr-v80-ma18b}. 

\indent In some recent works, task difficulty is additionally considered in modeling the fidelity matrix $\mathbf{F}$. In \cite{khetan2016achieving}, task difficulty is modeled by the parameter $c_i \in \left[ 1/2, 1 \right]$, which represents the probability that the $i$-th task is perceived correctly, and the fidelity matrix $\mathbf{F}$ is modeled by $F_{ij} = c_i r_j+ \left( 1 - c_i \right) \left( 1 - r_j \right)$. In \cite{zhou2014aggregating}, the difficulty of task $i$ is represented by the parameter $\tau_i(c,k)$, which shows how likely item $i$ in class $c$ is answered as class $k$ when a randomly chosen worker answers it. Similarly, the skill of worker $j$ is parameterized using a variable $\sigma_j(c,k)$, which represents how likely worker $j$ labels a randomly given task in class $c$ as class $k$. Then, the probability that worker $j$ answers the task $i$ having class $c$ as class $k$ is modeled as $\frac{1}{Z_{ij}} \exp \left \{ \sigma_i(c,k) + \tau_j(c,k) \right \}$, where $Z_{ij}$ is the normalization term given by $Z_{ij} = \sum_{k} \exp \left \{ \sigma_i(c,k) + \tau_j(c,k) \right \}$. In \cite{shah2020permutation}, a non-parametric crowdsourcing model, called the \emph{permutation-based model}, is proposed, where it is assumed that there are a fixed ordering of workers and a fixed ordering of tasks in terms of their skill levels and difficulty levels, respectively. This model includes the single-coin D\&S model and the model in \cite{khetan2016achieving} as special cases. 
In \cite{jeong2023recovering}, the task difficulty is modeled by confusion probability among top two plausible answers for the multiple-choice crowdsourcing problem. 
In all the aforementioned crowdsourcing models, there is a fixed ordering of workers in terms of their skill levels that does not change across tasks. Thus, a reasonable inference algorithm mainly tries to figure out a fixed subset of workers whose responses tend to be more reliable than the rest of the workers for all the tasks, which thus can be effectively utilized to infer the correct labels \cite{shah2020permutation}.
 
\indent These models, however, inherently impose a homogeneity condition on the task types, which can be easily violated in practical scenarios. In contrast, we consider a more general crowdsourcing model where the reliability of a worker can vary depending on the worker-task type pairs, and thus our model allows the case where the subset of highly skilled workers can vary across tasks. There exist some previous works \cite{shah2018reducing, ho2013adaptive, ho2012online, welinder2010multidimensional, zhou2012learning, zhou2015regularized, khetan2018learning} that assume the type structure of the tasks or the different expertise of the workers. However, the model has not been theoretically analyzed \cite{welinder2010multidimensional} yet or only a very specific case of the $d$-type specialization model is studied \cite{shah2018reducing}. In \cite{ho2013adaptive, ho2012online}, an additional assumption is considered so that the system has access to \emph{golden standard} tasks of each task type, for which the ground-truth labels are revealed. In this paper, on the other hand, we assume that each ground-truth label is unknown to the system, and investigate the general $d$-type specialization model.

\subsubsection{Other relevant problems} 

The crowdsourced labeling problem has intimate connections to other query-based data acquisition and inference problems, where the goal is to recover the true labels by using noisy measurements on the true labels. The type of the measurements is determined by the type of queries, designed to be suitable for a given application. The crowdsourcing systems usually rely on the most basic querying method, called the \emph{repetition querying}, that asks one label at a time repeatedly to many human workers. More sophisticated querying schemes such as the pairwise comparison \cite{mazumdar2017clustering} or ``triangle'' queries \cite{vinayak2016crowdsourced}, which query the homogeneity between two or more objects simultaneously, have also been considered in the crowdsourced labeling to examine the trade-offs between the sample-efficiency and the increased error probability of the answers to complex queries. The pairwise comparison has also been studied extensively in the context of community detection for random graphs \cite{abbe2015exact}, where the goal is to recover the labels of the nodes by using the pairwise measurements represented by the presence or absence of an edge between the node pairs. The homogeneity measurement, which measures the level of the homogeneity among a given set of nodes, extends the pairwise comparison to more than two nodes, and it has been investigated in the context of the hypergraph clustering \cite{kim2017community, ahn2019community, lee2020robust} or the group testing \cite{dorfman1943detection}. The binary classification with XOR queries has also been broadly studied in diverse applications, such as the channel coding \cite{gallager1962low, mackay2005fountain, arikan2009channel}, hypergraph clustering \cite{ahn2019community}, random XOR-constrained satisfaction problem \cite{pittel2016satisfiability}, and recently also for the crowdsourced labeling \cite{DBLP:journals/tit/KimC21, DBLP:conf/isit/KimC20,8437703}. Complex queries such as XOR queries increase the information efficiency but at the same time decreases the fidelity of the worker responses since XOR queries are relatively harder to answer for human workers and the corresponding error rate in the answers may increase in practice \cite{DBLP:journals/tit/KimC21}. Thus, it is an open question whether more complex queries can indeed bring budget-efficiency in real-world crowdsourcing systems. In this work, we focus on the simplest and thus most reliable and widely adopted query type, repetition query, under a very general error model where the error probability of the measurements (worker responses) changes depending on the worker-task type pair. On the other hand, other query-based data acquisition problems such as graph clustering or constrained satisfaction problem often assume a simple error model where the error probability is identical over all the measurements.

\subsection{Organization of the paper}
\label{subsec:organization}

\indent This article will be organized as follows. In Section \ref{sec:model_problem_formulation}, we first introduce our main observation model and then formulate the crowdsourced labeling problem for binary tasks. Throughout Section \ref{sec:info_limits_sample_complexity}, we establish the information-theoretically optimal sample complexity for correctly inferring the ground-truth labels within a target accuracy under the newly proposed crowdsourcing model. In Section \ref{sec:proposed_algorithms}, we present two inference algorithms: one is for the case where side information of the ground-truth task-type vector is available, and the other is for the case where it is not. In the sequel, we analyze the proposed algorithms theoretically and provide the model parameter regimes where each algorithm can achieve the order-wise optimal sample complexity. We also compare our proposed algorithms against several existing baseline algorithms in Section \ref{subsec:comparison_baseline_algorithms}. Section \ref{sec:empirical_results} includes empirical results both on synthetic and real-world datasets, and Section \ref{sec:discussion} concludes the paper. The proofs of the main results will be presented in Section \ref{sec:proofs_theoretical_results}, whose technical details are available in appendices.

\subsection{Notations}
\label{subsec:notations}

\indent We denote a vector and a matrix by a bold-face letter, \emph{e.g.}, $\mathbf{v} = \left( v_i : i \in [n] \right)$ and $\mathbf{M} = \left( M_{ij} : (i, j) \in [m] \times [n] \right)$; Given any matrix $\mathbf{M} \in \mathbb{R}^{m \times n}$, let $\mathbf{M}_{i*}$ and $\mathbf{M}_{*j}$ denote its $i$-th row vector and its $j$-th column vector, respectively; For any positive integers $m$ and $n$, we denote by $\mathbf{1}_{m \times n}$ the $m \times n$ all-one matrix and $\mathbf{I}_n$ the $n \times n$ identity matrix; Given any $n \times n$ real symmetric matrix $\mathbf{S}$, let $\lambda_i (\mathbf{S})$ denote the $i$-th largest eigenvalue of $\mathbf{S}$; For $n \in \mathbb{N}$, we use $[n]$ to denote $[n] := \left\{ 1, 2, \cdots, n \right\}$; Given any set $\mathcal{X}$ and any non-negative integer $l$, we set $\binom{\mathcal{X}}{l} := \left\{ \mathcal{Y} \subseteq \mathcal{X} : \left| \mathcal{Y} \right| = l \right\}$; 
Let $\Delta(\mathcal{A}) := \left\{ \varphi (\cdot) \in \mathbb{R}^{\mathcal{A}} : \varphi(a) \geq 0,\ \forall a \in \mathcal{A} \textnormal{ and } \sum_{a \in \mathcal{A}} \varphi (a) = 1 \right\}$ denote the probability simplex over the finite alphabet $\mathcal{A}$; For any $\mathbf{a}^n := \left( a_1, a_2, \cdots, a_n \right) \in \mathcal{A}^n$, we denote the \emph{empirical distribution of $\mathbf{a}^n$} by $\hat{\mathcal{P}}_{\mathbf{a}^n} (\cdot) \in \Delta (\mathcal{A})$, where $\hat{\mathcal{P}}_{\mathbf{a}^n} (a) := \frac{1}{n} \sum_{i=1}^{n} \mathbbm{1} \left( a_i = a \right)$ for every $a \in \mathcal{A}$.

\section{Model and problem formulation}
\label{sec:model_problem_formulation}

\indent Throughout this section, we propose our underlying crowdsourcing model and define the performance metric for the crowdsourced labeling problem.

\subsection{Observation model: the $d$-type specialization model} 
\label{subsec:observation_model}

We first introduce our main observation model.

\begin{defi} [Crowdsourcing system]
\label{defi:crowdsourcing_system}
\normalfont{
Let $m$ and $n$ denote the number of tasks and workers, respectively, and $\mathbf{a} \in \left\{ \pm 1 \right\}^m$ denote the ground-truth vector of unknown binary labels associated with $m$ tasks, and $\mathcal{A} \subseteq [m] \times [n]$ denote the \emph{worker-task assignment set}, \emph{i.e.},
\begin{equation*}
\begin{split}
    &\mathcal{A} := \\
    &\left\{ (i, j) \in [m] \times [n]: \textnormal{$i$-th task is assigned to $j$-th worker} \right\}.
\end{split}
\end{equation*}
The \emph{crowdsourcing system with fidelity matrix $\mathbf{F} \in \left[ 0, 1 \right]^{m \times n}$}, denoted by $\textnormal{CS}(\mathbf{F})$, is a generative probabilistic model which observes $\left( M_{ij} : (i, j) \in [m] \times [n] \right) \in \left\{ -1, 0, +1 \right\}^{[m] \times [n]}$ as follows: $M_{ij} = 0$ if $(i, j) \in \left( [m] \times [n] \right) \setminus \mathcal{A}$, and
\begin{equation}
    \label{eqn:crowdsourcing_system}
    M_{ij} =
    \begin{cases}
        a_{i} & \textnormal{with probability } F_{ij}; \\
        - a_{i} & \textnormal{with probability } 1 - F_{ij},
    \end{cases}
\end{equation}
otherwise. We further assume the independence of the aggregation of noisy answers $\mathbf{M} := \left( M_{ij} : (i, j) \in \mathcal{A} \right)$.
}
\end{defi}

\indent Our goal is to estimate the ground-truth label vector $\mathbf{a} \in \left\{ \pm 1 \right\}^m$ based on the observed data $\mathbf{M}$ at the minimum number of queries per task, $|\mathcal{A}|/m$. The task assignment rule for the specification of the worker-task assignment set $\mathcal{A}$ is discussed later in detail when we introduce our algorithms in Section \ref{sec:proposed_algorithms}.

\indent We now introduce our main framework, called the \emph{$d$-type specialization model}, where each worker and task is associated with a certain type from the set $[d]$ and the fidelity of the response from the $j$-th worker on the $i$-th task, denoted by $F_{ij}$,  is determined by the type of the $i$-th task and the type of the $j$-th worker. 
 
\begin{defi} [The $d$-type specialization model]
\label{defi:specialization_model}
\normalfont{
Let $\bm{\mu}$, $\bm{\nu} \in \Delta ([d])$ be any two probability distributions over $[d]$ and $\mathcal{Q}(\cdot, \cdot) : [d] \times [d] \rightarrow \left[ \frac{1}{2}, 1 \right]$. The \emph{d-type specialization model}, denoted by $\textnormal{SM} \left( d; \mathcal{Q}, \bm{\mu}, \bm{\nu} \right)$, is defined to be the crowdsourcing system $\textnormal{CS}(\mathbf{F})$ whose fidelity matrix $\mathbf{F}$ is generated as per the following prior distribution over $\left[ \frac{1}{2}, 1 \right]^{m \times n}$:
\begin{enumerate}
    \item A \emph{task-type vector} $\mathbf{t} = \left( t_i : i \in [m] \right) \in [d]^m$ and a \emph{worker-type vector} $\mathbf{w} = \left( w_j : j \in [n] \right) \in [d]^n$ are generated according to the following distributions: $\mathbf{t} \sim \bm{\mu}^{\otimes m}$ and $\mathbf{w} \sim \bm{\nu}^{\otimes n}$;
    \item The value of the fidelity $F_{ij}$ is determined by the pair of the $i$-th task type and the $j$-th worker type $\left( t_i, w_j \right)$: for each $(i, j) \in [m] \times [n]$,
    \begin{equation*}
        F_{ij} = \mathcal{Q} \left( t_i, w_j \right).
    \end{equation*}
\end{enumerate}
Here, the matrix $\mathcal{Q}(\cdot, \cdot) : [d] \times [d] \rightarrow \left[ \frac{1}{2}, 1 \right]$ is referred to as the \emph{reliability matrix} of the $d$-type specialization model $\textnormal{SM} \left( d; \mathcal{Q}, \bm{\mu}, \bm{\nu} \right)$.
}
\end{defi}

In this model, we assume that the task type $t_i$ and the worker type $w_j$ are generated from some prior distributions $\bm{\mu}$ and $\bm{\nu}$ over the set $[d]$ independently, respectively, and the task-worker type pair $\left( t_i, w_j \right)$ determines the fidelity of the $j$-th worker on the $i$-th task as $F_{ij} = \mathcal{Q} \left( t_i, w_j \right)$ where $\mathcal{Q}(\cdot, \cdot) : [d] \times [d] \rightarrow \left[ \frac{1}{2}, 1 \right]$ is the (unknown) reliability matrix. If $\mathcal{Q}$ is a full-rank matrix, the fidelity matrix $\mathbf{F}$ has rank $d$. The number $d$ of types can scale in the number $m$ of tasks or the number $n$ of workers. Even though the number $d$ of possible types is assumed to be the same for the tasks and the workers, by allowing arbitrary prior distributions $\bm{\mu}$ and $\bm{\nu}$ over $[d]$, which may take zero probability at some types, we effectively allow the case where the number of task types and that of worker types are different. 

\indent The $d$-type specialization model was first explored by \cite{shah2018reducing}, but with a specific assumption that both the task types and worker types are uniformly distributed over $[d]$ and the reliability matrix $\mathcal{Q}(\cdot, \cdot) : [d] \times [d] \rightarrow \left[ \frac{1}{2}, 1 \right]$ is given by $\mathcal{Q} \left( t, w \right) = p > 1/2$ if $t = w$; $\mathcal{Q} \left( t, w \right) = 1/2$ otherwise, that is, the workers provide more reliable responses than random guesses only when the worker type and the task type match. In this article, we consider a more general $d$-type specialization model together with some assumptions on the reliability matrix $\mathcal{Q}$, which will be addressed later where we design inference algorithms and analyze their statistical performances. One assumption we impose commonly throughout the paper is stated below.

\begin{assumption}
\label{assumption:strong_ass_cqcm}
We define a $d \times d$ matrix $\Phi \left( \mathcal{Q}, \bm{\mu}, \bm{\nu} \right) (\cdot, \cdot) : [d] \times [d] \rightarrow [0, 1]$, called the collective quality correlation matrix corresponding to the reliability matrix $\mathcal{Q}$, by
\begin{equation}
    \begin{split}
    \label{eqn:collective_qual}
        &\Phi \left( \mathcal{Q}, \bm{\mu}, \bm{\nu} \right) (a, b) := \sum_{t=1}^{d} \mu (t) \left\{ 2 \mathcal{Q} (t, a) - 1 \right\} \left\{ 2 \mathcal{Q} (t, b) - 1 \right\}
    \end{split}
\end{equation}
for $\forall (a, b) \in [d] \times [d]$.
 Also, let 
\begin{equation}
    \begin{split}\label{eqn:def_pm_pu}
        p_m &:= \min \left\{ \Phi \left( \mathcal{Q}, \bm{\mu}, \bm{\nu} \right) (a, a) : a \in [d] \right\}; \\
        p_u &:= \max \left\{ \Phi \left( \mathcal{Q}, \bm{\mu}, \bm{\nu} \right) (a, b) : a \neq b \textnormal{ in } [d] \right\}
    \end{split}
\end{equation}
be the minimum intra-cluster collective quality correlation and the maximum inter-cluster collective quality correlation, respectively. Then, it holds that
\begin{equation}\label{eqn:cond_pm_pu}
    p_m > p_u.
\end{equation}
\end{assumption}

In Assumption \ref{assumption:strong_ass_cqcm} (which will be referred to as the \emph{strong assortativity of $\Phi \left( \mathcal{Q}, \bm{\mu}, \bm{\nu} \right)$} from now on), the collective quality correlation matrix $\Phi \left( \mathcal{Q}, \bm{\mu}, \bm{\nu} \right)$ extends the notion of the collective intelligence of the crowd \cite{karger2014budget, khetan2016achieving} to the $d$-type specialization model. The diagonal entry $\Phi \left( \mathcal{Q}, \bm{\mu}, \bm{\nu} \right) (a, a) = \sum_{t=1}^{d} \mu(t) \left\{ 2 \mathcal{Q}(t,a) - 1 \right\}^2$ represents the average quality of the type-$a$ worker cluster in responding $d$-different task types distributed by $\bm{\mu}$. Since we assume $\mathcal{Q}(t,a) \in \left[ \frac{1}{2},1 \right]$, the value $\Phi \left( \mathcal{Q}, \bm{\mu}, \bm{\nu} \right) (a, a)$ increases as each reliability $\mathcal{Q}(t,a)$ increases. The off-diagonal entry $\Phi \left( \mathcal{Q}, \bm{\mu}, \bm{\nu} \right) (a, b)$, where $a \neq b$ in $[d]$, on the other hand, reads the correlation of the quality between the type-$a$ and the type-$b$ clusters of workers over all task types. If the collective quality of each worker cluster averaged over all task types is the same, \emph{i.e.}, $\Phi \left( \mathcal{Q}, \bm{\mu}, \bm{\nu} \right) (a, a) = \sum_{t=1}^{d} \mu (t) \left\{ 2 \mathcal{Q} (t, a) - 1 \right\}^2$ are the same for every $a \in [d]$, then the Cauchy-Schwarz inequality gives $p_m \geq p_u$. Assumption \ref{assumption:strong_ass_cqcm} describes a scenario that the collective quality correlation between any two workers of the same type is strictly higher than that between any two workers of different types. This assumption will be used when we design a worker clustering algorithm to cluster workers based on their estimated types by using similarity on the responses for a subset of tasks between every pair of workers. 

\subsection{Performance metric}

\indent To measure the quality of an estimator $\hat{\mathbf{a}}(\cdot) : \left\{ \pm 1 \right\}^{\mathcal{A}} \rightarrow \left\{ \pm 1 \right\}^m$, the loss function $\mathcal{L} : \left\{ \pm 1 \right\}^m \times \left\{ \pm 1 \right\}^m \rightarrow \mathbb{R}_{+}$ defined by the mismatch ratio between two label vectors will be considered:
\begin{equation*}
    \mathcal{L} \left( \mathbf{x}, \mathbf{y} \right) := \frac{1}{m} \sum_{i=1}^{m} \mathbbm{1}(x_i \neq y_i ) = \frac{d_{\textnormal{H}} \left( \mathbf{x}, \mathbf{y} \right)}{m},
\end{equation*}
where $d_{\textnormal{H}}(\cdot, \cdot)$ denotes the Hamming distance between two vectors. Then, one can evaluate the performance of the estimator $\hat{\mathbf{a}} (\cdot)$ by its expected loss. Formally, the \emph{risk function of an estimator $\hat{\mathbf{a}}(\cdot) : \left\{ \pm 1 \right\}^{\mathcal{A}} \rightarrow \left\{ \pm 1 \right\}^m$} is given by
\begin{equation*}
    \mathcal{R} \left( \mathbf{a}, \hat{\mathbf{a}} \right) := \mathbb{E}_{\mathbf{a}} \left[ \mathcal{L} \left( \mathbf{a}, \hat{\mathbf{a}} (\mathbf{M}) \right) \right] = \frac{1}{m} \sum_{i=1}^{m} \mathbb{P} \left\{ \hat{a}_i (\mathbf{M}) \neq a_{i} \right\}.
\end{equation*}
The main question is to find the minimal number of queries per task, $|\mathcal{A}|/m$, required for the recovery performance
\begin{equation}
    \label{eqn:expected_accuracy}
    \mathcal{R} \left( \mathbf{a}, \hat{\mathbf{a}} \right) =   \frac{1}{m} \sum_{i=1}^{m} \mathbb{P} \left\{ \hat{a}_i (\mathbf{M}) \neq a_{i} \right\}\leq \alpha,
\end{equation}
for any given recovery accuracy $\alpha \in (0, 1)$.

\section{Information-theoretic limits on the sample complexity}
\label{sec:info_limits_sample_complexity}

\indent We first establish the information-theoretic limits on the sample complexity to achieve the recovery performance \eqref{eqn:expected_accuracy}, under general $d$-type specialization model $\textnormal{SM} \left( d; \mathcal{Q}, \bm{\mu}, \bm{\nu} \right)$. The statistical optimality result will be characterized in terms of the minimax risk:
\begin{equation}
    \label{eqn:minimax_risk}
    \mathcal{R}^* (\mathcal{A}) := \inf_{\hat{\mathbf{a}}} \left( \sup_{\mathbf{a} \in \left\{ \pm 1 \right\}^m} \mathcal{R} \left( \mathbf{a}, \hat{\mathbf{a}} \right) \right),
\end{equation}
where $\hat{\mathbf{a}}$ ranges over all estimators based on the worker-task assignment set $\mathcal{A} \subseteq [m] \times [n]$, \emph{i.e.}, all functions from $\left\{ \pm 1 \right\}^{\mathcal{A}}$ to $\left\{ \pm 1 \right\}^m$. 

\indent In order to derive an information-theoretic guarantee for the achievability of the recovery performance \eqref{eqn:expected_accuracy}, we consider the maximum likelihood (ML) estimation for the ground-truth vector of labels $\mathbf{a} \in \left\{ \pm 1 \right\}^m$ under the crowdsourcing system $\textnormal{CS}(\mathbf{F})$. The likelihood function of observing the collection of responses $\mathbf{M} = \left( M_{ij} : (i, j) \in \mathcal{A} \right) \in \left\{ \pm 1 \right\}^{\mathcal{A}}$ is given by
\begin{equation*}
    \begin{split}
        \mathbb{P}_{\mathbf{a}} \left\{ \mathbf{M} \right\} &= \prod_{(i, j) \in \mathcal{A}} \mathbb{P}_{a_i} \left\{ M_{ij} \right\} \\
        &= \prod_{(i, j) \in \mathcal{A}} \left[ F_{ij}^{\frac{ 1 + a_i M_{ij}}{2}} \left( 1 - F_{ij} \right)^{\frac{ 1 - a_i M_{ij}}{2}} \right] \\
        &= \prod_{(i, j) \in \mathcal{A}} \left[ \sqrt{ F_{ij} \left( 1 - F_{ij} \right)} \left( \frac{F_{ij}}{1 - F_{ij}} \right)^{\frac{a_i M_{ij}}{2}} \right].
    \end{split}
\end{equation*}
So its log-likelihood function can be computed as
\begin{equation}
    \label{eqn:log_likelihood_v1}
    \begin{split}
        &\log \left( \mathbb{P}_{\mathbf{a}} \left\{ \mathbf{M} \right\} \right) \\
        = \ & \sum_{(i, j) \in \mathcal{A}} \left[ \frac{1}{2} \log \left\{ F_{ij} \left( 1 - F_{ij} \right) \right\} + \frac{a_i M_{ij}}{2} \log \left( \frac{F_{ij}}{1 - F_{ij}} \right) \right] \\
        = \ & \frac{1}{2} \sum_{(i, j) \in \mathcal{A}} \log \left\{ F_{ij} \left( 1 - F_{ij} \right) \right\} \\
        &+ \frac{1}{2} \sum_{i=1}^{m} a_i \left[ \sum_{j \in \mathcal{A}(i)} \log \left( \frac{F_{ij}}{1 - F_{ij}} \right) M_{ij} \right]
    \end{split}
\end{equation}
where $\mathcal{A}(i) := \left\{ j \in [n] : (i, j) \in \mathcal{A} \right\}$ denotes the set of workers assigned to the $i$-th task. Therefore, the ML estimator $\hat{\mathbf{a}}^{\textnormal{ML}}(\cdot) : \left\{ \pm 1 \right\}^{\mathcal{A}} \rightarrow \left\{ \pm 1 \right\}^{m}$ of the ground-truth label vector $\mathbf{a} \in \left\{ \pm 1 \right\}^m$ is given by
\begin{equation}
    \label{eqn:ml_estimator}
    \hat{a}_{i}^{\textnormal{ML}}(\mathbf{M}) = \textnormal{sign} \left( \sum_{j \in \mathcal{A}(i)} \log \left( \frac{F_{ij}}{1 - F_{ij}} \right) M_{ij} \right).
\end{equation}
As the computation of the ML estimator $\hat{\mathbf{a}}^{\textnormal{ML}}(\cdot) : \left\{ \pm 1 \right\}^{\mathcal{A}} \rightarrow \left\{ \pm 1 \right\}^{m}$ \eqref{eqn:ml_estimator} requires an exact knowledge of the fidelity matrix $\mathbf{F}$, the decision rule \eqref{eqn:ml_estimator} is also called the \emph{maximum likelihood oracle}. In practical crowdsourcing scenarios, the fidelity matrix $\mathbf{F}$ is unknown at the label inference algorithms, and we cannot devise the ML estimator. Nevertheless, our analysis of the ML estimator provides a benchmark for making comparisons against other estimators, to be presented in the subsequent sections, that do not require any knowledge of the fidelity matrix $\mathbf{F}$. We now present an information-theoretic guarantee for the desired recovery performance \eqref{eqn:expected_accuracy} achievable by the ML estimator \eqref{eqn:ml_estimator}. The proof of the following result can be found in Appendix \ref{sec:proof_thm:achievability_ml}.

\begin{thm} [Information-theoretic achievability]
\label{thm:achievability_ml}
For any given recovery accuracy $\alpha \in \left( 0, \frac{1}{2} \right]$, the ML estimator \eqref{eqn:ml_estimator} achieves the recovery performance \eqref{eqn:expected_accuracy}:
\[
    \inf_{\hat{\mathbf{a}}} \left( \sup_{\mathbf{a} \in \left\{ \pm 1 \right\}^m} \mathcal{R} \left( \mathbf{a}, \hat{\mathbf{a}} \right) \right) \leq \sup_{\bfa \in \left\{ \pm 1 \right\}^m} \calR \left( \bfa, \hat{\bfa}^{\textnormal{ML}} \right) \leq \alpha,
\]
under the $d$-type specialization model $\textnormal{SM} \left( d; \mathcal{Q}, \bm{\mu}, \bm{\nu} \right)$ if the following condition holds:
\begin{equation}
    \label{eqn:achievability_ml}
    \min_{i \in [m]} \left| \mathcal{A}(i) \right|
    \geq \frac{1}{\gamma_1 \left( d; \mathcal{Q}, \bm{\mu}, \bm{\nu} \right)} \log \left( \frac{1}{\alpha} \right),
\end{equation}
where the error exponent $\gamma_1 \left( d; \mathcal{Q}, \bm{\mu}, \bm{\nu} \right)$ is defined by
\begin{equation}
    \label{eqn:error_exponent_achievability_ml}
    \begin{split}
        &\gamma_1 \left( d; \mathcal{Q}, \bm{\mu}, \bm{\nu} \right)  :=  \\
        \ & \log \left( \frac{1}{2 \max_{t \in [d]} \left\{ \sum_{w=1}^{d} \nu (w) \sqrt{\mathcal{Q}(t, w) \left( 1 - \mathcal{Q}(t, w) \right)} \right\}} \right).
    \end{split}
\end{equation}
\end{thm}

\indent Theorem \ref{thm:achievability_ml} implies that the error probability of the ML estimator decreases exponentially in the number of responses $\left| \mathcal{A}(i) \right|$ to the $i$-th task in the rate of $\mathbb{P} \left\{ \hat{a}_{i}^{\textnormal{ML}}(\mathbf{M}) \neq a_{i} \right\} \leq \exp \left\{ - \left| \mathcal{A} (i) \right| \cdot \gamma_1 \left( d; \mathcal{Q}, \bm{\mu}, \bm{\nu} \right) \right\}$. Next, the statistical impossibility result can be summarized into the following form whose detailed proof is deferred to Appendix \ref{sec:proof_thm:statistical_impossibility}:

\begin{thm} [Statistical impossibility]
\label{thm:statistical_impossibility}
Given any recovery accuracy $\alpha \in \left( 0, \frac{1}{8} \right]$, no inference methods with the average number of queries per task, $\left| \mathcal{A} \right| / m$, satisfying
\begin{equation}
    \label{eqn:statistical_impossibility}
    \gamma_2 \left( d; \mathcal{Q}, \bm{\mu}, \bm{\nu} \right) \left( \frac{\left| \mathcal{A} \right|}{m} \right) + \Gamma (d;\mathcal{Q}) \sqrt{\frac{\left| \mathcal{A} \right|}{m}} < \log \left( \frac{1}{4 \alpha} \right),
\end{equation}
can achieve the desired statistical performance \eqref{eqn:expected_accuracy}, i.e., $\inf_{\hat{\mathbf{a}}} \left( \sup_{\mathbf{a} \in \left\{ \pm 1 \right\}^m} \mathcal{R} \left( \mathbf{a}, \hat{\mathbf{a}} \right) \right) > \alpha$, in the $d$-type specialization model $\textnormal{SM} \left( d; \mathcal{Q}, \bm{\mu}, \bm{\nu} \right)$.
  Here, the error exponent $\gamma_2 \left( d; \mathcal{Q}, \bm{\mu}, \bm{\nu} \right)$ is defined by
\begin{equation}
    \label{eqn:error_exponent_statistical_impossibility}
    \begin{split}
        &\gamma_2 \left( d; \mathcal{Q}, \bm{\mu}, \bm{\nu} \right):= \\
         \ & \log \left( \frac{1}{2 \sum_{(t, w) \in [d] \times [d]} \mu (t) \nu (w) \sqrt{\mathcal{Q}(t, w) \left( 1 - \mathcal{Q} (t, w) \right)}} \right),
    \end{split}
\end{equation}
and $\Gamma(d;\mathcal{Q})$ denotes the log-odds of the maximum reliability between tasks and workers, i.e., 
\begin{equation}
    \label{eqn:gamma_exponent}
    \Gamma(d;\mathcal{Q}) := \log \left( \frac{\max \left\{ \mathcal{Q}(t, w) : (t, w) \in [d] \times [d] \right\}}{1 - \max \left\{ \mathcal{Q}(t, w) : (t, w) \in [d] \times [d] \right\}} \right).
\end{equation}
\end{thm}

\indent In summary, the ML estimator \eqref{eqn:ml_estimator} succeeds to recover the ground-truth vector within the desired recovery accuracy \eqref{eqn:expected_accuracy} provided that
\begin{equation}
    \label{eqn:achievability_ml_simplified_v1}
    \frac{\left| \mathcal{A} \right|}{m} = \Omega \left( \frac{1}{\gamma_1 \left( d; \mathcal{Q}, \bm{\mu}, \bm{\nu} \right)} \log \left( \frac{1}{\alpha} \right) \right),
\end{equation}
while the recovery is impossible within the target accuracy \eqref{eqn:expected_accuracy} whenever
\begin{equation}
    \label{eqn:statistical_impossibility_simplified_v1}
    \begin{split}
    \frac{\left| \mathcal{A} \right|}{m} =
    &o \left( \min \left\{ \frac{1}{\gamma_2 \left( d; \mathcal{Q}, \bm{\mu}, \bm{\nu} \right)} \log \left( \frac{1}{\alpha} \right),\right.\right.\\
    &\qquad\qquad \left.\left. \left
    ( \frac{1}{\Gamma \left( d; \mathcal{Q} \right)} \log \left( \frac{1}{\alpha} \right) \right)^2 \right\} \right)
    \end{split}
\end{equation}
from Theorem \ref{thm:achievability_ml} and \ref{thm:statistical_impossibility}. Note that the error exponents for the information-theoretic upper bound $\gamma_1 \left( d; \mathcal{Q}, \bm{\mu}, \bm{\nu} \right)$ and the information-theoretic lower bound $\gamma_2 \left( d; \mathcal{Q}, \bm{\mu}, \bm{\nu} \right)$ coincide when the quantities 
\begin{equation*}
    \sum_{w=1}^{d} \nu (w) \sqrt{\mathcal{Q} (t, w) \left( 1 - \mathcal{Q} (t, w) \right)}
\end{equation*}
are equal for every $t \in [d]$, \emph{i.e.}, when all task types have the same difficulty when averaged over worker types.

\begin{remark} [Tightness of the information-theoretic bounds]
\label{rmk:tightness_info_bounds} 
\normalfont{
To see the tightness of the information-theoretic bounds, we now pay our attention to the parameter regime for which
\begin{equation}
    \label{eqn:specific_regime_v1}
    \begin{split}
        \gamma_1 \left( d; \mathcal{Q}, \bm{\mu}, \bm{\nu} \right) \asymp & \ \gamma_2 \left( d; \mathcal{Q}, \bm{\mu}, \bm{\nu} \right); \\
        \sup_{d \in \bbN} \max_{(t, w) \in [d] \times [d]} & \mathcal{Q} (t, w) < 1.
    \end{split}
\end{equation}
Here, it is worth noting that $\gamma_1 \left( d; \mathcal{Q}, \bm{\mu}, \bm{\nu} \right) \leq \gamma_2 \left( d; \mathcal{Q}, \bm{\mu}, \bm{\nu} \right)$ in general. In the parameter regime \eqref{eqn:specific_regime_v1}, it is evident that $\sup \left\{ \Gamma \left( d; \mathcal{Q} \right) : d \in \mathbb{N} \right\} < +\infty$ for $\Gamma \left( d; \mathcal{Q} \right)$ in \eqref{eqn:gamma_exponent}, thereby the recovery accuracy \eqref{eqn:expected_accuracy} is not achievable whenever
\begin{equation}
\begin{split}
    \label{eqn:statistical_impossibility_simplified_v2}
 &   \frac{\left| \mathcal{A} \right|}{m} =\\
 &   o \left( \min \left\{ \frac{1}{\gamma_2 \left( d; \mathcal{Q}, \bm{\mu}, \bm{\nu} \right)} \log \left( \frac{1}{\alpha} \right), \left( \log \left( \frac{1}{\alpha} \right) \right)^2 \right\} \right).
    \end{split}
\end{equation}
\begin{enumerate} [label=(\alph*)]
    \item $\gamma_1 \left( d; \mathcal{Q}, \bm{\mu}, \bm{\nu} \right) \asymp \gamma_2 \left( d; \mathcal{Q}, \bm{\mu}, \bm{\nu} \right) \gtrsim \left\{ \log \left( \frac{1}{\alpha} \right) \right\}^{-1}$: it is clear that the right-hand side of the equation \eqref{eqn:statistical_impossibility_simplified_v2} becomes $o \left( \frac{1}{\gamma_1 \left( d; \mathcal{Q}, \bm{\mu}, \bm{\nu} \right)} \log \left( \frac{1}{\alpha} \right) \right)$. Hence, the ML estimator \eqref{eqn:ml_estimator} is statistically optimal with the order-wise fundamental limit of the sample complexity:
    \begin{equation}
        \label{eqn:fundamental_limits_v1}
        \frac{|\mathcal{A}|}{m} = \Theta \left( \frac{1}{\gamma_1 \left( d; \mathcal{Q}, \bm{\mu}, \bm{\nu} \right)} \log \left( \frac{1}{\alpha} \right) \right).
    \end{equation}
    \item $\gamma_1 \left( d; \mathcal{Q}, \bm{\mu}, \bm{\nu} \right) \asymp \gamma_2 \left( d; \mathcal{Q}, \bm{\mu}, \bm{\nu} \right) \lesssim \left\{ \log \left( \frac{1}{\alpha} \right) \right\}^{-1}$: the equations \eqref{eqn:achievability_ml_simplified_v1} and \eqref{eqn:statistical_impossibility_simplified_v2} guarantee that the ML estimator \eqref{eqn:ml_estimator} achieves the recovery performance \eqref{eqn:expected_accuracy} when
    \begin{equation*}
        \frac{|\mathcal{A}|}{m} = \Omega \left( \frac{1}{\gamma_1 \left( d; \mathcal{Q}, \bm{\mu}, \bm{\nu} \right)} \log \left( \frac{1}{\alpha} \right) \right),
    \end{equation*}
    while no algorithm whatsoever can reach the recovery accuracy \eqref{eqn:expected_accuracy}
    \begin{equation*}
        \frac{|\mathcal{A}|}{m} = o \left( \left\{ \log \left( \frac{1}{\alpha} \right) \right\}^2 \right).
    \end{equation*}
    This result shows that there exists an order-wise gap between the information-theoretical achievability condition and the converse result in this regime.
\end{enumerate}
}
\end{remark}

\section{Algorithms and performance analysis}
\label{sec:proposed_algorithms}

\indent Throughout this section, we propose novel estimation algorithms for the ground-truth label vector, and provide their performance guarantees under the $d$-type specialization model. Here, we remind that in the $d$-type specialization model, the fidelity $F_{ij} = \mathcal{Q} \left( t_i, w_j \right)$ of the $j$-th worker on the $i$-th task is determined by the task type $t_i$, the worker type $w_j$ and the reliability matrix $\mathcal{Q}(\cdot, \cdot): [d] \times [d] \to \left[ \frac{1}{2}, 1 \right]$, which are all unknown in most practical scenarios. In Section \ref{subsec:worker_clustering}, we first introduce a worker clustering algorithm based on their types $\mathbf{w} \in [d]^n$. This algorithm is used as an initial stage of our final inference algorithms. In Section \ref{subsec:algorithm_with_side_information}, we introduce an inference algorithm to estimate the true labels $\mathbf{a} \in \left\{ \pm 1 \right\}^m$ under the presence of side information of the task-type vector $\mathbf{t} \in [d]^m$. In Section \ref{subsec:algorithm_without_side_information}, we return to the setting where both the worker types and the task types are not revealed and then present a provably sample-efficient algorithm. We provide theoretical guarantees of the proposed inference algorithms under some reasonable assumptions on the reliability matrix $\mathcal{Q}$ and demonstrate the order-wise optimality of these algorithms by comparing the results with the information-theoretic limits from Section \ref{sec:info_limits_sample_complexity}.

\subsection{An SDP-based worker clustering stage}
\label{subsec:worker_clustering}

\indent We first propose an algorithm for clustering workers based on their responses on a subset of tasks, by solving a semi-definite program (SDP). This algorithm plays a critical role at the very first stage of the inference algorithms we will present later.

\begin{alg} [The SDP-based worker clustering algorithm]
\label{alg:worker_clustering}
\normalfont{ \
\begin{enumerate} [label=\arabic*.]
    \item[] \textbf{Input}:  
    the number of types $d \in \mathbb{N}$, and the parameters $\left( r, \eta\right) \in [m] \times \left( 0, +\infty \right)$;
    \item Let $\mathcal{S} \subseteq [m]$ denote the set of randomly chosen $r$ tasks, and assign each task in $\mathcal{S}$ to all $n$ workers;
    \item Based on the responses $\mathbf{M}_{i*} = \left( M_{ij} : j \in [n] \right)$ for each $i \in \mathcal{S}$, we define the \emph{similarity matrix} $\mathbf{A} \in \mathbb{R}^{n \times n}$ by $\mathbf{A} := \mathcal{P}_{\textnormal{off-diag}} \left( \sum_{i \in \mathcal{S}} \mathbf{M}_{i*}^{\top} \mathbf{M}_{i*} \right)$, where $\mathcal{P}_{\textnormal{off-diag}}(\cdot) : \mathbb{R}^{n \times n} \rightarrow \mathbb{R}^{n \times n}$ defines the linear operator that zeros out all diagonal entries of a matrix;
    \item Solve the following semi-definite program:        \begin{equation}
        \label{eqn:clustering_SDP}
        \begin{split}
            \max_{\mathbf{X} \in \mathbb{R}^{n \times n}} \ & \left\langle \mathbf{A} - \eta \mathbf{1}_{n \times n}, \mathbf{X} \right\rangle \\
            \textnormal{subject to } & \mathbf{X} \succeq \mathbf{0}; 
            \quad\left\langle \mathbf{I}_n, \mathbf{X} \right\rangle = n; \\
            &0 \leq X_{ij} \leq 1,\ \forall (i, j) \in [n] \times [n],
        \end{split}
    \end{equation}
    where $\eta \geq 0$ is a tuning parameter that should be pre-determined. Let $\hat{\mathbf{X}}_{\textnormal{SDP}}\in[0,1]^{n\times n}$ denote the optimal solution to the SDP \eqref{eqn:clustering_SDP};
    \item Extract the worker clusters $\left\{ \hat{\mathcal{W}}_1, \hat{\mathcal{W}}_2, \cdots, \hat{\mathcal{W}}_d \right\}$ by performing the approximate $k$-medoids clustering algorithm (\emph{Algorithm 1} in \cite{fei2018exponential}) on the row vectors of $\hat{\mathbf{X}}_{\textnormal{SDP}}$ to cluster the $n$ rows into $d$ groups. Here, we note that the number of types $d$ is known to us;
    \item[] \textbf{Output}: the explicit clusters $\left\{ \hat{\mathcal{W}}_1, \hat{\mathcal{W}}_2, \cdots, \hat{\mathcal{W}}_d \right\}$ of $n$ workers.
\end{enumerate}
}
\end{alg}

\indent  The SDP-based worker clustering stage of our inference algorithm is motivated by a line of works on the SDP-based similarity clustering \cite{ames2014guaranteed, vinayak2016similarity, lee2020robust, kim2017community, chen2014improved, chen2016statistical, chen2018convexified, fei2018exponential, kim2022generalized}. The rationale of the proposed worker clustering algorithm can be elucidated as follows. In Step 1 of Algorithm \ref{alg:worker_clustering}, we randomly choose $r$ tasks and collect responses for each of the tasks from all $n$ workers. The collected responses are then used to compute the \emph{similarity matrix} $\mathbf{A}$ whose $(i, j)$-th entry represents the similarity between the $i$-th worker and the $j$-th worker. If we denote $\mathbf{A}^{(i)} := \mathcal{P}_{\textnormal{off-diag}} \left( \mathbf{M}_{i*}^{\top} \mathbf{M}_{i*} \right)$ for each $i \in \mathcal{S}$, it can be easily shown that 
\begin{equation*}
    \begin{split}
     &   \mathbb{E} \left[ \left. A_{jk}^{(i)} \right| \mathbf{t}, \mathbf{w} \right] 
        \\
       &= \begin{cases}
            0 & \textnormal{if } j = k; \\
            \left\{ 2 \mathcal{Q} \left( t_i, w_j \right) - 1 \right\} \left\{ 2 \mathcal{Q} \left( t_i, w_k \right) - 1 \right\} & \textnormal{otherwise.}
        \end{cases}
    \end{split}
\end{equation*}
By taking expectations with respect to $\mathbf{t} \sim \bm{\mu}^{\otimes m}$ to both sides, we reach
\begin{equation*}
    \begin{split} 
        \mathbb{E} \left[ \left. A_{jk}^{(i)} \right| \mathbf{w} \right] 
       & = \  \mathbb{E}_{\mathbf{t} \sim \bm{\mu}^{\otimes m}} \left[ \mathbb{E} \left[ \left. A_{jk}^{(i)} \right| \mathbf{t}, \mathbf{w} \right] \right] \\
       & = \ 
        \begin{cases}
            0 & \textnormal{if } j = k; \\
            \Phi \left( \mathcal{Q}, \bm{\mu}, \bm{\nu} \right) \left( w_j, w_k \right) & \textnormal{otherwise,} 
        \end{cases}
    \end{split}
\end{equation*}
for every $i \in \mathcal{S}$ where $\Phi \left( \mathcal{Q}, \bm{\mu}, \bm{\nu} \right)(\cdot,\cdot)$ is the collective quality correlation matrix defined in \eqref{eqn:collective_qual}. From the definition of $p_m$ and $p_u$ in \eqref{eqn:def_pm_pu} of Assumption \ref{assumption:strong_ass_cqcm}, one can observe that
\begin{equation*}
    \begin{split}
        \mathbb{E} \left[ \left. A_{jk} \right| \mathbf{w} \right] &= \sum_{i \in \mathcal{S}} \mathbb{E} \left[ \left. A_{jk}^{(i)} \right| \mathbf{w} \right] \\
        &= r \cdot \Phi \left( \mathcal{Q}, \bm{\mu}, \bm{\nu} \right) \left( w_j, w_k \right) \\
        &
        \begin{cases}
            \geq r p_m & \textnormal{if } j \neq k \textnormal{ and } w_j = w_k; \\
            \leq r p_u & \textnormal{if } w_j \neq w_k.
        \end{cases}
    \end{split}
\end{equation*}
In words, if any two workers have the same type, their expected similarity is higher than or equal to $r p_m$. On the other hand, when their worker types are different, their expected similarity is lower than or equal to $r p_u$. Since we have assumed $p_m > p_u$ in Assumption \ref{assumption:strong_ass_cqcm}, one can exploit the gap between the similarity values of pairs of workers of the same type and pairs of workers of different types to recover the community structure of workers from the similarity matrix $\mathbf{A}$. This is the fundamental reason why the strong assortativity assumption of the collective quality correlation matrix $\Phi \left( \mathcal{Q}, \bm{\mu}, \bm{\nu} \right)$ (Assumption \ref{assumption:strong_ass_cqcm}) is necessary for establishing a theoretical guarantee of Algorithm \ref{alg:worker_clustering}.

\indent If we choose the tuning parameter $\eta$ in \eqref{eqn:clustering_SDP} to lie between $rp_m$ and $rp_u$, say $\eta=\frac{r(p_m+p_u)}{2}$, we have
\begin{equation*}
    \mathbb{E} \left[ \left. A_{jk} \right| \mathbf{w} \right] - \eta    
    \begin{cases}
        \geq 0 & \textnormal{if } j \neq k \textnormal{ and } w_j = w_k; \\
        \leq 0 & \textnormal{if } w_j \neq w_k.
    \end{cases}
\end{equation*}
With the above observation in place, one can consider the following combinatorial optimization problem:
\begin{equation}
    \label{eqn:combinatorial_opt}
    \max_{\mathbf{X} \in \mathcal{X} (n, d)}
    \left\langle \mathbb{E} \left[ \left. \mathbf{A} \right| \mathbf{w} \right] - \eta \mathbf{1}_{n \times n}, \mathbf{X} \right\rangle,
\end{equation}
where the search space $\mathcal{X}(n, d)$ of the problem \eqref{eqn:combinatorial_opt} denotes the set of all worker cluster matrices with $n$ nodes and $d$ clusters. Formally, the worker cluster matrix $\mathbf{X} (\sigma) \in \left\{ 0, 1 \right\}^{n \times n}$ associated with the given community assignment $\sigma : [n] \rightarrow [d]$ is defined by
\begin{equation*}
    \begin{split}
        \left[ \mathbf{X} (\sigma) \right]_{jk} :=
        \begin{cases}
            1 & \textnormal{if } \sigma (j) = \sigma (k); \\
            0 & \textnormal{otherwise.}
        \end{cases}
    \end{split}
\end{equation*}
Then, the search space $\mathcal{X}(n, d)$ of the combinatorial optimization problem \eqref{eqn:combinatorial_opt} is given by
\begin{equation*}
    \mathcal{X} (n, d) := \left\{ \left. \mathbf{X} (\sigma) \in \left\{ 0, 1 \right\}^{n \times n} \right| \sigma : [n] \rightarrow [d] \right\}.
\end{equation*}
One can find that the optimal solution to the problem \eqref{eqn:combinatorial_opt} reveals the ground-truth worker cluster matrix $\mathbf{X} (\mathbf{w}) \in \mathcal{X} (n, d)$ (here, we consider the worker-type vector $\mathbf{w} \in [d]^n$ as the ground-truth community assignment of workers), since the optimal solution $\mathbf{X}^* \in \mathcal{X} (n, d)$ to \eqref{eqn:combinatorial_opt} should take $X_{jk}^* = 1$ provided that $\mathbb{E} \left[ \left. A_{jk} \right| \mathbf{w} \right] - \eta \geq 0$, \emph{i.e.}, when $j \neq k$ and $w_j = w_k$, and 0 otherwise. 

\indent In the SDP \eqref{eqn:clustering_SDP} of Algorithm \ref{alg:worker_clustering}, we make use of the \emph{sample-level similarity matrix} $\mathbf{A}$ in lieu of the \emph{population-level similarity matrix} $\mathbb{E} \left[ \left. \mathbf{A} \right| \mathbf{w} \right]$, which requires the exact knowledge of the true worker-type vector $\mathbf{w}$, and relax the non-convex constraint $\mathbf{X} \in \mathcal{X} (n, d)$ in \eqref{eqn:combinatorial_opt} into convex constraints. This enables us to address the computational intractability issue of the original problem \eqref{eqn:combinatorial_opt} since the SDPs are known to be solvable efficiently by using various algorithms including the interior-point method \cite{huang2021solving, jiang2020faster} and first-order methods \cite{wen2010alternating}. 

\indent We next establish a sufficient condition on $\left( r, \eta \right)$ to guarantee the exact recovery of the true worker clusters by solving the SDP \eqref{eqn:clustering_SDP}. Hereafter, let $s_{w} := \left| \mathcal{W}_w \right|$ denote the size of the worker cluster of type $w \in [d]$ and 
\begin{equation*}
\begin{split}
   & s_{\min} := \min \left\{ s_w : w \in [d] \right\} \quad \textnormal{and}\\
   & s_{\max} := \max \left\{ s_w : w \in [d] \right\}
    \end{split}
\end{equation*}
denote the minimum size and the maximum size of the worker clusters, respectively.

\begin{lemma} [Exact recovery guarantee of Algorithm \ref{alg:worker_clustering}]
\label{lemma:exact_recovery_alg:worker_clustering}
Consider the $d$-type specialization model $\textnormal{SM} \left( d; \mathcal{Q}, \bm{\mu}, \bm{\nu} \right)$ whose collective quality correlation matrix $\Phi \left( \mathcal{Q}, \bm{\mu}, \bm{\nu} \right)$ is strongly assortative (Assumption \ref{assumption:strong_ass_cqcm}). Then, Algorithm \ref{alg:worker_clustering} exactly recovers the worker clusters with probability at least $1 - 4 n^{-11}$, provided that the parameter $\eta \geq 0$ required in the SDP \eqref{eqn:clustering_SDP} satisfies
\begin{equation}
    \label{eqn:tuning_parameter_condition}
    r \left( \frac{1}{4} p_m + \frac{3}{4} p_u \right) \leq \eta \leq r \left( \frac{3}{4} p_m + \frac{1}{4} p_u \right),
\end{equation}
as well as the number of randomly chosen tasks $r$ in Step 1 of Algorithm \ref{alg:worker_clustering} is at least
\begin{equation}
    \label{eqn:r_condition}
    r \geq C_1 \cdot \frac{n^2 \left( \log n \right)^2}{\left( p_m - p_u \right)^2 s_{\min}^2}
\end{equation}
for some absolute constant $C_1 > 0$.
\end{lemma}

\indent The proof of Lemma \ref{lemma:exact_recovery_alg:worker_clustering} will be provided in Appendix \ref{sec:proof_lemma:exact_recovery_alg:worker_clustering}. Let $\mathcal{X} \subseteq \mathbb{R}^{n \times n}$ denote the feasible region of the SDP \eqref{eqn:clustering_SDP}, and $\mathbf{X}^* :=
\mathbf{X}(\mathbf{w}) \in \mathcal{X} (n, d)$ refer to the \emph{ground-truth worker cluster matrix} induced by the worker-type vector $\mathbf{w} \in [d]^n$. In order to establish Lemma \ref{lemma:exact_recovery_alg:worker_clustering}, it suffices to show that $\mathbf{X}^*$ is the unique optimal solution to the SDP \eqref{eqn:clustering_SDP}. Thus, the main conclusion of Lemma \ref{lemma:exact_recovery_alg:worker_clustering} reduces to the following claim: for any $\mathbf{X} \in \mathcal{X} \setminus \left\{ \mathbf{X}^* \right\}$,
\begin{equation}
    \label{eqn:main_claim_lemma:exact_recovery_alg:worker_clustering}
    \Delta (\mathbf{X}) := \left\langle \mathbf{A} - \eta \mathbf{1}_{n \times n}, \mathbf{X}^* - \mathbf{X} \right\rangle > 0.
\end{equation}

\indent One of the key components in the proof of the claim \eqref{eqn:main_claim_lemma:exact_recovery_alg:worker_clustering} is to develop a sharp concentration result for the spectral norm $\left\| \mathbf{A} - \mathbb{E} \left[ \left. \mathbf{A} \right| \mathbf{w} \right] \right\|$. Due to the strong dependency between entries of the similarity matrix $\mathbf{A}$, we cannot employ the standard techniques from the random matrix theory literature mostly assuming the independence between entries of the data matrix. In order to establish a tight concentration bound on the spectral norm $\left\| \mathbf{A} - \mathbb{E} \left[ \left. \mathbf{A} \right| \mathbf{w} \right] \right\|$, we utilize an extensively used matrix concentration inequality, known as the \emph{matrix Bernstein's inequality} \cite{tropp2012user}. Details of this result can be found in Lemma \ref{lemma:proof_lemma:exact_recovery_alg:worker_clustering_v2}.

\begin{remark} [Choice of the tuning parameter $\eta$ for the SDP \eqref{eqn:clustering_SDP}]
\label{rmk:choice_tuning_parameter}
Solving the SDP \eqref{eqn:clustering_SDP} requires us a suitable choice of the tuning parameter $\eta \geq 0$ such that it obeys the required bound \eqref{eqn:tuning_parameter_condition} for the exact recovery of worker clusters of Algorithm \ref{alg:worker_clustering}. The condition \eqref{eqn:tuning_parameter_condition} effectively means that the tuning parameter $\eta$ should be chosen to lie between the minimum within-cluster similarity $rp_m$ and the maximum cross-cluster similarity $rp_u$ to ensure the exact recovery of worker clusters. In fact, one can realize that the parameter $\eta$ determines the resolution of the clusters in the solution, since a higher $\eta$ detects finer clusters with similarity larger than $\eta$. Therefore, varying $\eta$ results in different solutions with the desired within-cluster similarity specified by $\eta$. Therefore, it is generally impossible to determine a unique choice of $\eta$ from the data for the recovery of the hidden membership structure. Similar arguments have been discussed for the SDP-based clustering algorithms for the graph case \cite{chen2014improved} and for the uniform weighted hypergraph case \cite{lee2020robust}. 
\end{remark}

\subsection{Inference algorithm with side information of the task-type vector}
\label{subsec:algorithm_with_side_information}

\indent Throughout this section, we propose an algorithm for the estimation of the ground-truth labels $\mathbf{a} \in \left\{ \pm 1 \right\}^m$ when an \emph{approximate task-type vector} $\hat{\mathbf{t}} := \left( \hat{t}_1, \hat{t}_2, \cdots, \hat{t}_m \right) \in [d]^m$ that matches the ground-truth task types fairly well is revealed as side information. The proposed algorithm consists of the following two stages: (\romannumeral 1) clustering workers via Algorithm \ref{alg:worker_clustering}, and (\romannumeral 2) for each task type, ruling out the unreliable worker clusters and estimating labels by performing standard majority voting only among reliable worker clusters.

\begin{alg} [Two-stage inference algorithm with the presence of side information of the task-type vector]
\label{alg:with_side_information}
\normalfont{ \
\begin{enumerate} [label=\arabic*.]
    \item[] \textbf{Input}: the number of types $d \in \mathbb{N}$, the parameters $\left( r, \eta, l \right) \in [m] \times \left( 0, +\infty \right) \times [n]$ and $\bm{\zeta} := \left( \zeta_{ab} : (a, b) \in [d] \times [d] \right) \in \left( 0, +\infty \right)^{d \times d}$, and an approximate task-type vector $\hat{\mathbf{t}} := \left( \hat{t}_1, \hat{t}_2, \cdots, \hat{t}_m \right) \in [d]^m$ of the ground-truth task-type vector $\mathbf{t} \in [d]^m$;
    \item \textbf{SDP-based worker clustering}:
    \begin{enumerate} [label=(\alph*)]
        \item Employ Algorithm \ref{alg:worker_clustering} and obtain an explicit worker clusters $\left\{ \hat{\mathcal{W}}_1, \hat{\mathcal{W}}_2, \cdots, \hat{\mathcal{W}}_d \right\}$;
        \item For each task $i \in [m] \setminus \mathcal{S}$ and worker cluster $w \in [d]$, assign the task $i$ to $l$ workers sampled uniformly at random from the inferred worker cluster $\hat{\mathcal{W}}_w$. Let $\mathcal{A}(i) := \left\{ j \in [n] : (i, j) \in \mathcal{A} \right\}$ denote the set of workers assigned to the $i$-th task;
    \end{enumerate}
    \item \textbf{Ruling out the worker clusters of spammers and label inference via standard majority voting}:
    \begin{enumerate} [label=(\alph*)]
        \item For every task $i \in [m]$, we select $\mathcal{A}_w (i) \in \binom{\mathcal{A} (i) \cap \hat{\mathcal{W}}_w}{l}$ for $w \in [d]$, and define $\mathcal{A}' (i) := \bigcup_{w=1}^{d} \mathcal{A}_w (i) \subseteq \mathcal{A}(i)$;
        \item Set $\hat{\mathcal{T}}_a := \left\{ i \in [m]: \hat{t}_i = a \right\}$ for each $a \in [d]$. Then, we determine weights in a data-driven way as follows:
        \begin{equation}
        \begin{split}
            \label{eqn:choice_weights_alg:with_side_information}
            &\hat{\theta}_{ab} (\mathbf{M}) :=\\
            &
            \begin{cases}
                1 & \textnormal{if } \frac{1}{\left| \hat{\mathcal{T}}_a \right|} \sum_{i \in \hat{\mathcal{T}}_a} \left| \sum_{j \in \mathcal{A}_b (i)} M_{ij} \right| \geq \zeta_{ab}; \\
                0 & \textnormal{otherwise},
            \end{cases}
            \end{split}
        \end{equation}
        where $\zeta_{ab} > 0$, $(a, b) \in [d] \times [d]$, is a tuning parameter that should be specified;
        \item Finally, we estimate the ground-truth label $a_i$ via the following decision rule:
        \begin{equation}
            \label{eqn:decision_rule_alg:with_side_information}
            \hat{a}_i (\mathbf{M}) := \textnormal{sign} \left\{ \sum_{w=1}^{d} \left( \sum_{j \in \mathcal{A}_w (i)} \hat{\theta}_{\hat{t}_i w} (\mathbf{M}) \cdot M_{ij} \right) \right\},
        \end{equation}
        for every $i \in [m]$;
    \end{enumerate}
    \item[] \textbf{Output}: $\hat{\mathbf{a}}(\cdot) := \left( \hat{a}_i (\cdot) : i \in [m] \right) : \left\{ \pm 1 \right\}^{\mathcal{A}} \rightarrow \left\{ \pm 1 \right\}^m$.
\end{enumerate}
}
\end{alg}

\indent The main takeaway of Algorithm \ref{alg:with_side_information} is to select a subset of worker clusters for each task type that are believed to be composed of reliable workers, and use the responses only from those reliable worker clusters to infer the ground-truth labels. To that end, in Step 1-(b) of Algorithm \ref{alg:with_side_information}, we use the worker clusters recovered from Algorithm \ref{alg:worker_clustering} in order to assign each task to $l$ randomly sampled workers from each inferred cluster. To examine the reliability of each worker cluster for each task type $a \in [d]$, we then measure the bias of responses provided by each worker cluster $\mathcal{A}_b (i)$, $b \in [d]$, averaged over every task $i$ having (approximate) type $\hat{t}_i = a$, and set the weight $\hat{\theta}_{ab} (\mathbf{M})=1$ only for the worker clusters with averaged bias larger than a certain threshold $\zeta_{ab} > 0$ as in \eqref{eqn:choice_weights_alg:with_side_information}. The estimation of the ground-truth label is then conducted by using only the answers from worker clusters $w \in [d]$ having $\hat{\theta}_{\hat{t}_i w}(\mathbf{M}) = 1$ and ignoring answers from worker clusters having $\hat{\theta}_{\hat{t}_i w}(\mathbf{M}) = 0$ as in \eqref{eqn:decision_rule_alg:with_side_information}.

\indent We now address an assumption on the $d$-type specialization model to provide theoretical guarantees of Algorithm \ref{alg:with_side_information}.

\begin{assumption}
\label{assumption:multiple_spammers}
There is a universal constant $\epsilon \in \left( 0, \frac{1}{2} \right)$ such that $\mathcal{Q} (t, w) \in \left\{ \frac{1}{2} \right\} \cup \left[ \frac{1}{2} + \epsilon, 1 \right]$ for all $(t, w) \in [d] \times [d]$, and there exists a function $\delta (\cdot; d): [d] \to \left\{ 0, \frac{1}{d}, \cdots, \frac{d-1}{d}, 1 \right\}$ such that $\left| \textnormal{spammer}_{\mathcal{Q}} (t) \right| = d \left\{ 1 - \delta (t; d) \right\}$ for every $t \in [d]$, where
\begin{equation*}
    \textnormal{spammer}_{\mathcal{Q}} (t) := \left\{ w \in [d]: \mathcal{Q} (t, w) = \frac{1}{2} \right\}
\end{equation*}
denotes the set of worker clusters of spammers for the task type $t \in [d]$ with respect to the reliability matrix $\mathcal{Q}$. Automatically, $[d] \setminus \textnormal{spammer}_{\mathcal{Q}}(t)$ is the set of clusters of reliable workers for the task type $t$ with respect to the reliability matrix $\mathcal{Q}$.
\end{assumption}

\indent For this setup, there exists a gap $\epsilon \in \left( 0, \frac{1}{2} \right)$ between the reliability of the clusters of reliable workers and that of the worker clusters of spammers with respect to the reliability matrix $\mathcal{Q}$. We further define
\begin{equation*}
    \delta_{\min} (d) := \min_{t \in [d]} \delta \left( t; d \right) \quad \textnormal{and} \quad \delta_{\max} (d) := \max_{t \in [d]} \delta \left( t; d \right).
\end{equation*}
It would be worth noting that the weighting scheme \eqref{eqn:choice_weights_alg:with_side_information} aims to recover the ground-truth signals $\left\{ \theta_{ab}: (a, b) \in [d] \times [d] \right\}$, where
\begin{equation}\label{eqn:ground_truth_signal}
    \theta_{ab} :=
    \begin{cases}
        1 & \textnormal{if } b \in [d] \setminus \textnormal{spammer}(a); \\
        0 & \textnormal{otherwise.}
    \end{cases}
\end{equation}

\noindent To establish a statistical guarantee for Algorithm \ref{alg:with_side_information}, we further impose the approximated balancedness assumption on the task types.

\begin{assumption} [The approximate balancedness on the task-type clusters]
\label{assumption:approximate_balancedness_task_type}
Let $\mathcal{T}_a := \left\{ i \in [m]: t_i = a \right\}$.
The ground-truth task-type vector $\mathbf{t} \in [d]^m$ satisfies
\begin{equation*}
    \frac{\max \left\{ \left| \mathcal{T}_a \right|: a \in [d] \right\}}{\min \left\{ \left| \mathcal{T}_a \right|: a \in [d] \right\}} = \Theta (1) \textnormal{ as } d \to \infty.
\end{equation*}
\end{assumption}

\indent The following theorem characterizes the statistical performance of Algorithm \ref{alg:with_side_information} with side information, \emph{i.e.}, an approximate task-type vector $\hat{\mathbf{t}}$ that recovers the ground-truth task-type vector $\mathbf{t}$ fairly well.

\begin{thm} [Statistical analysis of Algorithm \ref{alg:with_side_information}]
\label{thm:performance_alg:with_side_information}
Consider the $d$-type specialization model $\textnormal{SM} \left( d; \mathcal{Q}, \bm{\mu}, \bm{\nu} \right)$ satisfying Assumptions \ref{assumption:strong_ass_cqcm}--\ref{assumption:approximate_balancedness_task_type}, $\delta (t; d) \geq \frac{1}{d}$ for $t \in [d]$, and $\min_{w \in [d]} \nu (w) > 0$. Let $\alpha \in \left( 0, \frac{1}{2} \right]$ be any target accuracy. Assume that the given approximate task-type vector $\hat{\mathbf{t}}$ (side information of the ground-truth task-type vector $\mathbf{t} \in [d]^m$) satisfies the following two weak recovery assumptions on side information $\hat{\mathbf{t}}$:
\begin{equation}
    \label{eqn:sufficient_condition_alg:with_side_information_v1}
    \frac{\left| \hat{\mathcal{T}}_a \setminus \mathcal{T}_a \right|}{\left| \hat{\mathcal{T}}_a \right|} = o(1) \textnormal{ as } d \to \infty,
\end{equation}
and
\begin{equation}
    \label{eqn:sufficient_condition_alg:with_side_information_v2}
    \frac{\left| \left\{ i \in [m]: \hat{t}_i \neq t_i \right\} \right|}{m} = \frac{1}{m} \sum_{i=1}^{m} \mathbbm{1} \left( \hat{t}_i \neq t_i \right) \leq \frac{\alpha}{2}.
\end{equation}
Then, the recovery performance \eqref{eqn:expected_accuracy} is achievable via Algorithm \ref{alg:with_side_information} with the average number of queries per task
\begin{equation}
    \label{eqn:sufficient_condition_alg:with_side_information_v3}
    \begin{split}
        \frac{|\mathcal{A}|}{m} \geq \frac{1}{\epsilon^2 \cdot \delta_{\min} (d)} \log \left( \frac{8}{\alpha} \right)
    \end{split}
\end{equation}
for every sufficiently large $d$, where $m = \Omega \left( \frac{n^3 \left( \log n \right)^2}{\left( p_m - p_u \right)^2 s_{\min}^2} \right)$.
\end{thm}

\indent The detailed proof of Theorem \ref{thm:performance_alg:with_side_information} is given in Section \ref{subsec:proof_thm:performance_alg:with_side_information}.

\begin{remark} [Statistical optimality of Algorithm \ref{alg:with_side_information}]
\label{rmk:stat_opt_alg:with_side_information}
\normalfont{
An immediate yet remarkable consequence of Theorem \ref{thm:performance_alg:with_side_information} is the information-theoretic optimality of the sample complexity of Algorithm \ref{alg:with_side_information} for the $d$-type specialization model whose reliability matrix $\mathcal{Q}$ obeys Assumption \ref{assumption:multiple_spammers}. 
The information-theoretic achievability (Theorem \ref{thm:achievability_ml}) result and converse (Theorem \ref{thm:statistical_impossibility}) result imply that
under the $d$-type specialization model whose reliability matrix $\mathcal{Q}$ satisfies Assumption \ref{assumption:multiple_spammers}, the desired recovery performance \eqref{eqn:expected_accuracy} is achievable via the ML estimator \eqref{eqn:ml_estimator} if
\begin{equation}
    \label{eqn:Q_in_M2_info_upper_m}
    \frac{\left| \mathcal{A} \right|}{m} = \Omega \left( \frac{1}{\delta_{\min}(d)} \log \left( \frac{1}{\alpha} \right) \right),
\end{equation}
while is statistically impossible whenever
\begin{equation}
    \label{eqn:Q_in_M2_info_lower_m}
    \begin{split}
&    \frac{\left| \mathcal{A} \right|}{m} = o \left( \min \left\{ \frac{1 - \delta_{\max}(d)}{\delta_{\max}(d)} \log \left( \frac{1}{\alpha} \right),\right.\right.\\
&\qquad \qquad \qquad \left.\left.  \left( \frac{1}{\Gamma \left( d; \mathcal{Q} \right)} \log \left( \frac{1}{\alpha} \right) \right)^2
    \right\} \right).
    \end{split}
\end{equation}
The detailed derivation is stated as Corollary \ref{cor:IT_bounds_Q_in_M2} in Appendix \ref{sec:info_limits_special_cases}.
The sample complexity per task \eqref{eqn:sufficient_condition_alg:with_side_information_v3} of Algorithm \ref{alg:with_side_information} exactly matches the information-theoretic upper bound \eqref{eqn:Q_in_M2_info_upper_m} in the current setup. Recall that the information-theoretic bounds \eqref{eqn:Q_in_M2_info_upper_m} and \eqref{eqn:Q_in_M2_info_lower_m} become order-wise sharp results when (\romannumeral 1) $\delta_{\min}(d) \asymp \delta_{\max}(d)$ as $d \to \infty$, (\romannumeral 2) $\limsup_{d \to \infty} \delta_{\max}(d) < 1$, and (\romannumeral 3) the target accuracy $\alpha \in \left( 0, 1 \right]$ satisfies $\log \left( \frac{1}{\alpha} \right) \gtrsim \frac{\Gamma \left( d; \mathcal{Q} \right)^2}{\delta_{\max}(d)}$ as $d \to \infty$.
}
\end{remark}

\subsection{Inference algorithm without side information of the task-type vector}
\label{subsec:algorithm_without_side_information}

\indent Finally, we provide an algorithm that does not require the presence of side information $\hat{\mathbf{t}}$ of the ground-truth task type vector $\mathbf{t} \in [d]^m$. It consists of the following two stages: (\romannumeral 1) clustering workers via Algorithm \ref{alg:worker_clustering}, and (\romannumeral 2) matching task types and inferring the labels by running weighted majority voting with suitable weights different between the worker cluster of the matched type and the rest worker clusters.

\begin{alg} [Two-stage inference algorithm without side information of the task-type vector]
\label{alg:without_side_information}
\normalfont{ \
\begin{enumerate} [label=\arabic*.]
    \item[] \textbf{Input}: the number of types $d \in \mathbb{N}$, and the parameters $\left( r, \eta, l \right) \in [m] \times \left( 0, +\infty \right) \times [n]$;
    \item \textbf{SDP-based worker clustering stage}:
    \begin{enumerate} [label=(\alph*)]
        \item Employ Algorithm \ref{alg:worker_clustering} and obtain an explicit worker clusters $\left\{ \hat{\mathcal{W}}_1, \hat{\mathcal{W}}_2, \cdots, \hat{\mathcal{W}}_d \right\}$;
        \item For each task $i \in [m] \setminus \mathcal{S}$ and worker cluster $w \in [d]$, assign the task $i$ to $l$ workers sampled uniformly at random from the inferred worker cluster $\hat{\mathcal{W}}_w$. Let $\mathcal{A}(i) := \left\{ j \in [n] : (i, j) \in \mathcal{A} \right\}$ denote the set of workers assigned to the $i$-th task;
    \end{enumerate}
    \item \textbf{Task-type matching and label inference via weighted majority voting}:
    \begin{enumerate} [label=(\alph*)]
        \item For every task $i \in [m]$, we select $\mathcal{A}_w (i) \in \binom{\mathcal{A} (i) \cap \hat{\mathcal{W}}_w}{l}$ for $w \in [d]$, and define $\mathcal{A}' (i) := \bigcup_{w=1}^{d} \mathcal{A}_w (i) \subseteq \mathcal{A}(i)$;
        \item We infer the types associated with each task via the following decision rule: for each task $i \in [m]$, the type associated with the $i$-th task is inferred by finding the worker cluster whose response is the most biased:
        \begin{equation}         
            \label{eqn:task_type_matching}
            \hat{t}_i := \hat{t}_i (\mathbf{M}) = \argmax_{w \in [d]} \left| \sum_{j \in \mathcal{A}_{w} (i)} M_{ij} \right|.
        \end{equation}
        \item Determine weights $\bm{\theta}_{i*} = \left( \theta_{ij} : j \in \mathcal{A}' (i) \right)$ for each $i \in [m]$ as per the following rule:
        \begin{equation}
            \label{eqn:choice_weights_alg:without_side_information}
            \theta_{ij} :=
            \begin{cases}
                1 & \textnormal{if } j \in \mathcal{A}' (i) \cap \hat{\mathcal{W}}_{\hat{t}_i} = \mathcal{A}_{\hat{t}_i}(i); \\
                \frac{1}{\sqrt{d-1}} & \textnormal{otherwise.}
            \end{cases}
        \end{equation}
        \item Finally, we estimate the ground-truth label $a_i$ via the following weighted majority voting rule:
        \begin{equation}
            \label{eqn:decision_rule_alg:without_side_information}
            \hat{a}_i (\mathbf{M}) := \textnormal{sign} \left( \sum_{j \in \mathcal{A}' (i)} \theta_{ij} M_{ij} \right).
        \end{equation}
    \end{enumerate}
    \item[] \textbf{Output}: $\hat{\mathbf{a}}(\cdot) := \left( \hat{a}_i (\cdot) : i \in [m] \right) : \left\{ \pm 1 \right\}^{\mathcal{A}} \rightarrow \left\{ \pm 1 \right\}^m$.
\end{enumerate}
}
\end{alg}

\indent Step 1 in Algorithm \ref{alg:without_side_information} is the same as that of Algorithm \ref{alg:with_side_information}. The main difference from Algorithm \ref{alg:with_side_information} is that since we do not have side information of the task type vector, we first estimate the type of each individual task by identifying the worker cluster whose responses are the most biased as in \eqref{eqn:task_type_matching}. Our label inference stage then uses the weighted majority voting with weights \eqref{eqn:choice_weights_alg:without_side_information} that are differently assigned to the cluster of the matched type and the remaining clusters. 

\begin{remark} [Comparison between Algorithm \ref{alg:with_side_information} and Algorithm \ref{alg:without_side_information}]
\label{rmk:comparison_alg:with_side_information_alg:without_side_information_v1}
\normalfont{
Algorithm \ref{alg:with_side_information} makes use of side information of the ground-truth task-type vector in order to evaluate each worker cluster in terms of the bias of the responses averaged over all tasks of a given type as in \eqref{eqn:choice_weights_alg:with_side_information}. Algorithm \ref{alg:without_side_information}, on the other hand, matches the worker cluster which has the most biased answers without any averaging scheme as in \eqref{eqn:task_type_matching}. Even though one can utilize the approximate task-type vector $\hat{\mathbf{t}} = \left( \hat{t}_1, \hat{t}_2, \cdots, \hat{t}_m \right)$ obtained via \eqref{eqn:task_type_matching} as side information for Algorithm \ref{alg:without_side_information}, we did not make effort in this direction since the sample complexity for \emph{recovering the true task-type vector $\mathbf{t}$ within a desirable accuracy}, which is also necessary as a sufficient condition of the theoretical guarantee of Algorithm \ref{alg:with_side_information}, already dominates the sample complexity for achieving the desired recovery performance \eqref{eqn:expected_accuracy}.
}
\end{remark}

\indent We next provide a performance guarantee of Algorithm \ref{alg:without_side_information}. To establish a condition that the estimation of the task type through the task-type matching rule \eqref{eqn:task_type_matching} can recover the ground-truth task type for each task, we impose an assumption on the reliability matrix $\mathcal{Q}$ described below.

\begin{assumption}
\label{assumption:weak_ass_rm}
For any task type $t \in [d]$, let $p^* (t) := \mathcal{Q} (t, t)$ and $q^* (t) := \max_{w \in [d] \setminus \left\{ t \right\}} \mathcal{Q} (t, w)$ denote the matched reliability and the maximum mismatched reliability, respectively. Then, we have 
\begin{equation*}
    p^* (t) > q^* (t),\ \forall t \in [d].
\end{equation*}
\end{assumption}

\indent Assumption \ref{assumption:weak_ass_rm} (which we call the \emph{weak assortativity of the reliability matrix} $\mathcal{Q}$) implies that the workers whose types match the type of a given task responds more reliably than the workers of other types do. With this assumption, one can establish the following statistical guarantee of Algorithm \ref{alg:without_side_information}.

\begin{thm} [Statistical analysis of Algorithm \ref{alg:without_side_information}]
\label{thm:performance_alg:without_side_information}
Consider the $d$-type specialization model $\textnormal{SM} \left( d; \mathcal{Q}, \bm{\mu}, \bm{\nu} \right)$ whose reliability matrix $\mathcal{Q}$ satisfies Assumption \ref{assumption:strong_ass_cqcm} and \ref{assumption:weak_ass_rm}, and let $\alpha \in \left( 0, \frac{1}{2} \right]$ be any target accuracy. Then, it is possible to achieve the recovery performance \eqref{eqn:expected_accuracy} via Algorithm \ref{alg:without_side_information} with the average number of queries per task
\begin{equation}
    \label{eqn:sufficient_condition_alg:without_side_information}
    \begin{split}
        &\frac{|\mathcal{A}|}{m} \\
        \geq \ & \min \left\{ \frac{4d \log \left( \frac{6d+3}{\alpha} \right)}{ \min_{t \in [d]} \left\{ \left( p^{*} (t) - q^{*} (t) \right)^2 + \theta_1 \left( t; \mathcal{Q} \right) \right\}},  \right. \\
        &\left.\qquad\quad \frac{4d \log \left( \frac{3}{\alpha} \right)}{\min_{t \in [d]} \theta_1 \left( t; \mathcal{Q} \right)} \right \}
    \end{split}
\end{equation}
for all sufficiently large $d$, where $m = \Omega \left( \frac{n^3 \left( \log n \right)^2}{\left( p_m - p_u \right)^2 s_{\min}^2} \right)$, and the function $\theta_1 \left( \cdot; \mathcal{Q} \right) : [d] \rightarrow \mathbb{R}$ is given by
\begin{equation}
    \label{eqn:error_exponent_alg:without_side_information}
    \begin{split}
        \theta_1 \left( t; \mathcal{Q} \right)
        := \ & \frac{1}{2} \left[ \frac{1}{\sqrt{d-1}} \sum_{w=1}^{d} \left\{ 2 \mathcal{Q} (t, w) - 1 \right\} \right. \\
        &\left.+ \left( 1 - \frac{1}{\sqrt{d-1}} \right) \left \{ 2 \min_{w \in [d]} \mathcal{Q}(t, w)  - 1 \right \} \right]^2
    \end{split}
\end{equation}
for $t \in [d]$.
\end{thm}

\indent The detailed proof of Theorem \ref{thm:performance_alg:without_side_information} is provided in Section \ref{subsec:proof_thm:performance_alg:without_side_information}. Theorem \ref{thm:performance_alg:without_side_information} reveals a few intriguing remarks of Algorithm \ref{alg:without_side_information}:

\begin{remark} [Statistical optimality of Algorithm \ref{alg:without_side_information} in the special $(p, q)$ model]
\label{rmk:stat_opt_alg:without_side_information}
\normalfont{
Consider the following special case of the $d$-type specialization model:
\begin{equation}
    \label{eqn:special_model}
    \mathcal{Q} = q \mathbf{1}_{d \times d} + (p-q) \mathbf{I}_d \quad \textnormal{and} \quad \bm{\mu} = \bm{\nu} = \frac{1}{d} \mathbf{1}_d,
\end{equation}
where $\frac{1}{2} \leq q < p < 1$ are some universal constants and $\mathbf{1}_d \in \mathbb{R}^d$ denotes the $d$-dimensional all-one vector. We call this model the \emph{special $(p,q)$ model}. Under this model, each worker provides a correct response to the tasks of the matched type with probability $p \in \left( \frac{1}{2}, 1 \right)$, while providing a correct answer to any tasks of mismatched types  with probability $q \in \left[ \frac{1}{2}, p \right)$. It is straightforward to see that this model clearly satisfies Assumption \ref{assumption:strong_ass_cqcm} and \ref{assumption:weak_ass_rm}. This model was considered for $q = \frac{1}{2}$ in \cite{shah2018reducing}, and for general $q \in \left[ \frac{1}{2}, p \right)$ in the conference version \cite{kim2021crowdsourced} of this paper. 

\indent The information-theoretic achievability result (Theorem \ref{thm:achievability_ml}) and converse result (Theorem \ref{thm:statistical_impossibility}) imply that in the special $(p,q)$ model \eqref{eqn:special_model}, the desired recovery performance \eqref{eqn:expected_accuracy} is achievable via the ML estimator \eqref{eqn:ml_estimator} if
\begin{equation}
    \label{eqn:special_model_info_upper_m}
    \frac{\left| \mathcal{A} \right|}{m} =
    \begin{cases}
        \Omega \left( \log \left( \frac{1}{\alpha} \right) \right) & \textnormal{if } q > \frac{1}{2}; \\
        \Omega \left( d \log \left( \frac{1}{\alpha} \right) \right) & \textnormal{otherwise},
    \end{cases}
\end{equation}
while it is statistically impossible whenever
\begin{equation}
    \label{eqn:special_model_info_lower_m}
    \frac{\left| \mathcal{A} \right|}{m} =
    \begin{cases}
        o \left( \log \left( \frac{1}{\alpha} \right) \right) & \textnormal{if } q > \frac{1}{2}; \\
        o \left( d \log \left( \frac{1}{\alpha} \right) \right) & \textnormal{if } q = \frac{1}{2} \textnormal{ and } \log \left( \frac{1}{\alpha} \right) = \Omega (d); \\
        o \left( \left\{ \log \left( \frac{1}{\alpha} \right) \right\}^2 \right) & \textnormal{if } q = \frac{1}{2} \textnormal{ and } \log \left( \frac{1}{\alpha} \right) = o (d).
    \end{cases}
\end{equation}
The detailed derivation of \eqref{eqn:special_model_info_upper_m} and \eqref{eqn:special_model_info_lower_m} will be discussed in Corollary \ref{cor:special_model_IT_result} in Appendix \ref{sec:info_limits_special_cases}.

\indent One can realize that Algorithm \ref{alg:without_side_information} achieves the order-wise optimal sample complexity since for the special $(p, q)$ model \eqref{eqn:special_model}, the error exponent $\theta_1 \left( t; \mathcal{Q} \right)$ in \eqref{eqn:error_exponent_alg:without_side_information} equals
\[
    \theta_1 \left( t; \mathcal{Q} \right) = \frac{1}{2} \left[ \left( 1 + \sqrt{d-1} \right) (2q-1) + \frac{2}{\sqrt{d-1}} (p-q) \right]^2
\] 
for $\ \forall t \in [d]$.
Thus, the right-hand side of the bound \eqref{eqn:sufficient_condition_alg:without_side_information} is given by
\begin{equation}
    \label{eqn:special_model_sufficient_condition_alg:without_side_information_v1}
    \begin{split}
        &\min \Bigg\{ \quad \frac{4d \log \left( \frac{3}{\alpha} \right)}{\frac{1}{2} \left[ \left( 1 + \sqrt{d-1} \right) (2q-1) + \frac{2}{\sqrt{d-1}} (p-q) \right]^2},  \\
        & \frac{4d \log \left( \frac{6d+3}{\alpha} \right)}{(p-q)^2 + \frac{1}{2} \left[ \left( 1 + \sqrt{d-1} \right) (2q-1) + \frac{2}{\sqrt{d-1}} (p-q) \right]^2} \Bigg\}.
    \end{split}
\end{equation}
One can observe that
\begin{equation*}
   \begin{split}
        \eqref{eqn:special_model_sufficient_condition_alg:without_side_information_v1} =
        \begin{cases}
            \Theta \left( \log \left( \frac{1}{\alpha} \right) \right) & \textnormal{if } q > \frac{1}{2}; \\
            \Theta \left( d \log \left( \frac{d}{\alpha} \right) \right) & \textnormal{otherwise,}
        \end{cases}
    \end{split}
\end{equation*}
thereby the recovery accuracy \eqref{eqn:expected_accuracy} would be achievable via Algorithm \ref{alg:without_side_information} provided that
\begin{equation}
    \label{eqn:special_model_sufficient_condition_alg:without_side_information_v2}
    \begin{split}
        \frac{\left| \mathcal{A} \right|}{m} =
        \begin{cases}
            \Theta \left( \log \left( \frac{1}{\alpha} \right) \right) & \textnormal{if } q > \frac{1}{2}; \\
            \Theta \left( d \log \left( \frac{d}{\alpha} \right) \right) & \textnormal{otherwise.}
        \end{cases}
    \end{split}
\end{equation}
By comparing the performance guarantee \eqref{eqn:special_model_sufficient_condition_alg:without_side_information_v2} of Algorithm \ref{alg:without_side_information} with the sample complexity per task \eqref{eqn:special_model_info_upper_m} of the ML estimator \eqref{eqn:ml_estimator}, one can find that the bound \eqref{eqn:special_model_sufficient_condition_alg:without_side_information_v2} matches the information-theoretic upper bound \eqref{eqn:special_model_info_upper_m} (matches up to a logarithmic factor in the case for which $q = \frac{1}{2}$ and $\alpha = \omega \left( \frac{1}{d} \right)$). Consequently, Algorithm \ref{alg:without_side_information} is statistically optimal for achieving the recovery accuracy \eqref{eqn:expected_accuracy} for the regime where either $q > \frac{1}{2}$ or $q = \frac{1}{2}$ and $\log \left( \frac{1}{\alpha} \right) \gtrsim d$ as $d \to \infty$.
}
\end{remark}

\begin{remark} [How large can the number of types $d$ be?]
\label{rmk:range_of_d}
\normalfont{
The current paper copes with a crowdsourced labeling problem in a model with fidelity matrix $\mathbf{F}$ that can have high rank, the $d$-type specialization model. It is clear that $\textnormal{rank}(\mathbf{F}) \leq d$ and the equality holds if and only if $\mathcal{Q}$ has full rank. Therefore, in order to argue how large $\textnormal{rank}(\mathbf{F})$ can be to guarantee the desired recovery performance \eqref{eqn:expected_accuracy} under the $d$-type specialization model by using Algorithm \ref{alg:without_side_information}, it is necessary to identify the range of possible order for the number of types $d$ as a function of $\left( m, n, \alpha \right)$. Taking closer inspections on the proof of Theorem \ref{thm:performance_alg:without_side_information} and Lemma \ref{lemma:exact_recovery_alg:worker_clustering}, one can reveal that the following conditions are mandated for the model parameters: (\romannumeral 1) $\frac{n^2 \left( \log n \right)^2}{\left( p_m - p_u \right)^2 s_{\min}^2} \lesssim r$ and $\frac{nr}{m} = \calO (ld)$; (\romannumeral 2) $dl \leq n$. In order for $r$ that satisfies the condition (\romannumeral 1) to exist, the following condition should be guaranteed:
\begin{equation}
    \label{eqn:range_of_d_v1}
    \frac{n^2 \left( \log n \right)^2}{\left( p_m - p_u \right)^2 s_{\min}^2} = \calO \left( \frac{mld}{n} \right).
\end{equation}
Note that under the assumption $m = \Omega \left( \frac{n^3 \left( \log n \right)^2}{\left( p_m - p_u \right)^2 s_{\min}^2} \right)$ made in Theorem \ref{thm:performance_alg:without_side_information}, the property \eqref{eqn:range_of_d_v1} immediately follows. Thus, it suffices to choose the model parameters to satisfy the condition (\romannumeral 2). Under the special $(p, q)$ model \eqref{eqn:special_model}, our choice \eqref{eqn:proof_thm:performance_alg:without_side_information_v17} of the model parameters obeys
\begin{equation*}
    \begin{split}
        dl =
        \begin{cases}
            \Theta \left( \log \left( \frac{1}{\alpha} \right) \right) & \textnormal{if } q > \frac{1}{2}; \\
            \Theta \left( d \log \left( \frac{d}{\alpha} \right) \right) & \textnormal{otherwise.}
        \end{cases}
    \end{split}
\end{equation*}
So, the number of types $d$ only needs to satisfy $d \log \left( \frac{d}{\alpha} \right) = \mathcal{O}(n)$ for the desired recovery performance \eqref{eqn:expected_accuracy} in the special $(p, q)$ model \eqref{eqn:special_model}. To summarize, $\textnormal{rank}(\mathbf{F})$ can be as large as the number of types $d$ fulfilling the condition $d \log \left( \frac{d}{\alpha} \right) = \mathcal{O}(n)$, for instance, $d = n^{1 - \epsilon}$ for some constant $\epsilon \in \left( 0, 1 \right)$ provided that $\frac{1}{\alpha} = \textnormal{poly}(n)$ for the theoretical guarantee of the recovery performance \eqref{eqn:expected_accuracy} under the special $(p, q)$ model \eqref{eqn:special_model}. It is worth noting that the possible range of the rank of the fidelity matrix $\mathbf{F}$ is much higher than the existing models, which have mostly considered the rank-one cases for the ease of analysis \cite{dawid1979maximum, khetan2016achieving}.
}
\end{remark}

\section{Comparison with the existing baseline algorithms}
\label{subsec:comparison_baseline_algorithms}

\indent We now compare the statistical performances of Algorithm \ref{alg:with_side_information} and \ref{alg:without_side_information} with two existing baseline algorithms in the $d$-type specialization model. All the baseline algorithms that we consider throughout this section do not require any prior knowledge of the fidelity matrix $\mathbf{F}$. In Section \ref{subsubsec:standard_mv}, we review the standard majority voting algorithm, and in Section \ref{subsubsec:SS_algorithm}, we review the \emph{type-dependent subset-selection algorithm} \cite{shah2018reducing}, which is a previous algorithm developed for the type-based crowdsourcing model. We provide the performance analysis of these baselines under the $d$-type specialization model. In Section \ref{subsubsec:performance_comparison}, we compare these baselines with Algorithm \ref{alg:with_side_information} and \ref{alg:without_side_information} and show that Algorithm \ref{alg:with_side_information} and \ref{alg:without_side_information} achieve better performance guarantees than the previous algorithms, more robust against changes in model parameters.

\subsection{Standard majority voting estimation}
\label{subsubsec:standard_mv}

\indent We first consider the standard majority voting algorithm. To begin with, let $\mathcal{A}(i) := \left\{ j \in [n] : (i, j) \in \mathcal{A} \right\}$ denote the set of workers assigned to the $i$-th task. The standard majority voting (MV) estimator to infer the ground-truth label ${a}_{i}$ takes the same weight for all responses, and is given by $\hat{\mathbf{a}}^{\textnormal{MV}} (\cdot) := \left( \hat{a}_{1}^{\textnormal{MV}}(\cdot), \hat{a}_{2}^{\textnormal{MV}}(\cdot), \cdots, \hat{a}_{m}^{\textnormal{MV}}(\cdot) \right) : \left\{ \pm 1 \right\}^{\mathcal{A}} \rightarrow \left\{ \pm 1 \right\}^m$, where
\begin{equation}
    \label{eqn:standard_mv}
    \hat{a}_{i}^{\textnormal{MV}}(\mathbf{M}) := \textnormal{sign} \left( \sum_{j \in \mathcal{A}(i)} M_{ij} \right),\ \forall i \in [m].
\end{equation}
In Appendix \ref{sec:theoretical_analysis_standard_mv}, we provide a rigorous theoretical analysis of the standard majority voting estimator \eqref{eqn:standard_mv} under the $d$-type specialization model $\textnormal{SM} \left( d; \mathcal{Q}, \bm{\mu}, \bm{\nu} \right)$. Informally, we will show that the recovery performance \eqref{eqn:expected_accuracy} can be achieved via the standard majority voting estimator \eqref{eqn:standard_mv} provided that the worker-task assignment set $\mathcal{A} \subseteq [m] \times [n]$ satisfies
\begin{equation}
    \label{eqn:sufficient_condition_standard_mv}
    \min_{i \in [m]} \left| \mathcal{A}(i) \right| \geq \frac{1}{\min_{t \in [d]} \theta_2 \left( t; \mathcal{Q} \right)} \log \left( \frac{1}{\alpha} \right)
\end{equation}
for any given $\alpha \in \left( 0, \frac{1}{2} \right]$ ($\alpha$ may depend on $m$), where $\theta_2 \left( \cdot; \mathcal{Q} \right) : [d] \rightarrow \mathbb{R}_{+}$ is defined by
\begin{equation}
    \label{eqn:error_exponent_standard_mv}
    \theta_2 \left( t; \mathcal{Q} \right) := \frac{1}{8} \left[ \sum_{w=1}^{d} \nu (w) \left\{ 2 \mathcal{Q}(t, w) - 1 \right\} \right]^2,\ \forall t \in [d].
\end{equation}
Since all responses offered by the workers in $\mathcal{A}(i)$ are aggregated with the same weight to estimate the true label $a_i$, the error exponent $\theta_2 \left( t; \mathcal{Q} \right)$ is determined by the overall quality $\sum_{w=1}^d \nu (w) \mathcal{Q} (t,w)$, which takes account of all $d$-different types of workers in responding to tasks of type $t$. We emphasize that the sample complexity result \eqref{eqn:sufficient_condition_standard_mv} of the standard majority voting does not require any additional assumption on the reliability matrix $\mathcal{Q}$.

\subsection{Type-dependent subset-selection algorithm}
\label{subsubsec:SS_algorithm}

\indent The second baseline estimation algorithm we consider is the \emph{type-dependent subset-selection algorithm} \cite{shah2018reducing}. The basic idea of this algorithm is to exploit the responses only from the workers whose type matches the type of the given task, which are believed to be more reliable than the responses provided by the workers of mismatched types. Since neither task types nor worker types are known, this algorithm also estimates the worker clusters and the task types using the responses provided by workers. For the worker clustering, the similarity on the responses between every pair of workers on the subset $\mathcal{S}$ of tasks are sequentially compared, \emph{i.e.}, for the $j$-th worker, if there is a worker cluster $\mathcal{C} \subseteq \left[ j-1 \right]$ such that for every $j' \in \mathcal{C}$,
$\frac{1}{r} \sum_{i \in \mathcal{S}} \mathbbm{1} \left( M_{ij} = M_{ij'} \right) > \xi$ for some threshold $\xi > 0$ that should be pre-determined, then the $j$-th worker is assigned to the worker cluster $\mathcal{C}$. Otherwise, a new worker cluster $\left\{ j \right\}$ is created. After this sequential worker clustering process, the task type is estimated by the same rule as \eqref{eqn:task_type_matching}.
The main difference of the subset-selection algorithm compared to Algorithm \ref{alg:without_side_information} is that Algorithm \ref{alg:without_side_information} infers the ground-truth binary labels via weighted majority voting by using all responses with suitable weights based on the task-type matching result, while the subset-selection algorithm estimates the ground-truth by performing the standard majority voting using only answers provided by the matched workers. For the subset-selection algorithm, when the subset of workers among $\mathcal{A}(i)$ whose type match the inferred task type $\hat{t}_i := \hat{t}_i (\mathbf{M})$ is denoted by $\mathcal{A}_{\hat{t}_i}(i)$, the ground-truth label $a_i$ is inferred by running the standard majority voting using only the answers given by the \emph{workers of the matched type}:
\begin{equation}
    \label{eqn:SS_decision_rule_main} \hat{a}_{i}^{\textnormal{SS}}(\mathbf{M}) := \textnormal{sign} \left( \sum_{j \in \mathcal{A}_{\hat{t}_i}(i)} M_{ij} \right),\ \forall i \in [m].
\end{equation}
We refer to Appendix \ref{sec:detailed_description_SS_algorithm} for the detailed description of the subset-selection algorithm. The performance guarantee of this algorithm is established in \cite{shah2018reducing} only for the special $(p,q)$ model \eqref{eqn:special_model} with a fixed value of $q$, $q = 1/2$. We present a general statistical analysis of the subset-selection algorithm by extending the proof therein under a much more general $d$-type specialization model. See Appendix \ref{sec:proof_prop:performance_SS_Q_in_M1} for the detailed proof.

\begin{prop} [Statistical guarantee of the subset-selection algorithm \MakeUppercase{\romannumeral 1}]
\label{prop:performance_SS_Q_in_M1}
Consider the $d$-type specialization model $\textnormal{SM} \left( d; \mathcal{Q}, \bm{\mu}, \bm{\nu} \right)$ with $\mathcal{Q}$ satisfying Assumption \ref{assumption:strong_ass_cqcm} and \ref{assumption:weak_ass_rm}, and let $\alpha \in \left( 0, \frac{1}{2} \right]$ be any given target accuracy. Then, the subset-selection algorithm can achieve the recovery performance \eqref{eqn:expected_accuracy} with the average number of queries per task
\begin{equation}
    \label{eqn:sufficient_condition_SS_Q_in_M1}
    \begin{split}
        \frac{|\mathcal{A}|}{m} \geq \ & \min \left\{ \frac{4d \log \left( \frac{6d+3}{\alpha} \right)}{ \min_{t \in [d]} \left\{ \left( p^{*} (t) - q^{*} (t) \right)^2 + \theta_3 \left( t; \mathcal{Q} \right) \right\}},  \right. \\
        &\left.\qquad\quad \frac{4d \log \left( \frac{3}{\alpha} \right)}{\min_{t \in [d]} \theta_3 \left( t; \mathcal{Q} \right)} \right\}
    \end{split}
\end{equation}
for every sufficiently large $d$, where $m \geq C_2 \cdot \frac{n^{1 + \beta}}{\left( p_m - p_u \right)^2}$ for some universal constants $C_2, \beta \in \left( 0, +\infty \right)$, and the function $\theta_3 \left( \cdot; \mathcal{Q} \right) : [d] \rightarrow \mathbb{R}_{+}$ is defined by
\begin{equation}
    \label{eqn:error_exponent_SS_Q_in_M1}
    \theta_3 \left( t; \mathcal{Q} \right) := \left[ 2 \min \limits_{w \in [d]} \mathcal{Q} (t, w) - 1 \right]^2,\ \forall t \in [d].
\end{equation}
\end{prop}

\indent Next, we provide a performance guarantee of the subset-selection algorithm for the $d$-type specialization model whose reliability matrix $\mathcal{Q}$ satisfies Assumption \ref{assumption:strong_ass_cqcm} and \ref{assumption:multiple_spammers}. We refer the readers to Appendix \ref{sec:proof_prop:performance_SS_Q_in_M2} for the detailed proof:

\begin{prop} [Statistical guarantee of the subset-selection algorithm \MakeUppercase{\romannumeral 2}]
\label{prop:performance_SS_Q_in_M2}
Let us consider the $d$-type specialization model $\textnormal{SM} \left( d; \mathcal{Q}, \bm{\mu}, \bm{\nu} \right)$ with the reliability matrix $\mathcal{Q}$ satisfying Assumption \ref{assumption:strong_ass_cqcm} and \ref{assumption:multiple_spammers}, $\delta (t; d) \geq \frac{1}{d}$ for every $t \in [d]$, and $\min_{w \in [d]} \nu (w) > 0$. Let $\alpha \in \left( 0, \frac{1}{2} \right]$ be any target accuracy. Then, the subset-selection algorithm can achieve the recovery performance with the average number of queries per task
\begin{equation}
    \label{eqn:sufficient_condition_SS_Q_in_M2}
    \begin{split}
        \frac{|\mathcal{A}|}{m} \geq \frac{d}{\epsilon^2} \varphi^{-1} \left\{ \log \left( \frac{15 d^2 \delta_{\max}(d)}{\epsilon^2 \alpha} \right) \right\}
    \end{split}
\end{equation}
for all sufficiently large $d$, where $m \geq C_3 \cdot \frac{n^{1 + \beta}}{\left( p_m - p_u \right)^2}$ for some absolute constants $C_3, \beta \in \left( 0, +\infty \right)$, and the function $\varphi(\cdot): \left[ 1, +\infty \right) \to \left[ 1, +\infty \right)$ is defined by $\varphi (x) := x - \log x$.
\end{prop}

\subsection{Comparison of Algorithm \ref{alg:with_side_information} and \ref{alg:without_side_information} against the baseline algorithms}
\label{subsubsec:performance_comparison}

\indent We first compare the statistical performances of the standard majority voting rule, the subset-selection algorithm, and Algorithm \ref{alg:without_side_information} under the $d$-type specialization model whose reliability matrix $\mathcal{Q}$ obeys Assumption \ref{assumption:strong_ass_cqcm} and \ref{assumption:weak_ass_rm}. Let us consider the sample complexities of the subset-selection algorithm and Algorithm \ref{alg:without_side_information}. Note that the sample complexity results \eqref{eqn:sufficient_condition_SS_Q_in_M1} and \eqref{eqn:sufficient_condition_alg:without_side_information} are the same except that the error exponent $\theta_3 \left( t; \mathcal{Q} \right)$ \eqref{eqn:error_exponent_SS_Q_in_M1} for the subset-selection scheme is replaced by $\theta_1 \left( t; \mathcal{Q} \right)$ \eqref{eqn:error_exponent_alg:without_side_information} in Algorithm \ref{alg:without_side_information}. One can realize that
\begin{equation}
    \label{eqn:comparison_SS_alg:without_side_information}
    \theta_1 \left( t; \mathcal{Q} \right) \geq \frac{\left( 1 + \sqrt{d-1} \right)^2}{2} \theta_3 \left( t; \mathcal{Q} \right),\ \forall t \in [d],
\end{equation}
implying that the error exponent $\theta_1 \left( t; \mathcal{Q} \right)$ of Algorithm \ref{alg:without_side_information} is strictly larger than the error exponent $\theta_3 \left( t; \mathcal{Q} \right)$ of the subset-selection algorithm. Hence, the sample complexity \eqref{eqn:sufficient_condition_alg:without_side_information} of Algorithm \ref{alg:without_side_information} is always smaller than the sample complexity \eqref{eqn:sufficient_condition_SS_Q_in_M1} of the subset-selection algorithm. So, Algorithm \ref{alg:without_side_information} is a strict improvement over the subset-selection algorithm in the sense of the sample complexity required for the recovery performance \eqref{eqn:expected_accuracy}.

\indent Next, we compare the sample complexity of Algorithm \ref{alg:without_side_information} against that of the standard majority voting estimation. To this end, we focus on the case where the prior distribution $\bm{\nu}$ of worker types is \emph{approximately balanced}:
\begin{equation}
    \label{eqn:comparison_MV_alg:without_side_information_v1}
    \frac{\max_{w \in [d]} \nu (w)}{\min_{w \in [d]} \nu (w)} = \Theta (1) \textnormal{ as } d \to \infty.
\end{equation}
With this assumption, it holds that 
\[
    \max_{w \in [d]} \nu (w) \asymp \min_{w \in [d]} \nu (w) \asymp \frac{1}{d}
\]
as $d \to \infty$. Thus, we have
\begin{equation*}
    \theta_2 \left( t; \mathcal{Q} \right) \asymp \frac{1}{d^2} \left[ \sum_{w=1}^{d} \left\{ 2 \mathcal{Q} (t, w) - 1 \right\} \right]^2,\ \forall t \in [d],
\end{equation*}
for the error exponent $\theta_2 \left( t; \mathcal{Q} \right)$ of the majority voting estimator, and thus
\begin{equation}
    \label{eqn:comparison_MV_alg:without_side_information_v2}
    \theta_1 \left( t; \mathcal{Q} \right)
    \geq \frac{1}{2 (d-1)} \left[ \sum_{w=1}^{d} \left\{ 2 \mathcal{Q} (t, w) - 1 \right\} \right]^2
    \asymp d \theta_2 \left( t; \mathcal{Q} \right),
\end{equation}
for $ \forall t \in [d]$
as $d \to \infty$. Hence, it follows from \eqref{eqn:comparison_MV_alg:without_side_information_v2} that
\begin{equation}
    \label{eqn:comparison_MV_alg:without_side_information_v3}
    \frac{4d}{\min_{t \in [d]} \theta_1 \left( t; \mathcal{Q} \right)} \log \left( \frac{3}{\alpha} \right) 
    \lesssim \frac{1}{\min_{t \in [d]} \theta_2 \left( t; \mathcal{Q} \right)} \log \left( \frac{1}{\alpha} \right)
\end{equation}
as $d \to \infty$.
This shows that the sample efficiency of Algorithm \ref{alg:without_side_information} is always better than or equal to that of the standard majority voting estimator in terms of the order of $d$.

\indent To investigate the cases where only Algorithm \ref{alg:without_side_information} achieves the order-wise optimal sample complexity result robustly against the model parameter changes, we compare the performance guarantees of the algorithms under the special $(p, q)$ model \eqref{eqn:special_model}. For the standard majority voting estimator \eqref{eqn:standard_mv}, we obtain $\theta_2 \left( t; \mathcal{Q} \right) = \frac{\left\{ (2p-1) + (d-1)(2q-1) \right\}^2}{2d^2}$ for every $t \in [d]$, and thus the RHS of \eqref{eqn:sufficient_condition_standard_mv} equals to $\frac{2d^2}{\left\{ (2p-1) + (d-1)(2q-1) \right\}^2} \log \left( \frac{1}{\alpha} \right)$. This implies that the sufficient condition of the standard majority voting to achieve the desired recovery performance \eqref{eqn:expected_accuracy} is given by
\begin{equation}
    \label{eqn:special_standard_mv_performance}
    \frac{\left| \mathcal{A} \right|}{m} =
    \begin{cases}
         \Omega \left( \log \left( \frac{1}{\alpha} \right) \right) & \textnormal{if } q > \frac{1}{2}; \\
        \Omega \left( d^2 \log \left( \frac{1}{\alpha} \right) \right) & \textnormal{otherwise.}
    \end{cases}
\end{equation}
For the subset-selection algorithm, it is straightforward that $\theta_3 \left( t; \mathcal{Q} \right) = (2q-1)^2$ for all $t \in [d]$, and the RHS of \eqref{eqn:sufficient_condition_SS_Q_in_M1} equals to
$\min \left\{ \frac{4d}{(p-q)^2 + (2q-1)^2} \log \left( \frac{6d+3}{\alpha} \right), \frac{4d}{(2q-1)^2} \log \left( \frac{3}{\alpha} \right) \right\}$. So Proposition \ref{prop:performance_SS_Q_in_M1} implies that the subset-selection algorithm succeeds if
\begin{equation}
    \label{eqn:special_subset_selection_performance}
    \frac{\left| \mathcal{A} \right|}{m} =
    \begin{cases}
        \Omega \left( d \log \left( \frac{1}{\alpha} \right) \right) & \textnormal{if } q > \frac{1}{2}; \\
        \Omega \left( d \log \left( \frac{d}{\alpha} \right) \right) & \textnormal{otherwise.}
    \end{cases}
\end{equation}

\indent From \eqref{eqn:special_standard_mv_performance} and \eqref{eqn:special_subset_selection_performance}, one can find that the order-wise information-theoretic limit \eqref{eqn:special_model_info_upper} is achievable by the standard majority voting estimator \eqref{eqn:standard_mv} in the parameter regime where $q > \frac{1}{2}$, while it is achievable via the subset-selection algorithm in the parameter regime where $q = \frac{1}{2}$ and $\log \left( \frac{1}{\alpha} \right) \gtrsim d$. Hence, the standard majority voting rule \eqref{eqn:standard_mv} and the subset-selection algorithm do not consistently beat each other and the optimality of these algorithms highly depends on the model parameters. On the other hand, our algorithm, achieves the order-wise information-theoretic limit for both cases as shown in Remark \ref{rmk:stat_opt_alg:without_side_information}.

\indent In order to understand the fundamental reason behind this result, we consider the spammer-hammer model \cite{karger2014budget}: the $j$-th worker is called a \emph{hammer} for the $i$-th task if $F_{ij} = 1$; a \emph{spammer} for the $i$-th task if $F_{ij} = \frac{1}{2}$. If all the workers are almost hammers, that is, $\mathcal{Q} (t, w) \approx 1$ for all $(t, w) \in [d] \times [d]$, the standard majority voting using all responses outperforms the subset-selection algorithm since the subset-selection algorithm abandons $\left( \frac{d-1}{d} \right)$-fraction of answers that are provided by workers whose types do not match the type of the given task. On the other hand, if we consider the regime where $q^*(t) \approx \frac{1}{2}$ and $p^* (t) - q^*(t) = \Theta (1)$ for every $t \in [d]$, then all workers whose types are different from the given task type can almost be viewed as spammers. For this case, the subset-selection algorithm is much better than the standard majority voting estimator, since the standard majority voting does not rule out the dominant random noisy responses. Indeed, as demonstrated in \eqref{eqn:special_standard_mv_performance} and \eqref{eqn:special_subset_selection_performance}, the subset-selection algorithm requires $d$ times more queries compared to the standard majority voting if $q > 1/2$, while it requires only $1/d$ times queries if $q = 1/2$. On the contrary, our proposed algorithm, Algorithm \ref{alg:without_side_information}, achieves the order-wise optimal sample complexity in both cases by aggregating answers from different worker clusters with proper weights as defined in \eqref{eqn:choice_weights_alg:without_side_information}.

\indent Now, we make a comparison between the performances of the standard majority voting estimator \eqref{eqn:standard_mv}, the subset-selection algorithm, and Algorithm \ref{alg:with_side_information} under the $d$-type specialization model whose reliability matrix $\mathcal{Q}$ satisfies Assumption \ref{assumption:strong_ass_cqcm} and \ref{assumption:multiple_spammers}. Consider the sample complexities of the standard majority voting and Algorithm \ref{alg:with_side_information} provided that all the required assumptions of Theorem \ref{thm:performance_alg:with_side_information} together with the approximate balancedness condition \eqref{eqn:comparison_MV_alg:without_side_information_v1} of the prior distribution $\nu (\cdot) \in \Delta \left( [d] \right)$ of worker types hold. Then, we have
\begin{equation}
    \label{eqn:comparison_MV_alg:with_side_information_v1}
    \begin{split}
        \theta_2 \left( t; \mathcal{Q} \right) &\asymp \frac{1}{d^2} \left[ \sum_{w=1}^{d} \left\{ 2 \mathcal{Q} (t, w) - 1 \right\} \right]^2 \\
        &= \frac{1}{d^2} \left[ \sum_{w \in [d] \setminus \textnormal{spammer}_{\mathcal{Q}} (t)} \left\{ 2 \mathcal{Q} (t, w) - 1 \right\} \right]^2 \\
        &\geq \frac{\epsilon^2}{2} \delta (t; d)^2
    \end{split}
\end{equation}
for every $t \in [d]$, since $\mathcal{Q} (t, w) \geq \frac{1}{2} + \epsilon$ for all $w \in [d] \setminus \textnormal{spammer}_{\mathcal{Q}} (t)$ and $\left| [d] \setminus \textnormal{spammer}_{\mathcal{Q}} (t) \right| = d \cdot \delta (t; d)$. From \eqref{eqn:comparison_MV_alg:with_side_information_v1}, one can observe $\min_{t \in [d]} \theta_2 \left( t; \mathcal{Q} \right) \gtrsim \epsilon^2 \delta_{\min} (d)^2$ and the sample complexity \eqref{eqn:sufficient_condition_standard_mv} of the standard majority voting rule implies that if
\begin{equation}
    \label{eqn:comparison_MV_alg:with_side_information_v2}
    \min_{i \in [m]} \left| \mathcal{A} (i) \right| \gtrsim \frac{1}{\epsilon^2 \cdot \delta_{\min} (d)^2} \log \left( \frac{1}{\alpha} \right),
\end{equation}
then the recovery accuracy \eqref{eqn:expected_accuracy} is achievable by the standard majority voting estimator. Hence, the sample complexity \eqref{eqn:sufficient_condition_alg:with_side_information_v3} of Algorithm \ref{alg:with_side_information} is always smaller than the sample complexity \eqref{eqn:comparison_MV_alg:with_side_information_v2} of the standard majority voting order-wisely. 

\indent As a next stage, we take a closer look at the sample complexity of Algorithm \ref{alg:with_side_information} against that of the subset-selection algorithm. To this end, let us embark on our discussion with the following simple and straightforward observation:
\begin{equation}
    \label{eqn:comparison_SS_alg:with_side_information_v1}
    \frac{1}{2} \varphi^{-1} (x) < x < \varphi^{-1} (x),\ \forall x \in \left[ 2 - \log 2, +\infty \right).
\end{equation}
For a complete argument, we assume that the assumptions of Theorem \ref{thm:performance_alg:with_side_information} and Proposition \ref{prop:performance_SS_Q_in_M2} hold. Based on the observation \eqref{eqn:comparison_SS_alg:with_side_information_v1}, one can realize that the right-hand side of the sample complexity result \eqref{eqn:sufficient_condition_SS_Q_in_M2} has the order
\begin{equation}
    \label{eqn:comparison_SS_alg:with_side_information_v2}
    \begin{split}
        \frac{d}{\epsilon^2} \varphi^{-1} \left\{ \log \left( \frac{5 d^2 \delta_{\max}(d)}{\epsilon^2 \alpha} \right) \right\}
        \asymp \ & d \log \left( \frac{d^2 \delta_{\max} (d)}{\alpha} \right) \\
        \stackrel{\textnormal{(a)}}{\asymp} \ & d \log \left( \frac{d}{\alpha} \right),
    \end{split}
\end{equation}
where the step (a) follows from the fact
\begin{equation*}
    \frac{d}{\alpha} \leq \frac{d^2 \delta_{\max} (d)}{\alpha} \leq \frac{d^2}{\alpha^2}.
\end{equation*}
Thus, the desired recovery accuracy \eqref{eqn:expected_accuracy} can be achieved by the subset-selection algorithm when
\begin{equation}
    \label{eqn:comparison_SS_alg:with_side_information_v3}
    \frac{\left| \mathcal{A} \right|}{m} \gtrsim d \log \left( \frac{d}{\alpha} \right).
\end{equation}
From the condition $\delta_{\min} (d) \geq \frac{1}{d}$, it is clear that
\begin{equation*}
    d \log \left( \frac{d}{\alpha} \right) \gtrsim \frac{1}{\epsilon^2 \cdot \delta_{\min} (d)} \log \left( \frac{8}{\alpha} \right) \textnormal{ as } d \to \infty.
\end{equation*}
This elucidates that the sample efficiency of Algorithm \ref{alg:with_side_information} is always superior than or equal to that of the subset-selection algorithm in terms of the order of $d$.

\section{Empirical results}
\label{sec:empirical_results}

\indent Throughout this section, we provide numerical simulation results, both for the synthetic and real-world datasets, to illustrate the advantages of the proposed label inference algorithms compared to several baseline methods.

\subsection{Experiments with synthetic data}
\label{subsec:experiment_syn}

\indent In this section, we conduct synthetic simulations to compare the performance of proposed algorithms (Algorithms \ref{alg:with_side_information} and \ref{alg:without_side_information}) with two baseline algorithms, the standard majority voting (MV) \eqref{eqn:standard_mv} and the subset-selection (SS) algorithm \eqref{eqn:SS_decision_rule_main}, as well as with the maximum likelihood (ML) oracle \eqref{eqn:ml_estimator}.
In implementing Algorithm \ref{alg:with_side_information}, it is assumed that the estimator $\hat{\mathbf{t}}$ of the task type ${\mathbf{t}}$ is provided to the algorithm. To evaluate the performance of Algorithm \ref{alg:with_side_information} in terms of the accuracy of the estimator $\hat{\mathbf{t}}$, we compare the performances of Algorithm \ref{alg:with_side_information} for the case where we have an exact task-type vector estimate $\hat{\mathbf{t}} = \mathbf{t}$, denoted by Algorithm \ref{alg:with_side_information} (0\%), and for the case where we have an approximate task-type vector estimate $\hat{\mathbf{t}}$ whose elements are obtained by flipping the elements of the ground-truth task-type vector $\mathbf{t}$ with probability 0.1, denoted by Algorithm \ref{alg:with_side_information} (10\%). In implementing Algorithm \ref{alg:with_side_information}, we set the tuning parameter $\zeta_{ab}$  in \eqref{eqn:choice_weights_alg:with_side_information} for discriminating the spammer worker groups from non-spammer worker groups as $\zeta_{ab} = \zeta = 0.6l$ for all $(a,b) \in [d] \times [d]$. The tuning parameter $\eta$ for the worker clustering (the first stage of both Algorithms \ref{alg:with_side_information} and \ref{alg:without_side_information}) is set to be  $\eta = \frac{r \left( p_m + p_u \right)}{2}$. In our simulations, we set the numbers of tasks, workers and types as $\left( m, n, d \right) = \left( 15000, 60, 4 \right)$, respectively, and set the size of the subset of tasks for the worker clustering as $r = 200$. We change the number of queries assigned to each worker cluster per task as $l \in \left\{ 3, 4, 5, 6 \right\}$ and report the performance of the algorithms with respect to the varying $l$. The overall inference quality is measured by the fraction of labels that do not match with the ground-truth label, \emph{i.e.,} $\frac{1}{m }\sum_{i=1}^m  \mathbbm{1} \left( \hat{a}_i (\mathbf{M}) \neq a_i \right)$. The standard MV rule and the ML estimator use $ld$ answers in total to estimate the ground-truth label. On the other hand, Algorithm \ref{alg:with_side_information}, Algorithm \ref{alg:without_side_information} and the subset-selection algorithm additionally use $rn$ responses for the worker clustering and thus the total number of queries per task of these algorithms are $\frac{rn+(m-r)ld}{m}$. This overhead is counted in measuring the performances of the clustering-based algorithms.

\indent In the experiments using synthetic dataset, we generate two types of reliability matrices $\mathcal{Q}$, depending on assumptions imposed on $\mathcal{Q}$. The prior distributions for worker \& task types are set to be balanced as $\bm{\mu} = \bm{\nu} = \begin{bmatrix} \frac{1}{d} & \cdots & \frac{1}{d} \end{bmatrix}^{\top}$. For each fixed reliability matrix $\mathcal{Q}$ and a pair of the task-type vector and the worker-type vector $\left( \mathbf{t}, \mathbf{w} \right)$, the measurement matrix $\mathbf{M}$ is randomly generated 20 times according to the observation model \eqref{eqn:crowdsourcing_system}, and we report the empirical performances of each algorithm averaged over these 20 measurements.
\medskip

\subsubsection{Model assuming the existence of the most reliable worker cluster}
\label{subsubsec:model_m1}
\begin{table}[t]
    \begin{subtable}[h]{\columnwidth}
        \centering
        \begin{tabular}{|l | l | l | l |}
            \hline
            0.9365 & 0.5000 & 0.5000 & 0.5000 \\
            \hline
            0.5000 & 0.9835 & 0.5000 & 0.5000 \\
            \hline
            0.5000 & 0.5000 & 0.9128 & 0.5000 \\
            \hline
            0.5000 & 0.5000 & 0.5000 & 0.9749 \\
            \hline
        \end{tabular}
        \caption{$(p_{\sf min},q_{\sf max}) = (0.9,0.5)$}
    \end{subtable}
    \begin{subtable}[h]{\columnwidth}
        \centering
        \begin{tabular}{|l | l | l| l|}
            \hline
            0.9095 & 0.5270 & 0.6301 & 0.5495 \\
            \hline
            0.5658 & 0.9305 & 0.6421 & 0.5816 \\
            \hline
            0.6396 & 0.5670 & 0.9591 & 0.5705 \\
            \hline
            0.6026 & 0.6693 & 0.6648 & 0.9706 \\
            \hline
        \end{tabular}
        \caption{$(p_{\sf min},q_{\sf max}) = (0.9,0.7)$}
    \end{subtable}
    \caption{Sampled reliability matrices for (a) $\left( p_{\sf min}, q_{\sf max} \right) = (0.9, 0.5)$ and (b) $\left( p_{\sf min}, q_{\sf max} \right) = (0.9,0.7)$, satisfying Assumption \ref{assumption:strong_ass_cqcm} \& \ref{assumption:weak_ass_rm}. The minimum intra-cluster collective quality correlation and the maximum inter-cluster collective quality correlation are $(p_m,p_u) = (0.1704,0)$ for (a) and $(p_m,p_u) = (0.2021,0.1436)$ for (b).}
    \label{tab:tie_m1d4_relmatrix}
\end{table}

In the first case, we make two assumptions (Assumptions \ref{assumption:strong_ass_cqcm} and \ref{assumption:weak_ass_rm}) on the reliability matrix $\mathcal{Q}$, in order to evaluate the performance of the algorithms under the existence of the most reliable worker cluster for each task type. To generate the reliability matrix $\mathcal{Q}$ based on these assumptions, we define two variables $p_{\sf min}$ and $q_{\sf max}$, a lower bound on the reliability of workers to the tasks of the matched type, \emph{i.e.}, the diagonal entries of $\mathcal{Q}$, and an upper bound on the reliability of workers to the tasks of mismatched types, \emph{i.e.}, the off-diagonal entries of $\mathcal{Q}$, respectively. We set $\left( p_{\sf min}, q_{\sf max} \right) \in \left\{ (0.9, 0.5), (0.9, 0.7) \right\}$, and then generate the diagonal entries $\left\{ \mathcal{Q}(a,a): a \in [d] \right\}$ uniformly at random from the interval $\left[ p_{\sf min}, 1 \right]$ and the off-diagonal entries $\left\{ \mathcal{Q}(a,b): a \neq b \text{ in } [d] \right\}$ uniformly at random from the interval $\left[ 0.5, q_{\sf max} \right]$. So, there are two fixed reliability matrices $\mathcal{Q}$ each of which corresponds to $q_{\sf max} \in \left\{ 0.5, 0.7 \right\}$ for the fixed $p_{\sf min} =0.9$. As $q_{\sf max}$ increases, the quality of the responses from workers of mismatched types improves. By creating the reliability matrix $\mathcal{Q}$ in this way, Assumption \ref{assumption:weak_ass_rm} is clearly fulfilled due to the gap between $p_{\sf min}$ and $q_{\sf max}$. We also check that the  condition $p_m > p_u$ of Assumption \ref{assumption:strong_ass_cqcm} is satisfied for the generated $\mathcal{Q}$ matrices. Table \ref{tab:tie_m1d4_relmatrix} shows the two $\mathcal{Q}$ matrices used in the simulations. 

\begin{figure}[t]
    \centering
    \begin{subfigure}{0.9\columnwidth}   
    \includegraphics[width=\linewidth]{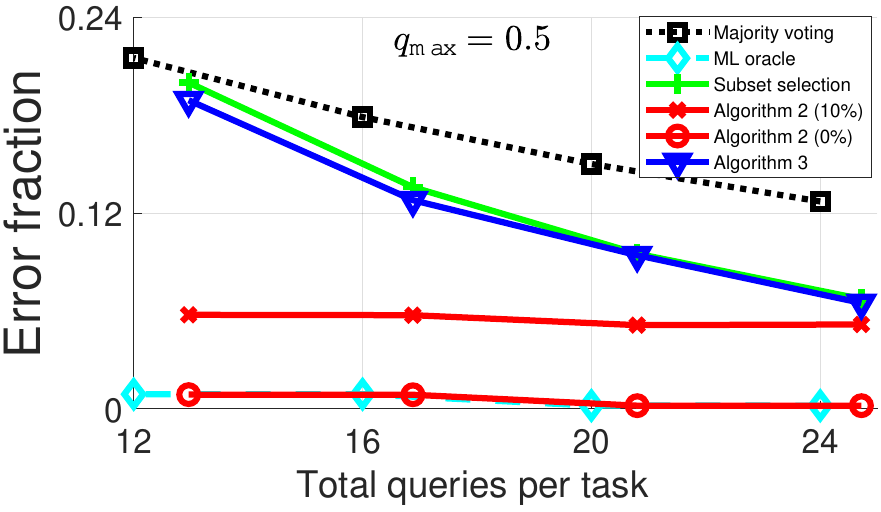}
    \caption{$(p_{\sf min},q_{\sf max})=(0.9,0.5)$}
    \label{fig:tie_m1d405}
    \end{subfigure}
    \begin{subfigure}{0.9\columnwidth}
    \includegraphics[width=\linewidth]{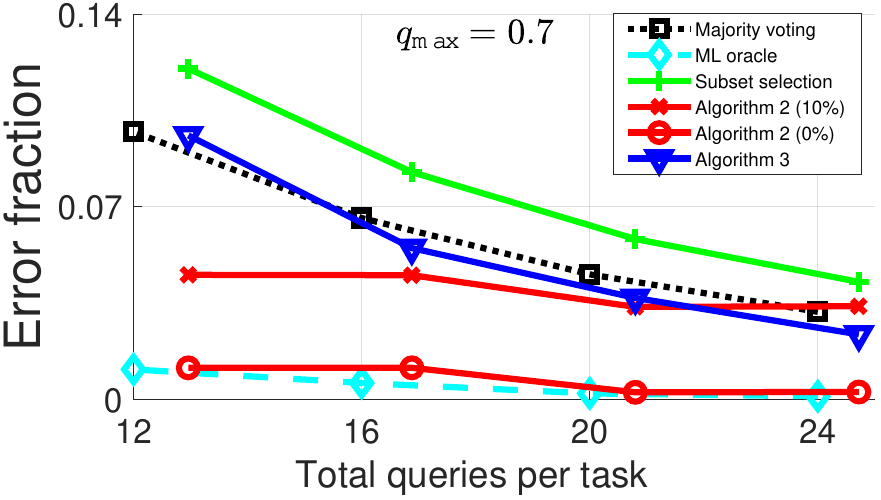}
    \caption{$(p_{\sf min},q_{\sf max})=(0.9,0.7)$}
    \label{fig:tie_m1d407}
    \end{subfigure}
    \caption{Comparison of the inference qualities (the error fraction, $\frac{1}{m}\sum_{i=1}^m  \mathbbm{1} \left( \hat{a}_i (\mathbf{M}) \neq a_i \right)$) between the algorithms for the cases $\left( p_{\sf min}, q_{\sf max} \right) = (0.9,0.5)$ (left) and $\left( p_{\sf min}, q_{\sf max} \right) = (0.9,0.7)$ (right). The corresponding reliability matrices $\mathcal{Q}$, satisfying Assumptions \ref{assumption:strong_ass_cqcm} and \ref{assumption:weak_ass_rm}, are shown in Table \ref{tab:tie_m1d4_relmatrix}. Algorithm \ref{alg:without_side_information} consistently shows a better performance than the standard MV estimator and the SS algorithm. Algorithm \ref{alg:with_side_information}, which makes use of side information of the ground-truth task-type vector, achieves the best empirical performance, near that of the ML estimator.
    }
    \label{fig:tie_m1d4}
\end{figure}

\indent In Fig. \ref{fig:tie_m1d4}, we compare the inference qualities of the proposed algorithms (Algorithm \ref{alg:with_side_information} and Algorithm \ref{alg:without_side_information}) with the existing algorithms. The performance of Algorithm \ref{alg:without_side_information} is always better than or comparable to that of the MV and the SS algorithm for all the parameter regimes under consideration. Note that the performance of the SS algorithm is superior than that of the MV estimator when $q_{\sf max} = 0.5$, \emph{i.e.,} when the reliability of workers having mismatched types is lower, while that of the MV estimator gets improved for a larger $q_{\sf max} = 0.7$. Since the SS algorithm uses only $1/d$ fraction of total answers, which are believed to come from the worker cluster whose type matches the type of the given task, its efficiency compared to the MV highly depends on the quality of workers having different types from the given task. As $q_{\sf max}$ increases, since the answers from the workers of mismatched types can have reliable information, the MV rule, which utilizes all the responses with the same weights, performs better than that of the SS algorithm. On the other hand, our proposed algorithm, Algorithm \ref{alg:without_side_information}, consistently demonstrates the better performance than the two baselines (the standard MV estimator and the SS algorithm) regardless of $q_{\sf max}$, since our algorithm assigns suitable weights in aggregating answers from matched and mismatched workers as specified in \eqref{eqn:choice_weights_alg:without_side_information}.

\indent One can also observe that side information of the ground-truth task-type vector $\mathbf{t}$ is indispensable in improving the empirical performances of the inference algorithms. Compared to Algorithm \ref{alg:without_side_information}, Algorithm \ref{alg:with_side_information}, which uses side information of the ground-truth task-type vector, achieves a better performance, near that of the ML estimator when $\hat{\mathbf{t}} = \mathbf{t}$. The performance of Algorithm \ref{alg:with_side_information} is slightly degraded for the case where the error fraction of the approximate task-type vector $\hat{\mathbf{t}}$ equals 10\%, but it still achieves the best empirical performance among the baseline algorithms in the comparison group.
\medskip

\subsubsection{Model with spammer and non-spammer worker clusters}
\label{subsubsec:model_m2}

Here, we make two assumptions (Assumptions \ref{assumption:strong_ass_cqcm} and \ref{assumption:multiple_spammers}) on the reliability matrix $\mathcal{Q}$ to evaluate the empirical performance of the inference algorithms under the existence of two different types of worker clusters, spammer and non-spammer groups, for each task type. We set the gap between the reliability of spammer and non-spammer clusters, defined in Assumption \ref{assumption:multiple_spammers},  as $\epsilon \in \{0.1,0.3 \}$. We randomly sample two reliability matrices $\mathcal{Q}$, as shown in Table \ref{tab:tie_m2d4_relmatrix} for each $\epsilon \in \left\{ 0.1, 0.3 \right\}$, such that they satisfy both Assumptions \ref{assumption:strong_ass_cqcm} and \ref{assumption:multiple_spammers}.

\begin{table}[t]
    \begin{subtable}[h]{\columnwidth}
        \centering
        \begin{tabular}{|l | l | l | l |}
            \hline
            0.7941 & 0.8240 & 0.5000 & 0.8970 \\
            \hline
            0.5000 & 0.5000 & 0.9236 & 0.8136 \\
            \hline
            0.5000 & 0.8875 & 0.7833 & 0.5000 \\
            \hline
            0.9143 & 0.5000 & 0.5000 & 0.5000 \\
            \hline
        \end{tabular}
        \caption{$\epsilon = 0.1$}
    \end{subtable}
    \begin{subtable}[h]{\columnwidth}
        \centering
        \begin{tabular}{|l | l | l| l|}
            \hline
            0.9623 & 0.5000 & 0.8946 & 0.5000 \\
            \hline
            0.5000 & 0.5000 & 0.5000 & 0.9380 \\
            \hline
            0.5000 & 0.9682 & 0.9197 & 0.5000 \\
            \hline
            0.8486 & 0.8419 & 0.5000 & 0.8754 \\
            \hline
        \end{tabular}
        \caption{$\epsilon = 0.3$}
    \end{subtable}
    \caption{The reliability matrices $\mathcal{Q}$ corresponding to Fig. \ref{fig:tie_m2d4}. As the value of $\epsilon$ varies from 0.1 to 0.3, the values of $\left( p_m, p_u \right)$ become $\left\{ (0.2551,0.1328), (0.3318,0.1965) \right\}$ (left to right), respectively.}
    \label{tab:tie_m2d4_relmatrix}
\end{table}

\begin{figure}[t]
    \centering
    \begin{subfigure}{0.9\columnwidth}    \includegraphics[width=\linewidth]{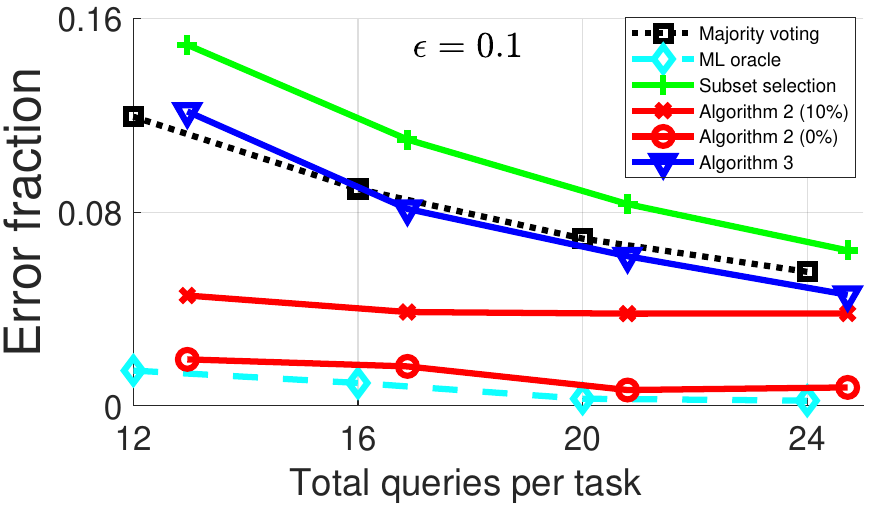}
    \caption{$\epsilon=0.1$}
    \label{fig:tie_m2d4_eps01}
    \end{subfigure}
    \begin{subfigure}{0.9\columnwidth} 
    \includegraphics[width=\linewidth]{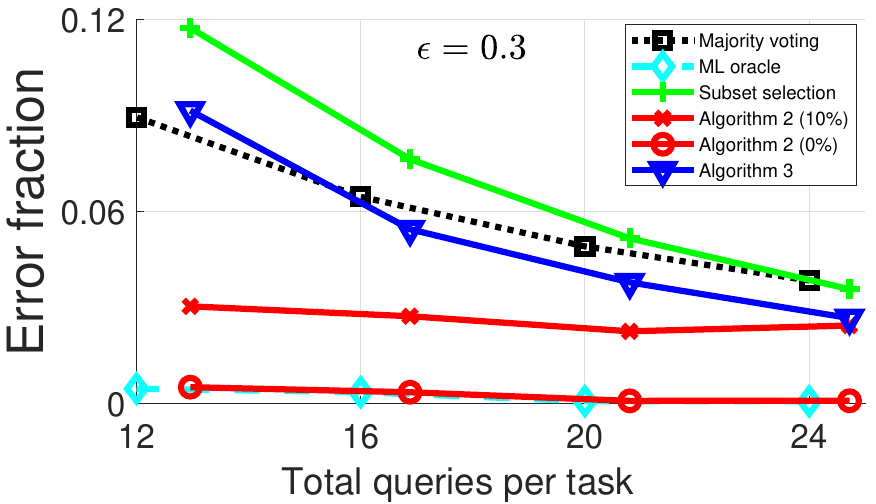}
    \caption{$\epsilon=0.3$}
    \label{fig:tie_m2d4_eps03}
    \end{subfigure}
    \caption{Comparison of the inference qualities (the error fraction, $\frac{1}{m}\sum_{i=1}^m  \mathbbm{1} \left( \hat{a}_i (\mathbf{M}) \neq a_i \right)$) between the algorithms for $\epsilon = 0.1$ (left) and $\epsilon = 0.3$ (right). The corresponding reliability matrices $\mathcal{Q}$, satisfying Assumption \ref{assumption:strong_ass_cqcm} and \ref{assumption:multiple_spammers}, are shown in Table \ref{tab:tie_m2d4_relmatrix}. Algorithm \ref{alg:without_side_information} consistently shows a better performance than the standard majority voting estimator and the subset-selection algorithm. Algorithm \ref{alg:with_side_information}, which uses side information of the ground-truth task-type vector, achieves the best empirical performance, near that of the ML estimator.
    }
    \label{fig:tie_m2d4}
\end{figure}

\indent In Fig. \ref{fig:tie_m2d4}, the inference qualities of the proposed algorithms, Algorithms \ref{alg:with_side_information} \ref{alg:without_side_information}, are compared against the other baseline algorithms for the above two reliability matrices $\mathcal{Q}$ with $\epsilon = 0.1$ (left) and $\epsilon = 0.3$ (right). For the model with spammer and non-spammer worker clusters, where the number of non-spammer worker clusters can be larger than one for each task type, the empirical performance of the SS algorithm, which uses only one matched worker cluster for estimating the ground-truth labels, becomes worse than that of the standard MV rule. Algorithm \ref{alg:without_side_information}, on the other hand, can achieve a better empirical performance than that of the standard MV estimator and the SS algorithm. Algorithm \ref{alg:with_side_information}, which uses side information of the ground-truth task types, achieves the best performance under this model as well, near that of the ML estimator, and the performance of Algorithm \ref{alg:with_side_information} is better when the given side information of the ground-truth task-type vector is more accurate. Comparing Fig. \ref{fig:tie_m2d4_eps01} and Fig. \ref{fig:tie_m2d4_eps03}, it can be found that the higher the $\epsilon$, the better the performance of Algorithm \ref{alg:without_side_information} than the MV estimator. If there is a larger difference between the reliability of the spammer clusters and that of the non-spammer clusters, there is more likely to be an improvement by giving appropriate weights to their responses.

\subsection{Experiments with real-world data}
\label{subsec:exp_real}

\begin{figure}[t]
\centering
\begin{subfigure}{\columnwidth}
    \centering
    \includegraphics[width=\linewidth]{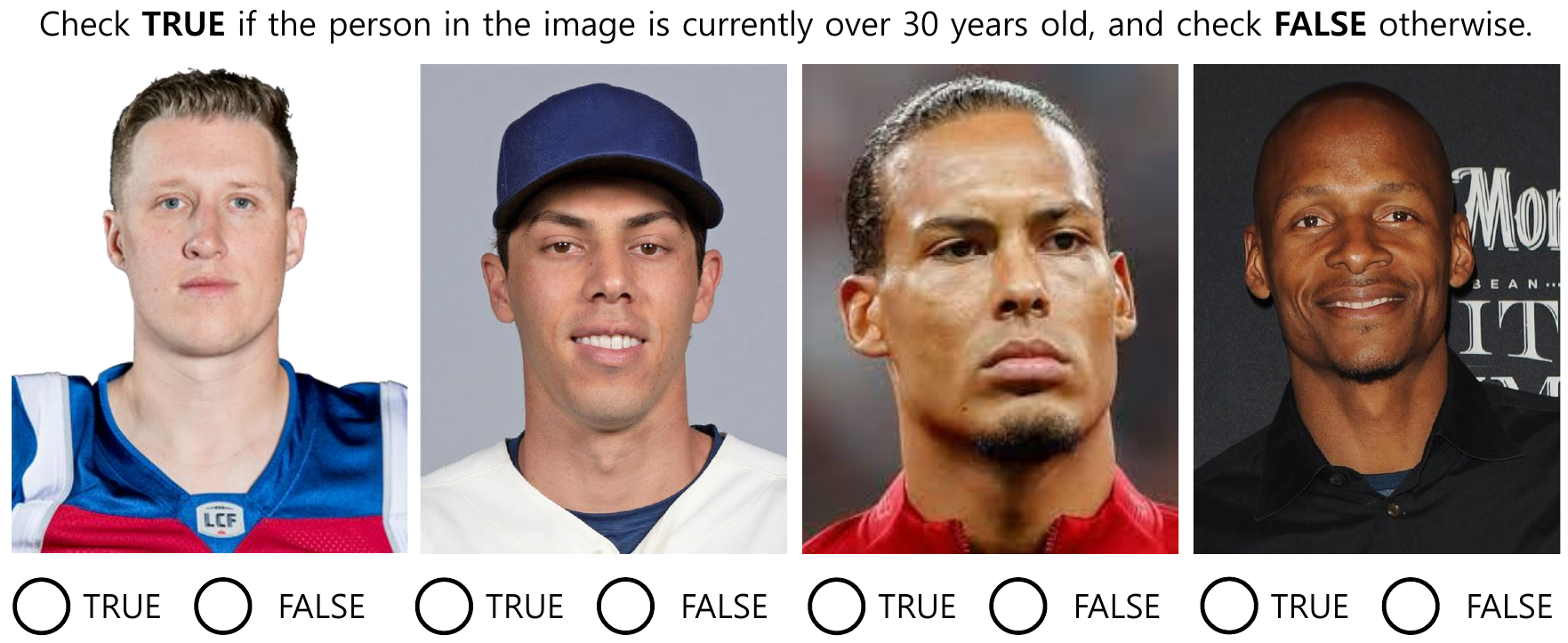}
    \caption{Tasks for athlete data.}
  \label{fig:mturk_athlete}
\end{subfigure}
\hspace{0.05\textwidth}
\begin{subfigure}{\columnwidth}
    \centering
    \includegraphics[width=\linewidth]{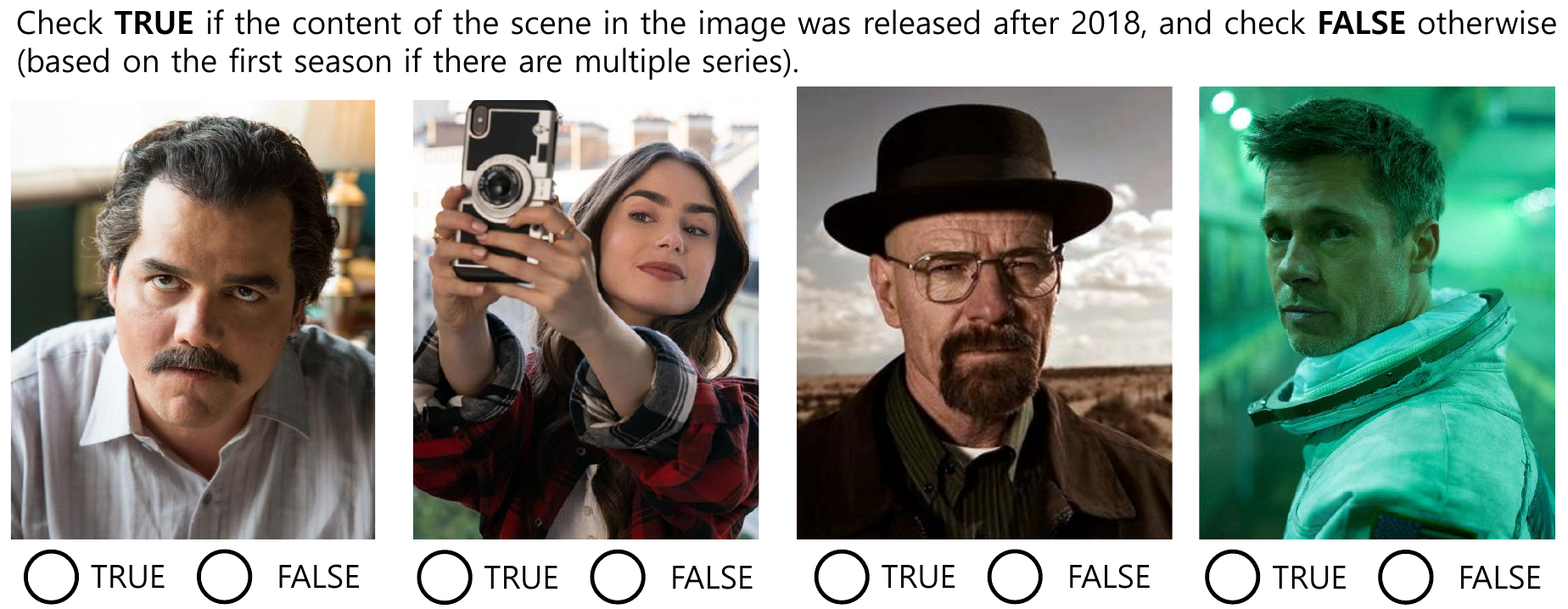}
    \caption{Tasks for Netflix data.}
  \label{fig:mturk_netflix}
\end{subfigure}
\caption{Examples of tasks posted on Amazon Mechanical Turk.}
\label{fig:mturk}
\end{figure}

\indent We also conduct experiments on the real-world data collected by the celebrated crowdsourcing platform, known as the \emph{Amazon Mechanical Turk (MTurk)}. The binary tasks using different types of images are designed to get binary responses, TRUE or FALSE, from the workers in the MTurk. Details for the two different datasets we have collected are described below.

\begin{enumerate}
    \item {\bf Athlete}: The experiment consists of 500 images of athletes where each a quarter of images is from one of four sports types: football, baseball, soccer, and basketball. Each human intelligent task (HIT) includes 400 images that contain 100 randomly sampled images of each type of sports. A HIT represents a single, self-contained, virtual task set that a worker can give answers, and recieve a reward for completing. We ask whether the athlete in each image is currently over 30 years old, and receive binary answers $\left\{ \text{TRUE}, \text{FALSE} \right\}$. We design total 50 HITs and assign them to 50 different workers;
    \item {\bf Netflix}: The experiment consists of 500 scenes each of which is from a content (movies, dramas, or series, etc.) at Netflix, where the 500 contents can be categorized into four types: action, romance, thriller, and science-fiction (SF). We select 125 scenes from each category. We ask if the content (movies, dramas, or series, etc.) corresponding to each scene was released after 2018, and obtain the binary responses $\{\text{TRUE}, \text{FALSE}\}$. Each HIT is designed to contain 400 scenes that contain 100 randomly sampled scenes from each category. Then, we design total 50 HITs and assign them to 50 individual workers.
\end{enumerate}

\indent We first check whether the collected real-world dataset indeed follows a \emph{type structure}. Since only the task types are known, we infer the ground-truth worker types based on the rates of correct responses of each worker on each task type, {which are evaluated using the ground-truth label information. For the experiments on the Athlete dataset, the composition of workers based on their \emph{empirically estimated} types is given by (football, baseball, soccer, basketball)$ = \left( 13, 10, 11, 16 \right)$. For the experiments on the Netflix dataset, we have the composition of workers such as (action, romance, thriller, SF)$ = \left( 14, 14, 11, 11 \right)$. The reliability matrices $\mathcal{Q}$ computed by averaging the rates of correct answers for each task-worker type pair $(t, w) \in [d] \times [d]$ for two different datasets are presented in TABLE \ref{tab:relmat_real} with their $\left( p_m, p_u \right)$ values.

\begin{table}[t]
	\begin{subtable}[h]{\columnwidth}
		\centering
		\begin{tabular}{|l | l | l | l |}
			\hline
			0.8914 & 0.4715 & 0.5429 & 0.5267 \\
			\hline
			0.6118 & 0.8978 & 0.5527 & 0.4514 \\
			\hline
			0.4782 & 0.5497 & 0.8952 & 0.5202 \\
			\hline
			0.5329 & 0.4450 & 0.5239 & 0.9493 \\
			\hline
		\end{tabular}
		\caption{Athlete: $(p_m,p_u)=(0.1429,0.0345)$}
		\label{tab:m1}
	\end{subtable}
	\begin{subtable}[h]{\columnwidth}
		\centering
		\begin{tabular}{|l | l | l| l|}
			\hline
			0.9456 & 0.8179 & 0.8181 & 0.5000 \\
			\hline
			0.4694 & 0.9404 & 0.5117 & 0.5127 \\
			\hline
			0.4490 & 0.5284 & 0.9430 & 0.7765 \\
			\hline
			0.8148 & 0.4648 & 0.4802 & 0.9406 \\
			\hline
		\end{tabular}
		\caption{Netflix: $(p_m,p_u)=(0.2741,0.1315)$}
		\label{tab:m2}
	\end{subtable}
	\caption{The reliability matrix $\mathcal{Q}$ evaluated by using the ground-truth task types and the estimated worker types for the Athlete dataset (left) and the Netflix dataset (right).}
	\label{tab:relmat_real}
\end{table}

\indent One can first observe that the diagonal entries are larger than the off-diagonal entries for every row in the reliability matrices provided in TABLE \ref{tab:relmat_real}. We can also realize that the strong assortativity assumption (Assumption \ref{assumption:strong_ass_cqcm}: $p_m > p_u$) holds with the above two real-world datasets, which will enable the clustering of workers based on their types. In TABLE \ref{tab:m1} (Athlete data), it can be seen that for each task type (each row), only a worker cluster of the matched type tends to have relatively high rates of correct responses, while the remaining worker clusters have the rates of correct responses near 0.5. On the other hand, in TABLE \ref{tab:m2}, one can observe that for each task type (each row), multiple worker clusters may respond more reliably with rates of correct answers higher than 0.8, \emph{i.e.}, for the SF task type (last row), both the worker cluster of the action type (the first entry) and the worker cluster of the SF type (the last entry) have the rates of correct responses higher than 0.8. This implies that the workers who provided responses on the Netflix dataset tend to have diverse interests on different movie types, or the four different movie types (action, romance, thriller, SF) share some common features. On the contrary, for the athlete dataset, it turns out that the sport types are rather independent from each other, and each worker tends to have interest on a unique sport type. Thus, the Athlete dataset turns out to satisfy the Assumption \ref{assumption:strong_ass_cqcm} and \ref{assumption:weak_ass_rm}, while the Netflix dataset satisfies the Assumption \ref{assumption:strong_ass_cqcm} and \ref{assumption:multiple_spammers}. One can also observe that the original Dawid-Skene model, where the order of workers in terms of their reliabilities is fixed for each task, cannot explain both the datasets, since the tasks are heterogeneous and the reliability of workers may change widely across different task types.

\begin{figure}[t]
	\begin{subfigure}{0.9\columnwidth}
		\includegraphics[width=\linewidth]{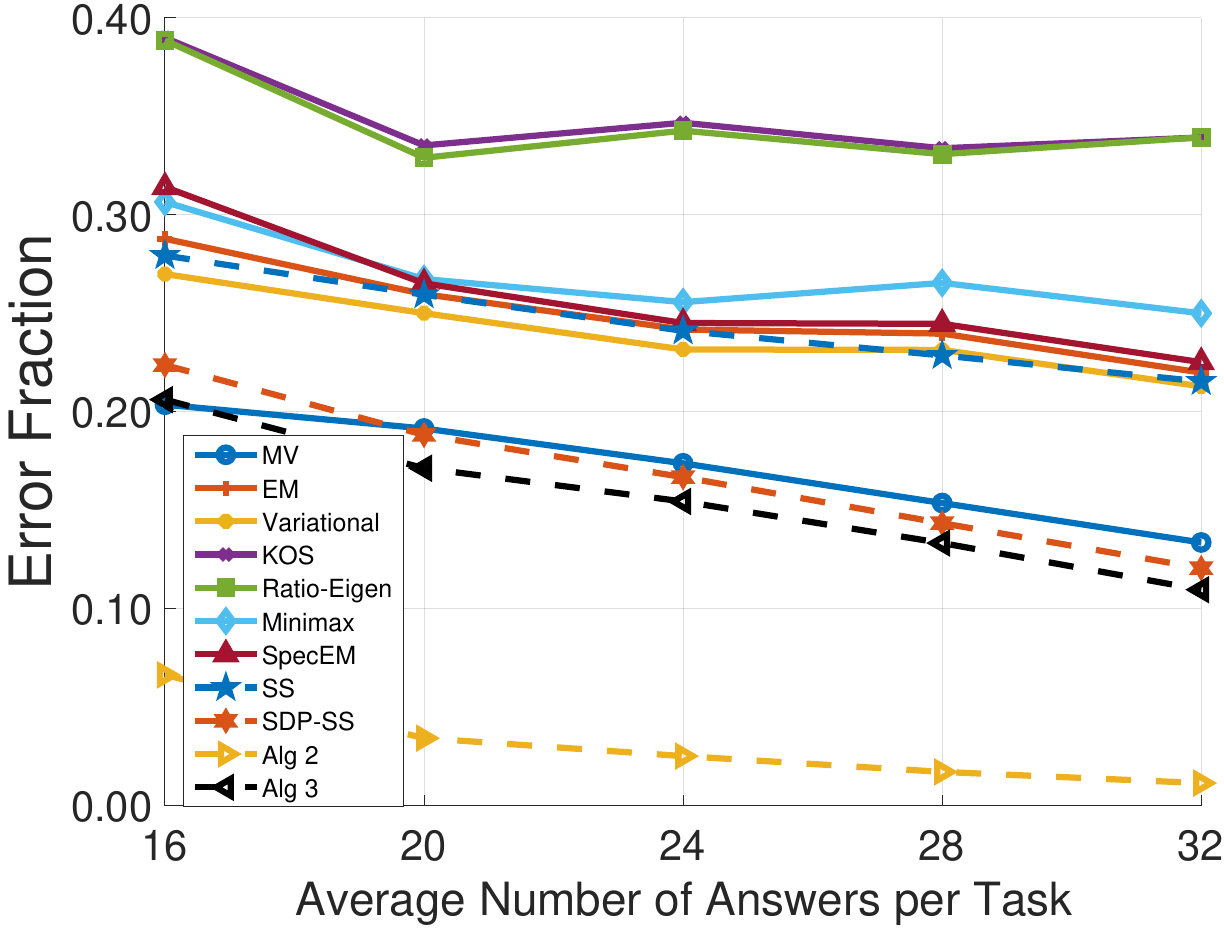}
		\caption{Athlete}
	\end{subfigure}
	\begin{subfigure}{0.9\columnwidth}
		\includegraphics[width=\linewidth]{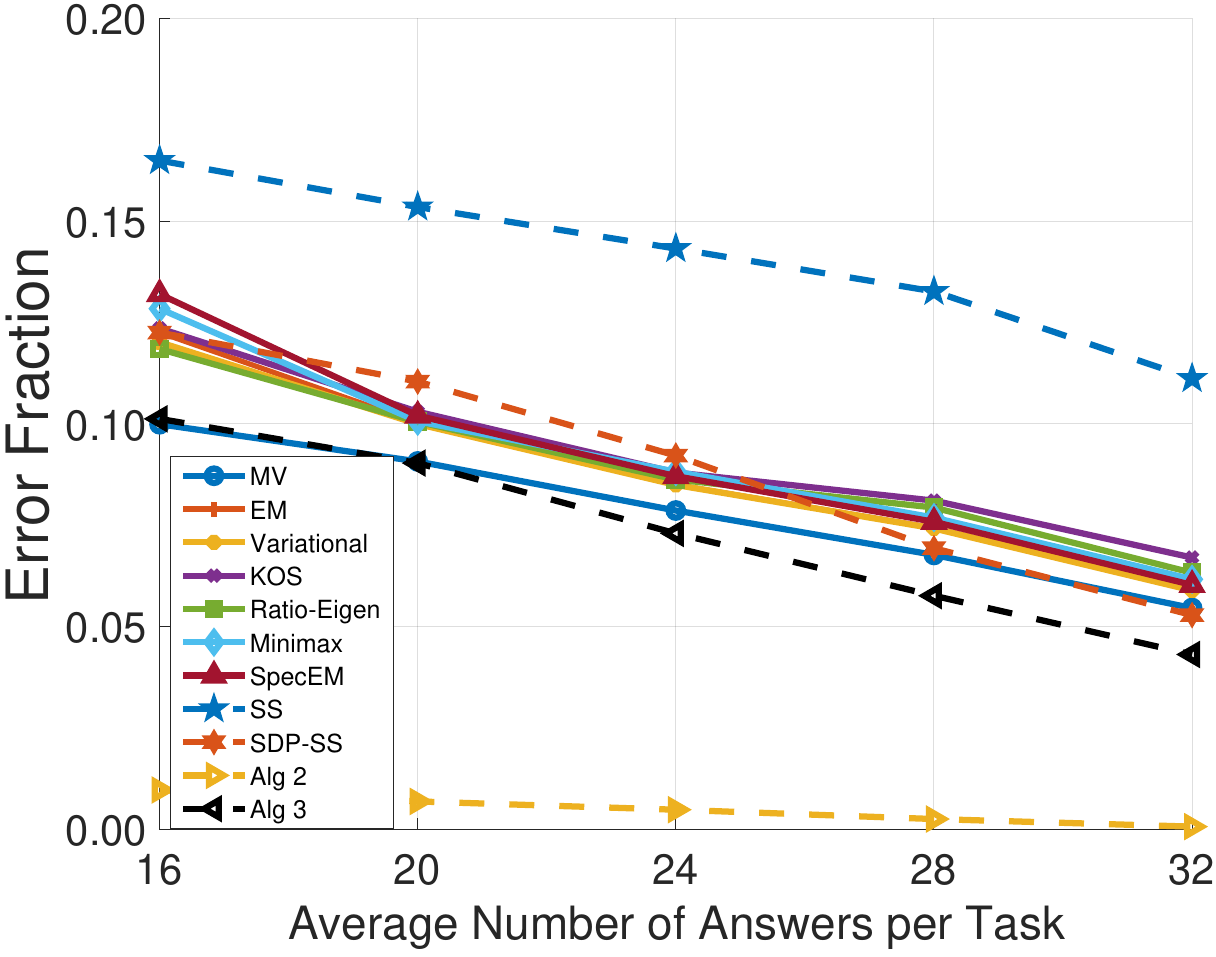}
		\caption{Netflix}
	\end{subfigure}
	\caption{Comparison of the inference qualities between different algorithms for (a) the Athelte dataset, and (b) the Netflix dataset, respectively. Among the baseline methods, not using side information of the ground-truth task types, Algorithm  \ref{alg:without_side_information} demonstrates the best empirical performance. When the task type is available, Algorithm \ref{alg:with_side_information} shows the best empirical performance with a large margin from the second-best algorithm.}
	\label{fig:real}
\end{figure}

\indent In Fig. \ref{fig:real}, we compare the empirical performances of the proposed algorithms, Algorithms \ref{alg:with_side_information} and \ref{alg:without_side_information}, against the state-of-the-art algorithms, including EM \cite{dawid1979maximum}, Variational \cite{liu2012variational}, KOS \cite{karger2014budget}, Ratio-Eigen \cite{dalvi2013aggregating}, and specEM \cite{zhang2016spectral}, all of which are developed under the model assumptions of the Dawid-Skene model. Also, Minimax \cite{zhou2014aggregating}, which assumes that the error probability of a worker's answer to a given task may vary depending on the task difficulty, is also compared to our proposed algorithms. The empirical performances of the standard MV estimator and the SS algorithm are also plotted. For the ablation study of our algorithms, which has two prominent differences from the SS algorithm, we also consider the scheme whose worker clustering stage is replaced by our SDP-based worker clustering method (SDP-SS). For the clustering-based algorithms including Alg. \ref{alg:with_side_information}, Alg. \ref{alg:without_side_information} and the subset-selection algorithm, the worker clustering stage is performed prior to the task label inference. In the worker clustering stage, the tuning parameters $\eta$ in \eqref{eqn:clustering_SDP} and $\xi$ in \eqref{eqn:SS_sequential_clustering} need to be chosen accordingly. Since there is no prior information about the model parameters, which may assist for the parameter tuning, in the real-world datasets, we instead do the exhaustive search over the proper ranges of the tuning parameters to choose the values that give the smallest weighted cross-edge sum between the clusters. In particular, we apply the normalized-cut method to measure the weighted cross-edge sum. Details about the normalized-cut method can be found in \cite{shi2000normalized,von2007tutorial}.

\indent In order to examine how the inference quality changes as the average number of queries per task increases, we sample the responses from each HIT, which are composed of 400 total answers, with probability $p_{s} \in \left\{ 0.4, 0.5, 0.6, 0.7, 0.8 \right\}$ for 30 times, and report the average inference quality in Fig. \ref{fig:real}. From the plots, we observe that Algorithm \ref{alg:with_side_information}, which uses side information of the ground-truth task types, outperforms all the other algorithms. In TABLE \ref{tab:m1:theta} and \ref{tab:m2:theta}, we report the mean of $\hat{\theta}_{ab} (\bf{M})$ in \eqref{eqn:choice_weights_alg:with_side_information} averaged over 30 random samplings for Athlete data and Netflix data. Remind that by using $\hat{\theta}_{ab} (\bf{M})$, we aim to recover the ground-truth signals $\theta_{ab}$ for $(a,b) \in [d] \times [d]$, defined in \eqref{eqn:ground_truth_signal}, which play a role as indicators of whether the cluster is spammer or non-spammer. By comparing the empirical reliability matrices in TABLE \ref{tab:relmat_real} and the estimated signals in TABLE \ref{tab:m1:theta} and \ref{tab:m2:theta}, one can observe that Algorithm \ref{alg:with_side_information} reliably filters out the responses from the unreliable (spammer) worker clusters. We can also see that as more samplers are used the accuracy of estimating the ground-truth signals $\theta_{ab}$ increases. 

\indent Among the algorithms that are not using side information of the ground-truth task types, Algorithm \ref{alg:without_side_information} achieves the best empirical performance. In particular, Algorithm \ref{alg:without_side_information} shows better performance than the algorithms which heavily rely upon the model assumptions of the Dawid-Skene model, since the strong model assumptions of the Dawid-Skene model, where the worker skill is fixed regardless of a given task, does not hold well for the real-world datasets we have collected where the tasks are heterogeneous and the worker reliabilities can vary with the type of a given task. Also, Algorithm \ref{alg:without_side_information} is better than the SS algorithm because it has gains both from the worker clustering stage, observed by the performance gap between SS and SDP-SS, and the weighting scheme for the label inference, observed by the performance gap between SDP-SS and Algorithm \ref{alg:without_side_information}. In particular, by comparing the performances of SDP-SS and SS, one can find that our worker clustering algorithm, Algorithm \ref{alg:worker_clustering}, performs much better than the na\"{i}ve sequential-clustering algorithm used in the SS algorithm. 

\begin{table}[t]
	\begin{subtable}[h]{\columnwidth}
		\centering
		\begin{tabular}{|l|l|l|l|}
			\hline
			1.000 & 0.667 & 0.033 & 0.033 \\
			\hline
			0.000 & 1.000 & 0.033 & 0.000 \\
			\hline
			0.067 & 0.100 & 1.000 & 0.167 \\
			\hline
			0.000 & 0.033 & 0.000 & 1.000 \\
			\hline
		\end{tabular}
		\caption{40\% answers are sampled}
	\end{subtable}
	\begin{subtable}[h]{\columnwidth}
		\centering
		\begin{tabular}{|l | l | l| l|}
			\hline
			1.000 & 0.000 & 0.000 & 0.000 \\
			\hline
			0.000 & 1.000 & 0.000 & 0.000 \\
			\hline
			0.000 & 0.033 & 1.000 & 0.033 \\
			\hline
			0.000 & 0.000 & 0.000 & 1.000 \\
			\hline
		\end{tabular}
		\caption{50\% answers are sampled}
	\end{subtable}
	\begin{subtable}[h]{\columnwidth}
		\centering
		\begin{tabular}{|l | l | l| l|}
			\hline
			1.000 & 0.000 & 0.000 & 0.000 \\
			\hline
			0.000 & 1.000 & 0.000 & 0.000 \\
			\hline
			0.000 & 0.000 & 1.000 & 0.000 \\
			\hline
			0.000 & 0.000 & 0.000 & 1.000 \\
			\hline
		\end{tabular}
		\caption{60\% answers are sampled}
	\end{subtable}
	\begin{subtable}[h]{\columnwidth}
		\centering
		\begin{tabular}{|l | l | l| l|}
			\hline
			1.000 & 0.000 & 0.000 & 0.000 \\
			\hline
			0.000 & 1.000 & 0.000 & 0.000 \\
			\hline
			0.000 & 0.000 & 1.000 & 0.000 \\
			\hline
			0.000 & 0.000 & 0.000 & 1.000 \\
			\hline
		\end{tabular}
		\caption{70\% answers are sampled}
	\end{subtable}
	\begin{subtable}[h]{\columnwidth}
		\centering
		\begin{tabular}{|l | l | l| l|}
			\hline
			1.000 & 0.000 & 0.000 & 0.000 \\
			\hline
			0.000 & 1.000 & 0.000 & 0.000 \\
			\hline
			0.000 & 0.000 & 1.000 & 0.000 \\
			\hline
			0.000 & 0.000 & 0.000 & 1.000 \\
			\hline
		\end{tabular}
		\caption{80\% answers are sampled}
	\end{subtable}
	\caption{The means of $\hat{\theta}_{ab} (\bf{M})$ in \eqref{eqn:choice_weights_alg:with_side_information} over 30 iterations for the Athlete dataset.}
	\label{tab:m1:theta}
\end{table}

\begin{table}[t]
	\begin{subtable}[h]{\columnwidth}
		\centering
		\begin{tabular}{|l|l|l|l|}
			\hline
			1.000 & 0.600 & 0.667 & 0.000 \\
			\hline
			0.000 & 1.000 & 0.000 & 0.000 \\
			\hline
			0.000 & 0.000 & 1.000 & 0.433 \\
			\hline
			0.833 & 0.000 & 0.000 & 1.000 \\
			\hline
		\end{tabular}
		\caption{40\% answers are sampled}
	\end{subtable}
	\begin{subtable}[h]{\columnwidth}
		\centering
		\begin{tabular}{|l | l | l| l|}
			\hline
			1.000 & 0.567 & 0.767 & 0.000 \\
			\hline
			0.000 & 1.000 & 0.000 & 0.000 \\
			\hline
			0.000 & 0.000 & 1.000 & 0.233 \\
			\hline
			0.867 & 0.000 & 0.000 & 1.000 \\
			\hline
		\end{tabular}
		\caption{50\% answers are sampled}
	\end{subtable}
	\begin{subtable}[h]{\columnwidth}
		\centering
		\begin{tabular}{|l | l | l| l|}
			\hline
			1.000 & 0.500 & 0.633 & 0.000 \\
			\hline
			0.000 & 1.000 & 0.000 & 0.000 \\
			\hline
			0.000 & 0.000 & 1.000 & 0.200 \\
			\hline
			0.833 & 0.000 & 0.000 & 1.000 \\
			\hline
		\end{tabular}
		\caption{60\% answers are sampled}
	\end{subtable}
	\begin{subtable}[h]{\columnwidth}
		\centering
		\begin{tabular}{|l | l | l| l|}
			\hline
			1.000 & 0.900 & 0.667 & 0.000 \\
			\hline
			0.000 & 1.000 & 0.000 & 0.000 \\
			\hline
			0.000 & 0.000 & 1.000 & 0.133 \\
			\hline
			0.933 & 0.000 & 0.000 & 1.000 \\
			\hline
		\end{tabular}
		\caption{70\% answers are sampled}
	\end{subtable}
	\begin{subtable}[h]{\columnwidth}
		\centering
		\begin{tabular}{|l | l | l| l|}
			\hline
			1.000 & 0.800 & 0.967 & 0.000 \\
			\hline
			0.000 & 1.000 & 0.000 & 0.000 \\
			\hline
			0.000 & 0.000 & 1.000 & 0.200 \\
			\hline
			1.000 & 0.000 & 0.000 & 1.000 \\
			\hline
		\end{tabular}
		\caption{80\% answers are sampled}
	\end{subtable}
	\caption{The means of $\hat{\theta}_{ab} (\bf{M})$ in \eqref{eqn:choice_weights_alg:with_side_information} over 30 iterations for the Netflix dataset.}
	\label{tab:m2:theta}
\end{table}

\section{Conclusion}
\label{sec:discussion}

\indent In this paper, we study the binary crowdsourced labeling problem for the $d$-type specialization model, where the error probability of a worker response changes depending on the worker-task type pair. Compared to the celebrated Dawid-Skene model, where it is assumed that the skilled worker set among the crowd worker is fixed regardless of the tasks, our model allows much general scenarios where the skilled worker set can vary over the tasks. In the $d$-type specialization model, we fully characterized the information-theoretic limits on the sample complexity required to correctly estimate the labels within a target accuracy, and designed two inference algorithms achieving the order-wise optimal sample complexity under some practical assumptions on the model, where the first algorithm utilizes side information of the ground-truth task-type vector and the second algorithm does not. For both the algorithms, we assumed that the worker types are unknown, and designed a worker clustering algorithm to extract the clusters of workers based on their types. The worker clustering stage has been critical in identifying reliable vs. unreliable workers for each task. By up-weighting the responses from the reliable worker clusters, while down-weighting the answers from the unreliable worker clusters, we showed that the proposed inference algorithms achieve the desired recovery accuracy at the order-wise optimal sample complexity, achievable via the maximum likelihood estimator, even when the exact fidelity matrix is unknown.
 
\indent We conclude this paper by highlighting some few interesting open directions. In most existing works on crowdsourcing where an inference algorithm that can beat the simple majority voting has been designed with theoretical guarantees, the structural assumption on the expected observation matrix has been restricted to be rank-one. In this work, we extended the model so that the expected observation matrix can have rank-$d$, where $d$ can scale in the number of tasks up to some limit, but we imposed a specific block structure based on the types of tasks and workers. Involving real-world dataset simulations, we demonstrated that when the tasks have some inherent types, such a block structure describes well the error probabilities of worker responses and our inference algorithms recovering the block structure outperforms the other algorithms assuming rank-one structure for the model. It would be an interesting open direction to consider a more general error model where the fidelity matrix has rank higher than one but without any block structure. Also, it would be worthwhile to investigate whether there is an inference algorithm that can achieve theoretically better performance than that of the standard majority voting estimator and whether such a general model allows the algorithm to perform more robustly in estimating the ground-truth labels from the real-world datasets.

\section{Proofs of theoretical results}
\label{sec:proofs_theoretical_results}

\subsection{Proof of Theorem \ref{thm:performance_alg:with_side_information}}
\label{subsec:proof_thm:performance_alg:with_side_information}

\indent To begin with, we assume that we are on the event $\mathcal{E}_1 \cap \mathcal{E}_2$, where
\begin{equation}
    \label{eqn:proof_thm:performance_alg:with_side_information_v1}
    \begin{split}
        \mathcal{E}_1 &:= \left\{ \hat{\mathcal{W}}_b = \mathcal{W}_b := \left\{ j \in [n]: w_j = b \right\} \textnormal{ for all } b \in [d] \right\}; \\
        \mathcal{E}_2 &:= \bigcap_{b=1}^{d} \left\{ \left| \hat{\mathcal{W}}_b \right| \geq l \right\}.
    \end{split}
\end{equation}
Here, the event $\mathcal{E}_1$ defines an event that the worker clustering algorithm (Algorithm \ref{alg:worker_clustering}) exactly recovers the true clusters and the event $\mathcal{E}_2$ refers to an event that the size of inferred worker clusters are all greater than or equal to $l$. Now, let us recall the weighting scheme \eqref{eqn:choice_weights_alg:with_side_information} for performing weighted majority voting rules in Algorithm \ref{alg:with_side_information}: for each $(a, b) \in [d] \times [d]$,
\begin{equation*}
    \hat{\theta}_{ab}(\mathbf{M}) :=
    \begin{cases}
        1 & \textnormal{if } \frac{1}{\left| \hat{\mathcal{T}}_a \right|} \sum_{i \in \hat{\mathcal{T}}_a} \left| \sum_{j \in \mathcal{A}_b (i)} M_{ij} \right| \geq \zeta_{ab}; \\
        0 & \textnormal{otherwise.}
    \end{cases}
\end{equation*}
Note that the weighting scheme \eqref{eqn:choice_weights_alg:with_side_information} is designed for recovering the ground-truth signals $\left\{ \theta_{ab}: (a, b) \in [d] \times [d] \right\}$, where
\begin{equation}
    \label{eqn:proof_thm:performance_alg:with_side_information_v2}
        \theta_{ab} :=
        \begin{cases}
            1 & \textnormal{if } b \in [d] \setminus \textnormal{spammer}(a); \\
            0 & \textnormal{otherwise.}
        \end{cases}
\end{equation}

\indent We first aim to bound the probability
\begin{equation*}
    \mathbb{P} \left\{ \left. \bigcup_{(a, b) \in [d] \times [d]} \left\{ \hat{\theta}_{ab}(\mathbf{M}) \neq \theta_{ab} \right\} \right| \mathcal{E}_1 \cap \mathcal{E}_2, \left( \mathbf{t}, \mathbf{w} \right) \right\}
\end{equation*}
that the weighting scheme \eqref{eqn:choice_weights_alg:with_side_information} fails to exactly recover the ground-truth signals $\left\{ \theta_{ab}: (a, b) \in [d] \times [d] \right\}$. 
Let $\mathcal{E}_3 := \bigcap_{(a, b) \in [d] \times [d]} \left\{ \hat{\theta}_{ab}(\mathbf{M}) = \theta_{ab} \right\}$. Due to the union bound, we have
\begin{equation}
    \label{eqn:proof_thm:performance_alg:with_side_information_v3}
    \begin{split}
        &\mathbb{P} \left\{ \left. \mathcal{E}_{3}^c \right| \mathcal{E}_1 \cap \mathcal{E}_2, \left( \mathbf{t}, \mathbf{w} \right) \right\} \\
        \leq \ & \sum_{(a, b) \in [d] \times [d]} \mathbb{P} \left\{ \left. \hat{\theta}_{ab}(\mathbf{M}) \neq \theta_{ab} \right| \mathcal{E}_1 \cap \mathcal{E}_2, \left( \mathbf{t}, \mathbf{w} \right) \right\}.
    \end{split}
\end{equation}
Thus, it suffices to take a closer inspection on the probability $\mathbb{P} \left\{ \left. \hat{\theta}_{ab}(\mathbf{M}) \neq \theta_{ab} \right| \mathcal{E}_1 \cap \mathcal{E}_2, \left( \mathbf{t}, \mathbf{w} \right) \right\}$ for each $(a, b) \in [d] \times [d]$. Here, we consider the following two cases:

\begin{enumerate} [label=(\roman*)]
    \item $b \in \textnormal{spammer}_{\mathcal{Q}}(a)$: we have $\mathcal{Q}(a, b) = \frac{1}{2}$ and $\theta_{ab} = 0$. If it is able to choose the threshold $\zeta_{ab} > 0$ such that
    \begin{equation}
        \label{eqn:proof_thm:performance_alg:with_side_information_v4}
            \zeta_{ab} > \frac{1}{\left| \hat{\mathcal{T}}_a \right|} \sum_{i \in \hat{\mathcal{T}}_a} \mathbb{E} \left[ \left. \left| \sum_{j \in \mathcal{A}_b (i)} M_{ij} \right| \right| \mathcal{E}_1 \cap \mathcal{E}_2, \left( \mathbf{t}, \mathbf{w} \right) \right],
    \end{equation}
    then the one-sided Hoeffding's inequality yields
    \begin{equation}
        \label{eqn:proof_thm:performance_alg:with_side_information_v5}
        \begin{split}
            &\mathbb{P} \left\{ \left. \hat{\theta}_{ab}(\mathbf{M}) \neq \theta_{ab} \right| \mathcal{E}_1 \cap \mathcal{E}_2, \left( \mathbf{t}, \mathbf{w} \right) \right\} \leq \\
            & \  \exp \left\{ - \frac{2 \left| \hat{\mathcal{T}}_a \right|}{l^2}\Bigg( \zeta_{ab}  \right.  \\
            &\left. - \frac{1}{\left| \hat{\mathcal{T}}_a \right|} \sum_{i \in \hat{\mathcal{T}}_a} \mathbb{E} \left[ \left. \left| \sum_{j \in \mathcal{A}_b (i)} M_{ij} \right| \right| \mathcal{E}_1 \cap \mathcal{E}_2, \left( \mathbf{t}, \mathbf{w} \right) \right] \Bigg)^2 \right\}.
        \end{split}
    \end{equation}
    \item $b \in [d] \setminus \textnormal{spammer}_{\mathcal{Q}}(a)$: it holds that $\mathcal{Q}(a, b) \geq \frac{1}{2} + \epsilon$ and $\theta_{ab} = 1$. If one can select the threshold $\zeta_{ab} > 0$ such that
    \begin{equation}
        \label{eqn:proof_thm:performance_alg:with_side_information_v6}
            \zeta_{ab} < \frac{1}{\left| \hat{\mathcal{T}}_a \right|} \sum_{i \in \hat{\mathcal{T}}_a} \mathbb{E} \left[ \left. \left| \sum_{j \in \mathcal{A}_b (i)} M_{ij} \right| \right| \mathcal{E}_1 \cap \mathcal{E}_2, \left( \mathbf{t}, \mathbf{w} \right) \right],
    \end{equation}
    then the one-sided Hoeffding's inequality yields
    \begin{equation}
        \label{eqn:proof_thm:performance_alg:with_side_information_v7}
        \begin{split}
            &\mathbb{P} \left\{ \left. \hat{\theta}_{ab}(\mathbf{M}) \neq \theta_{ab} \right| \mathcal{E}_1 \cap \mathcal{E}_2, \left( \mathbf{t}, \mathbf{w} \right) \right\} \\
            \leq \ & \exp \left\{ - \frac{2 \left| \hat{\mathcal{T}}_a \right|}{l^2}\Bigg( \zeta_{ab} \right. \\
            &\left. - \frac{1}{\left| \hat{\mathcal{T}}_a \right|} \sum_{i \in \hat{\mathcal{T}}_a} \mathbb{E} \left[ \left. \left| \sum_{j \in \mathcal{A}_b (i)} M_{ij} \right| \right| \mathcal{E}_1 \cap \mathcal{E}_2, \left( \mathbf{t}, \mathbf{w} \right) \right] \Bigg)^2 \right\}.
        \end{split}
    \end{equation}
\end{enumerate}

\indent We now define the function $\Delta (\cdot, \cdot): [d] \times [d] \to \mathbb{R}_{+}$ by
\begin{equation*}
    \Delta (a, b) := \frac{1}{\left| \hat{\mathcal{T}}_a \right|} \sum_{i \in \hat{\mathcal{T}}_a} \mathbb{E} \left[ \left. \left| \sum_{j \in \mathcal{A}_b (i)} M_{ij} \right| \right| \mathcal{E}_1 \cap \mathcal{E}_2, \left( \mathbf{t}, \mathbf{w} \right) \right]
\end{equation*}
for each $(a, b) \in [d] \times [d]$. Given any $(n, p) \in \mathbb{N} \times [0, 1]$, let $\mathbb{B}(n, p)$ denotes the binomial distribution with parameter $(n, p) \in \mathbb{N} \times [0, 1]$. Then, one can observe that
\begin{equation*}
    \left| \sum_{j \in \mathcal{A}_b (i)} M_{ij} \right| \stackrel{d}{=} \left| 2 \mathbb{B} \left( l, \mathcal{Q} \left( t_i, b \right) \right) - l \right|,\ \forall (i, b) \in [m] \times [d],
\end{equation*}
while being conditioned on the event $\mathcal{E}_1 \cap \mathcal{E}_2$ as well as the pair of type vectors $\left( \mathbf{t}, \mathbf{w} \right)$. Thus,
\begin{equation}
    \label{eqn:proof_thm:performance_alg:with_side_information_v8}
    \begin{split}
        \Delta (a, b) &= \frac{1}{\left| \hat{\mathcal{T}}_a \right|} \sum_{i \in \hat{\mathcal{T}}_a} \mathbb{E} \left[ \left| 2 \mathbb{B} \left( l, \mathcal{Q} \left( t_i, b \right) \right) - l \right| \right] \\
        &= \frac{1}{\left| \hat{\mathcal{T}}_a \right|} \sum_{i \in \hat{\mathcal{T}}_a \setminus \mathcal{T}_a} \mathbb{E} \left[ \left| 2 \mathbb{B} \left( l, \mathcal{Q} \left( t_i, b \right) \right) - l \right| \right] \\
        &\quad+ \frac{\left| \hat{\mathcal{T}}_a \cap \mathcal{T}_a \right|}{\left| \hat{\mathcal{T}}_a \right|} \mathbb{E} \left[ \left| 2 \mathbb{B} \left( l, \mathcal{Q} (a, b) \right) - l \right| \right]
    \end{split}
\end{equation}
for every $(a, b) \in [d] \times [d]$.

\indent For the implementation of the weighting scheme \eqref{eqn:choice_weights_alg:with_side_information}, we need to choose suitable thresholds $\left\{ \zeta_{ab} \in \left( 0, +\infty \right): b \in [d] \right\}$ such that they satisfy the inequality \eqref{eqn:proof_thm:performance_alg:with_side_information_v4} for the case $b \in \textnormal{spammer}_{\mathcal{Q}}(a)$ and the inequality \eqref{eqn:proof_thm:performance_alg:with_side_information_v6} otherwise. However, we do not know whether $b \in \textnormal{spammer}_{\mathcal{Q}}(a)$ or $b \in [d] \setminus \textnormal{spammer}_{\mathcal{Q}}(a)$ for any given pair $(a, b) \in [d] \times [d]$ due to the lack of prior knowledge of the reliability matrix $\mathcal{Q}$. So, our desideratum is to make a uniform choice $\zeta_{ab} \equiv \lambda_a$ for every $b \in [d]$ such that $\lambda_a > \Delta (a, b)$ if $b \in \textnormal{spammer}_{\mathcal{Q}} (a)$ and $\lambda_a < \Delta (a, b)$ otherwise. Therefore, it suffices to establish the following conclusion:
\begin{equation}
    \label{eqn:proof_thm:performance_alg:with_side_information_v9}
    \begin{split}
        &\max \left\{ \Delta \left( a, b_0 \right): b_0 \in \textnormal{spammer}_{\mathcal{Q}} (a) \right\} \\
        < \ & \min \left\{ \Delta \left( a, b_1 \right): b_1 \in [d] \setminus \textnormal{spammer}_{\mathcal{Q}} (a) \right\}.
    \end{split}
\end{equation}
Once we prove the claim \eqref{eqn:proof_thm:performance_alg:with_side_information_v9}, we specify the threshold $\zeta_{ab} > 0$ for $b \in [d]$ by  $\zeta_{ab} =  \lambda_a $ where
\begin{equation}
    \label{eqn:proof_thm:performance_alg:with_side_information_v10}
    \begin{split}
        & \lambda_a \\
        &:= \  \frac{1}{2} \left[ \max_{b_0 \in \textnormal{spammer}_{\mathcal{Q}} (a)} \Delta \left( a, b_0 \right) + \min_{b_1 \in [d] \setminus \textnormal{spammer}_{\mathcal{Q}} (a)} \Delta \left( a, b_1 \right) \right]
    \end{split}
\end{equation}
for each $a \in [d]$. Consequently, our main question is now: how can we establish the inequality \eqref{eqn:proof_thm:performance_alg:with_side_information_v9}? To this end, the following useful lemma, whose detailed proof can be found in Appendix \ref{subsec:proof_lemma:proof_thm:performance_alg:with_side_information_v1}, is necessary:

\begin{lemma}
\label{lemma:proof_thm:performance_alg:with_side_information_v1}
For any $(n, p) \in \mathbb{N} \times [0, 1]$, the following results hold:
\begin{enumerate} [label=(\roman*)]
    \item If $p \neq \frac{1}{2}$, then
    \begin{equation}
        \label{eqn:lemma:proof_thm:performance_alg:with_side_information_v1_v1}
        \begin{split}
            n \left| 2p - 1 \right| \leq \ & \mathbb{E} \left[ \left| 2 \mathbb{B} (n, p) - n \right| \right] \\
            \leq \ & n \left| 2p - 1 \right| + 2n \exp \left\{ - 2n \left( p - \frac{1}{2} \right)^2 \right\}.
        \end{split}
    \end{equation}
    In particular, if $p$ does not depend on $n$, then 
    \[
        \mathbb{E} \left[ \left| 2 \mathbb{B} (n, p) - n \right| \right] \asymp n \left| 2p-1 \right|
    \]
    as $n \to \infty$;
    \item If $p = \frac{1}{2}$, then it holds that
    \begin{equation}
        \label{eqn:lemma:proof_thm:performance_alg:with_side_information_v1_v2}
        \lim_{n \to \infty} \frac{\mathbb{E} \left[ \left| 2 \mathbb{B} \left( n, \frac{1}{2} \right) - n \right| \right]}{\sqrt{n}} = \sqrt{\frac{2}{\pi}}.
    \end{equation}
    Thus, $\mathbb{E} \left[ \left| 2 \mathbb{B} \left( n, \frac{1}{2} \right) - n \right| \right] \asymp \sqrt{\frac{2}{\pi}} \cdot \sqrt{n}$ as $n \to \infty$.
\end{enumerate}
\end{lemma}

\indent In order to prove the desired result \eqref{eqn:proof_thm:performance_alg:with_side_information_v9}, we first choose any $\left( b_0, b_1 \right) \in \textnormal{spammer}_{\mathcal{Q}}(a) \times \left( [d] \setminus \textnormal{spammer}_{\mathcal{Q}}(a) \right)$. With the above technical lemma in hand, one can establish a desirable lower bound on the term $\Delta \left( a, b_1 \right) - \Delta \left( a, b_0 \right)$ as follows: if $l = \omega (1)$ as $d \to \infty$, then
\begin{equation}
    \label{eqn:proof_thm:performance_alg:with_side_information_v11}
    \begin{split}
        &\Delta \left( a, b_1 \right) - \Delta \left( a, b_0 \right) \\
        \stackrel{\textnormal{(a)}}{=} \ &
        \frac{\left| \hat{\mathcal{T}}_a \cap \mathcal{T}_a \right|}{\left| \hat{\mathcal{T}}_a \right|} \left( \mathbb{E} \left[ \left| 2 \mathbb{B} \left( l, \mathcal{Q} \left( a, b_1 \right) \right) - l \right| \right] \right.\\
        &\qquad\qquad\left.- \mathbb{E} \left[ \left| 2 \mathbb{B} \left( l, \mathcal{Q} \left( a, b_0 \right) \right) - l \right| \right] \right) \\
        &+ \frac{1}{\left| \hat{\mathcal{T}}_a \right|} \sum_{i \in \hat{\mathcal{T}}_a \setminus \mathcal{T}_a} \left( \underbrace{\mathbb{E} \left[ \left| 2 \mathbb{B} \left( l, \mathcal{Q} \left( t_i, b_1 \right) \right) - l \right| \right]}_{\geq \ 0} \right. \\
        &\qquad\qquad\left. - \underbrace{\mathbb{E} \left[ \left| 2 \mathbb{B} \left( l, \mathcal{Q} \left( t_i, b_0 \right) \right) - l \right| \right]}_{\leq \ l} \right) \\
        \geq \ & \frac{\left| \hat{\mathcal{T}}_a \cap \mathcal{T}_a \right|}{\left| \hat{\mathcal{T}}_a \right|} \left( \mathbb{E} \left[ \left| 2 \mathbb{B} \left( l, \mathcal{Q} \left( a, b_1 \right) \right) - l \right| \right] \right. \\
        &\left. - \mathbb{E} \left[ \left| 2 \mathbb{B} \left( l, \mathcal{Q} \left( a, b_0 \right) \right) - l \right| \right] \right) - \frac{\left| \hat{\mathcal{T}}_a \setminus \mathcal{T}_a \right|}{\left| \hat{\mathcal{T}}_a \right|} l \\
        \stackrel{\textnormal{(b)}}{\geq} \ & \left( 1 - \frac{\left| \hat{\mathcal{T}}_a \setminus \mathcal{T}_a \right|}{\left| \hat{\mathcal{T}}_a \right|} \right) \left\{ 2 \epsilon l - \sqrt{\frac{2l}{\pi}} \cdot \left( 1 + o(1) \right) \right\} \\
        &\quad- \frac{\left| \hat{\mathcal{T}}_a \setminus \mathcal{T}_a \right|}{\left| \hat{\mathcal{T}}_a \right|} l \\
        \stackrel{\textnormal{(c)}}{=} \ & \left( 1 - o(1) \right) \underbrace{\left\{ 2 \epsilon l - \sqrt{\frac{2l}{\pi}} \cdot \left( 1 + o(1) \right) \right\}}_{\geq \ \frac{3}{2}\epsilon l} - o(1) \cdot l \\
        \geq \ & \frac{3}{2} \epsilon l - \underbrace{o(1) \cdot \left( \frac{3}{2} \epsilon + 1 \right) l}_{\leq \ \frac{1}{2} \epsilon l} \\
        \geq \ & \epsilon l
    \end{split}
\end{equation}
for all sufficiently large $d \in \mathbb{N}$, where the step (a) utilizes the identity \eqref{eqn:proof_thm:performance_alg:with_side_information_v8}, the step (b) holds due to Lemma \ref{lemma:proof_thm:performance_alg:with_side_information_v1}, and the step (c) follows from the condition \eqref{eqn:sufficient_condition_alg:with_side_information_v1}. Therefore, we obtain
\begin{equation}
    \label{eqn:proof_thm:performance_alg:with_side_information_v12}
    \min_{b_1 \in [d] \setminus \textnormal{spammer}_{\mathcal{Q}} (a)} \Delta \left( a, b_1 \right) - \max_{b_0 \in \textnormal{spammer}_{\mathcal{Q}} (a)} \Delta \left( a, b_0 \right) \geq \epsilon l
\end{equation}
for all sufficiently large $d$. From the bound \eqref{eqn:proof_thm:performance_alg:with_side_information_v12} together with the choice \eqref{eqn:proof_thm:performance_alg:with_side_information_v10} of the thresholds $\zeta_{ab}$, one can conclude from the bounds \eqref{eqn:proof_thm:performance_alg:with_side_information_v5} and \eqref{eqn:proof_thm:performance_alg:with_side_information_v7} that for every pair $(a, b) \in [d] \times [d]$, it holds that
\begin{equation}
    \label{eqn:proof_thm:performance_alg:with_side_information_v13}
    \begin{split}
        &\mathbb{P} \left\{ \left. \hat{\theta}_{ab}(\mathbf{M}) \neq \theta_{ab} \right| \mathcal{E}_1 \cap \mathcal{E}_2, \left( \mathbf{t}, \mathbf{w} \right) \right\} \\
        \leq \ & \exp \left\{ - \frac{2 \left| \hat{\mathcal{T}}_a \right|}{l^2} \cdot \left( \frac{1}{2} \epsilon l \right)^2 \right\} \\
        = \ & \exp \left( - \frac{1}{2} \epsilon^2 \left| \hat{\mathcal{T}}_a \right| \right)
    \end{split}
\end{equation}
for all sufficiently large $d$. From the approximated balancedness assumption on the clusters of tasks (Assumption \ref{assumption:approximate_balancedness_task_type}) and the weak recovery assumption \eqref{eqn:sufficient_condition_alg:with_side_information_v1} on side information $\hat{\mathbf{t}}$, it follows that $\left| \hat{\mathcal{T}}_a \right| = \Theta \left( \frac{m}{d} \right)$ for all $a \in [d]$ as $d \to \infty$. Hence, by putting the inequality \eqref{eqn:proof_thm:performance_alg:with_side_information_v13} into the bound \eqref{eqn:proof_thm:performance_alg:with_side_information_v3}, we reach
\begin{equation}
    \label{eqn:proof_thm:performance_alg:with_side_information_v14}
    \begin{split}
        &\mathbb{P} \left\{ \left. \mathcal{E}_{3}^c \right| \mathcal{E}_1 \cap \mathcal{E}_2, \left( \mathbf{t}, \mathbf{w} \right) \right\} \\
        \leq \ & \sum_{(a, b) \in [d] \times [d]} \exp \left( - \frac{1}{2} \epsilon^2 \left| \hat{\mathcal{T}}_a \right| \right) \\
        \leq \ & d^2 \exp \left\{ - \epsilon^2 \cdot \Theta \left( \frac{m}{d} \right) \right\}.
    \end{split}
\end{equation}

\indent Now we analyze the probability of error for the weighted majority voting rule \eqref{eqn:decision_rule_alg:with_side_information} while being conditioned on the event $\mathcal{E}_1 \cap \mathcal{E}_2 \cap \mathcal{E}_3$. Let $\left\{ \Lambda_{ij}: (i, j) \in \mathcal{A} \right\}$ be a collection of conditionally independent random variables given a pair of type vectors $\left( \mathbf{t}, \mathbf{w} \right)$ such that $\Lambda_{ij} \sim \textnormal{Bern} \left( F_{ij} \right) = \textnormal{Bern} \left( \mathcal{Q} \left( t_i, w_j \right) \right)$ for every $(i, j) \in \mathcal{A}$. Then, the weighted majority voting rule \eqref{eqn:decision_rule_alg:with_side_information} can be written as:
\begin{equation*}
    \begin{split}
        &\hat{a}_i (\mathbf{M}) = \textnormal{sign} \left\{ \sum_{w=1}^{d} \left( \sum_{j \in \mathcal{A}_w (i)} \hat{\theta}_{\hat{t}_i w} (\mathbf{M}) \cdot M_{ij} \right) \right\} \\
        &= \  \textnormal{sign} \left[ a_i \cdot \sum_{w=1}^{d} \left\{ \sum_{j \in \mathcal{A}_w (i)} \hat{\theta}_{\hat{t}_i w} (\mathbf{M}) \left( 2 \Lambda_{ij} - 1 \right) \right\} \right].
    \end{split}
\end{equation*}
So the error probability $\mathbb{P} \left\{ \left. \hat{a}_i (\mathbf{M}) \neq a_i \right| \mathcal{E}_1 \cap \mathcal{E}_2 \cap \mathcal{E}_3, \left( \mathbf{t}, \mathbf{w} \right) \right\}$ can be bounded by
\begin{equation}
    \label{eqn:proof_thm:performance_alg:with_side_information_v15}
    \begin{split}
        &\mathbb{P} \left\{ \left. \hat{a}_i (\mathbf{M}) \neq a_i \right| \mathcal{E}_1 \cap \mathcal{E}_2 \cap \mathcal{E}_3, \left( \mathbf{t}, \mathbf{w} \right) \right\} \\
        = \ & \mathbb{P} \Bigg\{  \sum_{w=1}^{d} \left\{ \sum_{j \in \mathcal{A}_w (i)} \hat{\theta}_{\hat{t}_i w} (\mathbf{M}) \left( 2 \Lambda_{ij} - 1 \right) \right\} \leq 0\Bigg|  \\
        &\qquad\mathcal{E}_1 \cap \mathcal{E}_2 \cap \mathcal{E}_3, \left( \mathbf{t}, \mathbf{w} \right) \Bigg\} \\
        \stackrel{\textnormal{(d)}}{\leq} \ & \exp \left[ - \frac{l}{2} \cdot \frac{\left\{ \sum_{w \in [d] \setminus \textnormal{spammer}_{\mathcal{Q}} \left( \hat{t}_i \right)} \left( 2 \mathcal{Q} \left( t_i, w \right) - 1 \right) \right\}^2}{d - \left| \textnormal{spammer}_{\mathcal{Q}} \left( \hat{t}_i \right) \right|} \right] \\
        \stackrel{\textnormal{(e)}}{\leq} \ &
        \exp \Bigg[ - \frac{2l \epsilon^2}{d \delta \left( \hat{t}_i; d \right)}\\
        &\times \left\{ d - \left| \textnormal{spammer}_{\mathcal{Q}} \left( t_i \right) \cup \textnormal{spammer}_{\mathcal{Q}} \left( \hat{t}_i \right) \right| \right\}^2 \Bigg],
    \end{split}
\end{equation}
where the step (d) is due to the Hoeffding's inequality and the step (e) follows from the assumptions $\left| \textnormal{spammer}_{\mathcal{Q}} (t) \right| = d \left\{ 1 - \delta (t; d) \right\}$ and $\mathcal{Q} (t, w) \geq \frac{1}{2} + \epsilon$, $\forall w \in [d] \setminus \textnormal{spammer}_{\mathcal{Q}}(t)$, for every $t \in [d]$. Taking two ingredients \eqref{eqn:proof_thm:performance_alg:with_side_information_v14} and \eqref{eqn:proof_thm:performance_alg:with_side_information_v15} collectively, we obtain
\begin{equation}
    \label{eqn:proof_thm:performance_alg:with_side_information_v16}
    \begin{split}
        &\mathbb{P} \left\{ \hat{a}_i (\mathbf{M}) \neq a_i \right\} \\
        \leq \ & \mathbb{P} \left\{ \left. \hat{a}_i (\mathbf{M}) \neq a_i \right| \mathcal{E}_1 \cap \mathcal{E}_2 \cap \mathcal{E}_3 \right\} \underbrace{\mathbb{P} \left\{ \mathcal{E}_1 \cap \mathcal{E}_2 \cap \mathcal{E}_3 \right\}}_{\leq \ 1} \\
        &+ \underbrace{\mathbb{P} \left\{ \left. \hat{a}_i (\mathbf{M}) \neq a_i \right| \mathcal{E}_1 \cap \mathcal{E}_2 \cap \mathcal{E}_{3}^c \right\}}_{\leq \ 1} \underbrace{\mathbb{P} \left\{ \mathcal{E}_1 \cap \mathcal{E}_2 \cap \mathcal{E}_{3}^c \right\}}_{\leq \ \mathbb{P} \left\{ \left. \mathcal{E}_{3}^c \right| \mathcal{E}_1 \cap \mathcal{E}_2 \right\}} \\
        &+ \underbrace{\mathbb{P} \left\{ \left. \hat{a}_i (\mathbf{M}) \neq a_i \right| \mathcal{E}_{1}^c \right\}}_{\leq \ 1} \mathbb{P} \left\{ \mathcal{E}_{1}^c \right\} \\
        &+ \underbrace{\mathbb{P} \left\{ \left. \hat{a}_i (\mathbf{M}) \neq a_i \right| \mathcal{E}_{2}^c \right\}}_{\leq \ 1} \mathbb{P} \left\{ \mathcal{E}_{2}^c \right\} \\
        \leq \ & \mathbb{P} \left\{ \left. \hat{a}_i (\mathbf{M}) \neq a_i \right| \mathcal{E}_1 \cap \mathcal{E}_2 \cap \mathcal{E}_3 \right\} \\
        &+ \mathbb{P} \left\{ \left. \mathcal{E}_{3}^c \right| \mathcal{E}_1 \cap \mathcal{E}_2 \right\} + \mathbb{P} \left\{ \mathcal{E}_{1}^c \right\} + \mathbb{P} \left\{ \mathcal{E}_{2}^c \right\} \\
        \leq \ & \exp \Bigg[ - \frac{2l \epsilon^2}{d \delta \left( \hat{t}_i; d \right)}\\
        &\qquad\times \left\{ d - \left| \textnormal{spammer}_{\mathcal{Q}} \left( t_i \right) \cup \textnormal{spammer}_{\mathcal{Q}} \left( \hat{t}_i \right) \right| \right\}^2 \Bigg] \\
        &+ d^2 \exp \left\{ - \epsilon^2 \cdot \Theta \left( \frac{m}{d} \right) \right\} + \mathbb{P} \left\{ \mathcal{E}_{1}^c \right\} + \mathbb{P} \left\{ \mathcal{E}_{2}^c \right\}.
    \end{split}
\end{equation}
Therefore, in view of the inequality \eqref{eqn:proof_thm:performance_alg:with_side_information_v16}, it only remains to establish upper bounds on the probabilities $\mathbb{P} \left\{ \mathcal{E}_{1}^c \right\}$ and $\mathbb{P} \left\{ \mathcal{E}_{2}^c \right\}$. We know from Lemma \ref{lemma:exact_recovery_alg:worker_clustering} that $\mathbb{P} \left\{ \mathcal{E}_{1}^c \right\} \leq 4 n^{-11}$ if we choose $r = C_1 \cdot \frac{n^2 \left( \log n \right)^2}{\left( p_m - p_u \right)^2 s_{\min}^2}$ workers randomly in Step 1 of the SDP-based worker clustering algorithm (Algorithm \ref{alg:worker_clustering}). Now, we should take account with the probability that the event $\mathcal{E}_2$ does not occur, \emph{i.e.}, $\mathbb{P} \left\{ \mathcal{E}_{2}^c \right\}$. While being conditioned on the event $\mathcal{E}_1$, it holds that $\hat{\mathcal{W}}_b = \mathcal{W}_b$ for every $b \in [d]$. Also since the type of each worker is randomly generated from the distribution $\bm{\nu} \in \Delta \left( [d] \right)$ independently, the number of workers of type $b \in [d]$ is given by $\left| \mathcal{W}_b \right| = \sum_{j=1}^{n} \mathbbm{1} \left( w_j = b \right) \sim \textnormal{Binomial} \left( n, \nu(b) \right)$. This gives:
\begin{equation}
    \begin{split}\label{eqn:cond_prob_E2_E1}
       & \mathbb{P} \left\{ \left. \mathcal{E}_{2}^c \right| \mathcal{E}_1 \right\}\\
        &= \mathbb{P} \left\{ \left. \bigcup_{b=1}^{d} \left\{ \left| \mathcal{W}_b \right| < l \right\} \right| \mathcal{E}_1 \right\} \\
        &\stackrel{\textnormal{(f)}}{\leq} \sum_{b=1}^{d} \mathbb{P} \left\{ \left. \sum_{j=1}^{n} \mathbbm{1} \left( w_j = b \right) < l \right| \mathcal{E}_1 \right\} \\
        &\leq \frac{1}{\mathbb{P} \left\{ \mathcal{E}_1 \right\}} \sum_{b=1}^{d} \mathbb{P} \left\{ \sum_{j=1}^{n} \left( \mathbbm{1} \left( w_j = b \right) - \nu (b) \right) < l - n \nu(b) \right\} \\
        &\stackrel{\textnormal{(g)}}{\leq} \frac{1}{\mathbb{P} \left\{ \mathcal{E}_1 \right\}} \sum_{b=1}^{d} \exp \left\{ - \frac{n \nu(b)}{2} \left( 1 - \frac{l}{n \nu (b)} \right)^2 \right\} \\
        &\leq \frac{d}{\mathbb{P} \left\{ \mathcal{E}_1 \right\}} \exp \left[ - \min_{b \in [d]} \left\{ \frac{n \nu(b)}{2} \left( 1 - \frac{l}{n \nu (b)} \right)^2 \right\} \right],
    \end{split}
\end{equation}
where the step (f) comes from the union bound and the step (g) holds due to the multiplicative form of the Chernoff's bound. Hence, we eventually arrive at
\begin{equation}
    \label{eqn:proof_thm:performance_alg:with_side_information_v17}
    \begin{split}
        &\mathbb{P} \left\{ \mathcal{E}_{2}^c \right\}\\
        &= \mathbb{P} \left\{ \left. \mathcal{E}_{2}^c \right| \mathcal{E}_1 \right\} \mathbb{P} \left\{ \mathcal{E}_1 \right\} + \underbrace{\mathbb{P} \left\{ \left. \mathcal{E}_{2}^c \right| \mathcal{E}_{1}^c \right\}}_{\leq \ 1} \mathbb{P} \left\{ \mathcal{E}_{1}^c \right\} \\
        &\leq d \exp \left[ - \min_{b \in [d]} \left\{ \frac{n \nu(b)}{2} \left( 1 - \frac{l}{n \nu (b)} \right)^2 \right\} \right] + 4n^{-11}.
    \end{split}
\end{equation}
By putting two bounds $\mathbb{P} \left\{ \mathcal{E}_{1}^c \right\} \leq 4 n^{-11}$ and \eqref{eqn:proof_thm:performance_alg:with_side_information_v17} into the inequality \eqref{eqn:proof_thm:performance_alg:with_side_information_v16}, it follows that
\begin{equation}
    \label{eqn:proof_thm:performance_alg:with_side_information_v18}
    \begin{split}
        &\mathbb{P} \left\{ \hat{a}_i (\mathbf{M}) \neq a_i \right\}\\
        \leq \ & \exp \Bigg[ - \frac{2l \epsilon^2}{d \delta \left( \hat{t}_i; d \right)}\\
       &\quad \times \left\{ d - \left| \textnormal{spammer}_{\mathcal{Q}} \left( t_i \right) \cup \textnormal{spammer}_{\mathcal{Q}} \left( \hat{t}_i \right) \right| \right\}^2 \Bigg] \\
        &+ d^2 \exp \left\{ - \epsilon^2 \cdot \Theta \left( \frac{m}{d} \right) \right\} \\
        &+ d \exp \left[ - \min_{b \in [d]} \left\{ \frac{n \nu(b)}{2} \left( 1 - \frac{l}{n \nu (b)} \right)^2 \right\} \right] + 8n^{-11}.
    \end{split}
\end{equation}
Note that if $\hat{t}_i = t_i$, then the bound \eqref{eqn:proof_thm:performance_alg:with_side_information_v18} can be simplified as
\begin{equation}
    \label{eqn:proof_thm:performance_alg:with_side_information_v19}
    \begin{split}
        &\mathbb{P} \left\{ \hat{a}_i (\mathbf{M}) \neq a_i \right\} \\
        \leq \ & \exp \left\{ - 2l \epsilon^2 \cdot d \delta \left( t_i; d \right) \right\} + d^2 \exp \left\{ - \epsilon^2 \cdot \Theta \left( \frac{m}{d} \right) \right\} \\
        &+ d \exp \left[ - \min_{b \in [d]} \left\{ \frac{n \nu(b)}{2} \left( 1 - \frac{l}{n \nu (b)} \right)^2 \right\} \right] + 8n^{-11}.
    \end{split}
\end{equation}
So one can upper bound the risk function of the output $\hat{\mathbf{a}}(\cdot): \left\{ \pm 1 \right\}^{\mathcal{A}} \to \left\{ \pm 1 \right\}^m$ of Algorithm \ref{alg:with_side_information} by
\begin{equation}
    \label{eqn:proof_thm:performance_alg:with_side_information_v20}
    \begin{split}
        &\mathcal{R} \left( \mathbf{a}, \hat{\mathbf{a}} \right) \\
        \leq \ & \frac{1}{m} \sum_{\substack{i \in [m]: \\\hat{t}_i = t_i}} \mathbb{P} \left\{ \hat{a}_i (\mathbf{M}) \neq a_i \right\} + \frac{\left| \left\{ i \in [m]: \hat{t}_i \neq t_i \right\} \right|}{m} \\
        \stackrel{\textnormal{(h)}}{\leq} \ & \exp \left\{ - 2l \epsilon^2 \cdot d \delta_{\min} (d) \right\} + d^2 \exp \left\{ - \epsilon^2 \cdot \Theta \left( \frac{m}{d} \right) \right\} \\
        &+ d \exp \left[ - \min_{b \in [d]} \left\{ \frac{n \nu(b)}{2} \left( 1 - \frac{l}{n \nu (b)} \right)^2 \right\} \right] \\
        &+ 8n^{-11} + \frac{\alpha}{2},
    \end{split}
\end{equation}
where the step (h) follows from the weak recovery assumption \eqref{eqn:sufficient_condition_alg:with_side_information_v2} of the given side information $\hat{\mathbf{t}}$ and the bound \eqref{eqn:proof_thm:performance_alg:with_side_information_v19}.

\indent Finally, it's time to complete the proof of Theorem \ref{thm:performance_alg:with_side_information}. In order to guarantee the target recovery accuracy \eqref{eqn:expected_accuracy}, one may choose parameters as follows:
\begin{equation}
    \label{eqn:proof_thm:performance_alg:with_side_information_v21}
    \begin{split}
        r &= C_1 \cdot \frac{n^2 \left( \log n \right)^2}{\left( p_m - p_u \right)^2 s_{\min}^2}; \\
        l &= \frac{1}{2 \epsilon^2 \cdot d \delta_{\min} (d)} \log \left( \frac{8}{\alpha} \right); \\
        n &\geq \max \Bigg\{ \frac{8}{\min_{b \in [d]} \nu(b)} \log \left( \frac{8d}{\alpha} \right),\\
        &\qquad\qquad \frac{2l}{\min_{b \in [d]} \nu(b)}, \left( \frac{64}{\alpha} \right)^{\frac{1}{11}} \Bigg\}.
    \end{split}
\end{equation}
With the above choice of parameters in hand, we can conclude that the sample complexity per task of Algorithm \ref{alg:with_side_information} for achieving the recovery performance \eqref{eqn:expected_accuracy} is bounded above by
\begin{equation}
    \label{eqn:proof_thm:performance_alg:with_side_information_v22}
    \begin{split}
        \frac{1}{m} \left\{ nr + ld(m-r) \right\} &\leq \frac{nr}{m} + ld \\
        &= C_1 \cdot \frac{n^3 \left( \log n \right)^2}{m \left( p_m - p_u \right)^2 s_{\min}^2} + ld.
    \end{split}
\end{equation}
The assumption $m \geq \Omega \left( \frac{n^3 \left( \log n \right)^2}{\left( p_m - p_u \right)^2 s_{\min}^2} \right)$ directly implies $C_1 \cdot \frac{n^3 \left( \log n \right)^2}{m \left( p_m - p_u \right)^2 s_{\min}^2} = \calO (1)$. Hence, from the upper bound \eqref{eqn:proof_thm:performance_alg:with_side_information_v22}, it holds that
\begin{equation*}
    \begin{split}
        \frac{1}{m} \left\{ nr + ld(m-r) \right\} \leq 2ld = \frac{1}{\epsilon^2 \cdot \delta_{\min} (d)} \log \left( \frac{8}{\alpha} \right)
    \end{split}
\end{equation*}
for all sufficiently large $d \in \mathbb{N}$, and this completes the proof of Theorem \ref{thm:performance_alg:with_side_information}.

\subsection{Proof of Theorem \ref{thm:performance_alg:without_side_information}}
\label{subsec:proof_thm:performance_alg:without_side_information}

\indent We first analyze the event that an error occurs for the estimation of the unknown labels in Step 2-(d) of Algorithm \ref{alg:without_side_information}, where the weight vectors $\bm{\theta}_{i*}$ are selected as \eqref{eqn:choice_weights_alg:without_side_information}, while being conditioned on the event $\mathcal{E}_1 \cap \mathcal{E}_2$. Recall the decision rule in Step 2-(d) of Algorithm \ref{alg:without_side_information}:
\begin{equation*}
    \begin{split}
        \hat{a}_i (\mathbf{M}) = \ & \textnormal{sign} \left( \sum_{j \in \mathcal{A}'(i)} \theta_{ij} M_{ij} \right) \\
        = \ & \textnormal{sign} \left( a_i \sum_{j \in \mathcal{A}'(i)} \theta_{ij} \left( 2 \Lambda_{ij} - 1 \right) \right),
    \end{split}
\end{equation*}
where $\left\{ \Lambda_{ij} : (i, j) \in \mathcal{A} \right\}$ is a collection of conditionally independent random variables whose probability distributions are given by $\Lambda_{ij} \sim \textnormal{Bern} (F_{ij}) = \textnormal{Bern} \left( \mathcal{Q} \left( t_i, w_j \right) \right)$ for every $(i, j) \in \mathcal{A}$, when a pair of type vectors $\left( \mathbf{t}, \mathbf{w} \right)$ is given.

\indent Being conditioned on the event $\left\{ \hat{t}_i (\mathbf{M}) = t_i \right\} \cap \left( \mathcal{E}_1 \cap \mathcal{E}_2 \right)$, the Hoeffding's inequality gives
\begin{equation}
    \label{eqn:proof_thm:performance_alg:without_side_information_v1}
    \begin{split}
        &\mathbb{P} \left\{ \left. \hat{a}_i (\mathbf{M}) \neq a_i \right| \left\{ \hat{t}_i (\mathbf{M}) = t_i \right\} \cap \left( \mathcal{E}_1 \cap \mathcal{E}_2 \right), \left( \mathbf{t}, \mathbf{w} \right) \right\} \\
        = \ & \mathbb{P} \Bigg\{\left. \sum_{j \in \mathcal{A}'(i)} \theta_{ij} \left( \Lambda_{ij} - F_{ij} \right) \leq - \sum_{j \in \mathcal{A}'(i)} \theta_{ij} \left( F_{ij} - \frac{1}{2} \right) \right| \\
        &\quad \left\{ \hat{t}_i = t_i \right\} \cap \left( \mathcal{E}_1 \cap \mathcal{E}_2 \right), \left( \mathbf{t}, \mathbf{w} \right) \Biggl\} \\
        \leq \ & \exp \left[ - \frac{\left\{ \sum_{j \in \mathcal{A}' (i)} \theta_{ij} \left( 2 F_{ij} - 1 \right) \right\}^2}{2 \sum_{j \in \mathcal{A}' (i)} \theta_{ij}^2} \right].
    \end{split}
\end{equation}
\noindent For this case, one can observe the following facts:
\begin{equation}
    \label{eqn:proof_thm:performance_alg:without_side_information_v2}
    \begin{split}
        &\sum_{j \in \mathcal{A}' (i)} \theta_{ij} \left( 2 F_{ij} - 1 \right) \\
        = \ & \sum_{j \in \mathcal{A}' (i)} \theta_{ij} \left\{ 2 \mathcal{Q} (t_i, w_j) - 1 \right\} \\
        \stackrel{\textnormal{(a)}}{=} \ & \sum_{j \in \mathcal{A}_{\hat{t}_i}(i)} \left\{ 2 \mathcal{Q} \left( t_i, \hat{t}_i (\mathbf{M}) \right) - 1 \right\} \\
        &+ \frac{1}{\sqrt{d-1}} \sum_{w \in [d] \setminus \left\{ \hat{t}_i (\mathbf{M}) \right\}} \left[ \sum_{j \in \mathcal{A}_w (i)} \left\{ 2 \mathcal{Q} (t_i, w_j) - 1 \right\} \right] \\
        = \ & l \left\{ 2 \mathcal{Q} \left( t_i, \hat{t}_i \right) - 1 \right\} \\
        &+ \frac{l}{\sqrt{d-1}} \sum_{w \in [d] \setminus \left\{ \hat{t}_i (\mathbf{M}) \right\}} \left\{ 2 \mathcal{Q} (t_i, w) - 1 \right\} \\
        = \ & \frac{l}{\sqrt{d-1}} \sum_{w=1}^{d} \left\{ 2 \mathcal{Q} (t_i, w) - 1 \right\} \\
        &+ l \left( 1 - \frac{1}{\sqrt{d-1}} \right) \left\{ 2 \mathcal{Q} \left( t_i, \hat{t}_i (\mathbf{M}) \right) - 1 \right\},
    \end{split}
\end{equation}
where the step (a) holds while being conditioned on the event $\left\{ \hat{t}_i = t_i \right\} \cap \left( \mathcal{E}_1 \cap \mathcal{E}_2 \right)$. Also, note that
\begin{equation}
    \label{eqn:proof_thm:performance_alg:without_side_information_v3}
    \begin{split}
         \sum_{j \in \mathcal{A}' (i)} \theta_{ij}^2
         = \ & \sum_{j \in \mathcal{A}_{\hat{t}_i}(i)} \theta_{ij}^2 + \sum_{w \in [d] \setminus \left\{ \hat{t}_i \right\}} \left[ \sum_{j \in \mathcal{A}_w (i)} \theta_{ij}^2 \right] \\
         = \ & 2l.
    \end{split}
\end{equation}
Substituting the computations \eqref{eqn:proof_thm:performance_alg:without_side_information_v2} and \eqref{eqn:proof_thm:performance_alg:without_side_information_v3} into the inequality \eqref{eqn:proof_thm:performance_alg:without_side_information_v1}, we find that
\begin{equation}
    \label{eqn:proof_thm:performance_alg:without_side_information_v4}
    \begin{split}
        &\mathbb{P} \left\{ \left. \hat{a}_i (\mathbf{M}) \neq a_i \right| \left\{ \hat{t}_i (\mathbf{M}) = t_i \right\} \cap \left( \mathcal{E}_1 \cap \mathcal{E}_2 \right), \left( \mathbf{t}, \mathbf{w} \right) \right\} \\
        \leq \ & \exp \left[ - \frac{l}{4} \left( \frac{1}{\sqrt{d-1}} \sum_{w=1}^{d} \left\{ 2 \mathcal{Q} (t_i, w) - 1 \right\} \right. \right. \\
        &\left. \left. + \left( 1 - \frac{1}{\sqrt{d-1}} \right) \left\{ 2 \mathcal{Q} (t_i, t_i) - 1 \right\} \right)^2 \right].
    \end{split}
\end{equation}
On the other hand, while being conditioned on the event $\left\{ \hat{t}_i(\mathbf{M}) \neq t_i \right\} \cap \left( \mathcal{E}_1 \cap \mathcal{E}_2 \right)$, the same argument above results in the bound
\begin{equation}
    \label{eqn:proof_thm:performance_alg:without_side_information_v5}
    \begin{split}
        &\mathbb{P} \left\{ \left. \hat{a}_i (\mathbf{M}) \neq a_i \right| \left\{ \hat{t}_i (\mathbf{M}) \neq t_i \right\} \cap \left( \mathcal{E}_1 \cap \mathcal{E}_2 \right), \left( \mathbf{t}, \mathbf{w} \right) \right\} \\
        \leq \ & \exp \left[ - \frac{l}{4} \left( \frac{1}{\sqrt{d-1}} \sum_{w=1}^{d} \left\{ 2 \mathcal{Q} (t_i, w) - 1 \right\} \right. \right. \\
        &\left. \left. + \left( 1 - \frac{1}{\sqrt{d-1}} \right) \left\{ 2 \min_{w \in [d]} \mathcal{Q} \left( t_i, w \right) - 1 \right\} \right)^2 \right].
    \end{split}
\end{equation}

\indent Next, while being conditioned on the event $\mathcal{E}_1 \cap \mathcal{E}_2$, we analyze the error probability of the task-type matching rule \eqref{eqn:task_type_matching}. To do so, we define an auxiliary random variable $S_{ib} := \sum_{j \in \mathcal{A}_b (i)} \mathbbm{1} \left( M_{ij} = +1 \right)$ for $(i, b) \in [m] \times [d]$. Then,
\begin{equation}
    \label{eqn:proof_thm:performance_alg:without_side_information_v6}
    S_{ib} \sim
    \begin{cases}
        \textnormal{Binomial} \left( l, \mathcal{Q} \left( t_i, b \right) \right) & \textnormal{if } a_i = +1; \\
        \textnormal{Binomial} \left( l, 1 - \mathcal{Q} \left( t_i, b \right) \right) & \textnormal{otherwise,}
    \end{cases}
\end{equation}
since $\left| \mathcal{A}_b (i) \right| = l$ and $\left\{ \mathbbm{1} \left( M_{ij} = +1 \right): j \in \mathcal{A}_b (i) \right\} \stackrel{\textnormal{i.i.d.}}{\sim} \textnormal{Bern} \left( \mathcal{Q} \left( t_i, b \right) \right)$ if $a_i = +1$; $\left\{ \mathbbm{1} \left( M_{ij} = +1 \right): j \in \mathcal{A}_b (i) \right\} \stackrel{\textnormal{i.i.d.}}{\sim} \textnormal{Bern} \left( 1 - \mathcal{Q} \left( t_i, b \right) \right)$ otherwise. Also from
\begin{equation*}
    S_{ib} = \sum_{j \in \mathcal{A}_b (i)} \frac{1 + M_{ij}}{2} = \frac{l}{2} + \frac{1}{2} \sum_{j \in \mathcal{A}_b (i)} M_{ij},
\end{equation*}
we have $\sum_{j \in \mathcal{A}_b (i)} M_{ij} = 2 \left( S_{ib} - \frac{l}{2} \right)$. So, it follows that $\hat{t}_i (\mathbf{M}) = t_i$ if $\left| S_{i t_i} - \frac{l}{2} \right| > \left| S_{ib} - \frac{l}{2} \right|$ for every $b \in [d] \setminus \left\{ t_i \right\}$. The following lemma, whose proof is deferred to Appendix \ref{subsec:proof_lemma:proof_thm:performance_alg:without_side_information_v1}, provides a sufficient condition for the accurate estimation of the true task-types of the task-type matching rule \eqref{eqn:task_type_matching}:

\begin{lemma}
\label{lemma:proof_thm:performance_alg:without_side_information_v1}
Conditioned on the event $\mathcal{E}_1 \cap \mathcal{E}_2$, it holds that
\begin{equation}
    \label{eqn:lemma:proof_thm:performance_alg:without_side_information_v1_v1}
    \bigcap_{b=1}^{d} \left\{ \left| S_{ib} - \mathbb{E} \left[ S_{ib} \right] \right| < \frac{p^* \left( t_i \right) - q^* \left( t_i \right)}{2} l \right\} \subseteq \left\{ \hat{t}_i (\mathbf{M}) = t_i \right\}.
\end{equation}
\end{lemma}

\indent With the above supporting lemma in hand, it follows that
\begin{equation}
    \label{eqn:proof_thm:performance_alg:without_side_information_v7}
    \begin{split}
        &\mathbb{P} \left\{ \left. \hat{t}_i (\mathbf{M}) \neq t_i \right| \mathcal{E}_1 \cap \mathcal{E}_2, \left( \mathbf{t}, \mathbf{w} \right) \right\} \\
        \leq \ & \mathbb{P} \Bigg\{ \left. \bigcup_{b=1}^{d} \left\{ \left| S_{ib} - \mathbb{E} \left[ S_{ib} \right] \right| \geq \frac{p^* \left( t_i \right) - q^* \left( t_i \right)}{2} l \right\}  \right|\\
        &\quad\mathcal{E}_1 \cap \mathcal{E}_2, \left( \mathbf{t}, \mathbf{w} \right) \Bigg\} \\
        \stackrel{\textnormal{(b)}}{\leq} \ & \sum_{b=1}^{d} \mathbb{P} \Bigg\{ \left. \left| S_{ib} - \mathbb{E} \left[ S_{ib} \right] \right| \geq \frac{p^* \left( t_i \right) - q^* \left( t_i \right)}{2} l \right| \\
        &\qquad\qquad \mathcal{E}_1 \cap \mathcal{E}_2, \left( \mathbf{t}, \mathbf{w} \right) \Bigg\} \\
        \stackrel{\textnormal{(c)}}{\leq} \ & 2d \exp \left\{ - \frac{l}{2} \left( p^* \left( t_i \right) - q^* \left( t_i \right) \right)^2 \right\},
    \end{split}
\end{equation}
where the step (b) holds due to the union bound, and the step (c) follows from the Chernoff-Hoeffding's theorem. Combining the previous three inequalities \eqref{eqn:proof_thm:performance_alg:without_side_information_v4}, \eqref{eqn:proof_thm:performance_alg:without_side_information_v5}, and \eqref{eqn:proof_thm:performance_alg:without_side_information_v7}, we find that
\begin{equation}
    \label{eqn:proof_thm:performance_alg:without_side_information_v8}
    \begin{split}
        &\mathbb{P} \left\{ \left. \hat{a}_i (\mathbf{M}) \neq a_i \right| \mathcal{E}_1 \cap \mathcal{E}_2, \left( \mathbf{t}, \mathbf{w} \right) \right\} \\
        = \ & \mathbb{P} \left\{ \left. \hat{a}_i (\mathbf{M}) \neq a_i \right| \left\{ \hat{t}_i (\mathbf{M}) = t_i \right\} \cap \left( \mathcal{E}_1 \cap \mathcal{E}_2 \right), \left( \mathbf{t}, \mathbf{w} \right) \right\} \\
        &\times \mathbb{P} \left\{ \left. \hat{t}_i (\mathbf{M}) = t_i \right| \mathcal{E}_1 \cap \mathcal{E}_2, \left( \mathbf{t}, \mathbf{w} \right) \right\} \\
        &+ \mathbb{P} \left\{ \left. \hat{a}_i (\mathbf{M}) \neq a_i \right| \left\{ \hat{t}_i (\mathbf{M}) \neq t_i \right\} \cap \left( \mathcal{E}_1 \cap \mathcal{E}_2 \right), \left( \mathbf{t}, \mathbf{w} \right) \right\} \\
        &\times \mathbb{P} \left\{ \left. \hat{t}_i (\mathbf{M}) \neq t_i \right| \mathcal{E}_1 \cap \mathcal{E}_2, \left( \mathbf{t}, \mathbf{w} \right) \right\} \\
        \leq \ & \exp \Biggl[- \frac{l}{4} \Biggl( \frac{1}{\sqrt{d-1}} \sum_{w=1}^{d} \left\{ 2 \mathcal{Q} (t_i, w) - 1 \right\} \\
        &\quad + \left( 1 - \frac{1}{\sqrt{d-1}} \right) \left\{ 2 \mathcal{Q} (t_i, t_i) - 1 \right\} \Biggr)^2 \Biggr] \\
        &+ 2d \exp \Biggl[ - \frac{l}{2} \biggl\{ \left( p^* \left( t_i \right) - q^* \left( t_i \right) \right)^2  \\
        &\quad+ \ \frac{1}{2}\biggl(  \frac{1}{\sqrt{d-1}} \sum_{w=1}^{d} \left\{ 2 \mathcal{Q} (t_i, w) - 1 \right\}  \\
        & \quad + \left( 1 - \frac{1}{\sqrt{d-1}} \right) \left\{ 2 \min_{w \in [d]} \mathcal{Q} \left( t_i, w \right) - 1 \right\} \biggr)^2  \biggr\} \Biggr].
    \end{split}
\end{equation}
Using the shorthand
\begin{equation*}
    \begin{split}
        \theta_{1}' \left( t; \mathcal{Q}\right) := \ & \frac{1}{2} \left( \frac{1}{\sqrt{d-1}} \sum_{w=1}^{d} \left\{ 2 \mathcal{Q} (t, w) - 1 \right\} \right. \\
        &+ \left. \left( 1 - \frac{1}{\sqrt{d-1}} \right) \left\{ 2 \mathcal{Q} (t, t) - 1 \right\} \right)^2
    \end{split}
\end{equation*}
it leads to the following simpler bound:
\begin{equation}
    \label{eqn:proof_thm:performance_alg:without_side_information_v9}
    \begin{split}
        &\mathbb{P} \left\{ \left. \hat{a}_i (\mathbf{M}) \neq a_i \right| \mathcal{E}_1 \cap \mathcal{E}_2, \left( \mathbf{t}, \mathbf{w} \right) \right\} \\
        \leq \ & \exp \left\{ - \frac{l}{2} \theta_{1}' \left( t_i; \mathcal{Q} \right) \right\} \\
        &+ 2d \exp \left[ - \frac{l}{2} \left\{ \left( p^* \left( t_i \right) - q^* \left( t_i \right) \right)^2  + \theta_{1} \left( t_i; \mathcal{Q} \right) \right\} \right].
    \end{split}
\end{equation}
where $\theta_{1} \left( \cdot; \mathcal{Q} \right) : [d] \rightarrow \mathbb{R}$ is given by 
\begin{equation*}
    \begin{split}
        \theta_{1} \left( t; \mathcal{Q} \right) := \ &
        \frac{1}{2} \left[ \frac{1}{\sqrt{d-1}} \sum_{w=1}^{d} \left\{ 2 \mathcal{Q} (t, w) - 1 \right\} \right. \\
        &\left. + \left( 1 - \frac{1}{\sqrt{d-1}} \right) \left\{ 2 \min_{w \in [d] } \mathcal{Q}(t, w) - 1 \right\} \right]^2.
    \end{split}
\end{equation*}
Taking expectation with respect to $\left( \mathbf{t}, \mathbf{w} \right) \sim \bm{\mu}^{\otimes m} \otimes \bm{\nu}^{\otimes n}$ on both sides of \eqref{eqn:proof_thm:performance_alg:without_side_information_v9}, we find that
\begin{equation}
    \label{eqn:proof_thm:performance_alg:without_side_information_v10}
    \begin{split}
        &\mathbb{P} \left\{ \left. \hat{a}_i (\mathbf{M}) \neq a_i \right| \mathcal{E}_1 \cap \mathcal{E}_2 \right\} \\
        = \ & \mathbb{E}_{\left( \mathbf{t}, \mathbf{w} \right) \sim \bm{\mu}^{\otimes m} \otimes \bm{\nu}^{\otimes n}} \left[ \mathbb{P} \left\{ \left. \hat{a}_i (\mathbf{M}) \neq a_i \right| \mathcal{E}_1 \cap \mathcal{E}_2, \left( \mathbf{t}, \mathbf{w} \right) \right\} \right] \\
        \leq \ & \mathbb{E}_{\left( \mathbf{t}, \mathbf{w} \right) \sim \bm{\mu}^{\otimes m} \otimes \bm{\nu}^{\otimes n}} \left[ \exp \left\{ - \frac{l}{2} \theta_{1}' \left( t_i ; \mathcal{Q} \right) \right\} \right. \\
        &\left. + 2d \exp \left[ - \frac{l}{2} \left\{ \left( p^* \left( t_i \right) - q^* \left( t_i \right) \right)^2  + \theta_1 \left( t_i ; \mathcal{Q} \right) \right\} \right] \right] \\
        = \ & \sum_{t=1}^{d} \mu(t) \left( \exp \left\{ - \frac{l}{2} \theta_{1}' \left( t ; \mathcal{Q} \right) \right\} \right. \\
        &\left. + 2d \exp \left[ - \frac{l}{2} \left\{ \left( p^* (t) - q^* (t) \right)^2  + \theta_1 \left( t; \mathcal{Q} \right) \right\} \right] \right) \\
        \stackrel{\textnormal{(d)}}{\leq} 
        \ & \sum_{t=1}^{d} \mu(t)  (2d+1)\\
        &\quad\times \exp \left[ - \frac{l}{2} \left\{ \left( p^* (t) - q^* (t) \right)^2 + \theta_1 \left( t; \mathcal{Q} \right) \right\} \right] \\
        \leq \ & (2d+1) \exp \left[ - \frac{l}{2} \min_{t \in [d]} \left\{ \left( p^* (t) - q^* (t) \right)^2 + \theta_1 \left( t; \mathcal{Q} \right) \right\} \right],
    \end{split}
\end{equation}
where the step (d) holds due to the following simple fact
\begin{equation}
    \label{eqn:proof_thm:performance_alg:without_side_information_v11}
    \theta_{1}' \left( t; \mathcal{Q} \right) \geq \left( p^* (t) - q^* (t) \right)^2 + \theta_1 \left( t; \mathcal{Q} \right),\ \forall t \in [d],
\end{equation}
for all $d \geq 3$, which can be justified by doing some straightforward algebra.

\indent On the other hand, by taking two inequalities \eqref{eqn:proof_thm:performance_alg:without_side_information_v4} and \eqref{eqn:proof_thm:performance_alg:without_side_information_v5} collectively, we arrive at
\begin{equation}
    \label{eqn:proof_thm:performance_alg:without_side_information_v12}
    \begin{split}
        &\mathbb{P} \left\{ \left. \hat{a}_i (\mathbf{M}) \neq a_i \right| \mathcal{E}_1 \cap \mathcal{E}_2, \left( \mathbf{t}, \mathbf{w} \right) \right\} \\
        = \ & \mathbb{P} \left\{ \left. \hat{a}_i (\mathbf{M}) \neq a_i \right| \left\{ \hat{t}_i (\mathbf{M}) = t_i \right\} \cap \left( \mathcal{E}_1 \cap \mathcal{E}_2 \right), \left( \mathbf{t}, \mathbf{w} \right) \right\} \\
        &\mathbb{P} \left\{ \left. \hat{t}_i = t_i \right| \mathcal{E}_1 \cap \mathcal{E}_2, \left( \mathbf{t}, \mathbf{w} \right) \right\} \\
        &+ \mathbb{P} \left\{ \left. \hat{a}_i (\mathbf{M}) \neq a_i \right| \left\{ \hat{t}_i (\mathbf{M}) \neq t_i \right\} \cap \left( \mathcal{E}_1 \cap \mathcal{E}_2 \right), \left( \mathbf{t}, \mathbf{w} \right) \right\} \\
        &\mathbb{P} \left\{ \left. \hat{t}_i (\mathbf{M}) \neq t_i \right| \mathcal{E}_1 \cap \mathcal{E}_2, \left( \mathbf{t}, \mathbf{w} \right) \right\} \\
        \leq \ & \exp \left\{ - \frac{l}{2} \theta_{1}' \left( t_i ; \mathcal{Q} \right) \right\} \mathbb{P} \left\{ \left. \hat{t}_i (\mathbf{M}) = t_i \right| \mathcal{E}_1 \cap \mathcal{E}_2, \left( \mathbf{t}, \mathbf{w} \right) \right\} \\
        &+ \exp \left\{ - \frac{l}{2} \theta_{1} \left( t_i ; \mathcal{Q} \right) \right\} \mathbb{P} \left\{ \left. \hat{t}_i (\mathbf{M}) \neq t_i \right| \mathcal{E}_1 \cap \mathcal{E}_2, \left( \mathbf{t}, \mathbf{w} \right) \right\} \\
        \stackrel{\textnormal{(e)}}{\leq} \ & \exp \left\{ - \frac{l}{2} \theta_{1} \left( t_i ; \mathcal{Q} \right) \right\},
    \end{split}
\end{equation}
where the step (e) holds due to the fact \eqref{eqn:proof_thm:performance_alg:without_side_information_v11}. By taking expectation with respect to $\left( \mathbf{t}, \mathbf{w} \right) \sim \bm{\mu}^{\otimes m} \otimes \bm{\nu}^{\otimes n}$ to the bound \eqref{eqn:proof_thm:performance_alg:without_side_information_v9} yields
\begin{equation}
    \label{eqn:proof_thm:performance_alg:without_side_information_v13}
    \begin{split}
        &\mathbb{P} \left\{ \left. \hat{a}_i (\mathbf{M}) \neq a_i \right| \mathcal{E}_1 \cap \mathcal{E}_2 \right\} \\
        = \ & \mathbb{E}_{\left( \mathbf{t}, \mathbf{w} \right) \sim \bm{\mu}^{\otimes m} \otimes \bm{\nu}^{\otimes n}} \left[ \mathbb{P} \left\{ \left. \hat{a}_i (\mathbf{M}) \neq a_i \right| \mathcal{E}_1 \cap \mathcal{E}_2, \left( \mathbf{t}, \mathbf{w} \right) \right\} \right] \\
        \leq \ & \mathbb{E}_{\left( \mathbf{t}, \mathbf{w} \right) \sim \bm{\mu}^{\otimes m} \otimes \bm{\nu}^{\otimes n}} \left[ \exp \left\{ - \frac{l}{2} \theta_{1} \left( t_i ; \mathcal{Q}  \right) \right\} \right] \\
        = \ & \sum_{t=1}^{d} \nu (t) \exp \left\{ - \frac{l}{2} \theta_{1} \left( t; \mathcal{Q} \right) \right\} \\
        \leq \ & \exp \left\{ - \frac{l}{2} \min_{t \in [d]} \theta_{1} \left( t; \mathcal{Q} \right) \right\}.
    \end{split}
\end{equation}
So by combining the bounds \eqref{eqn:proof_thm:performance_alg:without_side_information_v10} and \eqref{eqn:proof_thm:performance_alg:without_side_information_v13} together, we obtain
\begin{equation}
    \label{eqn:proof_thm:performance_alg:without_side_information_v14}
    \begin{split}
        &\mathbb{P} \left\{ \left. \hat{a}_i (\mathbf{M}) \neq a_i \right| \mathcal{E}_1 \cap \mathcal{E}_2 \right\} \leq \\
        &
        \min \left\{ \exp \left\{ - \frac{l}{2} \min_{t \in [d]} \theta_1 \left( t; \mathcal{Q} \right) \right\},  \right. \\
        &\quad\left. (2d+1) \exp \left[ - \frac{l}{2} \min_{t \in [d]} \left\{ \left( p^* (t) - q^* (t) \right)^2 + \theta_1 \left( t; \mathcal{Q} \right) \right\} \right] \right\}.
    \end{split}
\end{equation}

\noindent We already know from the proof of Theorem \ref{thm:performance_alg:with_side_information} (Section \ref{subsec:proof_thm:performance_alg:with_side_information}) that
\begin{equation}
    \label{eqn:proof_thm:performance_alg:without_side_information_v15}
    \begin{split}
        \mathbb{P} \left\{ \mathcal{E}_{1}^c \right\} &\leq 4 n^{-11}; \\
        \mathbb{P} \left\{ \mathcal{E}_{2}^c \right\} &\leq d \exp \left[ - \min_{b \in [d]} \left\{ \frac{n \nu(b)}{2} \left( 1 - \frac{l}{n \nu (b)} \right)^2 \right\} \right] \\
        &\quad+ 4n^{-11}.
    \end{split}
\end{equation}
By combining two pieces \eqref{eqn:proof_thm:performance_alg:without_side_information_v14} and \eqref{eqn:proof_thm:performance_alg:without_side_information_v15} together, we now have
\begin{equation}
    \label{eqn:proof_thm:performance_alg:without_side_information_v16}
    \begin{split}
        &\mathbb{P} \left\{ \hat{a}_i (\mathbf{M}) \neq a_i \right\} \\
        &\leq  \underbrace{\mathbb{P} \left\{ \left. \hat{a}_i (\mathbf{M}) \neq a_i \right| \mathcal{E}_{1}^c \right\}}_{\leq \ 1} \mathbb{P} \left\{ \mathcal{E}_{1}^c \right\} + \underbrace{\mathbb{P} \left\{ \left. \hat{a}_i (\mathbf{M}) \neq a_i \right| \mathcal{E}_{2}^c \right\}}_{\leq \ 1} \mathbb{P} \left\{ \mathcal{E}_{2}^c \right\} \\
        &+ \mathbb{P} \left\{ \left. \hat{a}_i (\mathbf{M}) \neq a_i \right| \mathcal{E}_{1} \cap \mathcal{E}_2 \right\} \underbrace{\mathbb{P} \left\{ \mathcal{E}_{1} \cap \mathcal{E}_2 \right\}}_{\leq \ 1} \\
        &\leq 8 n^{-11} + d \exp \left[ - \min_{b \in [d]} \left\{ \frac{n \nu(b)}{2} \left( 1 - \frac{l}{n \nu (b)} \right)^2 \right\} \right]+ \\
 &     \min \left\{ \exp \left\{ - \frac{l}{2} \min_{t \in [d]} \theta_1 \left( t; \mathcal{Q} \right) \right\},  \right. \\
        &\quad\left. (2d+1) \exp \left[ - \frac{l}{2} \min_{t \in [d]} \left\{ \left( p^* (t) - q^* (t) \right)^2 + \theta_1 \left( t; \mathcal{Q} \right) \right\} \right] \right\}.
    \end{split}
\end{equation}

\indent We are now ready to finish the proof of Theorem \ref{thm:performance_alg:without_side_information}. In order to achieve the desired recovery accuracy \eqref{eqn:expected_accuracy}, one may choose
\begin{equation}
    \label{eqn:proof_thm:performance_alg:without_side_information_v17}
    \begin{split}
        r = \ & C_1 \cdot \frac{n^2 \left( \log n \right)^2}{\left( p_m - p_u \right)^2 s_{\min}^2}; \\
        l = \ & \min \left\{ \frac{2 \log \left( \frac{6d+3}{\alpha} \right)}{ \min_{t \in [d]} \left\{ \left( p^{*} (t) - q^{*} (t) \right)^2 + \theta_1 \left( t; \mathcal{Q} \right) \right\}},  \right. \\
        &\left.\qquad\quad \frac{2 \log \left( \frac{3}{\alpha} \right)}{\min_{t \in [d]} \theta_1 \left( t; \mathcal{Q} \right)} \right \}; \\
        n \geq \ & \max \left\{ \frac{8 \log \left( \frac{8d}{\alpha} \right)}{\min_{b \in [d]} \nu(b)}, \frac{2l}{\min_{b \in [d]} \nu(b)}, \left( \frac{64}{\alpha} \right)^{\frac{1}{11}} \right\}.
    \end{split}
\end{equation}
With the above choice of parameters in hand, it follows that the sample complexity per task that Algorithm \ref{alg:without_side_information} calls for the recovery accuracy \eqref{eqn:expected_accuracy} can be bounded above by
\begin{equation}
    \label{eqn:proof_thm:performance_alg:without_side_information_v18}
    \begin{split}
        &\frac{1}{m} \left\{ nr + ld (m-r) \right\} \\
        \leq \ & \frac{nr}{m} + ld \\
        = \ & C_1 \cdot \frac{n^3 \left( \log n \right)^2}{m \left( p_m - p_u \right)^2 s_{\min}^2} + ld
    \end{split}
\end{equation}
Akin to the proof of Theorem \ref{thm:performance_alg:with_side_information}, from the assumption $m = \Omega \left( \frac{n^3 \left( \log n \right)^2}{\left( p_m - p_u \right)^2 s_{\min}^2} \right)$, it follows that $C_1 \cdot \frac{n^3 \left( \log n \right)^2}{m \left( p_m - p_u \right)^2 s_{\min}^2} = \calO (1)$. Hence,
\begin{equation}
    \label{eqn:proof_thm:performance_alg:without_side_information_v19}
    \begin{split}
        &\frac{1}{m} \left\{ nr + ld (m-r) \right\} \\
        \leq \ & 2ld \\
        = \ & \min \left\{ \frac{4d \log \left( \frac{6d+3}{\alpha} \right)}{ \min_{t \in [d]} \left\{ \left( p^{*} (t) - q^{*} (t) \right)^2 + \theta_1 \left( t; \mathcal{Q} \right) \right\}},  \right. \\
        &\left.\qquad\quad \frac{4d \log \left( \frac{3}{\alpha} \right)}{\min_{t \in [d]} \theta_1 \left( t; \mathcal{Q} \right)} \right \}
    \end{split}
\end{equation}
for all sufficiently large $d$. This finishes the proof of Theorem \ref{thm:performance_alg:without_side_information}.


\appendices

\section{Proof of Theorem \ref{thm:achievability_ml}}
\label{sec:proof_thm:achievability_ml}

\indent Let $\left\{ \Lambda_{ij} : (i, j) \in \mathcal{A} \right\}$ be a collection of random variables such that $\Lambda_{ij} \sim \textnormal{Bern} \left( F_{ij} \right)$, $(i, j) \in \mathcal{A}$, are conditionally independent given a pair of type vectors $\left( \mathbf{t}, \mathbf{w} \right)$. Then, the following bound holds: for any $\lambda \geq 0$,
\begin{equation}
    \label{eqn:proof_thm:achievability_ml_v1}
    \begin{split}
        &\mathbb{P} \left\{ \left.  \hat{a}_{i}^{\textnormal{ML}} (\mathbf{M}) \neq a_{i} \right| \mathbf{t}, \mathbf{w} \right\} \\
        = \ & \mathbb{P} \left\{ \left. \sum_{j \in \mathcal{A}(i)} \log \left( \frac{F_{ij}}{1 - F_{ij}} \right) \left( 2 \Lambda_{ij} - 1 \right) \leq 0 \right| \mathbf{t}, \mathbf{w} \right\} \\
        \overset{\textnormal{(a)}}{\leq} \ & \mathbb{E} \left[ \left. \exp \left( \lambda \left( \sum_{j \in \mathcal{A}(i)} \log \left( \frac{1 - F_{ij}}{F_{ij}} \right) \left( 2 \Lambda_{ij} - 1 \right) \right) \right) \right| \mathbf{t}, \mathbf{w} \right] \\
        \overset{\textnormal{(b)}}{=} \ & \prod_{j \in \mathcal{A}(i)} \mathbb{E} \left[ \left. \exp \left( \lambda \log \left( \frac{1 - F_{ij}}{F_{ij}} \right) \left( 2 \Lambda_{ij} - 1 \right) \right) \right| \mathbf{t}, \mathbf{w} \right] \\
        = \ & \prod_{j \in \mathcal{A}(i)} \left[ \left( 1 -  F_{ij} \right)^{\lambda} F_{ij}^{1 - \lambda} + \left( 1 - F_{ij} \right)^{1 - \lambda} F_{ij}^{\lambda}\right],
    \end{split}
\end{equation}
where the step (a) follows from the Markov inequality, and the step (b) holds due to the conditional independence of $\left\{ \Lambda_{ij} : (i, j) \in \mathcal{A} \right\}$ given a pair of type vectors $\left( \mathbf{t}, \mathbf{w} \right) \in [d]^m \times [d]^n$. Given any $\theta \in [0, 1]$, we define the function $\varphi_{\theta}(\lambda) := \theta^{1 - \lambda} (1 - \theta)^{\lambda} + \theta^{\lambda} (1 - \theta)^{1 - \lambda}$ for $\lambda \in [0, 1]$. It can be easily shown that $\frac{1}{2} \in \argmin_{\lambda \in [0, 1]} \varphi_{\theta}(\lambda)$ for every $\theta \in [0, 1]$. Putting $\lambda = \frac{1}{2}$ into the inequality \eqref{eqn:proof_thm:achievability_ml_v1} yields
\begin{equation}
    \label{eqn:proof_thm:achievability_ml_v2}
    \begin{split}
        \mathbb{P} \left\{ \left.  \hat{a}_{i}^{\textnormal{ML}}(\mathbf{M}) \neq a_{i} \right| \mathbf{t}, \mathbf{w} \right\} &\leq \prod_{j \in \mathcal{A}(i)} \left\{ 2 \sqrt{F_{ij} \left( 1 - F_{ij} \right)} \right\}.
    \end{split}
\end{equation}
Taking expectations to both sides of the inequality \eqref{eqn:proof_thm:achievability_ml_v2} with respect to $\mathbf{w} \sim \bm{\nu}^{\otimes n}$ yields
\begin{equation}
    \label{eqn:proof_thm:achievability_ml_v3}
    \begin{split}
        &\mathbb{P} \left\{ \left.  \hat{a}_{i}^{\textnormal{ML}}(\mathbf{M}) \neq a_{i} \right| \mathbf{t} \right\} \\
        = \ & \mathbb{E}_{\mathbf{w} \sim \bm{\nu}^{\otimes n}} \left[ \mathbb{P} \left\{ \left.  \hat{a}_{i}^{\textnormal{ML}}(\mathbf{M}) \neq a_{i} \right| \mathbf{t}, \mathbf{w} \right\} \right] \\
        \leq \ & \mathbb{E}_{\mathbf{w} \sim \bm{\nu}^{\otimes n}} \left[ \prod_{j \in \mathcal{A}(i)} \left\{ 2 \sqrt{F_{ij} \left( 1 - F_{ij} \right)} \right\} \right] \\
        \overset{\textnormal{(c)}}{=} \ & \prod_{j \in \mathcal{A}(i)} \mathbb{E}_{w_j \sim \bm{\nu}} \left[ 2 \sqrt{F_{ij} \left( 1 - F_{ij} \right)} \right] \\
        &\overset{\textnormal{(d)}}{=} \prod_{j \in \mathcal{A}(i)} \left\{ \sum_{w=1}^{d} 2 \nu (w) \sqrt{\mathcal{Q} \left( t_i, w \right) \left\{ 1 - \mathcal{Q} \left( t_i, w \right) \right\}} \right\} \\
        = \ & \left\{ \sum_{w=1}^{d} 2 \nu(w)  \sqrt{\mathcal{Q} \left( t_i, w \right) \left\{ 1 - \mathcal{Q} \left( t_i, w \right) \right\}} \right\}^{\left| \mathcal{A}(i) \right|}
    \end{split}
\end{equation}
where the step (c) holds from the fact that given a type $t_i \in [d]$ of the $i$-th task, $F_{ij}$ is completely determined based on the type $w_j \in [d]$ of the $j$-th worker for $j \in \mathcal{A}(i)$ and $\left\{ w_j : j \in \mathcal{A}(i) \right\}$ are mutually independent, and the step (d) follows from the fact that given a type $t_i \in [d]$ associated with the $i$-th task,
\begin{equation*}
    F_{ij} =
    \mathcal{Q} \left( t_i, w \right) \textnormal{ with probability } \nu (w)
\end{equation*}
for each $w \in [d]$. Finally, taking expectations to the bound \eqref{eqn:proof_thm:achievability_ml_v3} with respect to $\mathbf{t} \sim \bm{\mu}^{\otimes m}$ gives
\begin{equation}
    \label{eqn:proof_thm:achievability_ml_v4}
    \begin{split}
        &\mathbb{P} \left\{ \hat{a}_{i}^{\textnormal{ML}}(\mathbf{M}) \neq a_{i} \right\} \\
        = \ & \mathbb{E}_{\mathbf{t} \sim \bm{\mu}^{\otimes m}} \left[ \mathbb{P} \left\{ \left. \hat{a}_{i}^{\textnormal{ML}}(\mathbf{M}) \neq a_{i} \right| \mathbf{t} \right\} \right] \\
        \leq \ & \mathbb{E}_{t_i \sim \bm{\mu}} \left[ \left\{ \sum_{w=1}^{d} 2 \nu (w) \sqrt{\mathcal{Q} \left( t_i, w \right) \left\{ 1 - \mathcal{Q} \left( t_i, w \right) \right\}} \right\}^{\left| \mathcal{A}(i) \right|} \right] \\
        = \ & \sum_{t=1}^{d} \mu (t) \left\{  \sum_{w=1}^{d} 2 \nu (w) \sqrt{\mathcal{Q} \left( t, w \right) \left\{ 1 - \mathcal{Q} \left( t, w \right) \right\}} \right\}^{\left| \mathcal{A}(i) \right|} \\
        = \ & \exp \left\{ - \left| \mathcal{A} (i) \right| \cdot \gamma_1 \left( d; \mathcal{Q}, \bm{\mu}, \bm{\nu} \right) \right\}
    \end{split}
\end{equation}
for every $i \in [m]$. So in order to achieve the desired bound on the risk function $\mathcal{R} \left( \mathbf{a}, \hat{\mathbf{a}}_{\textnormal{ML}} \right) \leq \alpha$, where $\alpha \in \left( 0, \frac{1}{2} \right]$, it suffices to assign $\left| \mathcal{A}(i) \right|$ workers to the $i$-th task, where
\begin{equation}
    \label{eqn:proof_thm:achievability_ml_v5}
    \left| \mathcal{A}(i) \right| \geq \frac{1}{\gamma_1 \left( d; \mathcal{Q}, \bm{\mu}, \bm{\nu} \right)} \log \left( \frac{1}{\alpha} \right)
\end{equation}
for all $i \in [m]$, and this completes the proof of Theorem \ref{thm:achievability_ml}.

\section{Proof of Theorem \ref{thm:statistical_impossibility}}
\label{sec:proof_thm:statistical_impossibility}

\indent We begin the proof with the following basic inequality:
\begin{equation}
    \label{eqn:proof_thm:statistical_impossibility_v1}
    \begin{split}
       &\inf_{\hat{\mathbf{a}}} \left( \sup_{\mathbf{a} \in \left\{ \pm 1 \right\}^m} \mathcal{R} \left( \mathbf{a}, \hat{\mathbf{a}} \right) \right) \\
       \overset{\textnormal{(a)}}{\geq} \ & \inf_{\hat{\mathbf{a}}} \left(  \mathbb{E}_{\mathbf{a} \sim \textnormal{Unif} \left( \left\{ \pm 1 \right\}^m \right)} \left[ \mathcal{R} \left( \mathbf{a}, \hat{\mathbf{a}} \right) \right] \right) \\
       = \ & \frac{1}{m} \inf_{\hat{\mathbf{a}}} \left( \sum_{i=1}^{m} \mathbb{E}_{a_i \sim \textnormal{Unif} \left( \left\{ \pm 1 \right\} \right)} \left[ \mathbb{P} \left\{ \hat{a}_i (\bfM) \neq a_i \right\} \right] \right) \\
       = \ & \frac{1}{m} \sum_{i=1}^{m} \inf_{\hat{a}_i} \left( \mathbb{E}_{a_i \sim \textnormal{Unif} \left( \left\{ \pm 1 \right\} \right)} \left[ \mathbb{P} \left\{ \hat{a}_i (\mathbf{M}) \neq a_i \right\} \right] \right) \\
       \overset{\textnormal{(b)}}{=} \ & \frac{1}{m} \sum_{i=1}^{m} \mathbb{E}_{a_i \sim \textnormal{Unif} \left( \left\{ \pm 1 \right\} \right)} \left[ \mathbb{P} \left\{  \hat{a}_{i}^{\textnormal{ML}} (\mathbf{M}) \neq a_i \right\} \right],
    \end{split}
\end{equation}
where the step (a) uses a simple ``max$\geq$mean'' argument, and the step (b) follows from the well-known fact that the ML estimator is \emph{optimal} under the uniform prior together with an observation that the ML estimator of the ground-truth label $a_i$ associated with the $i$-th task equals to the $i$-th coordinate of the ML estimator of the vector of the ground-truth labels $\mathbf{a}$. This fact relies on the computation \eqref{eqn:log_likelihood_v1} of the log-likelihood function of observing the responses $\mathbf{M} = \left( M_{ij} : (i, j) \in \mathcal{A} \right)$, which gives
\begin{equation*}
    \begin{split}
        &\log \mathbb{P}_{\mathbf{a}} \left\{ \mathbf{M} \right\}\\
         &=  \sum_{k=1}^{m} a_k \left[ \sum_{j \in \mathcal{A}(k)} M_{kj} \log \left( \frac{F_{kj}}{1 - F_{kj}} \right) \right] \\
        &=  a_i \left[ \sum_{j \in \mathcal{A}(i)} M_{ij} \log \left( \frac{F_{ij}}{1 - F_{ij}} \right) \right] \\
        &\quad+ \sum_{k \in [m] \setminus \{ i \}} a_k \left[ \sum_{j \in \mathcal{A}(k)} M_{kj} \log \left( \frac{F_{kj}}{1 - F_{kj}} \right) \right].
    \end{split}
\end{equation*}

\indent Next, we analyze the error probability $\mathbb{P} \left\{ \hat{a}_{i}^{\textnormal{ML}}(\mathbf{M}) \neq a_i \right\}$ of the ML estimator. While being conditioned with a pair of type vectors $\left( \mathbf{t}, \mathbf{w} \right)$, we obtain from the definition of the ML estimator \eqref{eqn:ml_estimator} that
\begin{equation}
    \label{eqn:proof_thm:statistical_impossibility_v2}
    \begin{split}
        &\mathbb{P} \left\{ \left. \hat{a}_{i}^{\textnormal{ML}} (\mathbf{M}) \neq a_{i} \right| \mathbf{t}, \mathbf{w} \right\} \\
        = \ & \mathbb{P} \left\{ \left. \sum_{j \in \mathcal{A}(i)} \log \left( \frac{1 - F_{ij}}{F_{ij}} \right) \left( 2 \Lambda_{ij} - 1 \right) \geq 0 \right| \mathbf{t}, \mathbf{w} \right\},
    \end{split}
\end{equation}
where $\left\{ \Lambda_{ij} : (i, j) \in \mathcal{A} \right\}$ is a collection of conditionally independent random variables given a pair of type vectors $\left( \mathbf{t}, \mathbf{w} \right)$, with $\Lambda_{ij} \sim \textnormal{Bern} \left( F_{ij} \right)$ for each $(i, j) \in \mathcal{A}$. It would be good to keep in mind that \eqref{eqn:proof_thm:achievability_ml_v1} and \eqref{eqn:proof_thm:achievability_ml_v2} establish an upper bound on the right-hand side of \eqref{eqn:proof_thm:statistical_impossibility_v2}. Now, we provide its lower bound by employing a well-known technical argument adopted in the proof of \emph{Cram\'{e}r-Chernoff Theorem} \cite{van2000asymptotic}. Let
\begin{equation*}
    \lambda_i := \frac{1}{2} \sum_{j \in \mathcal{A}(i)} \log \left( \frac{1 - F_{ij}}{F_{ij}} \right).
\end{equation*}
Then, the right-hand side of \eqref{eqn:proof_thm:statistical_impossibility_v2} becomes
\begin{equation}
    \label{eqn:proof_thm:statistical_impossibility_v3}
    \mathbb{P} \left\{ \left. \sum_{j \in \mathcal{A}(i)} \log \left( \frac{1 - F_{ij}}{F_{ij}} \right) \Lambda_{ij} \geq \lambda_i \right| \mathbf{t}, \mathbf{w} \right\}.
\end{equation}
Let $X_{ij} := \log \left( \frac{1 - F_{ij}}{F_{ij}} \right) \Lambda_{ij}$ and $\mathcal{X}_{ij}$ refer to the state space of $X_{ij}$, \emph{i.e.}, $\mathcal{X}_{ij} := \left\{ 0, \log \left( \frac{1 - F_{ij}}{F_{ij}} \right) \right\}$ for $j \in \mathcal{A}(i)$. Now, we introduce new random variables $Y_{ij}$ for each $j \in \mathcal{A}(i)$ satisfying the following properties:
\begin{enumerate}
    \item $\left( X_{ij} : j \in \mathcal{A}(i) \right)$ and $\left( Y_{ij} : j \in \mathcal{A}(i) \right)$ are conditionally independent random vectors given a pair of type vectors $\left( \mathbf{t}, \mathbf{w} \right)$;
    \item $\left\{ Y_{ij} : j \in \mathcal{A}(i) \right\}$ are conditionally independent given a pair of type vectors $\left( \mathbf{t}, \mathbf{w} \right)$;
    \item $Y_{ij}$ has the same support as $X_{ij}$, and the conditional distribution of $Y_{ij}$ is given by
    \begin{equation}
        \label{eqn:proof_thm:statistical_impossibility_v4}
        \mathbb{P} \left\{ \left. Y_{ij} = x \right| \mathbf{t}, \mathbf{w} \right\} = \frac{\exp(x) \cdot \mathbb{P} \left\{ \left. X_{ij} = x \right| \mathbf{t}, \mathbf{w} \right\}}{\mathbb{E} \left[ \left. \exp (X_{ij}) \right| \mathbf{t}, \mathbf{w} \right]}
    \end{equation}
\end{enumerate}
for $\forall x \in \mathcal{X}_{ij}$.
 Since $\mathbb{P} \left\{ \left. X_{ij} = \log \left( \frac{1 - F_{ij}}{F_{ij}} \right) \right| \mathbf{t}, \mathbf{w} \right\} = F_{ij}$, we have
\begin{equation}
    \label{eqn:proof_thm:statistical_impossibility_v5}
    \begin{split}
        &\mathbb{E} \left[ \left. \exp (X_{ij}) \right| \mathbf{t}, \mathbf{w} \right] \\
        = \ & F_{ij} \cdot \exp \left\{ \log \left( \frac{1 - F_{ij}}{F_{ij}} \right) \right\} + \left( 1 - F_{ij} \right) \cdot \exp (0) \\
        = \ & 2 \left( 1 - F_{ij} \right),
    \end{split}
\end{equation}
and thus
\begin{equation}
    \label{eqn:proof_thm:statistical_impossibility_v6}
    \mathbb{P} \left\{ \left. Y_{ij} = \log \left( \frac{1 - F_{ij}}{F_{ij}} \right) \right| \mathbf{t}, \mathbf{w} \right\} = \mathbb{P} \left\{ \left. Y_{ij} = 0 \right| \mathbf{t}, \mathbf{w} \right\} = \frac{1}{2}.
\end{equation}
Therefore, we reach
\begin{equation}
    \label{eqn:proof_thm:statistical_impossibility_v7}
    \begin{split}
        &\mathbb{P} \left\{ \left. \hat{a}_{i}^{\textnormal{ML}} (\mathbf{M}) \neq a_{i} \right| \mathbf{t}, \mathbf{w} \right\} \\
        &=  \mathbb{P} \left\{ \left. \sum_{j \in \mathcal{A}(i)} X_{ij} \geq \lambda_i \right| \mathbf{t}, \mathbf{w} \right\} \\
        &=  \sum_{\substack{\mathbf{x}_{i*} \in \mathcal{X}_{i*} \\ : \sum_{j \in \mathcal{A}(i)} x_{ij} \geq \lambda_i}} \left[ \prod_{j \in \mathcal{A}(i)} \mathbb{P} \left\{ \left. X_{ij} = x_{ij} \right| \mathbf{t}, \mathbf{w} \right\} \right] \\
        &=  \sum_{\substack{\mathbf{y}_{i*} \in \mathcal{X}_{i*} \\ : \sum_{j \in \mathcal{A}(i)} y_{ij} \geq \lambda_i}} 
        \Biggl[ \prod_{j \in \mathcal{A}(i)} \bigg\{ \mathbb{E} \left[ \left. \exp (X_{ij}) \right| \mathbf{t}, \mathbf{w} \right] \cdot \exp(- y_{ij}) \\
        &\qquad\qquad\qquad \times \mathbb{P} \left\{ \left. Y_{ij} = y_{ij} \right| \mathbf{t}, \mathbf{w} \right\}\bigg\} \Biggr]\\
        &=  \left( \prod_{j \in \mathcal{A}(i)} \mathbb{E} \left[ \left. \exp (X_{ij}) \right| \mathbf{t}, \mathbf{w} \right] \right)\times \\
        &\sum_{\substack{\mathbf{y}_{i*} \in \mathcal{X}_{i*} \\ : \sum_{j \in \mathcal{A}(i)} y_{ij} \geq \lambda_i}} \left[ \exp \left( - \sum_{j \in \mathcal{A}(i)} y_{ij} \right) \mathbb{P} \left\{ \left. \mathbf{Y}_{i*} = \mathbf{y}_{i*} \right| \mathbf{t}, \mathbf{w} \right\} \right] \\
        &=  \left( \prod_{j \in \mathcal{A}(i)} \mathbb{E} \left[ \left. \exp (X_{ij}) \right| \mathbf{t}, \mathbf{w} \right] \right)\times \\
        &\mathbb{E} \left[ \left. \mathbbm{1}_{\left\{ \sum_{j \in \mathcal{A}(i)} Y_{ij} \geq \lambda_i \right\}} \exp \left( - \sum_{j \in \mathcal{A}(i)} Y_{ij} \right) \right| \mathbf{t}, \mathbf{w} \right],
    \end{split}
\end{equation}
where $\mathcal{X}_{i*} := \prod_{j \in \mathcal{A}(i)} \mathcal{X}_{ij}$, $\mathbf{x}_{i*} := \left( x_{ij} : j \in \mathcal{A}(i) \right)$, $\mathbf{y}_{i*} := \left( y_{ij} : j \in \mathcal{A}(i) \right)$, and $\mathbf{Y}_{i*} := \left( Y_{ij} : j \in \mathcal{A}(i) \right)$. Here, we note that $\mathbb{E} \left[ \left. Y_{ij} \right| \mathbf{t}, \mathbf{w} \right] = \frac{1}{2} \log \left( \frac{1 - F_{ij}}{F_{ij}} \right)$ for each $j \in \mathcal{A}(i)$, thereby
\begin{equation*}
    \lambda_i = \frac{1}{2} \sum_{j \in \mathcal{A}(i)} \log \left( \frac{1 - F_{ij}}{F_{ij}} \right) = \sum_{j \in \mathcal{A}(i)} \mathbb{E} \left[ \left. Y_{ij} \right| \mathbf{t}, \mathbf{w} \right],
\end{equation*}
and for every $j \in \mathcal{A}(i)$,
\begin{equation}
    \label{eqn:proof_thm:statistical_impossibility_v8}
        Y_{ij} - \mathbb{E} \left[ \left. Y_{ij} \right| \mathbf{t}, \mathbf{w} \right] =
        \begin{cases}
            \frac{1}{2} \log \left( \frac{1 - F_{ij}}{F_{ij}} \right) & \textnormal{with probability } \frac{1}{2}; \\
            - \frac{1}{2} \log \left( \frac{1 - F_{ij}}{F_{ij}} \right) & \textnormal{with probability } \frac{1}{2}.
        \end{cases}
\end{equation}
In particular, we find that $Y_{ij} - \mathbb{E} \left[ \left. Y_{ij} \right| \mathbf{t}, \mathbf{w} \right]$ is symmetrically distributed with center at $0$, \emph{i.e.},
\begin{equation*}
    Y_{ij} - \mathbb{E} \left[ \left. Y_{ij} \right| \mathbf{t}, \mathbf{w} \right] \overset{d}{=} - \left( Y_{ij} - \mathbb{E} \left[ \left. Y_{ij} \right| \mathbf{t}, \mathbf{w} \right] \right).
\end{equation*}
Due to the conditional independence of the random variables $\left\{ Y_{ij} - \mathbb{E} \left[ \left. Y_{ij} \right| \mathbf{t}, \mathbf{w} \right] : j \in \mathcal{A}(i) \right\}$ given a pair of type vectors $\left( \mathbf{t}, \mathbf{w} \right)$, their sum is also symmetrically distributed with center at $0$, \emph{i.e.},
\begin{equation*}
    \left( \sum_{j \in \mathcal{A}(i)} Y_{ij} \right) - \lambda_i \overset{d}{=} - \left\{ \left( \sum_{j \in \mathcal{A}(i)} Y_{ij} \right) - \lambda_i \right\}.
\end{equation*}
Consequently, we obtain
\begin{equation}
    \label{eqn:proof_thm:statistical_impossibility_v9}
    \mathbb{P} \left\{ \left. \left( \sum_{j \in \mathcal{A}(i)} Y_{ij} \right) \geq \lambda_i \right| \mathbf{t}, \mathbf{w} \right\} \geq \frac{1}{2}.
\end{equation}
 Now, we establish a lower bound the last term of the equation \eqref{eqn:proof_thm:statistical_impossibility_v7}: given any $\mu_i > 0$, one has
\begin{equation}
    \label{eqn:proof_thm:statistical_impossibility_v10}
    \begin{split}
        &\mathbb{E} \left[ \left. \mathbbm{1}_{\left\{ \sum_{j \in \mathcal{A}(i)} Y_{ij} \geq \lambda_i \right\}} \exp \left( - \sum_{j \in \mathcal{A}(i)} Y_{ij} \right) \right| \mathbf{t}, \mathbf{w} \right] \\
        \geq \ & \mathbb{E} \left[ \left. \mathbbm{1}_{\left\{0 \leq \left( \sum_{j \in \mathcal{A}(i)} Y_{ij} \right) - \lambda_i < \mu_i \right\}} \exp \left( - \sum_{j \in \mathcal{A}(i)} Y_{ij} \right) \right| \mathbf{t}, \mathbf{w} \right] \\
        \geq \ & \mathbb{E} \left[ \left. \mathbbm{1}_{\left\{0 \leq \left( \sum_{j \in \mathcal{A}(i)} Y_{ij} \right) - \lambda_i < \mu_i \right\}} \exp \left( - \mu_i - \lambda_i \right) \right| \mathbf{t}, \mathbf{w} \right] \\
        = \ & \exp \left( - \mu_i - \lambda_i \right) \mathbb{P} \left\{ \left. 0 \leq \left( \sum_{j \in \mathcal{A}(i)} Y_{ij} \right) - \lambda_i < \mu_i \right| \mathbf{t}, \mathbf{w} \right\}.
    \end{split}
\end{equation}
Using the fact \eqref{eqn:proof_thm:statistical_impossibility_v9} together with the Markov inequality yields
\begin{equation}
    \label{eqn:proof_thm:statistical_impossibility_v11}
    \begin{split}
        &\mathbb{P} \left\{ \left. 0 \leq \left( \sum_{j \in \mathcal{A}(i)} Y_{ij} \right) - \lambda_i < \mu_i \right| \mathbf{t}, \mathbf{w} \right\} \\
        = \ & \mathbb{P} \left\{ \left. \left( \sum_{j \in \mathcal{A}(i)} Y_{ij} \right) - \lambda_i \geq 0 \right| \mathbf{t}, \mathbf{w} \right\} \\
        &- \mathbb{P} \left\{ \left. \left( \sum_{j \in \mathcal{A}(i)} Y_{ij} \right) - \lambda_i \geq \mu_i \right| \mathbf{t}, \mathbf{w} \right\} \\
        \geq \ & \frac{1}{2} - \mathbb{P} \left\{ \left. \sum_{j \in \mathcal{A}(i)} \left( Y_{ij} - \mathbb{E} \left[ \left. Y_{ij} \right| \mathbf{t}, \mathbf{w} \right] \right) \geq \mu_i \right| \mathbf{t}, \mathbf{w} \right\} \\
        \geq \ &  \frac{1}{2} - \mu_{i}^{-2} \left( \sum_{j \in \mathcal{A}(i)} \mathbb{E} \left[ \left. \left( Y_{ij} - \mathbb{E} \left[ \left. Y_{ij} \right| \mathbf{t}, \mathbf{w} \right] \right)^2 \right| \mathbf{t}, \mathbf{w} \right] \right) \\
        \overset{\textnormal{(c)}}{=} \ & \frac{1}{2} - \mu_{i}^{-2} \left( \sum_{j \in \mathcal{A}(i)} \left\{ \frac{1}{2} \log \left( \frac{1 - F_{ij}}{F_{ij}} \right) \right\}^2 \right) \\
        = \ & \frac{1}{2} - \frac{1}{4 \mu_{i}^2} \left( \sum_{j \in \mathcal{A}(i)} \left\{ \log \left( \frac{1 - F_{ij}}{F_{ij}} \right) \right\}^2 \right)
    \end{split}
\end{equation}
where the step (c) follows from the fact \eqref{eqn:proof_thm:statistical_impossibility_v8}. By combining three bounds \eqref{eqn:proof_thm:statistical_impossibility_v7}, \eqref{eqn:proof_thm:statistical_impossibility_v10}, and \eqref{eqn:proof_thm:statistical_impossibility_v11}, we arrive at
\begin{equation}
    \label{eqn:proof_thm:statistical_impossibility_v12}
    \begin{split}
        &\mathbb{P} \left\{ \left.  \hat{a}_{i}^{\textnormal{ML}}(\mathbf{M}) \neq a_{i} \right| \mathbf{t}, \mathbf{w} \right\} \\
        \geq \ & \left[ \prod_{j \in \mathcal{A}(i)} 2 \sqrt{F_{ij} \left( 1 - F_{ij} \right)} \right] \exp \left( - \mu_i \right) \\
        &\times \left[ \frac{1}{2} - \frac{1}{4 \mu_{i}^2} \left( \sum_{j \in \mathcal{A}(i)} \left\{ \log \left( \frac{F_{ij}}{1 - F_{ij}} \right) \right\}^2 \right) \right]
    \end{split}
\end{equation}
for any $\mu_i > 0$. Now, we put $\mu_i = \Gamma (\mathcal{Q}) \sqrt{\left| \mathcal{A}(i) \right|}$ for $i \in [m]$. Since
\begin{equation*}
    \begin{split}
        &\frac{1}{4 \mu_{i}^2} \left( \sum_{j \in \mathcal{A}(i)} \left\{ \log \left( \frac{F_{ij}}{1 - F_{ij}} \right) \right\}^2 \right) \\
        \stackrel{\textnormal{(d)}}{\leq} \ & \frac{1}{4 \mu_{i}^2} \left\{ \log \left( \frac{\max \left\{ \mathcal{Q}(a, b) : (a, b) \in [d] \times [d] \right\}}{1 - \max \left\{ \mathcal{Q}(a, b) : (a, b) \in [d] \times [d] \right\}} \right) \right\}^2 \\
        &\times\left| \mathcal{A}(i) \right| \\
        = \ & \frac{1}{4},
    \end{split}
\end{equation*}
where the step (d) follows since the log-odds function $x \in \left[ \frac{1}{2}, 1 \right) \mapsto \log \left( \frac{x}{1-x} \right) \in \mathbb{R}$ is a non-negative, strictly increasing function, we deduce from the bound \eqref{eqn:proof_thm:statistical_impossibility_v12} that
\begin{equation}
    \label{eqn:proof_thm:statistical_impossibility_v13}
    \begin{split}
        &\mathbb{P} \left\{ \left. \hat{a}_{i}^{\textnormal{ML}}(\mathbf{M}) \neq a_{i} \right| \mathbf{t}, \mathbf{w} \right\} \\
        \geq \ & \frac{1}{4} \exp \left\{ - \Gamma (d;\mathcal{Q}) \sqrt{\left| \mathcal{A}(i) \right|} \right\} \left[ \prod_{j \in \mathcal{A}(i)} 2 \sqrt{F_{ij} \left( 1 - F_{ij} \right)} \right].
    \end{split}
\end{equation}
At this point, we recall from the computation in the upper bound \eqref{eqn:proof_thm:achievability_ml_v3} on the conditional error probability given a task type vector $\mathbf{t}$ that
\begin{equation}
    \label{eqn:proof_thm:statistical_impossibility_v14}
    \begin{split}
        &\mathbb{E}_{\mathbf{w} \sim \bm{\nu}^{\otimes n}} \left[ \prod_{j \in \mathcal{A}(i)} \left\{ 2 \sqrt{F_{ij} \left( 1 - F_{ij} \right)} \right\} \right] \\
        = \ & \left\{ \sum_{w=1}^{d} 2 \nu (w) \sqrt{\mathcal{Q} \left( t_i, w \right) \left\{ 1 - \mathcal{Q} \left( t_i, w \right) \right\}} \right\}^{\left| \mathcal{A}(i) \right|},
    \end{split}
\end{equation}
thereby
\begin{equation}
    \label{eqn:proof_thm:statistical_impossibility_v15}
    \begin{split}
        &\mathbb{E}_{\left( \mathbf{t}, \mathbf{w} \right) \sim \bm{\mu}^{\otimes m} \otimes \bm{\nu}^{\otimes n}} \left[ \prod_{j \in \mathcal{A}(i)} \left\{ 2 \sqrt{F_{ij} \left( 1 - F_{ij} \right)} \right\} \right] \\
        = \ & \mathbb{E}_{\mathbf{t} \sim \bm{\mu}^{\otimes m}} \left[ \mathbb{E}_{\mathbf{w} \sim \bm{\nu}^{\otimes n}} \left[ \prod_{j \in \mathcal{A}(i)} \left\{ 2 \sqrt{F_{ij} \left( 1 - F_{ij} \right)} \right\} \right] \right] \\
        = \ & \mathbb{E}_{t_i \sim \bm{\mu}} \left[ \left\{ \sum_{w=1}^{d} 2 \nu (w) \sqrt{\mathcal{Q} \left( t_i, w \right) \left\{ 1 - \mathcal{Q} \left( t_i, w \right) \right\}} \right\}^{\left| \mathcal{A}(i) \right|} \right] \\
        \stackrel{\textnormal{(e)}}{\geq} \ & \left\{ \mathbb{E}_{t_i \sim \bm{\mu}} \left[ \sum_{w=1}^{d} 2 \nu (w) \sqrt{\mathcal{Q} \left( t_i, w \right) \left\{ 1 - \mathcal{Q} \left( t_i, w \right) \right\}} \right] \right\}^{\left| \mathcal{A}(i) \right|} \\
        = \ & \left\{ 2\sum_{(t, w) \in [d] \times [d]} \mu (t) \nu (w) \sqrt{\mathcal{Q}(t, w) \left\{ 1 - \mathcal{Q}(t, w) \right\}} \right\}^{\left| \mathcal{A}(i) \right|} \\
        = \ & \exp \left\{ - \left| \mathcal{A} (i) \right| \gamma_2 \left( d; \mathcal{Q}, \bm{\mu}, \bm{\nu} \right) \right\},
    \end{split}
\end{equation}
where the step (e) follows by the Jensen's inequality. Therefore, we finally achieve the bound
\begin{equation}
    \label{eqn:proof_thm:statistical_impossibility_v16}
    \begin{split}
        &\mathbb{P} \left\{ \hat{a}_{i}^{\textnormal{ML}}(\mathbf{M}) \neq a_{i} \right\} \\
        = \ & \mathbb{E}_{\left( \mathbf{t}, \mathbf{w} \right) \sim \bm{\mu}^{\otimes m} \otimes \bm{\nu}^{\otimes n}} \left[ \mathbb{P} \left\{ \left. \hat{a}_{i}^{\textnormal{ML}}(\mathbf{M}) \neq a_{i} \right| \mathbf{t}, \mathbf{w} \right\} \right] \\
        \stackrel{\textnormal{(f)}}{\geq} \ & \frac{1}{4} \exp \left\{ - \Gamma (d;\mathcal{Q}) \sqrt{\left| \mathcal{A}(i) \right|} \right\} \\
        &\mathbb{E}_{\left( \mathbf{t}, \mathbf{w} \right) \sim \bm{\mu}^{\otimes m} \otimes \bm{\nu}^{\otimes n}} \left[ \prod_{j \in \mathcal{A}(i)} \left\{ 2 \sqrt{F_{ij} \left( 1 - F_{ij} \right)} \right\} \right] \\
        \stackrel{\textnormal{(g)}}{\geq} \ & \frac{1}{4} \exp \left[ - \left\{ \gamma_2 \left( d; \mathcal{Q}, \bm{\mu}, \bm{\nu} \right) \left| \mathcal{A}(i) \right| + \Gamma \left( d;\mathcal{Q} \right) \sqrt{\left| \mathcal{A}(i) \right|} \right\} \right],
    \end{split}
\end{equation}
where the step (f) and (g) make use of the inequalities \eqref{eqn:proof_thm:statistical_impossibility_v13} and \eqref{eqn:proof_thm:statistical_impossibility_v15}, respectively, and note that the \eqref{eqn:proof_thm:statistical_impossibility_v16} holds for any ground-truth label $a_{i} \in \left\{ \pm 1 \right\}$ associated with the $i$-th task.

\indent As the final step, it remains to establish a minimax lower bound. Based on the lower bound \eqref{eqn:proof_thm:statistical_impossibility_v1} of the minimax risk, we find that
\begin{equation}
    \label{eqn:proof_thm:statistical_impossibility_v17}
    \begin{split}
        &\mathcal{R}^* = \inf_{\hat{\mathbf{a}}} \left( \sup_{\mathbf{a} \in \left\{ \pm 1 \right\}^m} \mathcal{R} \left( \mathbf{a}, \hat{\mathbf{a}} \right) \right) \\
        &\geq \frac{1}{m} \sum_{i=1}^{m} \mathbb{E}_{a_i \sim \textnormal{Unif} \left( \left\{ \pm 1 \right\} \right)} \left[ \mathbb{P} \left\{ \hat{a}_{i}^{\textnormal{ML}}(\mathbf{M}) \neq a_i \right\} \right] \\
        &\stackrel{\textnormal{(h)}}{\geq}\frac{1}{4m} \sum_{i=1}^{m} \exp \bigg[ - \bigg\{ \gamma_2 \left( d; \mathcal{Q}, \bm{\mu}, \bm{\nu} \right) \left| \mathcal{A}(i) \right|\\
        &\qquad \qquad\qquad\qquad+ \Gamma (d;\mathcal{Q}) \sqrt{\left| \mathcal{A}(i) \right|} \bigg\} \bigg] \\
        &\stackrel{\textnormal{(i)}}{\geq} \frac{1}{4} \exp \left[ - \left\{ \gamma_2 \left( d; \mathcal{Q}, \bm{\mu}, \bm{\nu} \right) \left( \frac{\left| \mathcal{A} \right|}{m} \right) + \Gamma (d;\mathcal{Q}) \sqrt{\frac{\left| \mathcal{A} \right|}{m}} \right\} \right],
    \end{split}
\end{equation}
where the step (h) follows from the bound \eqref{eqn:proof_thm:statistical_impossibility_v16}, and the step (i) is due to Jensen's inequality together with the convexity of the function $x \in \left[ 0, +\infty \right) \mapsto \exp \left\{ - \left( \mu x + \nu \sqrt{x} \right) \right\} \in \mathbb{R}$ for any constants $\mu, \nu \geq 0$. This fact can be confirmed by computing the second-order derivative of the function directly. So in order to enforce the following conclusion holds
\begin{equation}
    \label{eqn:proof_thm:statistical_impossibility_v18}
    \mathcal{R}^* = \inf_{\hat{\mathbf{a}}} \left( \sup_{\mathbf{a} \in \left\{ \pm 1 \right\}^m} \mathcal{R} \left( \mathbf{a}, \hat{\mathbf{a}} \right) \right) > \alpha,
\end{equation}
one can see from the bound \eqref{eqn:proof_thm:statistical_impossibility_v17} that it suffices to make the following inequality holds:
\begin{equation}
    \label{eqn:proof_thm:statistical_impossibility_v19}
    \frac{1}{4} \exp \left[ - \left\{ \gamma_2 \left( d; \mathcal{Q}, \bm{\mu}, \bm{\nu} \right) \left( \frac{\left| \mathcal{A} \right|}{m} \right) + \Gamma (d;\mathcal{Q}) \sqrt{\frac{\left| \mathcal{A} \right|}{m}} \right\} \right] > \alpha,
\end{equation}
and this bound is equivalent to the condition \eqref{eqn:statistical_impossibility}. In other words, our desired order-wise lower bound on the minimax risk \eqref{eqn:proof_thm:statistical_impossibility_v19} follows, when the condition \eqref{eqn:statistical_impossibility} holds, and this completes the proof.

\section{Information-theoretic limits for special cases}
\label{sec:info_limits_special_cases}

\indent In this section, we aim to establish the information-theoretic results in the $d$-type worker-task specialization model under specific assumptions on $\mathcal{Q}$. More specifically, by leveraging Theorem \ref{thm:achievability_ml} and \ref{thm:statistical_impossibility}, we derive the order-wise information-theoretic limits under the model following Assumption \ref{assumption:multiple_spammers} in Corollary \ref{cor:IT_bounds_Q_in_M2}, and  under the special $(p,q)$ model  \eqref{eqn:special_model} in Corollary \ref{cor:special_model_IT_result}.

\indent Here, we first recall that if $\mathcal{Q}$ follows Assumption \ref{assumption:multiple_spammers}, then
\begin{enumerate} [label=(\roman*)]
    \item there is an absolute constant $\epsilon \in \left( 0, \frac{1}{2} \right)$ such that $\mathcal{Q} (t, w) \in \left\{ \frac{1}{2} \right\} \cup \left[ \frac{1}{2} + \epsilon, 1 \right]$ for all $(t, w) \in [d] \times [d]$;
    \item there exists a function $\delta (\cdot; d): [d] \to \left\{ 0, \frac{1}{d}, \cdots, \frac{d-1}{d}, 1 \right\}$ such that
    \begin{equation*}
        \left| \textnormal{spammer}_{\mathcal{Q}} (t) \right| = d \left\{ 1 - \delta (t; d) \right\},\ \forall t \in [d].
    \end{equation*}
\end{enumerate}
For simplicity, we consider the uniform distribution $\textnormal{Unif} \left( [d] \right)$ for the prior distributions of the task types and worker types, \emph{i.e.}, $\bm{\mu} = \bm{\nu} = \frac{1}{d} \mathbf{1}_d$. 
}

\begin{cor}
\label{cor:IT_bounds_Q_in_M2}
Under the $d$-type worker-task specialization model whose reliability matrix $\mathcal{Q}$ satisfies Assumption \ref{assumption:multiple_spammers}, the desired recovery performance \eqref{eqn:expected_accuracy} is achievable via the ML estimator \eqref{eqn:ml_estimator} if
\begin{equation}
    \label{eqn:Q_in_M2_info_upper}
    \frac{\left| \mathcal{A} \right|}{m} = \Omega \left( \frac{1}{\delta_{\min}(d)} \log \left( \frac{1}{\alpha} \right) \right),
\end{equation}
while it is statistically impossible whenever
\begin{equation}
    \label{eqn:Q_in_M2_info_lower}
    \begin{split}
        &\frac{\left| \mathcal{A} \right|}{m} = o \Bigg( \min\Bigg\{ \frac{1 - \delta_{\max}(d)}{\delta_{\max}(d)} \log \left( \frac{1}{\alpha} \right), \\
        &\qquad\qquad\left( \frac{1}{\Gamma \left( d; \mathcal{Q} \right)} \log \left( \frac{1}{\alpha} \right)
    \right) \Bigg\}^2 \Bigg).
    \end{split}
\end{equation}
\end{cor}

\begin{proof} [Proof of Corollary \ref{cor:IT_bounds_Q_in_M2}]
In view of the theoretical guarantees for the information-theoretic achievability \eqref{eqn:achievability_ml_simplified_v1} and the statistical impossibility \eqref{eqn:statistical_impossibility_simplified_v1}, it suffices to show that
\begin{equation}
    \label{eqn:proof_cor:IT_bounds_Q_in_M2_v1}
    \begin{split}
        \gamma_1 \left( d; \mathcal{Q}, \frac{1}{d} \mathbf{1}_d, \frac{1}{d} \mathbf{1}_d \right) &\gtrsim \delta_{\min} (d); \\
        \gamma_2 \left( d; \mathcal{Q}, \frac{1}{d} \mathbf{1}_d, \frac{1}{d} \mathbf{1}_d \right) &\lesssim \frac{\delta_{\max} (d)}{1 - \delta_{\max} (d)}.
    \end{split}
\end{equation}
Note from the definition of the class $\mathcal{M}_2$ of reliability matrices that
\begin{equation}
    \label{eqn:proof_cor:IT_bounds_Q_in_M2_v2}
    \begin{split}
        \frac{d \left\{ 1 - \delta \left( t; d \right) \right\}}{2} \leq \ & \sum_{w=1}^{d} \sqrt{\mathcal{Q} (t, w) \left( 1 - \mathcal{Q}(t, w) \right)} \\
        \leq \ & \frac{d \left\{ 1 - \delta \left( t; d \right) \right\}}{2} + \frac{\sqrt{1 - 4 \epsilon^2} \cdot d \delta \left( t; d \right)}{2}
    \end{split}
\end{equation}
for every $t \in [d]$. Thus, we reach
\begin{equation*}
    \begin{split}
        &\gamma_1 \left( d; \mathcal{Q}, \frac{1}{d} \mathbf{1}_d, \frac{1}{d} \mathbf{1}_d \right) \\
        = \ & \log \left( \frac{d}{2 \max_{t \in [d]} \sum_{w=1}^{d} \sqrt{\mathcal{Q} (t, w) \left( 1 - \mathcal{Q}(t, w) \right)}} \right) \\
        \stackrel{\textnormal{(a)}}{\geq} \ & \log \left( \frac{1}{\max_{t \in [d]} \left\{ 1 - \left( 1 - \sqrt{1 - 4 \epsilon^2} \right) \delta \left( t; d \right) \right\}} \right) \\
        = \ & \log \left( \frac{1}{1 - \left( 1 - \sqrt{1 - 4 \epsilon^2} \right) \delta_{\min}(d)} \right) \\
        \stackrel{\textnormal{(b)}}{\geq} \ & \left( 1 - \sqrt{1 - 4 \epsilon^2} \right) \delta_{\min}(d),
    \end{split}
\end{equation*}
where the step (a) follows from the observation \eqref{eqn:proof_cor:IT_bounds_Q_in_M2_v2}, and the step (b) is due to the fact $\log \left( \frac{1}{1 - x} \right) \geq x$ for every $x \geq 0$, as desired. On the other hand, we have
\begin{equation*}
    \begin{split}
        &\gamma_2 \left( d; \mathcal{Q}, \frac{1}{d} \mathbf{1}_d, \frac{1}{d} \mathbf{1}_d \right) \\
        &=
        \log \left( \frac{d^2}{2 \sum_{(t, w) \in [d] \times [d]} \sqrt{\mathcal{Q} (t, w) \left( 1 - \mathcal{Q}(t, w) \right)}} \right) \\
        &\stackrel{\textnormal{(c)}}{\leq} \log \left( \frac{1}{1 - \delta_{\max} (d)} \right) \\
        &\stackrel{\textnormal{(d)}}{\leq} \frac{\delta_{\max}(d)}{1 - \delta_{\max}(d)},
    \end{split}
\end{equation*}
where the step (c) makes use of the observation \eqref{eqn:proof_cor:IT_bounds_Q_in_M2_v2}, and the step (d) holds by the inequality
\begin{equation*}
    \log \left( \frac{1}{1 - x} \right) = \log \left( 1 + \frac{x}{1-x} \right) \leq \frac{x}{1-x},\ \forall x \geq 0,
\end{equation*}
and this establishes the desired result \eqref{eqn:proof_cor:IT_bounds_Q_in_M2_v1}.
\end{proof}

\indent One can observe from Corollary \ref{cor:IT_bounds_Q_in_M2} that the ML estimator \eqref{eqn:ml_estimator} becomes sample-optimal if (\romannumeral 1) $\delta_{\min}(d) \asymp \delta_{\max}(d)$ as $d \to \infty$, (\romannumeral 2) $\limsup_{d \to \infty} \delta_{\max}(d) < 1$, and (\romannumeral 3) the target accuracy $\alpha \in \left( 0, 1 \right]$ satisfies $\log \left( \frac{1}{\alpha} \right) 
\gtrsim \frac{\Gamma \left( d; \mathcal{Q} \right)^2 \left\{ 1 - \delta_{\max}(d) \right\}}{\delta_{\max}(d)}$ as $d \to \infty$. We next derive the order-wise information-theoretic limits under the special $(p,q)$ model  \eqref{eqn:special_model}.

\begin{cor}
\label{cor:special_model_IT_result}
Under the special $(p,q)$ model \eqref{eqn:special_model}, the recovery accuracy \eqref{eqn:expected_accuracy} is achievable via the ML estimator \eqref{eqn:ml_estimator} if
\begin{equation}
    \label{eqn:special_model_info_upper}
    \frac{\left| \mathcal{A} \right|}{m} =
    \begin{cases}
        \Omega \left( \log \left( \frac{1}{\alpha} \right) \right) & \textnormal{if } q > \frac{1}{2}; \\
        \Omega \left( d \log \left( \frac{1}{\alpha} \right) \right) & \textnormal{otherwise},
    \end{cases}
\end{equation}
while it is statistically impossible whenever
\begin{equation}
    \label{eqn:special_model_info_lower}
    \frac{\left| \mathcal{A} \right|}{m} =
    \begin{cases}
        o \left( \log \left( \frac{1}{\alpha} \right) \right) & \textnormal{if } q > \frac{1}{2}; \\
        o \left( d \log \left( \frac{1}{\alpha} \right) \right) & \textnormal{if } q = \frac{1}{2} \textnormal{ and } \log \left( \frac{1}{\alpha} \right) = \Omega (d); \\
        o \left( \left\{ \log \left( \frac{1}{\alpha} \right) \right\}^2 \right) & \textnormal{if } q = \frac{1}{2} \textnormal{ and } \log \left( \frac{1}{\alpha} \right) = o (d).
    \end{cases}
\end{equation}

\end{cor}

\begin{proof} [Proof of Corollary \ref{cor:special_model_IT_result}]
One can realize that the reliability matrix \eqref{eqn:special_model} of the special $(p, q)$ model obeys the bounded reliability assumption \eqref{eqn:specific_regime_v1} together with the error exponent 
\begin{equation*}
    \begin{split}
        \gamma^* (d) := \ & \log \left( \frac{d}{2 \sqrt{p (1-p)} + 2(d-1) \sqrt{q (1-q)}} \right) \\
        = \ &\gamma_1 \left( d; \mathcal{Q}, \bm{\mu}, \bm{\nu} \right) = \gamma_2 \left( d; \mathcal{Q}, \bm{\mu}, \bm{\nu} \right),
    \end{split}
\end{equation*}
where $\gamma_1 \left( d; \mathcal{Q}, \bm{\mu}, \bm{\nu} \right)$ is the error exponent defined in \eqref{eqn:error_exponent_achievability_ml} and $\gamma_2 \left( d; \mathcal{Q}, \bm{\mu}, \bm{\nu} \right)$ is the error exponent defined in \eqref{eqn:statistical_impossibility}. From Remark \ref{rmk:tightness_info_bounds}, we conclude that it is possible to achieve the recovery performance \eqref{eqn:expected_accuracy} via the ML estimator \eqref{eqn:ml_estimator} if
\begin{equation}
    \label{eqn:achievability_ml_simplified_v3}
    \frac{|\mathcal{A}|}{m} = \Omega \left( \frac{1}{\gamma^* (d)} \log \left( \frac{1}{\alpha} \right) \right).
\end{equation}
On the other hand, the recovery performance \eqref{eqn:expected_accuracy} cannot be achieved whenever the size of the worker-task assignment set $\mathcal{A}$ satisfies
\begin{equation}
    \label{eqn:statistical_impossibility_simplified_v3}
    \frac{|\mathcal{A}|}{m} = o \left( \min \left\{ \frac{1}{\gamma^* (d)} \log \left( \frac{1}{\alpha} \right), \left\{ \log \left( \frac{1}{\alpha} \right) \right\}^2 \right\} \right).
\end{equation}

\indent We mainly consider the following two parameter regimes of the special $(p,q)$ model \eqref{eqn:special_model}:
\begin{itemize}
    \item $q > \frac{1}{2}$: it holds that 
    \begin{equation*}
    \gamma^* (d) = \Theta (1),
    \end{equation*} thereby we obtain
    $\log \left( \frac{1}{\alpha} \right) \gtrsim 1 \asymp \frac{1}{\gamma^*(d)}$.
    So this regime corresponds to the case (a) of Remark \ref{rmk:tightness_info_bounds} and thus we may conclude that the ML estimator \eqref{eqn:ml_estimator} achieves desired recovery \eqref{eqn:expected_accuracy} if and only if
    $\frac{|\mathcal{A}|}{m} = \Theta \left( \log \left( \frac{1}{\alpha} \right) \right)$.
    \item $p > \frac{1}{2} = q$: from the fact $\lim_{x \to 0} \frac{\log (1 + x)}{x} = 1$, the order of the optimal error exponent $\gamma^* (d)$ is
    \begin{equation*}
        \begin{split}
            \gamma^* (d) = \ & \log \left\{ 1 + \frac{\left( \sqrt{p} - \sqrt{1-p} \right)^2}{d - 1 + 2\sqrt{p(1-p)}} \right\} \\
            \asymp \ & \frac{\left( \sqrt{p} - \sqrt{1-p} \right)^2}{d - 1 + 2\sqrt{p(1-p)}} \\
            = \ & \Theta \left( \frac{1}{d} \right),
        \end{split}
    \end{equation*}
    thereby the ML estimator \eqref{eqn:ml_estimator} succeeds in the achievement of the recovery performance \eqref{eqn:expected_accuracy} if
    $\frac{|\mathcal{A}|}{m} = \Omega \left( d \log \left( \frac{1}{\alpha} \right) \right)$. We note that the scaling order of the impossibility condition \eqref{eqn:statistical_impossibility_simplified_v3} relies heavily on the decay rate of the recovery accuracy $\alpha$ (or simply the growth rate of $\log \left( \frac{1}{\alpha} \right)$) as a function of $d$. If $\alpha$ is exponentially decaying in terms of $d$, \emph{i.e.}, $\log \left( \frac{1}{\alpha} \right) \gtrsim d$, then the impossibility condition reads $\frac{|\mathcal{A}|}{m} = o \left(  d \log \left( \frac{1}{\alpha} \right) \right)$. Otherwise, \emph{i.e.}, $\log \left( \frac{1}{\alpha} \right) = o (d)$, the impossibility condition simplifies to $\frac{|\mathcal{A}|}{m} = o \left( \left\{ \log \left( \frac{1}{\alpha} \right) \right\}^2 \right)$.
\end{itemize}
\end{proof}

\section{Proof of Lemma \ref{lemma:exact_recovery_alg:worker_clustering}}
\label{sec:proof_lemma:exact_recovery_alg:worker_clustering}

\indent The proof of Lemma \ref{lemma:exact_recovery_alg:worker_clustering} is rather technically involved as it requires a number of additional settings. We first define the \emph{normalized worker type matrix} $\mathbf{U} := \left[ U_{jz} \right]_{(i, z) \in [n] \times [d]} \in \mathbb{R}^{n \times d}$ by
\begin{equation*}
    U_{iz} :=
    \begin{cases}
        \frac{1}{\sqrt{s_z}} & \textnormal{if } i \in \mathcal{W}_z; \\
        0 & \textnormal{otherwise.}
    \end{cases}
\end{equation*}
Let $\mathcal{U}$ denote the linear subspace of $\mathbb{R}^{n \times n}$ spanned by elements of the form $\mathbf{U}_{*z} \cdot \mathbf{x}^{\top}$ and $\mathbf{y} \cdot \mathbf{U}_{*z}^{\top}$, where $z \in [d]$ and $\mathbf{x}$, $\mathbf{y}$ are arbitrary vectors in $\mathbb{R}^n$, and $\mathcal{U}^{\perp}$ refer to its orthogonal complement in $\mathbb{R}^{n \times n}$. Then, the linear subspace $\mathcal{U}$ of $\mathbb{R}^{n \times n}$ can be written explicitly as
\begin{equation*}
    \mathcal{U} = \left\{ \mathbf{U} \mathbf{A}^{\top} + \mathbf{B} \mathbf{U}^{\top}: \mathbf{A}, \mathbf{B} \in \mathbb{R}^{n \times d} \right\}.
\end{equation*}
The orthogonal projections $\mathcal{P}_{\mathcal{U}}$ and $\mathcal{P}_{\mathcal{U}^{\perp}}$ of $\mathbb{R}^{n \times n}$ onto $\mathcal{U}$ and $\mathcal{U}^{\perp}$, respectively, are given by
\begin{equation*}
    \begin{split}
        \mathcal{P}_{\mathcal{U}}(\mathbf{X}) &:= \mathbf{U} \mathbf{U}^{\top} \mathbf{X} + \mathbf{X} \mathbf{U} \mathbf{U}^{\top} - \mathbf{U} \mathbf{U}^{\top} \mathbf{X} \mathbf{U} \mathbf{U}^{\top}; \\
        \mathcal{P}_{\mathcal{U}^{\perp}}(\mathbf{X}) &:= \left( \mathcal{I} - \mathcal{P}_{\mathcal{U}} \right)(\mathbf{X}) = \left( \mathbf{I}_n - \mathbf{U} \mathbf{U}^{\top} \right) \mathbf{X} \left( \mathbf{I}_n - \mathbf{U} \mathbf{U}^{\top} \right),
    \end{split}
\end{equation*}
where $\mathcal{I} : \mathbb{R}^{n \times n} \rightarrow \mathbb{R}^{n \times n}$ denotes the identity map on $\mathbb{R}^{n \times n}$. 

\indent Let $\mathcal{X} \subseteq \mathbb{R}^{n \times n}$ be the feasible region of the SDP \eqref{eqn:clustering_SDP} and $\mathbf{X}^* \in \mathbb{R}^{n \times n}$ denote the \emph{ground-truth worker cluster matrix} induced by worker types:
\begin{equation*}
    X_{jk}^* :=
    \begin{cases}
        1 & \textnormal{if $j$ and $k$ belong to the same cluster}; \\
        0 & \textnormal{otherwise.}
    \end{cases}
\end{equation*}
Then, it can be readily observed that $\mathbf{X}^*$ has a rank-$d$ singular value decomposition $\mathbf{X}^* = \mathbf{U} \mathbf{\Sigma} \mathbf{U}^{\top}$, where $\mathbf{\Sigma} := \textnormal{diag} \left( s_1, s_2, \cdots, s_d \right) \in \mathbb{R}^{d \times d}$, where $s_z := \left| \mathcal{W}_z \right|$ for $z \in [d]$. In order to establish Lemma \ref{lemma:exact_recovery_alg:worker_clustering}, it suffices to prove that $\mathbf{X}^*$ is the unique optimal solution to the SDP \eqref{eqn:clustering_SDP}. Thus the main conclusion of Lemma \ref{lemma:exact_recovery_alg:worker_clustering} reduces to the following claim: for any $\mathbf{X} \in \mathcal{X} \setminus \left\{ \mathbf{X}^* \right\}$,
\begin{equation}
    \label{eqn:proof_lemma:exact_recovery_alg:worker_clustering_v1}
    \Delta (\mathbf{X}) := \left\langle \mathbf{A} - \eta \mathbf{1}_{n \times n}, \mathbf{X}^* - \mathbf{X} \right\rangle > 0.
\end{equation}
From the definition of the orthogonal projections $\mathcal{P}_{\mathcal{U}}(\cdot) : \mathbb{R}^{n \times n} \rightarrow \mathbb{R}^{n \times n}$ and $\mathcal{P}_{\mathcal{U}^{\perp}}(\cdot) : \mathbb{R}^{n \times n} \rightarrow \mathbb{R}^{n \times n}$, we obtain the following decomposition of the quantity in \eqref{eqn:proof_lemma:exact_recovery_alg:worker_clustering_v1}:
\begin{equation}
    \label{eqn:proof_lemma:exact_recovery_alg:worker_clustering_v2}
    \begin{split}
        \Delta (\mathbf{X}) = \ & \underbrace{\left\langle \mathcal{P}_{\mathcal{U}} \left( \mathbf{A} - \mathbb{E} \left[ \left. \mathbf{A} \right| \mathbf{w} \right] \right), \mathbf{X}^* - \mathbf{X} \right\rangle}_{=: \ \textnormal{(T1)}} \\
        &+ \underbrace{\left\langle \mathcal{P}_{\mathcal{U}^{\perp}} \left( \mathbf{A} - \mathbb{E} \left[ \left. \mathbf{A} \right| \mathbf{w} \right] \right), \mathbf{X}^* - \mathbf{X} \right\rangle}_{=: \ \textnormal{(T2)}} \\
        &+ \underbrace{\left\langle \mathbb{E} \left[ \left. \mathbf{A} \right| \mathbf{w} \right] - \eta \mathbf{1}_{n \times n}, \mathbf{X}^* - \mathbf{X} \right\rangle}_{=: \ \textnormal{(T3)}}.
    \end{split}
\end{equation}

\noindent Here, it would be worth noting that the ensuing arguments for bounding (T1), (T2), and (T3) resemble ones in \cite{chen2018convexified, chen2016statistical, lee2020robust}, and the entries of the similarity matrix $\mathbf{A}$ are not conditionally independent given a worker type vector $\mathbf{w}$. \\

\subsubsection{Lower bound of (T1)}
The following lemma offers a tight concentration bound on the $l^{\infty}$-norm of $\mathcal{P}_{\mathcal{U}} \left( \mathbf{A} - \mathbb{E} \left[ \left. \mathbf{A} \right| \mathbf{w} \right] \right)$.

\begin{lemma}
\label{lemma:proof_lemma:exact_recovery_alg:worker_clustering_v1}
Under $\textnormal{SM} \left( d; \mathcal{Q}, \bm{\mu}, \bm{\nu} \right)$, there exists a universal constant $\gamma_1 > 0$ such that with probability greater than $1 - 2 n^{-11}$, we have
\begin{equation}
    \label{eqn:lemma:proof_lemma:exact_recovery_alg:worker_clustering_v1_v1}
    \left\| \mathcal{P}_{\mathcal{U}} \left( \mathbf{A} - \mathbb{E} \left[ \left. \mathbf{A} \right| \mathbf{w} \right] \right) \right\|_{\infty} \leq \gamma_1 \cdot \sqrt{r} \log n.
\end{equation}
\end{lemma}

\noindent The proof of Lemma \ref{lemma:proof_lemma:exact_recovery_alg:worker_clustering_v1} is relegated to Appendix \ref{subsec:proof_lemma:proof_lemma:exact_recovery_alg:worker_clustering_v1}. Thanks to Lemma \ref{lemma:proof_lemma:exact_recovery_alg:worker_clustering_v1} together with the H\"{o}lder's inequality, we obtain the following conclusion: with probability exceeding $1 - 2 n^{-11}$,
\begin{equation}
    \label{eqn:proof_lemma:exact_recovery_alg:worker_clustering_v3}
    \begin{split}
        \textnormal{(T1)} \geq \ & - \left\| \mathcal{P}_{\mathcal{U}} \left( \mathbf{A} - \mathbb{E} \left[ \left. \mathbf{A} \right| \mathbf{w} \right] \right) \right\|_{\infty} \cdot \left\| \mathbf{X}^* - \mathbf{X} \right\|_1 \\
        \geq \ & - \gamma_1 \cdot \sqrt{r} \log n \cdot \left\| \mathbf{X}^* - \mathbf{X} \right\|_1.
    \end{split}
\end{equation}

\subsubsection{Lower bound of (T2)} We first remark that the ground-truth worker cluster matrix $\mathbf{X}^*$ has a rank-$d$ singular value decomposition $\mathbf{X}^* = \mathbf{U} \mathbf{\Sigma} \mathbf{U}^{\top}$, where $\mathbf{\Sigma}$ is the $d \times d$ diagonal matrix whose entries are given by $\Sigma_{zz} = s_z$ for $z \in [d]$. By making use of \cite[Example 2]{watson1992characterization}, the sub-differential of the nuclear norm $\left\| \cdot \right\|_*$ at $\mathbf{X}^*$ can be written as
\begin{equation}
    \label{eqn:proof_lemma:exact_recovery_alg:worker_clustering_v4}
    \begin{split}
    &\partial \left\| \mathbf{X}^* \right\|_* = \\
    &\left\{ \mathbf{M} \in \mathbb{R}^{n \times n}: \mathcal{P}_{\mathcal{U}} (\mathbf{M}) = \mathbf{U} \mathbf{U}^{\top} \textnormal{ and } \left\| \mathcal{P}_{\mathcal{U}^{\perp}} (\mathbf{M}) \right\| \leq 1 \right\}.
    \end{split}
\end{equation}
It follows that for every $\mathbf{X} \in \mathcal{X}$,
\begin{equation}
    \label{eqn:proof_lemma:exact_recovery_alg:worker_clustering_v5}
    \begin{split}
        0 &= \textnormal{Trace}(\mathbf{X}) - \textnormal{Trace} \left( \mathbf{X}^* \right) \\
        &\stackrel{\textnormal{(a)}}{=} \left\| \mathbf{X} \right\|_* - \left\| \mathbf{X}^* \right\|_* \\
        &\stackrel{\textnormal{(b)}}{\geq} \left\langle \mathbf{U} \mathbf{U}^{\top} + \mathcal{P}_{\mathcal{U}^{\perp}} \left( \frac{ \mathbf{A} - \mathbb{E} \left[ \left. \mathbf{A} \right| \mathbf{w} \right]}{\left\| \mathbf{A} - \mathbb{E} \left[ \left. \mathbf{A} \right| \mathbf{w} \right] \right\|} \right), \mathbf{X} - \mathbf{X}^* \right\rangle,
    \end{split}
\end{equation}
where the step (a) holds since both $\mathbf{X}$ and $\mathbf{X}^*$ are $n \times n$ positive semi-definite matrices, and the step (b) follows from the fact
\begin{equation*}
    \mathbf{U} \mathbf{U}^{\top} + \mathcal{P}_{\mathcal{U}^{\perp}} \left( \frac{ \mathbf{A} - \mathbb{E} \left[ \left. \mathbf{A} \right| \mathbf{w} \right]}{\left\| \mathbf{A} - \mathbb{E} \left[ \left. \mathbf{A} \right| \mathbf{w} \right] \right\|} \right) \in \partial \left\| \mathbf{X}^* \right\|_*,
\end{equation*}
which can easily be observed from the result \eqref{eqn:proof_lemma:exact_recovery_alg:worker_clustering_v4}. Hence, we obtain the following lower bound on (T2):
\begin{equation}
    \label{eqn:proof_lemma:exact_recovery_alg:worker_clustering_v6}
    \begin{split}
        \textnormal{(T2)} &= \left\langle \mathcal{P}_{\mathcal{U}^{\perp}} \left( \mathbf{A} - \mathbb{E} \left[ \left. \mathbf{A} \right| \mathbf{w} \right] \right), \mathbf{X}^* - \mathbf{X} \right\rangle \\
        &\stackrel{\textnormal{(c)}}{\geq} - \left\| \mathbf{A} - \mathbb{E} \left[ \left. \mathbf{A} \right| \mathbf{w} \right] \right\| \cdot \left\langle \mathbf{U} \mathbf{U}^{\top}, \mathbf{X}^* - \mathbf{X} \right\rangle \\
        &\stackrel{\textnormal{(d)}}{\geq} - \left\| \mathbf{A} - \mathbb{E} \left[ \left. \mathbf{A} \right| \mathbf{w} \right] \right\| \cdot \left\| \mathbf{U} \mathbf{U}^{\top} \right\|_{\infty} \cdot \left\| \mathbf{X}^* - \mathbf{X} \right\|_{1} \\
        &\stackrel{\textnormal{(e)}}{\geq} - \frac{1}{s_{\min}} \left\| \mathbf{A} - \mathbb{E} \left[ \left. \mathbf{A} \right| \mathbf{w} \right] \right\| \cdot \left\| \mathbf{X}^* - \mathbf{X} \right\|_{1},
    \end{split}
\end{equation}
where $s_{\min} := \min \left\{ s_z : z \in [d] \right\}$ denotes the minimum size of the worker clusters. We note that the step (c) follows from the bound \eqref{eqn:proof_lemma:exact_recovery_alg:worker_clustering_v5}, the step (d) holds by the H\"{o}lder's inequality, and the step (e) is a consequence of the fact
\begin{equation*}
    \left[ \mathbf{U} \mathbf{U}^{\top} \right]_{jk} = 
    \begin{cases}
        \frac{1}{s_z} & \textnormal{if } j, k \in \mathcal{W}_z,\ z \in [d]; \\
        0 & \textnormal{otherwise.}
    \end{cases}
\end{equation*}
From \eqref{eqn:proof_lemma:exact_recovery_alg:worker_clustering_v6}, it suffices to develop a sharp concentration result for the spectral norm $\left\| \mathbf{A} - \mathbb{E} \left[ \left. \mathbf{A} \right| \mathbf{w} \right] \right\|$ of the centered similarity matrix. Due to the strong dependency between entries of the similarity matrix $\mathbf{A}$, we cannot employ the standard techniques from the random matrix theory literature mostly assuming the independence between entries of the data matrix. In order to establish a tight probabilistic bound on the spectral norm $\left\| \mathbf{A} - \mathbb{E} \left[ \left. \mathbf{A} \right| \mathbf{w} \right] \right\|$, we utilize an extensively used matrix concentration inequality, known as the \emph{matrix Bernstein's inequality} \cite{tropp2012user}. Now, we present a desirable concentration bound on the spectral norm of the centered similarity matrix whose proof will be presented in Appendix \ref{subsec:proof_lemma:proof_lemma:exact_recovery_alg:worker_clustering_v2}:

\begin{lemma}
\label{lemma:proof_lemma:exact_recovery_alg:worker_clustering_v2}
For the $d$-type worker-task specialization model $\textnormal{SM} \left( d; \mathcal{Q}, \bm{\mu}, \bm{\nu} \right)$, there exists an absolute constant $\gamma_2 > 0$ such that with probability at least $1 - 2 n^{-11}$, the similarity matrix $\mathbf{A}$ obeys the spectral norm bound
\begin{equation}
    \label{eqn:lemma:proof_lemma:exact_recovery_alg:worker_clustering_v2_v1}
    \left\| \mathbf{A} - \mathbb{E} \left[ \left. \mathbf{A} \right| \mathbf{w} \right] \right\| \leq \gamma_2 \cdot \sqrt{r} n \log n.
\end{equation}
\end{lemma}

\indent Applying Lemma \ref{lemma:proof_lemma:exact_recovery_alg:worker_clustering_v2} to the lower bound \eqref{eqn:proof_lemma:exact_recovery_alg:worker_clustering_v6} of the second term (T2) yields
\begin{equation}
    \label{eqn:proof_lemma:exact_recovery_alg:worker_clustering_v7}
    \textnormal{(T2)} \geq - \gamma_2 \cdot \sqrt{r} \left( \frac{n}{s_{\min}} \right) \log n \cdot \left\| \mathbf{X}^* - \mathbf{X} \right\|_1,
\end{equation}
with probability higher than $1 - 2 n^{-11}$.
\medskip

\subsubsection{Lower bound of (T3)}
One can easily see that for each $i \in \mathcal{S}$,
\begin{equation}
    \label{eqn:proof_lemma:exact_recovery_alg:worker_clustering_v8}
    \begin{split}
       & \mathbb{E} \left[ \left. A_{jk}^{(i)} \right| \mathbf{t}, \mathbf{w} \right] \\
      &=  \begin{cases}
            0 & \textnormal{if } j = k; \\
            ( 2 \mathcal{Q} \left( t_i, w_j \right) - 1 ) ( 2 \mathcal{Q} \left( t_i, w_k \right) - 1 ) & \textnormal{otherwise.}
        \end{cases}
    \end{split}
\end{equation}
By taking expectations with respect to $\mathbf{t} \sim \bm{\mu}^{\otimes m}$ to both sides of \eqref{eqn:proof_lemma:exact_recovery_alg:worker_clustering_v8}, we reach
\begin{equation}
    \label{eqn:proof_lemma:exact_recovery_alg:worker_clustering_v9}
    \begin{split}
        &\mathbb{E} \left[ \left. A_{jk}^{(i)} \right| \mathbf{w} \right] = \mathbb{E}_{\mathbf{t} \sim \bm{\mu}^{\otimes m}} \left[ \mathbb{E} \left[ \left. A_{jk}^{(i)} \right| \mathbf{t}, \mathbf{w} \right] \right] \\
        &= 
        \begin{cases}
            0 & \textnormal{if } j = k; \\
            \sum_{t=1}^{d} \mu (t) ( 2 \mathcal{Q} \left( t, w_j \right) - 1)( 2 \mathcal{Q} \left( t, w_k \right) - 1 )& \textnormal{o.w.} 
        \end{cases}
        \\
        &= 
        \begin{cases}
            0 & \textnormal{if } j = k; \\
            \Phi \left( \mathcal{Q}, \bm{\mu}, \bm{\nu} \right) \left( w_j, w_k \right) & \textnormal{o.w.,} 
        \end{cases}
    \end{split}
\end{equation}
for every $i \in \mathcal{S}$. From the definition of $p_m$ and $p_u$ described in Assumption \ref{assumption:strong_ass_cqcm}, one can observe that
\begin{equation}
    \label{eqn:proof_lemma:exact_recovery_alg:worker_clustering_v10}
    \begin{split}
        \mathbb{E} \left[ \left. A_{jk} \right| \mathbf{w} \right]
        &= r \cdot \Phi \left( \mathcal{Q}, \bm{\mu}, \bm{\nu} \right) \left( w_j, w_k \right) \\
        &
        \begin{cases}
            \geq r p_m & \textnormal{if } j \neq k \textnormal{ and } w_j = w_k; \\
            \leq r p_u & \textnormal{if } w_j \neq w_k.
        \end{cases}
    \end{split}
\end{equation}
On top of that, we know $X_{jk}^* = 1$ if and only if $w_j = w_k$ due to the definition of the ground-truth worker cluster matrix $\mathbf{X}^*$. So it can be shown that
\begin{equation}
    \label{eqn:proof_lemma:exact_recovery_alg:worker_clustering_v11}
    \begin{split}
        \textnormal{(T3)} = \ & \sum_{\substack{j, k \in [n]: \\ j \neq k}} \left( \mathbb{E} \left[ \left. A_{jk} \right| \mathbf{w} \right] - \eta \right) \left( X_{jk}^* - X_{jk} \right) \\
        &+ \sum_{j=1}^{n} \left( - \eta \right) \underbrace{\left( X_{jj}^* - X_{jj} \right)}_{= \ 0} \\
        \stackrel{\textnormal{(f)}}{\geq} \ & \sum_{\substack{j, k \in [n]: \\ j \neq k,\ X_{jk}^* = 1}} \left( rp_m - \eta \right) \left( 1 - X_{jk} \right) \\
        &+ \sum_{\substack{j, k \in [n]: \\ j \neq k,\ X_{jk}^* = 0}} \left( rp_u - \eta \right) \left( - X_{jk} \right) \\
        \stackrel{\textnormal{(g)}}{\geq} \ & \frac{1}{4} r \left( p_m - p_u \right) \sum_{\substack{j, k \in [n]: \\ j \neq k}} \left| X_{jk}^* - X_{jk} \right| \\
        \stackrel{\textnormal{(h)}}{=} \ & \frac{1}{4} r \left( p_m - p_u \right) \left\| \mathbf{X}^* - \mathbf{X} \right\|_1,
    \end{split}
\end{equation}
where the step (f) follows from the observation \eqref{eqn:proof_lemma:exact_recovery_alg:worker_clustering_v10} together with the fact $X_{jj} = 1$, $j \in [n]$, the step (g) is due to the condition \eqref{eqn:tuning_parameter_condition}, and the step (h) holds since $X_{jj} = 1$, $j \in [n]$.

\indent Taking three results \eqref{eqn:proof_lemma:exact_recovery_alg:worker_clustering_v3}, \eqref{eqn:proof_lemma:exact_recovery_alg:worker_clustering_v7}, and \eqref{eqn:proof_lemma:exact_recovery_alg:worker_clustering_v11} collectively into the decomposition \eqref{eqn:proof_lemma:exact_recovery_alg:worker_clustering_v2}, the union bound leads to the following conclusion: with probability greater than $1 - 4 n^{-11}$, 
\begin{equation}
    \label{eqn:proof_lemma:exact_recovery_alg:worker_clustering_v12}
    \begin{split}
        \Delta (\mathbf{X})
        \geq \ & \left\{ \frac{1}{4} r \left( p_m - p_u \right) - \gamma_1 \cdot \sqrt{r} \log n \right. \\
        &\left. - \gamma_2 \cdot \sqrt{r} \left( \frac{n}{s_{\min}} \right) \log n \right\} \left\| \mathbf{X}^* - \mathbf{X} \right\|_1.
    \end{split}
\end{equation}
Due to the main condition \eqref{eqn:r_condition} of Lemma \ref{lemma:exact_recovery_alg:worker_clustering}, it can be seen that
\begin{equation}
    \label{eqn:proof_lemma:exact_recovery_alg:worker_clustering_v13}
    \begin{split}
        &\gamma_1 \cdot \sqrt{r} \log n + \gamma_2 \cdot \sqrt{r} \left( \frac{n}{s_{\min}} \right) \log n \\
        \leq \ & \left( \gamma_1 + \gamma_2 \right) \cdot \sqrt{r} \frac{n \log n}{s_{\min}} \\
        \leq \ & \frac{\gamma_1 + \gamma_2}{\sqrt{C_2}} \cdot r \left( p_m - p_u \right).
    \end{split}
\end{equation}
Therefore, plugging \eqref{eqn:proof_lemma:exact_recovery_alg:worker_clustering_v13} into \eqref{eqn:proof_lemma:exact_recovery_alg:worker_clustering_v12}, we get
\begin{equation}
    \label{eqn:proof_lemma:exact_recovery_alg:worker_clustering_v14}
    \Delta (\mathbf{X}) \geq  \left( \frac{1}{4} - \frac{\gamma_1 + \gamma_2}{\sqrt{C_2}} \right) r \left( p_m - p_u \right) \left\| \mathbf{X}^* - \mathbf{X} \right\|_1
\end{equation}
with probability greater than $1 - 4 n^{-11}$. By choosing the universal constant $C_2 > 0$ to be sufficiently large so that
\begin{equation*}
    C_2 \geq 64 \left( \gamma_1 + \gamma_2 \right)^2
\end{equation*}
we may conclude that with probability higher than $1 - 4 n^{-11}$, 
\begin{equation}
    \label{eqn:proof_lemma:exact_recovery_alg:worker_clustering_v15}
    \Delta (\mathbf{X}) \geq \frac{1}{8} r \left( p_m - p_u \right) \left\| \mathbf{X}^* - \mathbf{X} \right\|_1
\end{equation}
for every $\mathbf{X} \in \mathcal{X}$, thereby the final inequality \eqref{eqn:proof_lemma:exact_recovery_alg:worker_clustering_v15} implies $\Delta (\mathbf{X}) > 0$ for every $\mathbf{X} \in \mathcal{X} \setminus \left\{ \mathbf{X}^* \right\}$ as desired.

\section{Theoretical analysis of the standard majority voting estimator \eqref{eqn:standard_mv}}
\label{sec:theoretical_analysis_standard_mv}

Throughout this section, we present a performance guarantee of the standard majority voting estimator \eqref{eqn:standard_mv} under the $d$-type worker-task specialization model $\textnormal{SM}\left( d; \mathcal{Q}, \bm{\mu}, \bm{\nu} \right)$. If we consider the empirical distribution $\hat{\mathcal{P}}_{\mathbf{w}_{\mathcal{A}(i)}} (\cdot) \in \Delta ([d])$ of the vector $\mathbf{w}_{\mathcal{A}(i)} := \left( w_j : j \in \mathcal{A}(i) \right) \in [d]^{\mathcal{A}(i)}$, then
\begin{equation*}
    \begin{split}
        \mathbb{E} \left[ \hat{\mathcal{P}}_{\mathbf{w}_{\mathcal{A}(i)}} (w) \right]
        = \ & \mathbb{E} \left[ \frac{1}{\left| \mathcal{A}(i) \right|} \sum_{j \in \mathcal{A}(i)}
        \mathbbm{1} \left( w_j = w \right) \right] \\
        = \ & \frac{1}{\left| \mathcal{A}(i) \right|} \sum_{j \in \mathcal{A}(i)} \mathbb{P} \left\{ w_j = w \right\} \\
        = \ & \nu(w)
    \end{split}
\end{equation*}
for every $w \in [d]$. With the above observation in place, one can expect that the value of $\hat{\mathcal{P}}_{\mathbf{w}_{\mathcal{A}(i)}} (w)$ is concentrated around its expectation $\mathbb{E} \left[ \hat{\mathcal{P}}_{\mathbf{w}_{\mathcal{A}(i)}} (w) \right] = \nu (w)$ for all $w \in [d]$ with high probability. This argument can be established via a powerful technique in large deviation theory, known as the method of types \cite{csiszar1998method, csiszar2011information}, which serves as a bridge between information theory and statistics. To this end, we consider the set
\begin{equation*}
    \mathcal{D}(\bm{\nu}) := \left\{ \bm{\delta} \in \Delta ([d]) : \delta (w) \geq \frac{1}{2} \nu (w),\ \forall w \in [d] \right\}.
\end{equation*}
Then, it can be easily justified that the empirical distribution of the vector $\mathbf{w}_{\mathcal{A}(i)} := \left( w_j : j \in \mathcal{A}(i) \right)$ belongs to 
the set $\mathcal{D}(\bm{\nu})$, \emph{i.e.}, $\hat{\mathcal{P}}_{\mathbf{w}_{\mathcal{A}(i)}} (\cdot) \in \mathcal{D}(\bm{\nu})$, with high probability by applying the Sanov's theorem \cite{sanov1958probability}. So, we will continue the performance analysis of the standard majority voting estimator \eqref{eqn:standard_mv} under the $d$-type worker-task specialization model under the condition $\hat{\mathcal{P}}_{\mathbf{w}_{\mathcal{A}(i)}} (\cdot) \in \mathcal{D}(\bm{\nu})$, which holds with high probability.

\begin{prop} [Statistical analysis of the standard majority voting rule]
\label{prop:performance_standard_mv}
For the $d$-type worker-task specialization model $\textnormal{SM} \left( d; \mathcal{Q}, \bm{\mu}, \bm{\nu} \right)$, the desired recovery performance \eqref{eqn:expected_accuracy} is achievable via the standard majority voting rule \eqref{eqn:standard_mv} provided that:
\begin{enumerate}
    \item the empirical distribution of the vector $\mathbf{w}_{\mathcal{A}(i)} := \left( w_j : j \in \mathcal{A}(i) \right)$ belongs to the set $\mathcal{D}(\bm{\nu})$, i.e., $\hat{\mathcal{P}}_{\mathbf{w}_{\mathcal{A}(i)}} (\cdot) \in \mathcal{D}(\bm{\nu})$ for every $i \in [m]$;
    \item the worker-task assignment set $\mathcal{A} \subseteq [m] \times [n]$ satisfies
    \begin{equation}
        \label{eqn:prop:performance_standard_mv_v1}
        \min_{i \in [m]} \left| \mathcal{A}(i) \right| \geq \frac{1}{\min_{t \in [d]} \theta_2 \left( t; \mathcal{Q} \right)} \log \left( \frac{1}{\alpha} \right)
    \end{equation}
    for any given $\alpha \in \left( 0, \frac{1}{2} \right]$ ($\alpha$ may depend on $m$), where $\theta_2 \left( \cdot; \mathcal{Q} \right) : [d] \rightarrow \mathbb{R}_{+}$ is defined by
    \begin{equation}
        \label{eqn:prop:performance_standard_mv_v2}
        \theta_2 \left( t; \mathcal{Q} \right) := \frac{1}{8} \left[ \sum_{w=1}^{d} \nu (w) \left\{ 2 \mathcal{Q}(t, w) - 1 \right\} \right]^2,\ \forall t \in [d].
    \end{equation}
\end{enumerate}
\end{prop}

\begin{proof} [Proof of Proposition \ref{prop:performance_standard_mv}]
\indent Let $\hat{\mathbf{a}}^{\textnormal{MV}} = \hat{\mathbf{a}}^{\textnormal{MV}} (\mathbf{M}) : \left\{ \pm 1 \right\}^{\mathcal{A}} \rightarrow \left\{ \pm 1 \right\}^m$ be the standard majority voting estimator:
\begin{equation}
    \label{eqn:proof_prop:performance_standard_mv_v1}
    \begin{split}
        \hat{a}_{i}^{\textnormal{MV}} (\mathbf{M}) := \ & \textnormal{sign} \left( \sum_{j \in \mathcal{A}(i)} M_{ij} \right) \\
        &= \textnormal{sign} \left( a_i \sum_{j \in \mathcal{A}(i)} \left( 2 \Lambda_{ij} - 1 \right) \right),
    \end{split}
\end{equation}
where $\left\{ \Lambda_{ij} : (i, j) \in \mathcal{A} \right\}$ is a collection of conditionally independent random variables given a pair of type vectors $\left( \mathbf{t}, \mathbf{w} \right)$ such that $\Lambda_{ij} \sim \textnormal{Bern} \left( F_{ij} \right)$ for every $(i, j) \in \mathcal{A}$. Then for each $i \in [m]$, 
\begin{equation}
    \label{eqn:proof_prop:performance_standard_mv_v2}
    \begin{split}
        &\mathbb{P} \left\{ \left. \hat{a}_{i}^{\textnormal{MV}} (\mathbf{M}) \neq a_i \right| \left( \mathbf{t}, \mathbf{w} \right) \right\} \\
        = \ & \mathbb{P} \left\{ \left. \sum_{j \in \mathcal{A}(i)} \left( 2 \Lambda_{ij} - 1 \right) \leq 0 \right| \left( \mathbf{t}, \mathbf{w} \right) \right\} \\
        = \ & \mathbb{P} \left\{ \left. \sum_{j \in \mathcal{A}(i)} \left( \Lambda_{ij} - F_{ij} \right) \leq - \sum_{j \in \mathcal{A}(i)} \left( F_{ij} - \frac{1}{2} \right) \right| \left( \mathbf{t}, \mathbf{w} \right) \right\} \\
        \stackrel{\textnormal{(a)}}{\leq} \ & 
        \exp \left[ - \frac{\left\{ \sum_{j \in \mathcal{A}(i)} \left( 2 F_{ij} - 1 \right) \right\}^2}{2 \left| \mathcal{A}(i) \right|} \right] \\
        = \ & \exp \left[ - \frac{\left| \mathcal{A}(i) \right|}{2} \cdot \left\{ \frac{\sum_{j \in \mathcal{A}(i)} \left( 2 \mathcal{Q} \left( t_i, w_j \right) - 1 \right)}{\left| \mathcal{A} (i) \right|} \right\}^2 \right] \\
        = \ & \exp \left[ - \frac{\left| \mathcal{A}(i) \right|}{2} \cdot \left\{ \sum_{w=1}^{d} \hat{\mathcal{P}}_{\mathbf{w}_{\mathcal{A}(i)}} (w) \left( 2 \mathcal{Q} \left( t_i, w \right) - 1 \right) \right\}^2 \right] \\
        \stackrel{\textnormal{(b)}}{\leq} \ & \exp \left[ - \frac{\left| \mathcal{A}(i) \right|}{8} \cdot \left\{ \sum_{w=1}^{d} \nu(w) \left( 2 \mathcal{Q} \left( t_i, w \right) - 1 \right) \right\}^2 \right],
    \end{split}
\end{equation}
where the step (a) holds by the Hoeffding bound and the step (b) is due to the condition (\romannumeral 1): $\hat{\mathcal{P}}_{\mathbf{w}_{\mathcal{A}(i)}} (\cdot) \in \mathcal{D}$ of Proposition \ref{prop:performance_standard_mv}.
Thus, we obtain from \eqref{eqn:proof_prop:performance_standard_mv_v2} that
\begin{equation}
    \label{eqn:proof_prop:performance_standard_mv_v3}
    \begin{split}
        &\mathbb{P} \left\{ \left. \hat{a}_{i}^{\textnormal{MV}} (\mathbf{M}) \neq a_i \right| \left( \mathbf{t}, \mathbf{w} \right) \right\} \\
        \leq \ & \exp \left[ - \frac{\left| \mathcal{A}(i) \right|}{8} \cdot \left\{ \sum_{w=1}^{d} \nu(w) \left( 2 \mathcal{Q} \left( t_i, w \right) - 1 \right) \right\}^2 \right] \\
        = \ & \exp \left\{ - \left| \mathcal{A}(i) \right| \cdot \theta_2 \left( t_i; \mathcal{Q} \right) \right\}.
    \end{split}
\end{equation}
By taking expectation to the inequality \eqref{eqn:proof_prop:performance_standard_mv_v3} with respect to $\left( \mathbf{t}, \mathbf{w} \right) \sim \bm{\mu}^{\otimes m} \otimes \bm{\nu}^{\otimes n}$, we find that
\begin{equation}
    \label{eqn:proof_prop:performance_standard_mv_v4}
    \begin{split}
        &\mathbb{P} \left\{ \hat{a}_{i}^{\textnormal{MV}} (\mathbf{M}) \neq a_i \right\} \\
        = \ & \mathbb{E}_{\left( \mathbf{t}, \mathbf{w} \right) \sim \bm{\mu}^{\otimes m} \otimes \bm{\nu}^{\otimes n}} \left[ \mathbb{P} \left\{ \left. \hat{a}_{i}^{\textnormal{MV}}(\mathbf{M}) \neq a_i \right| \left( \mathbf{t}, \mathbf{w} \right) \right\} \right] \\
        \leq \ & \mathbb{E}_{\left( \mathbf{t}, \mathbf{w} \right) \sim \bm{\mu}^{\otimes m} \otimes \bm{\nu}^{\otimes n}} \left[ \exp \left\{ - \left| \mathcal{A}(i) \right| \cdot \theta_2 \left( t_i; \mathcal{Q}, \bm{\mu}, \bm{\nu} \right) \right\} \right] \\
        = \ & \sum_{t=1}^{d} \mu (t) \exp \left\{ - \left| \mathcal{A}(i) \right| \cdot \theta_2 \left( t; \mathcal{Q} \right) \right\} \\
        \leq \ & \exp \left\{ - \left| \mathcal{A}(i) \right| \cdot \min_{t \in [d]} \theta_2 \left( t; \mathcal{Q} \right) \right\}.
    \end{split}
\end{equation}
So in order to achieve the desired recovery performance
\begin{equation*}
    \mathcal{R} \left( \mathbf{a}, \hat{\mathbf{a}}^{\textnormal{MV}}(\mathbf{M}) \right) =
    \frac{1}{m} \sum_{i=1}^{m} \mathbb{P} \left\{ \hat{a}_{i}^{\textnormal{MV}}(\mathbf{M}) \neq a_i \right\} \leq \alpha
\end{equation*}
for any target recovery accuracy $\alpha \in \left( 0, \frac{1}{2} \right]$, it suffices to have $\left| \mathcal{A}(i) \right|$ workers to the $i$-th task, where
\begin{equation*}
    \left| \mathcal{A}(i) \right| \geq \frac{1}{\min_{t \in [d]} \theta_2 \left( t; \mathcal{Q} \right)} \log \left( \frac{1}{\alpha} \right),
\end{equation*}
for all $i \in [m]$. This completes the proof of Proposition \ref{prop:performance_standard_mv}.

\end{proof}

\section{Detailed description of the subset-selection algorithm}
\label{sec:detailed_description_SS_algorithm}

\indent The second baseline estimation we consider is the type-dependent subset-selection algorithm \cite{shah2018reducing}. The basic idea is to exploit the responses only from the workers whose type matches the type of the given task, which are believed to be more reliable than the rest of the answers provided by the workers of mismatched types by Assumption \ref{assumption:weak_ass_rm} in the $d$-type specialization model. Since neither task types nor worker types are known, the main challenge is to estimate the type $t_i$ associated with the $i$-th task and the subset of workers among $\mathcal{A}(i)$ whose type match the inferred task type $\hat{t}_i := \hat{t}_i (\mathbf{M})$, denoted by $\mathcal{A}_{\hat{t}_i}(i)$. Then, the ground-truth label $a_i$ is inferred by running the standard majority voting using only the answers given by the \emph{workers of the matched type}:
    \begin{equation}
        \label{eqn:SS_decision_rule} \hat{a}_{i}^{\textnormal{SS}}(\mathbf{M}) := \textnormal{sign} \left( \sum_{j \in \mathcal{A}_{\hat{t}_i}(i)} M_{ij} \right),\ \forall i \in [m].
    \end{equation}
The algorithm from \cite{shah2018reducing} for revealing $\mathcal{A}_{\hat{t}_i}(i)$ is summarized below.

\begin{alg*} [The subset-selection algorithm \cite{shah2018reducing}]
\label{alg:subset_selection}
\normalfont{ \
\begin{enumerate} [label=\arabic*.]
    \item[] \textbf{Input}: the noisy answers $\mathbf{M} := \left( M_{ij}: (i, j) \in \mathcal{A} \right)$, the number of types $d \in \mathbb{N}$, and the parameters $\left( r, \eta, l \right) \in [m] \times \left( 0, +\infty \right) \times [n]$;
    \item \textbf{Sequential worker clustering}:
    \begin{enumerate} [label=(\alph*)]
        \item Choose a set of $r$ tasks $\mathcal{S} \subseteq [m]$ and assign each task in $\mathcal{S}$ to all $n$ workers;
        \item Next, cluster workers sequentially by comparing the similarity on the responses between every pair of workers: for the $j$-th worker, if there is a cluster of workers $\mathcal{C} \subseteq \left[ j-1 \right]$ such that for every $j' \in \mathcal{C}$,
        \begin{equation}
            \label{eqn:SS_sequential_clustering}
            \frac{1}{r} \sum_{i \in \mathcal{S}} \mathbbm{1} \left( M_{ij} = M_{ij'} \right) > \xi,
        \end{equation}
        where $\xi > 0$ is a certain threshold that should be pre-determined, then assign the $j$-th worker to the cluster $\mathcal{C}$; otherwise, we create a new cluster containing $\left\{ j \right\}$;
        \item Let $\left\{ \hat{\mathcal{W}}_1, \hat{\mathcal{W}}_2, \cdots, \hat{\mathcal{W}}_c \right\}$ denote the resulting worker clusters, which form a disjoint partition of $[n]$. Assign task $i \in [m] \setminus \mathcal{S}$ to $l$ randomly sampled workers from each inferred worker cluster;
    \end{enumerate}
    \item \textbf{Task-type matching and label inference via standard majority voting}:
    \begin{enumerate} [label=(\alph*)]
        \item For every task $i \in [m]$, we select $\mathcal{A}_z (i) \in \binom{\mathcal{A} (i) \cap \hat{\mathcal{W}}_z}{l}$ for $z \in [c]$;
        \item We infer the types associated with each task via the following decision rule: for each task $i \in [m]$, the type associated with the $i$-th task is inferred by finding the cluster whose response is the most biased:
        \begin{equation}         
            \label{eqn:SS_task_type_matching}
            \hat{t}_i := \hat{t}_i (\mathbf{M}) = \argmax_{z \in [c]} \left| \sum_{j \in \mathcal{A}_{z} (i)} M_{ij} \right|.
        \end{equation}
        \item Finally, we infer the ground-truth label $a_{i}$ of the $i$-th task by performing the standard majority voting rule solely based on the answers provided by workers belonging to the set $\mathcal{A}_{\hat{t}_i}(i) \subseteq \hat{\mathcal{W}}_{\hat{t}_i}$ as in \eqref{eqn:SS_decision_rule};
    \end{enumerate}
    \item[] \textbf{Output}: $\hat{\mathbf{a}}^{\textnormal{SS}} (\cdot): \left\{ \pm 1 \right\}^{\mathcal{A}} \rightarrow \left\{ \pm 1 \right\}^m$.
\end{enumerate}
}
\end{alg*}

\section{Proof of Proposition \ref{prop:performance_SS_Q_in_M1}}
\label{sec:proof_prop:performance_SS_Q_in_M1}

\indent We proceed in a similar way as the proof of \emph{Theorem 3.1} in \cite{shah2018reducing} but extend it to a general $d$-type specialization model whose reliability matrix $\mathcal{Q}$  satisfy \ref{assumption:strong_ass_cqcm} and \ref{assumption:weak_ass_rm}, rather than it is restricted to the special $(p,q)$ model with $p>q=1/2$. First we discuss the consequence of the sequential clustering stage of workers by their types. Let $\mathcal{E}_1$ denote the event that Step 1 of the subset-selection algorithm (see Section \ref{sec:detailed_description_SS_algorithm} for detailed descriptions) exactly recovers the worker clusters, \emph{i.e.},
\begin{equation*}
    \mathcal{E}_1 := \left\{ c = d \textnormal{ and } \hat{\mathcal{W}}_z = \mathcal{W}_z \textnormal{ for every } z \in [d] \right\}.
\end{equation*}
For any $i \in \mathcal{S}$ and $a \neq b$ in $[n]$, we know 
\begin{equation}
    \label{eqn:proof_prop:performance_SS_Q_in_M1_v1}
    \begin{split}
       & \mathbb{P} \left\{ \left. M_{ia} = M_{ib} \right| \mathbf{t}, \mathbf{w} \right\} \\
       &=  \mathcal{Q} (t_i, w_a) \mathcal{Q} (t_i, w_b)+ \left\{ 1 - \mathcal{Q} (t_i, w_a) \right\} \left\{ 1 - \mathcal{Q} (t_i, w_b) \right\},
    \end{split}
\end{equation}
so we obtain
\begin{equation}
    \label{eqn:proof_prop:performance_SS_Q_in_M1_v2}
    \begin{split}
        &\mathbb{P} \left\{ \left. M_{ia} = M_{ib} \right|  \mathbf{w} \right\} \\
        = \ & \mathbb{E}_{\mathbf{t} \sim \bm{\mu}^{\otimes m}} \left[ \mathbb{P} \left\{ \left. M_{ia} = M_{ib} \right| \mathbf{t}, \mathbf{w} \right\} \right] \\
        = \ & \sum_{t=1}^{d} \mu (t) \big[ \mathcal{Q} (t, w_a) \mathcal{Q} (t, w_b) \\
        &\quad+ \left\{ 1 - \mathcal{Q} (t, w_a) \right\} \left\{ 1 - \mathcal{Q} (t, w_b) \right\} \big].
    \end{split}
\end{equation}
Given any reliability matrix $\mathcal{Q} (\cdot, \cdot) : [d] \times [d] \rightarrow \left[ \frac{1}{2}, 1 \right]$, we define $\Lambda \left( \mathcal{Q}, \bm{\mu},\bm{\nu} \right) (\cdot, \cdot): [d] \times [d] \rightarrow \mathbb{R}_{+}$ by
\begin{equation*}
    \begin{split}
        &\Lambda \left( \mathcal{Q}, \bm{\mu},\bm{\nu} \right) \left( w, w' \right):=  \\
        & \sum_{t=1}^{d} \mu (t) \left[ \mathcal{Q} (t, w) \mathcal{Q} (t, w') + \{ 1 - \mathcal{Q} (t, w) \} \{ 1 - \mathcal{Q} (t, w')\} \right].
    \end{split}
\end{equation*}
As a next stage, we establish the conditional independence of $\left\{ \mathbf{M}_{i*} := \left( M_{ij} : j \in [n] \right) : i \in \mathcal{S} \right\}$ given a worker type vector $\mathbf{w}$, and we defer its proof to Appendix \ref{subsec:proof_lemma:proof_prop:performance_SS_Q_in_M1_v1}.

\begin{lemma}
\label{lemma:proof_prop:performance_SS_Q_in_M1_v1}
Let $\mathbf{M}_{i*} := \left( M_{ij} : j \in [n] \right)$ be the aggregation of answers provided by all workers for each $i \in \mathcal{S}$. Then, $\left\{ \mathbf{M}_{i*} : i \in \mathcal{S} \right\}$ are conditionally independent random vectors given a worker type vector $\mathbf{w} \in [d]^n$.
\end{lemma}

\indent With Lemma \ref{lemma:proof_prop:performance_SS_Q_in_M1_v1}, one can see that $\left\{ \mathbbm{1} \left( M_{ia} = M_{ib} \right) : i \in \mathcal{S} \right\}$ are independent and identically distributed, conditionally given a worker type vector $\mathbf{w}$. So we arrive at
\begin{equation}
    \label{eqn:proof_prop:performance_SS_Q_in_M1_v3}
    \begin{split}
        &\mathbb{P} \left\{ \left. \mathcal{E}_1 \right| \mathbf{w} \right\} \\
        = \ & \mathbb{P} \Biggl\{ \Bigg[ \bigcap_{\substack{\left\{ a, b \right\} \in \binom{[n]}{2} \\ : w_a = w_b}} \left\{ \frac{1}{r} \sum_{i \in \mathcal{S}} \mathbbm{1} \left( M_{ia} = M_{ib} \right) > \xi \right\} \Bigg]  \\
        &\quad\cap \Bigg[ \bigcap_{\substack{\left\{ a, b \right\} \in \binom{[n]}{2} \\ : w_a \neq w_b}} \left\{ \frac{1}{r} \sum_{i \in \mathcal{S}} \mathbbm{1} \left( M_{ia} = M_{ib} \right) \leq \xi \right\} \Bigg] \Bigg| \mathbf{w} \Biggr\} \\
        \stackrel{\textnormal{(a)}}{\geq} \ & 1 - \sum_{\substack{\left\{ a, b \right\} \in \binom{[n]}{2} \\ : w_a = w_b}} \mathbb{P} \left\{ \left. \frac{1}{r} \sum_{i \in \mathcal{S}} \mathbbm{1} \left( M_{ia} = M_{ib} \right) \leq \xi \right| \mathbf{w} \right\} \\
        &- \sum_{\substack{\left\{ a, b \right\} \in \binom{[n]}{2} \\ : w_a \neq w_b}} \mathbb{P} \left\{ \left. \frac{1}{r} \sum_{i \in \mathcal{S}} \mathbbm{1} \left( M_{ia} = M_{ib} \right) > \xi \right| \mathbf{w} \right\} \\
        \stackrel{\textnormal{(b)}}{\geq} \ & 1 - \sum_{\substack{\left\{ a, b \right\} \in \binom{[n]}{2} \\ : w_a = w_b}} \exp \left[ - 2r \left\{ \Lambda \left( \mathcal{Q}, \bm{\mu}, \bm{\nu} \right) \left( w_a, w_b \right) - \xi \right\}^2 \right] \\
        &- \sum_{\substack{\left\{ a, b \right\} \in \binom{[n]}{2} \\ : w_a \neq w_b}} \exp \left[ - 2r \left\{ \xi - \Lambda \left( \mathcal{Q}, \bm{\mu}, \bm{\nu} \right) \left( w_a, w_b \right) \right\}^2 \right],
    \end{split}
\end{equation}
where the step (a) follows from the union bound, and the step (b) comes from the Chernoff-Hoeffding theorem. One can easily see that
\begin{equation}
    \label{eqn:proof_prop:performance_SS_Q_in_M1_v4}
    \Lambda \left( \mathcal{Q}, \bm{\mu}, \bm{\nu} \right) \left( w, w' \right) = \frac{1}{2} \left\{ \Phi \left( \mathcal{Q}, \bm{\mu}, \bm{\nu} \right) \left( w, w' \right) + 1 \right\}
\end{equation}
for every $\left( w, w' \right) \in [d] \times [d]$, thereby we obtain the following fact from the strong assortativity condition of the collective quality correlation matrix $\Phi \left( \mathcal{Q}, \bm{\mu}, \bm{\nu} \right)$:
\begin{equation*}
    \begin{split}
        &\min \left\{ \Lambda \left( \mathcal{Q}, \bm{\mu}, \bm{\nu} \right) (a, a) : a \in [d] \right\} = \frac{1}{2} \left( p_m + 1 \right) \\
        &> \frac{1}{2} \left( p_u + 1 \right) = \max \left\{ \Lambda \left( \mathcal{Q}, \bm{\mu}, \bm{\nu} \right) (a, b) : a \neq b \textnormal{ in } [d] \right\}.
    \end{split}
\end{equation*}
With the aid of the above fact, we can make the following choice of tuning parameter $\xi$ to be
\begin{equation}
    \label{eqn:proof_prop:performance_SS_Q_in_M1_v5}
    \xi = \frac{1}{2} \left\{ \frac{1}{2} \left( 1 + p_m \right) + \frac{1}{2} \left( 1 + p_u \right) \right\},
\end{equation}
and accordingly the probability that every worker is exactly clustered can be bounded below by
\begin{equation}
    \label{eqn:proof_prop:performance_SS_Q_in_M1_v6}
    \mathbb{P} \left\{ \left. \mathcal{E}_1 \right| \mathbf{w} \right\} \geq 1 - \binom{n}{2} \exp \left\{ - \frac{r}{8} \left( p_m - p_u \right)^2 \right\},
\end{equation}
thereby we arrive at
\begin{equation}
    \label{eqn:proof_prop:performance_SS_Q_in_M1_v7}
    \begin{split}
        \mathbb{P} \left\{ \mathcal{E}_{1}^{c} \right\} = \ & \mathbb{E}_{\mathbf{w} \sim \bm{\nu}^{\otimes n}} \left[ \mathbb{P} \left\{ \left. \mathcal{E}_{1}^{c} \right| \mathbf{w} \right\} \right] \\
        \leq \ & \binom{n}{2} \exp \left\{ - \frac{r}{8} \left( p_m - p_u \right)^2 \right\}.
    \end{split}
\end{equation}
\medskip

\indent In order to assign each task $i \in [m] \setminus \mathcal{S}$ to $l$ uniformly and randomly sampled workers from each inferred cluster $\hat{\mathcal{W}}_z$, $z \in [c]$, it is necessary to take a closer inspection on the event that the size of $\hat{\mathcal{W}}_z$ is greater than or equal to $l$ for every $z \in [c]$. Let $\mathcal{E}_2$ denote the event that the size of $\hat{\mathcal{W}}_z$ is greater than or equal to $l$ for every $z \in [c]$, \emph{i.e.},
\begin{equation*}
    \mathcal{E}_2 := \bigcap_{z=1}^{c} \left\{ \left| \hat{\mathcal{W}}_z \right| \geq l \right\}.
\end{equation*}
While being conditioned on the event $\mathcal{E}_1$, we have $c = d$ and $\hat{\mathcal{W}}_z = \mathcal{W}_z$ for every $z \in [d]$. 
By using the bound on $ \mathbb{P} \left\{ \left. \mathcal{E}_{2}^c \right| \mathcal{E}_1 \right\}$ from \eqref{eqn:cond_prob_E2_E1},
we conclude that 
\begin{equation}
    \label{eqn:proof_prop:performance_SS_Q_in_M1_v9}
    \begin{split}
        \mathbb{P} \left\{ \mathcal{E}_{2}^c \right\} = \ & \mathbb{P} \left\{ \left. \mathcal{E}_{2}^c \right| \mathcal{E}_1 \right\} \mathbb{P} \left\{ \mathcal{E}_1 \right\} + \underbrace{\mathbb{P} \left\{ \left. \mathcal{E}_{2}^c \right| \mathcal{E}_{1}^c \right\}}_{\leq \ 1} \mathbb{P} \left\{ \mathcal{E}_{1}^c \right\} \\
        \leq \ & d \exp \left[ - \min_{b \in [d]} \left \{ \frac{n \nu (b)}{2} \left( 1 - \frac{l}{n \nu(b)} \right)^2 \right\} \right] \\
        &+ \binom{n}{2} \exp \left\{ - \frac{r}{8} \left( p_m - p_u \right)^2 \right\}.
    \end{split} 
\end{equation}

\indent Next, being conditioned on the event $\mathcal{E}_{1} \cap \mathcal{E}_{2}$, we analyze the probability of task-type matching error. To this end, we reuse auxiliary random variables $S_{ib} := \sum_{j \in \mathcal{A}_{b} (i)} \mathbbm{1} \left( M_{ij} = +1 \right)$ for each $(i, b) \in [m] \times [d]$, defined in the proof of Theorem \ref{thm:performance_alg:without_side_information} (see Section \ref{subsec:proof_thm:performance_alg:with_side_information}). Then, by following the arguments in Section \ref{subsec:proof_thm:performance_alg:without_side_information} meticulously, one can obtain the following sufficient condition for the exact recovery of the ground-truth task-types of the task-type matching rule \eqref{eqn:SS_task_type_matching} of the subset-selection algorithm. 

\begin{lemma}
    \label{lemma:proof_prop:performance_SS_Q_in_M1_v2} 
    Conditioned on the event $\mathcal{E}_1 \cap \mathcal{E}_2$, we have
    \begin{equation*}
        \begin{split}
            &\bigcap_{b=1}^{d} \left\{ \left| S_{ib} - \mathbb{E} \left[ S_{ib} \right] \right| < \frac{p^{*} \left( t_i \right) - q^{*} \left( t_i \right)}{2} l \right\} \\
            \subseteq \ & \left\{ \hat{t}_i (\mathbf{M}) = t_i \textnormal{ and } \hat{a}_{i}^{\textnormal{SS}}(\mathbf{M}) = a_i \right\}
        \end{split}
    \end{equation*}
    for every $i \in [m]$.
\end{lemma}

\indent The proof of Lemma \ref{lemma:proof_prop:performance_SS_Q_in_M1_v2} is entirely analogous to the proof of Lemma \ref{lemma:proof_thm:performance_alg:without_side_information_v1} and hence is omitted. Due to Lemma \ref{lemma:proof_prop:performance_SS_Q_in_M1_v2}, it can be shown that
\begin{equation*}
    \begin{split}
        &\mathbb{P} \left\{ \left. \hat{t}_i (\mathbf{M}) \neq t_i \right| \mathcal{E}_1 \cap \mathcal{E}_2, \left( \mathbf{t}, \mathbf{w} \right) \right\} \\
        &\leq  \mathbb{P} \Bigg\{ \bigcup_{b=1}^{d} \left\{ \left. \left| S_{ib} - \mathbb{E} \left[ S_{ib} \right] \right| \geq \frac{p^{*} \left( t_i \right) - q^{*} \left( t_i \right)}{2} l \right\} \right|\\
        &\qquad \mathcal{E}_1 \cap \mathcal{E}_2, \left( \mathbf{t}, \mathbf{w} \right) \Bigg\} \\
       & \stackrel{\textnormal{(e)}}{\leq}  \sum_{b=1}^{d} \mathbb{P} \Bigg\{ \left. \left| S_{ib} - \mathbb{E} \left[ S_{ib} \right] \right| \geq \frac{p^{*} \left( t_i \right) - q^{*} \left( t_i \right)}{2} l \right| \\
       &\qquad\qquad\mathcal{E}_1 \cap \mathcal{E}_2, \left( \mathbf{t}, \mathbf{w} \right) \Bigg\} \\
      &  \stackrel{\textnormal{(f)}}{\leq}  2d \exp \left\{ - \frac{l}{2} \left( p^{*} \left( t_i \right) - q^{*} \left( t_i \right) \right)^2 \right\},
    \end{split}
\end{equation*}
where the step (e) makes use of the union bound, and the step (f) follows from the Chernoff-Hoeffding theorem. So, by taking expectations to the above bound with respect to the pair of type vectors $\left( \mathbf{t}, \mathbf{w} \right) \sim \bm{\mu}^{\otimes m} \otimes \bm{\nu}^{\otimes n}$,
\begin{equation}
    \label{eqn:proof_prop:performance_SS_Q_in_M1_v10}
    \begin{split}
        &\mathbb{P} \left\{ \left. \hat{t}_i (\mathbf{M}) \neq t_i \right| \mathcal{E}_1 \cap \mathcal{E}_2 \right\} \\
        = \ & \mathbb{E}_{\left( \mathbf{t}, \mathbf{w} \right) \sim \bm{\mu}^{\otimes m} \otimes \bm{\nu}^{\otimes n}} \left[ \mathbb{P} \left\{ \left. \hat{t}_i (\mathbf{M}) \neq t_i \right| \mathcal{E}_1 \cap \mathcal{E}_2, \left( \mathbf{t}, \mathbf{w} \right) \right\} \right] \\
        \leq \ & 2d \exp \left\{ - \frac{l}{2} \left( p^{*} \left( t_i \right) - q^{*} \left( t_i \right) \right)^2 \right\}.
    \end{split}
\end{equation}

\indent Furthermore, given a pair of type vectors $\left( \mathbf{t}, \mathbf{w} \right)$, we have
\begin{equation*}
    M_{ij} = a_i \left( 2 \Lambda_{ij} - 1 \right),\ \forall (i, j) \in \mathcal{A},
\end{equation*}
where $\left\{ \Lambda_{ij} : (i, j) \in \mathcal{A} \right\}$ are conditionally independent random variables with $\Lambda_{ij} \sim \textnormal{Bern} \left( F_{ij} \right)$, $\forall (i, j) \in \mathcal{A}$, given a pair of type vectors $\left( \mathbf{t}, \mathbf{w} \right)$. Recall the definition of $\hat{a}_{i}^{\textnormal{SS}}(\cdot): \left\{ \pm 1 \right\}^{\mathcal{A}} \to \left\{ \pm 1 \right\}^m$:
\begin{equation*}
\begin{split}
    \hat{a}_{i}^{\textnormal{SS}}(\mathbf{M}) &:= \textnormal{sign} \left( \sum_{j \in \mathcal{A}_{\hat{t}_i} (i)} M_{ij} \right) \\
    &= \textnormal{sign} \left( a_i \sum_{j \in \mathcal{A}_{\hat{t}_i} (i)} \left( 2 \Lambda_{ij} - 1 \right) \right).
    \end{split}
\end{equation*}
Conditioned on the event $\mathcal{E}_1 \cap \mathcal{E}_2$, we have 
\begin{equation}
    \label{eqn:proof_prop:performance_SS_Q_in_M1_v11}
    F_{ij} = \mathcal{Q} \left( t_i, \hat{t}_i (\mathbf{M}) \right)
    \begin{cases}
        \geq p^{*}(t_i) & \textnormal{if } \hat{t}_i (\mathbf{M}) = t_i; \\
        \leq q^{*}(t_i) & \textnormal{otherwise,}
    \end{cases}
\end{equation}
for all $j \in \mathcal{A}_{\hat{t}_i} (i)$. Applying the Chernoff-Hoeffding theorem, we reach
\begin{equation}
    \label{eqn:proof_prop:performance_SS_Q_in_M1_v12}
    \begin{split}
        &\mathbb{P} \left\{ \left. \hat{a}_{i}^{\textnormal{SS}} (\mathbf{M}) \neq a_i \right| \left\{ \hat{t}_i (\mathbf{M}) = t_i \right\} \cap \left( \mathcal{E}_1 \cap \mathcal{E}_2 \right), (\mathbf{t}, \mathbf{w}) \right\} \\
        \leq \ & \exp \left[ - \frac{l}{2} \left\{ 2 \mathcal{Q} (t_i, t_i) - 1 \right\}^2 \right]; \\
        &\mathbb{P} \left\{ \left. \hat{a}_{i}^{\textnormal{SS}} (\mathbf{M}) \neq a_i \right| \left\{ \hat{t}_i (\mathbf{M}) \neq t_i \right\} \cap \left( \mathcal{E}_1 \cap \mathcal{E}_2 \right), (\mathbf{t}, \mathbf{w}) \right\} \\
        \leq \ & \exp \left[ - \frac{l}{2} \left\{ 2 \mathcal{Q} \left( t_i, \hat{t}_i (\mathbf{M}) \right) - 1 \right\}^2 \right] \\
        \leq \ & \exp \left\{ - \frac{l}{2} \theta_3 \left( t_i; \mathcal{Q} \right) \right\}
    \end{split}
\end{equation}
Thus, we can derive the following upper bound on the probability $\mathbb{P} \left\{ \left. \hat{a}_{i}^{\textnormal{SS}} (\mathbf{M}) \neq a_i \right| \mathcal{E}_1 \cap \mathcal{E}_2, (\mathbf{t}, \mathbf{w}) \right\}$:
\begin{equation}
    \label{eqn:proof_prop:performance_SS_Q_in_M1_v13}
    \begin{split}
        &\mathbb{P} \left\{ \left. \hat{a}_{i}^{\textnormal{SS}} (\mathbf{M}) \neq a_i \right| \mathcal{E}_1 \cap \mathcal{E}_2, \left( \mathbf{t}, \mathbf{w} \right) \right\} \\
        = \ & \mathbb{P} \left\{ \left. \hat{a}_{i}^{\textnormal{SS}} (\mathbf{M}) \neq a_i \right| \left\{ \hat{t}_i (\mathbf{M}) = t_i \right\} \cap \left( \mathcal{E}_1 \cap \mathcal{E}_2 \right), \left( \mathbf{t}, \mathbf{w} \right) \right\} \\
        &\mathbb{P} \left\{ \left. \hat{t}_i (\mathbf{M}) = t_i \right| \mathcal{E}_1 \cap \mathcal{E}_2, \left( \mathbf{t}, \mathbf{w} \right) \right\} \\
        &+ \mathbb{P} \left\{ \left. \hat{a}_{i}^{\textnormal{SS}} (\mathbf{M}) \neq a_i \right| \left\{ \hat{t}_i (\mathbf{M}) \neq t_i \right\} \cap \left( \mathcal{E}_1 \cap \mathcal{E}_2 \right), \left( \mathbf{t}, \mathbf{w} \right) \right\} \\
        &\mathbb{P} \left\{ \left. \hat{t}_i (\mathbf{M}) \neq t_i \right| \mathcal{E}_1 \cap \mathcal{E}_2, \left( \mathbf{t}, \mathbf{w} \right) \right\} \\
        \stackrel{\textnormal{(g)}}{\leq} \ & \exp \left[ - \frac{l}{2} \left\{ 2 \mathcal{Q} (t_i, t_i) - 1 \right\}^2 \right] \\
        &+ 2d \exp \left[ - \frac{l}{2} \left\{ \left( p^{*}(t_i) - q^{*}(t_i) \right)^2 + \theta_3 \left( t_i; \mathcal{Q} \right) \right\} \right] \\
        \stackrel{\textnormal{(h)}}{\leq} \ & (2d+1) \exp \left[ - \frac{l}{2} \left\{ \left( p^{*}(t_i) - q^{*}(t_i) \right)^2 + \theta_3 \left( t_i; \mathcal{Q} \right) \right\} \right],
    \end{split}
\end{equation}
where the step (g) can be obtained by putting two pieces \eqref{eqn:proof_prop:performance_SS_Q_in_M1_v10} and \eqref{eqn:proof_prop:performance_SS_Q_in_M1_v12}, and the step (h) can be justified with the following simple computation:
\begin{equation}
    \label{eqn:proof_prop:performance_SS_Q_in_M1_v14}
    \begin{split}
        &\left\{ 2 \mathcal{Q} (t_i, t_i) - 1 \right\}^2 - \left\{ \left( p^{*}(t_i) - q^{*}(t_i) \right)^2 + \theta_3 \left( t_i; \mathcal{Q} \right) \right\} \\
        &\geq \left( 2 p^{*}(t_i) - 1 \right)^2 - \left\{ \left( p^{*}(t_i) - q^{*}(t_i) \right)^2 + \left( 2 q^{*}(t_i) - 1 \right)^2 \right\} \\
        &= \left\{ p^{*}(t_i) - q^{*}(t_i) \right\} \left\{ 3 p^{*}(t_i) + 5 q^{*}(t_i) - 4 \right\} \\
        &\stackrel{\textnormal{(i)}}{>} 0,
    \end{split}
\end{equation}
where the step (i) holds by the assumption that the reliability matrix $\mathcal{Q}$ is weakly assortative. (Assumption \ref{assumption:weak_ass_rm})

\indent On the other hand, we reach from two inequalities in \eqref{eqn:proof_prop:performance_SS_Q_in_M1_v12} that
\begin{equation}
    \label{eqn:proof_prop:performance_SS_Q_in_M1_v15}
    \begin{split}
        &\mathbb{P} \left\{ \left. \hat{a}_{i}^{\textnormal{SS}} (\mathbf{M}) \neq a_i \right| \mathcal{E}_1 \cap \mathcal{E}_2, \left( \mathbf{t}, \mathbf{w} \right) \right\} \\
        = \ & \mathbb{P} \left\{ \left. \hat{a}_{i}^{\textnormal{SS}} (\mathbf{M}) \neq a_i \right| \left\{ \hat{t}_i (\mathbf{M}) = t_i \right\} \cap \left( \mathcal{E}_1 \cap \mathcal{E}_2 \right), \left( \mathbf{t}, \mathbf{w} \right) \right\} \\
        &\mathbb{P} \left\{ \left. \hat{t}_i (\mathbf{M}) = t_i \right| \mathcal{E}_1 \cap \mathcal{E}_2, \left( \mathbf{t}, \mathbf{w} \right) \right\} \\
        &+ \mathbb{P} \left\{ \left. \hat{a}_{i}^{\textnormal{SS}} (\mathbf{M}) \neq a_i \right| \left\{ \hat{t}_i (\mathbf{M}) \neq t_i \right\} \cap \left( \mathcal{E}_1 \cap \mathcal{E}_2 \right), (\mathbf{t}, \mathbf{w}) \right\} \\
        &\mathbb{P} \left\{ \left. \hat{t}_i (\mathbf{M}) \neq t_i \right| \mathcal{E}_1 \cap \mathcal{E}_2, \left( \mathbf{t}, \mathbf{w} \right) \right\} \\
        \leq \ & \exp \left[ - \frac{l}{2} \left\{ 2 \mathcal{Q} (t_i, t_i) - 1 \right\}^2 \right]\\
        &\times \mathbb{P} \left\{ \left. \hat{t}_i (\mathbf{M}) = t_i \right| \mathcal{E}_1 \cap \mathcal{E}_2, \left( \mathbf{t}, \mathbf{w} \right) \right\} \\
        &+ \exp \left\{ - \frac{l}{2} \theta_3 \left( t_i; \mathcal{Q} \right) \right\} \mathbb{P} \left\{ \left. \hat{t}_i (\mathbf{M}) \neq t_i \right| \mathcal{E}_1 \cap \mathcal{E}_2, \left( \mathbf{t}, \mathbf{w} \right) \right\} \\
        \stackrel{\textnormal{(j)}}{\leq} \ &  \exp \left\{ - \frac{l}{2} \theta_3 \left( t_i; \mathcal{Q} \right) \right\},
    \end{split}
\end{equation}
where the step (j) makes use of the fact $\left\{ 2 \mathcal{Q} (t_i, t_i) - 1 \right\}^2 > \theta_3 \left( t_i; \mathcal{Q} \right)$, which immediately follows from \eqref{eqn:proof_prop:performance_SS_Q_in_M1_v14}. 

\indent Combining two pieces \eqref{eqn:proof_prop:performance_SS_Q_in_M1_v13} and \eqref{eqn:proof_prop:performance_SS_Q_in_M1_v15} together yields the following bound
\begin{equation}
    \label{eqn:proof_prop:performance_SS_Q_in_M1_v16}
    \begin{split}
        &\mathbb{P} \left\{ \left. \hat{a}_{i}^{\textnormal{SS}} (\mathbf{M}) \neq a_i \right| \mathcal{E}_1 \cap \mathcal{E}_2, (\mathbf{t}, \mathbf{w}) \right\} \leq\\
         & \min \left\{ (2d+1) \exp \left[ - \frac{l}{2} \left\{ \left( p^{*}(t_i) - q^{*}(t_i) \right)^2 + \theta_3 \left( t_i; \mathcal{Q} \right) \right\} \right], \right. \\
        &\qquad\left. \exp \left\{ - \frac{l}{2} \theta_3 \left( t_i; \mathcal{Q} \right) \right\} \right\}.
    \end{split}
\end{equation}
Taking expectation with respect to $\left( \mathbf{t}, \mathbf{w} \right) \sim \bm{\mu}^{\otimes m} \otimes \bm{\nu}^{\otimes n}$ leads to
\begin{equation}
    \label{eqn:proof_prop:performance_SS_Q_in_M1_v17}
    \begin{split}
        &\mathbb{P} \left\{ \left. \hat{a}_{i}^{\textnormal{SS}} (\mathbf{M}) \neq a_i \right| \mathcal{E}_1 \cap \mathcal{E}_2 \right\} \\
        &=  \mathbb{E}_{\left( \mathbf{t}, \mathbf{w} \right) \sim \bm{\mu}^{\otimes m} \otimes \bm{\nu}^{\otimes n}} \left[ \mathbb{P} \left\{ \left. \hat{a}_{i}^{\textnormal{SS}} (\mathbf{M}) \neq a_i \right| \mathcal{E}_1 \cap \mathcal{E}_2, \left( \mathbf{t}, \mathbf{w} \right) \right\} \right] \\
        &\leq  \min \left\{ \exp \left\{ - \frac{l}{2} \min_{t \in [d]} \theta_3 \left( t; \mathcal{Q} \right) \right\},  \right. \\ 
        &\left. (2d+1) \exp \left[ - \frac{l}{2} \min_{t \in [d]} \left\{ \left( p^{*}(t) - q^{*}(t) \right)^2 + \theta_3 \left( t; \mathcal{Q} \right) \right\} \right]\right\}
    \end{split}
\end{equation}

\indent To sum up, we obtain the following upper bound of the error probability $\mathbb{P} \left\{ \hat{a}_{i}^{\textnormal{SS}} (\mathbf{M}) \neq a_i \right\}$:
\begin{equation}
    \label{eqn:proof_prop:performance_SS_Q_in_M1_v18}
    \begin{split}
        &\mathbb{P} \left\{ \hat{a}_{i}^{\textnormal{SS}} (\mathbf{M}) \neq a_i \right\} \\
        &=  \mathbb{P} \left\{ \left. \hat{a}_{i}^{\textnormal{SS}} (\mathbf{M}) \neq a_i \right| \mathcal{E}_{1}^c \right\} \mathbb{P} \left\{ \mathcal{E}_{1}^c \right\} \\
        &+ \mathbb{P} \left\{ \left. \hat{a}_{i}^{\textnormal{SS}} (\mathbf{M}) \neq a_i \right| \mathcal{E}_{2}^c \right\} \mathbb{P} \left\{ \mathcal{E}_{2}^c \right\} \\
        &+ \mathbb{P} \left\{ \left. \hat{a}_{i}^{\textnormal{SS}} (\mathbf{M}) \neq a_i \right| \mathcal{E}_{1} \cap \mathcal{E}_2 \right\} \mathbb{P} \left\{ \mathcal{E}_{1} \cap \mathcal{E}_2 \right\} \\
        &\leq  \mathbb{P} \left\{ \mathcal{E}_{1}^c \right\} + \mathbb{P} \left\{ \mathcal{E}_{2}^c \right\} + \mathbb{P} \left\{ \left. \hat{a}_{i}^{\textnormal{SS}} (\mathbf{M}) \neq a_i \right| \mathcal{E}_1 \cap \mathcal{E}_{2} \right\} \\
        &\stackrel{\textnormal{(k)}}{\leq} 
        2 \binom{n}{2} \exp \left\{ - \frac{r}{8} \left( p_m - p_u \right)^2 \right\} \\
        &+ d \exp \left [ - \min_{b \in [d]} \left \{ \frac{n \nu (b)}{2} \left( 1 - \frac{l}{n \nu(b)} \right)^2 \right\} \right] \\
        &+  \min \left\{ \exp \left\{ - \frac{l}{2} \min_{t \in [d]} \theta_3 \left( t; \mathcal{Q} \right) \right\},  \right. \\ 
        &\left. (2d+1) \exp \left[ - \frac{l}{2} \min_{t \in [d]} \left\{ \left( p^{*}(t) - q^{*}(t) \right)^2 + \theta_3 \left( t; \mathcal{Q} \right) \right\} \right]\right\}
    \end{split}
\end{equation}
where the step (k) is deduced by taking three pieces \eqref{eqn:proof_prop:performance_SS_Q_in_M1_v7}, \eqref{eqn:proof_prop:performance_SS_Q_in_M1_v9}, and \eqref{eqn:proof_prop:performance_SS_Q_in_M1_v17} collectively. So, we arrive at
\begin{equation}
    \label{eqn:proof_prop:performance_SS_Q_in_M1_v19}
    \begin{split}
        &\mathcal{R} \left( \mathbf{a}, \hat{\mathbf{a}}^{\textnormal{SS}} \right) = \frac{1}{m} \sum_{i=1}^{m} \mathbb{P} \left\{ \hat{a}_{i}^{\textnormal{SS}} (\mathbf{M}) \neq a_i \right\} \\
        &\leq  2 \binom{n}{2} \exp \left\{ - \frac{r}{8} \left( p_m - p_u \right)^2 \right\} \\
        &+ d \exp \left [ - \min_{b \in [d]} \left \{ \frac{n \nu (b)}{2} \left( 1 - \frac{l}{n \nu(b)} \right)^2 \right\} \right] \\
	&+  \min \left\{ \exp \left\{ - \frac{l}{2} \min_{t \in [d]} \theta_3 \left( t; \mathcal{Q} \right) \right\},  \right. \\ 
        &\left. (2d+1) \exp \left[ - \frac{l}{2} \min_{t \in [d]} \left\{ \left( p^{*}(t) - q^{*}(t) \right)^2 + \theta_3 \left( t; \mathcal{Q} \right) \right\} \right]\right\}.
    \end{split}
\end{equation}

\indent In order to achieve the desired recovery accuracy \eqref{eqn:expected_accuracy}, we may choose
\begin{equation}
    \label{eqn:proof_prop:performance_SS_Q_in_M1_v20}
    \begin{split}
        r = \ & \frac{8}{\left( p_m - p_u \right)^2} \log \left\{ \frac{3 n(n-1)}{\alpha} \right\}; \\
        l = \ & \min \left\{ \frac{2 \log \left( \frac{6d+3}{\alpha} \right)}{ \min_{t \in [d]} \left\{ \left( p^{*}(t) - q^{*}(t) \right)^2 + \theta_3 \left( t; \mathcal{Q} \right) \right\} }, \right. \\ 
        &\left. \frac{2 \log \left( \frac{3}{\alpha} \right)}{\min_{t \in [d]} \theta_3 \left( t; \mathcal{Q} \right)} \right\}; \\
        n \geq \ & \max \left\{ 8 \log \left( \frac{3d}{\alpha} \right), 2l \right\} \frac{1}{\min_{b \in [d]} \nu (b)}.
    \end{split}
\end{equation}
Then, the average number of required queries per task can be bounded above by
\begin{equation}
    \label{eqn:proof_prop:performance_SS_Q_in_M1_v21}
    \begin{split}
        &\frac{1}{m} \left\{ nr + ld(m-r) \right\} \\
        \leq \ & \frac{nr}{m} + ld \\
        = \ & \frac{8n}{m \left( p_m - p_u \right)^2} \log \left\{ \frac{3n(n-1)}{\alpha} \right\} + ld \\
        \stackrel{\textnormal{(l)}}{\leq} \ & \underbrace{\frac{8}{C_2} \cdot \frac{1}{n^{\beta}} \log \left\{ \frac{3n(n-1)}{\alpha} \right\}}_{=: \ \textnormal{(T1)}} + \underbrace{ld}_{=: \ \textnormal{(T2)}},
    \end{split}
\end{equation}
where the step (l) holds because $m \left( p_m - p_u \right)^2 \geq C_2 \cdot n^{1 + \beta}$. 

\begin{claim} 
    \label{claim:proof_prop:performance_SS_Q_in_M1_v1}
    $\textnormal{(T2)} = \omega \left( \textnormal{(T1)} \right)$ as $d \to \infty$.
\end{claim}

\begin{proof} [Proof of Claim \ref{claim:proof_prop:performance_SS_Q_in_M1_v1}]
\ \\
\indent Since the function
\begin{equation*}
    x \in \left[ 1 + \exp \left( \frac{3}{2 \epsilon} \right), +\infty \right) \mapsto x^{-\beta} \log \left\{ \frac{3x(x-1)}{\alpha} \right\}
\end{equation*}
is a strictly decreasing function, and $n \geq \frac{8}{\min_{b \in [d]} \nu (b)} \log \left( \frac{3d}{\alpha} \right) \geq 8 d \log \left( \frac{3d}{\alpha} \right)$, one has
\begin{equation}
    \label{eqn:proof_prop:performance_SS_Q_in_M1_v22}
    \begin{split}
        \textnormal{(T1)} &\leq \frac{8}{C_2} \cdot \left\{ 8d \log \left( \frac{3d}{\alpha} \right) \right\}^{- \beta}\\
        &\quad\times \log \left[ 192 \left( \frac{d}{\alpha} \right)^2 \left\{ \log \left( \frac{3d}{\alpha} \right) \right\}^2 \right] \\
        &= \mathcal{O} \left( d^{- \beta} \left\{ \log \left( \frac{d}{\alpha} \right) \right\}^{1 - \beta} \right) \\
        &= o \left( \left\{ \log \left( \frac{d}{\alpha} \right) \right\}^{1 - \beta} \right).
    \end{split}
\end{equation}
On the other hand, one can see that
\begin{equation*}
    \begin{split}
        l = \ & \min \left\{\frac{2 \log \left( \frac{3}{\alpha} \right)}{\min_{t \in [d]} \theta_3 \left( t; \mathcal{Q} \right)},  \right. \\ 
        &\left. \frac{2} \log \left( \frac{6d+3}{\alpha} \right){ \min_{t \in [d]} \left\{ \left( p^{*}(t) - q^{*}(t) \right)^2 + \theta_3 \left( t; \mathcal{Q} \right) \right\} }  \right\} \\
        \stackrel{\textnormal{(m)}}{\geq} \ & 
        \min \left\{2 \log \left( \frac{3}{\alpha} \right), \log \left( \frac{6d+3}{\alpha} \right) \right\} \\
        = \ & \Theta \left( \log \left( \frac{1}{\alpha} \right) \right),
    \end{split}
\end{equation*}
where the step (m) holds since $\theta_3 \left( t; \mathcal{Q}, \bm{\mu}, \bm{\nu} \right) \leq 1$ for every $t \in [d]$. Therefore, we arrive at
\begin{equation}
    \label{eqn:proof_prop:performance_SS_Q_in_M1_v23}
    \textnormal{(T2)} = dl = \Omega \left( d \log \left( \frac{1}{\alpha} \right) \right).
\end{equation}
Combining two pieces \eqref{eqn:proof_prop:performance_SS_Q_in_M1_v22} and \eqref{eqn:proof_prop:performance_SS_Q_in_M1_v23} together yields $\textnormal{(T1)} = o \left( \textnormal{(T2)} \right)$ as $d$ tends to infinity and this completes the proof of Claim \ref{claim:proof_prop:performance_SS_Q_in_M1_v1}.

\end{proof}

\indent Due to Claim \ref{claim:proof_prop:performance_SS_Q_in_M1_v1}, we obtain for all sufficiently large $d$ that
\begin{equation*}
    \begin{split}
        &\frac{1}{m} \left\{ nr + ld(m-r) \right\} \\
        &\leq  2 \cdot \textnormal{(T2)} \\
        &=  \min \left\{ \frac{4d}{\min_{t \in [d]} \theta_3 \left( t; \mathcal{Q} \right)} \log \left( \frac{3}{\alpha} \right),  \right. \\ 
        &\left.\frac{4d}{ \min_{t \in [d]} \left\{ \left( p^{*}(t) - q^{*}(t) \right)^2 + \theta_3 \left( t; \mathcal{Q} \right) \right\} } \log \left( \frac{6d+3}{\alpha} \right)  \right\},
    \end{split}
\end{equation*}
which establishes our desired result.

\section{Proof of Proposition \ref{prop:performance_SS_Q_in_M2}}
\label{sec:proof_prop:performance_SS_Q_in_M2}

\indent We embark on the proof of Proposition \ref{prop:performance_SS_Q_in_M2} by establishing the following preliminary inequality controlling the error probability $\mathbb{P} \left\{ \left. \hat{a}_{i}^{\textnormal{SS}} (\mathbf{M}) \neq a_i \right| \mathcal{E}_1 \cap \mathcal{E}_2, \left( \mathbf{t}, \mathbf{w} \right) \right\}$, where we sometimes omit the conditioning $\{\mathcal{E}_1 \cap \mathcal{E}_2, \left( \mathbf{t}, \mathbf{w} \right)\}$ for simplicity:
\begin{equation}
    \label{eqn:proof_prop:performance_SS_Q_in_M2_v1}
    \begin{split}
        &\mathbb{P} \left\{ \left. \hat{a}_{i}^{\textnormal{SS}} (\mathbf{M}) \neq a_i \right| \mathcal{E}_1 \cap \mathcal{E}_2, \left( \mathbf{t}, \mathbf{w} \right) \right\} \\
        &=   \mathbb{P} \big\{ \left. \hat{a}_{i}^{\textnormal{SS}} (\mathbf{M}) \neq a_i \right| \left\{ \hat{t}_i (\mathbf{M}) \in [d] \setminus \textnormal{spammer}_{\mathcal{Q}} \left( t_i \right) \right\} \big\} \\
        &\quad\times \mathbb{P} \left\{ \left. \hat{t}_i (\mathbf{M}) \in [d] \setminus \textnormal{spammer}_{\mathcal{Q}} \left( t_i \right) \right| \mathcal{E}_1 \cap \mathcal{E}_2, \left( \mathbf{t}, \mathbf{w} \right) \right\} \\
        &\quad+ \mathbb{P} \big\{ \left. \hat{a}_{i}^{\textnormal{SS}} (\mathbf{M}) \neq a_i \right| \left\{ \hat{t}_i (\mathbf{M}) \in \textnormal{spammer}_{\mathcal{Q}} \left( t_i \right) \right\} \big\} \\
        &
        \quad\times \mathbb{P} \left\{ \left. \hat{t}_i (\mathbf{M}) \in \textnormal{spammer}_{\mathcal{Q}} \left( t_i \right) \right| \mathcal{E}_1 \cap \mathcal{E}_2, \left( \mathbf{t}, \mathbf{w} \right) \right\} \\
        &\stackrel{\textnormal{(a)}}{\leq}  \exp \left[ - \frac{l}{2} \left\{ 2 \mathcal{Q} \left( t_i, \hat{t}_i (\mathbf{M}) \right) - 1 \right\}^2 \right] \\
        &\quad\times\mathbb{P} \left\{ \left. \hat{t}_i (\mathbf{M}) \in [d] \setminus \textnormal{spammer}_{\mathcal{Q}} \left( t_i \right) \right| \mathcal{E}_1 \cap \mathcal{E}_2, \left( \mathbf{t}, \mathbf{w} \right) \right\} \\
        &\quad+ \mathbb{P} \left\{ \left. \hat{t}_i (\mathbf{M}) \in \textnormal{spammer}_{\mathcal{Q}} \left( t_i \right) \right| \mathcal{E}_1 \cap \mathcal{E}_2, \left( \mathbf{t}, \mathbf{w} \right) \right\} \\
        &\stackrel{\textnormal{(b)}}{\leq}  \exp \left( - 2l \epsilon^2 \right)\\
        &\quad\times \underbrace{\mathbb{P} \left\{ \left. \hat{t}_i (\mathbf{M}) \in [d] \setminus \textnormal{spammer}_{\mathcal{Q}} \left( t_i \right) \right| \mathcal{E}_1 \cap \mathcal{E}_2, \left( \mathbf{t}, \mathbf{w} \right) \right\}}_{=: \ \textnormal{(P1)}} \\
        &\quad+ \underbrace{\mathbb{P} \left\{ \left. \hat{t}_i (\mathbf{M}) \in \textnormal{spammer}_{\mathcal{Q}} \left( t_i \right) \right| \mathcal{E}_1 \cap \mathcal{E}_2, \left( \mathbf{t}, \mathbf{w} \right) \right\}}_{=: \ \textnormal{(P2)}},
    \end{split}
\end{equation}
where the step (a) can be obtained with the same technique as the argument used to establish the bound \eqref{eqn:proof_prop:performance_SS_Q_in_M1_v12} that uses the Chernoff-Hoeffding theorem, and the step (b) holds since $\mathcal{Q} \left( t_i, \hat{t}_i (\mathbf{M}) \right) \geq \frac{1}{2} + \epsilon$ when $\hat{t}_i (\mathbf{M}) \in [d] \setminus \textnormal{spammer}_{\mathcal{Q}} \left( t_i \right)$.

\indent As we discussed earlier in Section \ref{subsubsec:SS_algorithm}, we are going to focus on controlling the term $\textnormal{(P2)}$.
\begin{equation}
    \label{eqn:proof_prop:performance_SS_Q_in_M2_v2}
    \begin{split}
        &\textnormal{(P2)} \\
        = \ & \mathbb{P} \left\{ \left. \hat{t}_i (\mathbf{M}) \in \textnormal{spammer}_{\mathcal{Q}} \left( t_i \right) \right| \mathcal{E}_1 \cap \mathcal{E}_2, \left( \mathbf{t}, \mathbf{w} \right) \right\} \\
        \leq \ & \mathbb{P} \Bigg\{  \max_{u \in \textnormal{spammer}_{\mathcal{Q}} \left( t_i \right)} \bigg| \sum_{j \in \mathcal{A}_u (i)} M_{ij} \bigg|  \\
        &\quad \geq \min_{v \in [d] \setminus \textnormal{spammer}_{\mathcal{Q}} \left( t_i \right)} \bigg| \sum_{k \in \mathcal{A}_v (i)} M_{ik} \bigg|
        \Bigg| \mathcal{E}_1 \cap \mathcal{E}_2, \left( \mathbf{t}, \mathbf{w} \right) \Bigg\} \\
        \stackrel{\textnormal{(c)}}{\leq} \ & \sum_{(u, v) \in \textnormal{spammer}_{\mathcal{Q}} \left( t_i \right) \times \left\{ [d] \setminus \textnormal{spammer}_{\mathcal{Q}} \left( t_i \right) \right\}} \\
        &\mathbb{P} \left\{  \bigg| \sum_{j \in \mathcal{A}_u (i)} M_{ij} \bigg| \geq \bigg| \sum_{k \in \mathcal{A}_v (i)} M_{ik} \bigg|
        \Bigg| \mathcal{E}_1 \cap \mathcal{E}_2, \left( \mathbf{t}, \mathbf{w} \right) \right\},
    \end{split}
\end{equation}
where the step (c) follows by the union bound. Now, we reuse auxiliary random variables $S_{ib} := \sum_{j \in \mathcal{A}_{b} (i)} \mathbbm{1} \left( M_{ij} = +1 \right)$ for each $(i, b) \in [m] \times [d]$, defined in the proof of Theorem \ref{thm:performance_alg:without_side_information} (see Section \ref{subsec:proof_thm:performance_alg:with_side_information}). It's clear that $\sum_{j \in \mathcal{A}_b (i)} M_{ij} = 2 S_{ib} - l$ for all pairs $(i, b) \in [m] \times [d]$. Then, the bound \eqref{eqn:proof_prop:performance_SS_Q_in_M2_v2} can be simplified as
\begin{equation}
    \label{eqn:proof_prop:performance_SS_Q_in_M2_v3}
    \begin{split}
        \textnormal{(P2)}
        \leq \ & \sum_{(u, v) \in \textnormal{spammer}_{\mathcal{Q}} \left( t_i \right) \times \left\{ [d] \setminus \textnormal{spammer}_{\mathcal{Q}} \left( t_i \right) \right\}} \\
        &\mathbb{P} \left\{ \left. \left| 2 S_{iu} - l \right| \geq \left| 2 S_{iv} - l \right|
        \right| \mathcal{E}_1 \cap \mathcal{E}_2, \left( \mathbf{t}, \mathbf{w} \right) \right\}.
    \end{split}
\end{equation}
Hereafter, we assume that $a_i = +1$ without loss of generality. Then, we have $S_{ib} \sim \textnormal{Binomial} \left( l, \mathcal{Q} \left( t_i, b \right) \right)$ for every $b \in [d]$. For each $(u, v) \in \textnormal{spammer}_{\mathcal{Q}} \left( t_i \right) \times \left\{ [d] \setminus \textnormal{spammer}_{\mathcal{Q}} \left( t_i \right) \right\}$, it holds that
\begin{equation}
    \label{eqn:proof_prop:performance_SS_Q_in_M2_v4}
    \begin{split}
        &\mathbb{P} \left\{ \left. \left| 2 S_{iu} - l \right| \geq \left| 2 S_{iv} - l \right|
        \right| \mathcal{E}_1 \cap \mathcal{E}_2, \left( \mathbf{t}, \mathbf{w} \right) \right\} \\
        \stackrel{\textnormal{(d)}}{\leq} \ & \sum_{k=0}^{l} \mathbb{P} \left\{ \left. \left| 2 S_{iu} - l \right| \geq k \geq \left| 2 S_{iv} - l \right| \right| \mathcal{E}_1 \cap \mathcal{E}_2, \left( \mathbf{t}, \mathbf{w} \right) \right\} \\
        \stackrel{\textnormal{(e)}}{=} \ & \sum_{k=0}^{l} \mathbb{P} \left\{ \left. \left| 2 S_{iu} - l \right| \geq k \right| \mathcal{E}_1 \cap \mathcal{E}_2, \left( \mathbf{t}, \mathbf{w} \right) \right\} \\
        &\mathbb{P} \left\{ \left. \left| 2 S_{iv} - l \right| \leq k \right| \mathcal{E}_1 \cap \mathcal{E}_2, \left( \mathbf{t}, \mathbf{w} \right) \right\} \\
        \stackrel{\textnormal{(f)}}{\leq} \ & \sum_{k=0}^{l} \mathbb{P} \left\{ \left. \left| S_{iu} - \mathbb{E} \left[ S_{iu} \right] \right| \geq \frac{k}{2} \right| \mathcal{E}_1 \cap \mathcal{E}_2, \left( \mathbf{t}, \mathbf{w} \right) \right\} \\
        &\mathbb{P} \left\{ \left. \left| S_{iv} - \mathbb{E} \left[ S_{iv} \right] \right| \geq \frac{l \left\{ 2 \mathcal{Q} \left( t_i, v \right) - 1 \right\} - k}{2} \right| \mathcal{E}_1 \cap \mathcal{E}_2\right\} \\
        \stackrel{\textnormal{(g)}}{\leq} \ & 4 \sum_{k=0}^{\left\lfloor l \left\{ 2 \mathcal{Q} \left( t_i, v \right) - 1 \right\} \right\rfloor}\\
        &\qquad\quad \exp \left[ - 2l \left\{ \left( \frac{k}{2l} \right)^2 + \left( 2 \mathcal{Q} \left( t_i, v \right) - 1 - \frac{k}{2l} \right)^2 \right\} \right] \\
        \stackrel{\textnormal{(h)}}{\leq} \ & 4 (l+1) \exp \left[ - l \left\{ 2 \mathcal{Q} \left( t_i, v \right) - 1 \right\}^2 \right] \\
        \stackrel{\textnormal{(i)}}{\leq} \ & 4 (l+1) \exp \left( - 4l \epsilon^2 \right),
    \end{split}
\end{equation}
where the steps (d)--(i) can be justified due to the following reasons:
\begin{enumerate} [label=(\alph*)]
    \setcounter{enumi}{3}
    \item The union bound;
    \item The conditional independence of $S_{iu}$ and $S_{iv}$ given the event $\mathcal{E}_1 \cap \mathcal{E}_2$ and the pair $\left( \mathbf{t}, \mathbf{w} \right)$ of the task-type vector and the worker-type vector;
    \item We know that $\mathbb{E} \left[ S_{iu} \right] = \frac{l}{2}$ for every $u \in \textnormal{spammer}_{\mathcal{Q}} \left( t_i \right)$. On the other hand, if $\left| 2 S_{iv} - l \right| \leq k$, then the triangle inequality yields
    \begin{equation*}
        \begin{split}
            &l \left\{ 2 \mathcal{Q} \left( t_i, v \right) - 1 \right\} \\
            &= \left| 2 \mathbb{E} \left[ S_{iv} \right] - l \right| \\
            &\leq \left| \left( 2 \mathbb{E} \left[ S_{iv} \right] - l \right) - \left( 2 S_{iv} - l \right) \right| + \left| 2 S_{iv} - l \right| \\
            &\leq 2 \left| S_{iv} - \mathbb{E} \left[ S_{iv} \right] \right| + k,
        \end{split}
    \end{equation*}
    which leads to the relation
    \begin{equation*}
    \begin{split}
        &\left\{ \left| 2 S_{iv} - l \right| \leq k \right\} \\
        & \subseteq\left\{ \left| S_{iv} - \mathbb{E} \left[ S_{iv} \right] \right| \geq \frac{l \left\{ 2 \mathcal{Q} \left( t_i, v \right) - 1 \right\} - k}{2} \right\}.
        \end{split}
    \end{equation*}
    \item The Chernoff-Hoeffding theorem;
    \item The Cauchy-Schwarz inequality gives
    \begin{equation*}
        \begin{split}
            &\left( \frac{k}{2l} \right)^2 + \left( 2 \mathcal{Q} \left( t_i, v \right) - 1 - \frac{k}{2l} \right)^2 \\
            &\geq  \frac{1}{2} \left\{ \left( \frac{k}{2l} \right) + \left( 2 \mathcal{Q} \left( t_i, v \right) - 1 - \frac{k}{2l} \right) \right\} \\
            &= \frac{1}{2} \left\{ 2 \mathcal{Q} \left( t_i, v \right) - 1 \right\}^2.
        \end{split}
    \end{equation*}
    \item It holds that $2 \mathcal{Q} \left( t_i, v \right) - 1 \geq 2 \epsilon$ for all $v \in [d] \setminus \textnormal{spammer}_{\mathcal{Q}} \left( t_i \right)$.
\end{enumerate}
By putting the inequality \eqref{eqn:proof_prop:performance_SS_Q_in_M2_v4} into \eqref{eqn:proof_prop:performance_SS_Q_in_M2_v3}, we arrive at
\begin{equation}
    \label{eqn:proof_prop:performance_SS_Q_in_M2_v5}
    \begin{split}
        \textnormal{(P2)} &\leq d^2 \delta \left( t_i; d \right) \left\{ 1 - \delta \left( t_i; d \right) \right\} \cdot 4 (l+1) \exp \left( - 4l \epsilon^2 \right) \\
        &\leq 4 d^2 (l+1) \cdot \delta \left( t_i; d \right) \exp \left( - 4l \epsilon^2 \right). 
    \end{split}
\end{equation}
Now, it's time to leverage the upper bound \eqref{eqn:proof_prop:performance_SS_Q_in_M2_v5} of $\textnormal{(P2)}$ in the preliminary bound \eqref{eqn:proof_prop:performance_SS_Q_in_M2_v1} to establish an inequality controlling the error probability $\mathbb{P} \left\{ \left. \hat{a}_{i}^{\textnormal{SS}} (\mathbf{M}) \neq a_i \right| \mathcal{E}_1 \cap \mathcal{E}_2, \left( \mathbf{t}, \mathbf{w} \right) \right\}$:
\begin{equation}
    \label{eqn:proof_prop:performance_SS_Q_in_M2_v6}
    \begin{split}
        &\mathbb{P} \left\{ \left. \hat{a}_{i}^{\textnormal{SS}} (\mathbf{M}) \neq a_i \right| \mathcal{E}_1 \cap \mathcal{E}_2, \left( \mathbf{t}, \mathbf{w} \right) \right\} \\
        \leq \ & \exp \left( - 2l \epsilon^2 \right) + 4 d^2 (l+1) \cdot \delta \left( t_i; d \right) \exp \left( - 4l \epsilon^2 \right) \\
        \leq \ & \left\{ 1 + 4 d^2 (l+1) \cdot \delta \left( t_i; d \right) \right\} \exp \left( - 2l \epsilon^2 \right) \\
        \stackrel{\textnormal{(j)}}{\leq} \ & 5 d^2 (l+1) \cdot \delta \left( t_i; d \right) \exp \left( - 2l \epsilon^2 \right),
    \end{split}
\end{equation}
where the step (j) holds due to the assumption $\delta (t; d) \geq \frac{1}{d}$ for every $t \in [d]$.

\indent To summarize, one can deduce the following upper bound of the error probability $\mathbb{P} \left\{ \hat{a}_{i}^{\textnormal{SS}} (\mathbf{M}) \neq a_i \right\}$:
\begin{equation}
    \label{eqn:proof_prop:performance_SS_Q_in_M2_v7}
    \begin{split}
        &\mathbb{P} \left\{ \hat{a}_{i}^{\textnormal{SS}} (\mathbf{M}) \neq a_i \right\} \\
        \leq \ & \mathbb{P} \left\{ \left. \hat{a}_{i}^{\textnormal{SS}} (\mathbf{M}) \neq a_i \right| \mathcal{E}_1 \cap \mathcal{E}_2 \right\} \mathbb{P} \left\{ \mathcal{E}_1 \cap \mathcal{E}_2 \right\} \\
        &+ \mathbb{P} \left\{ \left. \hat{a}_{i}^{\textnormal{SS}} (\mathbf{M}) \neq a_i \right| \mathcal{E}_{1}^c \right\} \mathbb{P} \left\{ \mathcal{E}_{1}^c \right\} \\
        &+ \mathbb{P} \left\{ \left. \hat{a}_{i}^{\textnormal{SS}} (\mathbf{M}) \neq a_i \right| \mathcal{E}_{2}^c \right\} \mathbb{P} \left\{ \mathcal{E}_{2}^c \right\} \\
        \leq \ & \mathbb{P} \left\{ \left. \hat{a}_{i}^{\textnormal{SS}} (\mathbf{M}) \neq a_i \right| \mathcal{E}_1 \cap \mathcal{E}_2 \right\} + \mathbb{P} \left\{ \mathcal{E}_{1}^c \right\} + \mathbb{P} \left\{ \mathcal{E}_{2}^c \right\} \\
        \leq \ & 2 \binom{n}{2} \exp \left\{ - \frac{r}{8} \left( p_m - p_u \right)^2 \right\} \\
        &+ d \exp \left [ - \min_{b \in [d]} \left \{ \frac{n \nu (b)}{2} \left( 1 - \frac{l}{n \nu(b)} \right)^2 \right\} \right] \\
        &+ 5 d^2 (l+1) \cdot \delta \left( t_i; d \right) \exp \left( - 2l \epsilon^2 \right).
    \end{split}
\end{equation}
To achieve the recovery accuracy \eqref{eqn:expected_accuracy}, one may choose
\begin{equation}
    \label{eqn:proof_prop:performance_SS_Q_in_M2_v8}
    \begin{split}
        r &= \frac{8}{\left( p_m - p_u \right)^2} \log \left\{ \frac{3n(n-1)}{\alpha} \right\}; \\
        l &= \frac{1}{2 \epsilon^2} \varphi^{-1} \left\{ \log \left( \frac{15 d^2 \delta_{\max} (d)}{\epsilon^2 \alpha} \right) \right\}; \\
        n &\geq \max \left\{ 8 \log \left( \frac{3d}{\alpha} \right), 2l \right\} \frac{1}{\min_{b \in [d]} \nu (b)}.
    \end{split}
\end{equation}
So the sample complexity per task that the subset-selection algorithm requires to achieve the recovery accuracy \eqref{eqn:expected_accuracy} can be bounded above by
\begin{equation}
    \label{eqn:proof_prop:performance_SS_Q_in_M2_v9}
    \begin{split}
        &\frac{1}{m} \left\{ nr + ld(m-r) \right\} \\
        \leq \ & \frac{nr}{m} + ld \\
        = \ & \frac{8n}{m \left( p_m - p_u \right)^2} \log \left\{ \frac{3n(n-1)}{\alpha} \right\} + ld \\
        \stackrel{\textnormal{(k)}}{\leq} \ & \underbrace{\frac{8}{C_3} \cdot \frac{1}{n^{\beta}} \log \left\{ \frac{3n(n-1)}{\alpha} \right\}}_{=: \ \textnormal{(T1)}} + \underbrace{ld}_{=: \ \textnormal{(T2)}},
    \end{split}
\end{equation}
where the step (k) holds by the assumption $m \left( p_m - p_u \right)^2 \geq C_3 \cdot n^{1 + \beta}$. Here, by mimicking the proof of Claim \ref{claim:proof_prop:performance_SS_Q_in_M1_v1}, it can be easily shown that $\textnormal{(T1)} = o \left( \textnormal{(T2)} \right)$ as $d \to \infty$. Therefore, we may conclude that for every sufficiently large $d$,
\begin{equation*}
    \begin{split}
        \frac{1}{m} \left\{ nr + ld(m-r) \right\} &\leq 2 \cdot \textnormal{(T2)} \\
        &= \frac{d}{\epsilon^2} \varphi^{-1} \left\{ \log \left( \frac{15 d^2 \delta_{\max} (d)}{\epsilon^2 \alpha} \right) \right\},
    \end{split}
\end{equation*}
and this finishes the proof of Proposition \ref{prop:performance_SS_Q_in_M2}.

\section{Deferred proofs of technical lemmas}
\label{sec:deferred_proofs_intro}

\indent This section will be devoted to provide you detailed proofs of technical lemmas which play significant roles in the proofs of main theorems.

\subsection{Proof of Lemma \ref{lemma:proof_thm:performance_alg:with_side_information_v1}}
\label{subsec:proof_lemma:proof_thm:performance_alg:with_side_information_v1}

\indent (\romannumeral 1) Without loss of generality, one can assume $p \in \left( \frac{1}{2}, 1 \right]$. After proving the desired claim for this case, it is possible to recover the same result for the case $p \in \left[ 0, \frac{1}{2} \right)$ by replacing $p$ by $1-p$.
\begin{equation}
    \label{eqn:proof_lemma:proof_thm:performance_alg:with_side_information_v1_v1}
    \begin{split}
        &\mathbb{E} \left[ \left| 2 \mathbb{B} (n, p) - n \right| \right] \\
        = \ & \mathbb{E} \left[ \left| 2 \mathbb{B} (n, p) - n \right| \cdot \mathbbm{1}_{\left\{ \mathbb{B} (n, p) \geq \frac{n}{2} \right\}} \right] \\
        &+ \mathbb{E} \left[ \left| 2 \mathbb{B} (n, p) - n \right| \cdot \mathbbm{1}_{\left\{ \mathbb{B} (n, p) < \frac{n}{2} \right\}} \right] \\
        = \ & \mathbb{E} \left[ 2 \mathbb{B} (n, p) - n  \right] + 2 \underbrace{\mathbb{E} \left[ \left| 2 \mathbb{B} (n, p) - n \right| \cdot \mathbbm{1}_{\left\{ \mathbb{B} (n, p) < \frac{n}{2} \right\}} \right]}_{=: \ \mathcal{R}_n (p)} \\
        = \ & n (2p-1) + 2 \mathcal{R}_n (p).
    \end{split}
\end{equation}
We now take a closer inspection on the term $\mathcal{R}_n (p)$:
\begin{equation}
    \label{eqn:proof_lemma:proof_thm:performance_alg:with_side_information_v1_v2}
    \begin{split}
        0 \leq \ & \mathcal{R}_n (p) \\
        \leq \ & n \cdot \mathbb{P} \left\{ \mathbb{B} (n, p) - np < - n \left( p - \frac{1}{2} \right) \right\} \\
        \stackrel{\textnormal{(a)}}{\leq} \ & n \exp \left\{ - 2n \left( p - \frac{1}{2} \right)^2 \right\},
    \end{split}
\end{equation}
where the step (a) follows from the Hoeffding bound. Putting the bound \eqref{eqn:proof_lemma:proof_thm:performance_alg:with_side_information_v1_v2} into \eqref{eqn:proof_lemma:proof_thm:performance_alg:with_side_information_v1_v1}, we obtain
\begin{equation*}
    \begin{split}
        n (2p-1) \leq \ & \mathbb{E} \left[ \left| 2 \mathbb{B} (n, p) - n \right| \right] \\
        \leq \ & n (2p-1) + 2n \exp \left\{ - 2n \left( p - \frac{1}{2} \right)^2 \right\},
    \end{split}
\end{equation*}
as desired.
\medskip

\indent (\romannumeral 2) Let $\left\{ \varepsilon_{k} \right\}_{k=1}^{\infty}$ be a sequence of independent and identically distributed Rademacher random variables. By the central limit theorem, we have
\begin{equation}
    \label{eqn:proof_lemma:proof_thm:performance_alg:with_side_information_v1_v3}
    \frac{2 \mathbb{B} \left( n, \frac{1}{2} \right) - 1}{\sqrt{n}} \stackrel{d}{=} \frac{1}{\sqrt{n}} \sum_{k=1}^{n} \varepsilon_k \stackrel{d}{\rightarrow} \mathcal{N} (0, 1) \quad \textnormal{as } n \to \infty.
\end{equation}
Let $X_n := \frac{1}{\sqrt{n}} \sum_{k=1}^{n} \varepsilon_k$ for $n \in \mathbb{N}$. It's clear that $\mathbb{E} \left[ X_n \right] = 0$ and
\begin{equation*}
    \mathbb{E} \left[ \left| X_n \right|^2 \right] = \textnormal{Var} \left[ X_n \right] = \frac{1}{n} \sum_{k=1}^{n} \textnormal{Var} \left[ \varepsilon_k \right] = 1
\end{equation*}
for every $n \in \mathbb{N}$. Due to \emph{Corollary 6.21} in \cite{klenke2013probability}, $\left\{ X_n : n \in \mathbb{N} \right\}$ is a uniformly integrable family of random variables. Thus if we let $Z \stackrel{d}{=} \mathcal{N} (0, 1)$, then it follows from \eqref{eqn:proof_lemma:proof_thm:performance_alg:with_side_information_v1_v3} that $X_n \stackrel{n \to \infty}{\longrightarrow} Z$ in $L^1$ by \emph{Theorem 6.25} in \cite{klenke2013probability}. Hence, we arrive at
\begin{equation*}
    \begin{split}
        &\lim_{n \to \infty} \mathbb{E} \left[ \left| \frac{2 \mathbb{B} \left( n, \frac{1}{2} \right) - 1}{\sqrt{n}} \right| \right]\\
        &= \lim_{n \to \infty} \mathbb{E} \left[ \left| X_n \right| \right] \\
        &\stackrel{\textnormal{(b)}}{=} \mathbb{E} \left[ \left| Z \right| \right] \\
        &= \int_{-\infty}^{\infty} \left| x \right| \cdot \frac{1}{\sqrt{2 \pi}} \exp \left( - \frac{1}{2} x^2 \right) \mathrm{d}x \\
        &= 2 \int_{0}^{\infty} x \cdot \frac{1}{\sqrt{2 \pi}} \exp \left( - \frac{1}{2} x^2 \right) \mathrm{d}x \\
        &= \sqrt{\frac{2}{\pi}} \left[ - \exp \left( - \frac{1}{2} x^2 \right) \right]_{0}^{\infty} \\
        &= \sqrt{\frac{2}{\pi}},
    \end{split}
\end{equation*}
where the step (b) follows due to the $L^1$-convergence of the sequence $\left\{ X_n \right\}_{n=1}^{\infty}$ to $Z \stackrel{d}{=} \mathcal{N} (0, 1)$. This completes the proof of Lemma \ref{lemma:proof_thm:performance_alg:with_side_information_v1}.

\subsection{Proof of Lemma \ref{lemma:proof_thm:performance_alg:without_side_information_v1}}
\label{subsec:proof_lemma:proof_thm:performance_alg:without_side_information_v1}

\indent Let us focus on the case for which $a_i = +1$; the other case follows similarly. Assume that we are lying on the event
\begin{equation*}
    \left[ \bigcap_{b=1}^{d} \left\{ \left| S_{ib} - \mathbb{E} \left[ S_{ib} \right] \right| < \frac{p^* \left( t_i \right) - q^* \left( t_i \right)}{2} l \right\} \right] \cap \left( \mathcal{E}_1 \cap \mathcal{E}_2 \right).
\end{equation*}
We find from \eqref{eqn:proof_thm:performance_alg:without_side_information_v6} that $\mathbb{E} \left[ S_{ib} \right] = l \cdot \mathcal{Q} \left( t_i, b \right)$ for all pairs $(i, b) \in [m] \times [d]$. So it can be shown that
\begin{equation}
    \label{eqn:proof_lemma:proof_thm:performance_alg:without_side_information_v1_v1}
    \begin{split}
        S_{i t_i} - \frac{l}{2} &= \left( S_{i t_i} - \mathbb{E} \left[ S_{i t_i} \right] \right) + \left( \mathbb{E} \left[ S_{i t_i} \right] - \frac{l}{2} \right) \\
        &> - \frac{p^* \left( t_i \right) - q^* \left( t_i \right)}{2} l + \frac{2 \mathcal{Q} \left( t_i, t_i \right) - 1}{2} l \\
        &= - \frac{p^* \left( t_i \right) - q^* \left( t_i \right)}{2} l + \frac{2 p^* \left( t_i \right) - 1}{2} l \\
        &= \frac{p^* \left( t_i \right) + q^* \left( t_i \right) - 1}{2} l \\
        &> 0.
    \end{split}
\end{equation}
On the other hand, for every $b \in [d] \setminus \left\{ t_i \right\}$, one has
\begin{equation}
    \label{eqn:proof_lemma:proof_thm:performance_alg:without_side_information_v1_v2}
    \begin{split}
        \left| S_{ib} - \frac{l}{2} \right|
        &= \left| S_{ib} - \mathbb{E} \left[ S_{ib} \right] \right| + \left| \mathbb{E} \left[ S_{ib} \right] - \frac{l}{2} \right| \\
        &< \frac{p^* \left( t_i \right) - q^* \left( t_i \right)}{2} l + \frac{2 \mathcal{Q} \left( t_i, b \right) - 1}{2} l \\
        &\leq \frac{p^* \left( t_i \right) - q^* \left( t_i \right)}{2} l + \frac{2 q^* \left( t_i \right) - 1}{2} l \\
        &= \frac{p^* \left( t_i \right) + q^* \left( t_i \right) - 1}{2} l.
    \end{split}
\end{equation}
Combining two bounds \eqref{eqn:proof_lemma:proof_thm:performance_alg:without_side_information_v1_v1} and \eqref{eqn:proof_lemma:proof_thm:performance_alg:without_side_information_v1_v2} together leads to the desired conclusion
\begin{equation*}
    S_{i t_i} - \frac{l}{2} = \left| S_{i t_i} - \frac{l}{2} \right| > \left| S_{ib} - \frac{l}{2} \right|,\ \forall b \in [d] \setminus \left\{ t_i \right\},
\end{equation*}
which yields $\hat{t}_i (\mathbf{M}) = t_i$.

\subsection{Proof of Lemma \ref{lemma:proof_lemma:exact_recovery_alg:worker_clustering_v1}}
\label{subsec:proof_lemma:proof_lemma:exact_recovery_alg:worker_clustering_v1}

\indent Based on the definition of the orthogonal projection $\mathcal{P}_{\mathcal{U}} (\cdot) : \mathbb{R}^{n \times n} \rightarrow \mathbb{R}^{n \times n}$ together with the triangle inequality, it can be easily shown that
\begin{equation}
    \label{eqn:proof_lemma:proof_lemma:exact_recovery_alg:worker_clustering_v1_v1}
    \begin{split}
        &\left\| \mathcal{P}_{\mathcal{U}} \left( \mathbf{A} - \mathbb{E} \left[ \left. \mathbf{A} \right| \mathbf{w} \right] \right) \right\|_{\infty} \\
        &\leq  3 \big( \left\| \mathbf{U} \mathbf{U}^{\top} \left( \mathbf{A} - \mathbb{E} \left[ \left. \mathbf{A} \right| \mathbf{w} \right] \right) \right\|_{\infty}\\
        &\quad \vee \left\| \left( \mathbf{A} - \mathbb{E} \left[ \left. \mathbf{A} \right| \mathbf{w} \right] \right) \mathbf{U} \mathbf{U}^{\top} \right\|_{\infty} \big) \\
        &\stackrel{\textnormal{(a)}}{\leq}  3 \left\| \mathbf{U} \mathbf{U}^{\top} \left( \mathbf{A} - \mathbb{E} \left[ \left. \mathbf{A} \right| \mathbf{w} \right] \right) \right\|_{\infty},
    \end{split}
\end{equation}
where the step (a) follows since $\left\| \mathbf{U} \mathbf{U}^{\top} \left( \mathbf{A} - \mathbb{E} \left[ \left. \mathbf{A} \right| \mathbf{w} \right] \right) \right\|_{\infty} = \left\| \left( \mathbf{A} - \mathbb{E} \left[ \left. \mathbf{A} \right| \mathbf{w} \right] \right) \mathbf{U} \mathbf{U}^{\top} \right\|_{\infty}$.
\medskip

\indent In order to establish a concentration bound on the random variable $\left\| \mathbf{U} \mathbf{U}^{\top} \left( \mathbf{A} - \mathbb{E} \left[ \left. \mathbf{A} \right| \mathbf{w} \right] \right) \right\|_{\infty}$, we compute the $(j, k)$-th entry of $\mathbf{U} \mathbf{U}^{\top} \left( \mathbf{A} - \mathbb{E} \left[ \left. \mathbf{A} \right| \mathbf{w} \right] \right)$: by setting $z := w_j \in [d]$, \emph{i.e.}, the type of the $j$-th worker, one has
\begin{equation}
    \label{eqn:proof_lemma:proof_lemma:exact_recovery_alg:worker_clustering_v1_v2}
    \begin{split}
        &\left[ \mathbf{U} \mathbf{U}^{\top} \left( \mathbf{A} - \mathbb{E} \left[ \left. \mathbf{A} \right| \mathbf{w} \right] \right) \right]_{jk} \\
        = \ & \sum_{l=1}^{n} \left[ \mathbf{U} \mathbf{U}^{\top} \right]_{jl} \left( A_{lk} - \mathbb{E} \left[ \left. A_{lk} \right| \mathbf{w} \right] \right) \\
        \stackrel{\textnormal{(b)}}{=} \ & \frac{1}{s_z} \sum_{l \in \mathcal{W}_z \setminus \{ k \}} \left( A_{lk} - \mathbb{E} \left[ \left. A_{lk} \right| \mathbf{w} \right] \right) \\
        = \ & \frac{1}{s_z} \sum_{l \in \mathcal{W}_z \setminus \{ k \}} \left[ \sum_{i \in \mathcal{S}} \left( A_{lk}^{(i)} - \mathbb{E} \left[ \left. A_{lk}^{(i)} \right| \mathbf{w} \right] \right) \right] \\
        = \ & \frac{1}{s_z} \sum_{i \in \mathcal{S}} \left[ \sum_{l \in \mathcal{W}_z \setminus \{ k \}} \left( A_{lk}^{(i)} - \mathbb{E} \left[ \left. A_{lk}^{(i)} \right| \mathbf{w} \right] \right) \right],
    \end{split}
\end{equation}
where the step (b) makes use of the fact
\begin{equation*}
    \left[ \mathbf{U} \mathbf{U}^{\top} \right]_{jl} =
    \begin{cases}
        \frac{1}{s_z} & \textnormal{if } l \in \mathcal{W}_z; \\
        0 & \textnormal{otherwise.}
    \end{cases}
\end{equation*}
Here, we remind the setting
\begin{equation*}
    \mathbf{A}^{(i)} := \mathcal{P}_{\textnormal{off-diag}} \left( \mathbf{M}_{i*}^{\top} \mathbf{M}_{i*} \right),\ \forall i \in \mathcal{S},
\end{equation*}
which gives the decomposition $\mathbf{A} = \sum_{i \in \mathcal{S}} \mathbf{A}^{(i)}$ into the sum of $r = \left| \mathcal{S} \right|$ conditionally independent $n \times n$ random matrices given a worker type vector $\mathbf{w}$, due to Lemma \ref{lemma:proof_prop:performance_SS_Q_in_M1_v1}. We point out that this decomposition of the similarity matrix $\mathbf{A}$ plays a key role in the proof of Lemma \ref{lemma:proof_lemma:exact_recovery_alg:worker_clustering_v2}. Let
\begin{equation*}
    V_i := \sum_{l \in \mathcal{W}_z \setminus \{ k \}} \left( A_{lk}^{(i)} - \mathbb{E} \left[ \left. A_{lk}^{(i)} \right| \mathbf{w} \right] \right),\ \forall i \in \mathcal{S}.
\end{equation*}
Then, $\left\{ V_i : i \in \mathcal{S} \right\}$ are conditionally independent random variables given a worker type vector $\mathbf{w}$ by Lemma \ref{lemma:proof_prop:performance_SS_Q_in_M1_v1}, and we have
\begin{equation}
    \label{eqn:proof_lemma:proof_lemma:exact_recovery_alg:worker_clustering_v1_v3}
    s_z \left[ \mathbf{U} \mathbf{U}^{\top} \left( \mathbf{A} - \mathbb{E} \left[ \left. \mathbf{A} \right| \mathbf{w} \right] \right) \right]_{jk} = \sum_{i \in \mathcal{S}} V_i.
\end{equation}
Here, one can make the following observations:
\begin{enumerate} [label=(\roman*)]
    \item $\left| V_i \right| \leq \sum_{l \in \mathcal{W}_z \setminus \{ k \}} \left| \left( A_{lk}^{(i)} - \mathbb{E} \left[ \left. A_{lk}^{(i)} \right| \mathbf{w} \right] \right) \right| \leq 2 s_z$ for every $i \in \mathcal{S}$;
    \item The sum of second-order moments of $V_i$'s is bounded by
    \begin{equation*}
        \begin{split}
            \sum_{i \in \mathcal{S}} \mathbb{E} \left[ \left. V_{i}^2 \right| \mathbf{w} \right]
            = \ & \sum_{i \in \mathcal{S}} \textnormal{Var} \left[ \left. \sum_{l \in \mathcal{W}_z \setminus \{ k \}} A_{lk}^{(i)} \right| \mathbf{w} \right] \\
            \leq \ & \sum_{i \in \mathcal{S}} \mathbb{E} \left[ \left. \left( \sum_{l \in \mathcal{W}_z \setminus \{ k \}} A_{lk}^{(i)} \right)^2 \right| \mathbf{w} \right] \\
            \stackrel{\textnormal{(c)}}{\leq} \ & r \cdot s_{z}^2,
        \end{split}
    \end{equation*}
\end{enumerate}
where the step (c) holds because $\left| \sum_{l \in \mathcal{W}_z \setminus \{ k \}} A_{lk}^{(i)} \right| \leq s_z$. The Bernstein inequality together with the facts (\romannumeral 1) and (\romannumeral 2) implies that for any universal constant $\gamma_1 > 0$, we have
\begin{equation}
    \label{eqn:proof_lemma:proof_lemma:exact_recovery_alg:worker_clustering_v1_v4}
    \begin{split}
        &\mathbb{P} \left\{ \left. \left| \sum_{i \in \mathcal{S}} V_i \right| > \frac{\gamma_1}{3} \cdot s_z \sqrt{r} \log n \right| \mathbf{w} \right\} \\
        \leq \ & 2 \exp \left\{ - \frac{ \left( \frac{\gamma_1}{3} \right)^2 \cdot s_{z}^2 r \left( \log n \right)^2}{2 s_{z}^2 r + \frac{4 \gamma_1}{9} s_{z}^2 \sqrt{r} \log n} \right\} \\
        \leq \ & 2 \exp \left\{ - \frac{ \left( \frac{\gamma_1}{3} \right)^2 \cdot s_{z}^2 r \left( \log n \right)^2}{2 s_{z}^2 r \log n + \frac{4 \gamma_1}{9} s_{z}^2 r \log n} \right\} \\
        = \ & 2 \exp \left\{ - \frac{\left( \frac{\gamma_1}{3} \right)^2}{2 + \frac{4 \gamma_1}{9}} \log n \right\}.
    \end{split}
\end{equation}
So by taking the universal constant $\gamma_1$ to be sufficiently large so that $\frac{\left( \frac{\gamma_1}{3} \right)^2}{2 + \frac{4 \gamma_1}{9}} \geq 13$, we deduce from \eqref{eqn:proof_lemma:proof_lemma:exact_recovery_alg:worker_clustering_v1_v3} that with probability at least $1 - 2n^{-13}$,
\begin{equation*}
    s_z \left| \left[ \mathbf{U} \mathbf{U}^{\top} \left( \mathbf{A} - \mathbb{E} \left[ \left. \mathbf{A} \right| \mathbf{w} \right] \right) \right]_{jk} \right| = \left| \sum_{i \in \mathcal{S}} V_i \right| \leq \frac{\gamma_1}{3} \cdot s_z \sqrt{r} \log n.
\end{equation*}
By the union bound, the following result holds: with probability greater than $1 - 2 n^{-11}$, we have
\begin{equation}
    \label{eqn:proof_lemma:proof_lemma:exact_recovery_alg:worker_clustering_v1_v5}
    \left\| \mathbf{U} \mathbf{U}^{\top} \left( \mathbf{A} - \mathbb{E} \left[ \left. \mathbf{A} \right| \mathbf{w} \right] \right) \right\|_{\infty} \leq \frac{\gamma_1}{3} \cdot \sqrt{r} \log n.
\end{equation}
So, plugging \eqref{eqn:proof_lemma:proof_lemma:exact_recovery_alg:worker_clustering_v1_v5} into \eqref{eqn:proof_lemma:proof_lemma:exact_recovery_alg:worker_clustering_v1_v1} completes the proof.

\subsection{Proof of Lemma \ref{lemma:proof_lemma:exact_recovery_alg:worker_clustering_v2}}
\label{subsec:proof_lemma:proof_lemma:exact_recovery_alg:worker_clustering_v2}

\indent We begin with the following decomposition of $\mathbf{A} - \mathbb{E} \left[ \left. \mathbf{A} \right| \mathbf{w} \right]$ into the sum of $r = \left| \mathcal{S} \right|$ centered and conditionally independent $n \times n$ random matrices given a worker type vector $\mathbf{w}$:
\begin{equation}
    \label{eqn:proof_lemma:proof_lemma:exact_recovery_alg:worker_clustering_v2_v1}
    \mathbf{A} - \mathbb{E} \left[ \left. \mathbf{A} \right| \mathbf{w} \right] = \sum_{i \in \mathcal{S}} \left( \mathbf{A}^{(i)} - \mathbb{E} \left[ \left. \mathbf{A}^{(i)} \right| \mathbf{w} \right] \right).
\end{equation}
For reader's convenience, let
\[
    \sigma^2 := \left\| \sum_{i \in \mathcal{S}} \mathbb{E} \left[ \left. \left( \mathbf{A}^{(i)} - \mathbb{E} \left[ \left. \mathbf{A}^{(i)} \right| \mathbf{w} \right] \right)^2 \right| \mathbf{w} \right] \right\|.
\]
Then we may observe the following property: for every $i \in \calS$
\begin{equation}
    \label{eqn:proof_lemma:proof_lemma:exact_recovery_alg:worker_clustering_v2_v2}
    \begin{split}
        \left\| \mathbf{A}^{(i)} - \mathbb{E} \left[ \left. \mathbf{A}^{(i)} \right| \mathbf{w} \right] \right\| \leq \ & n \left\| \mathbf{A}^{(i)} - \mathbb{E} \left[ \left. \mathbf{A}^{(i)} \right| \mathbf{w} \right] \right\|_{\infty} \\
        \leq \ & 2n.
    \end{split}
\end{equation}

\indent Now, it's time to bound $\sigma^2$. Let
\begin{equation*}
    \begin{split}
        \mathbf{M}^{(i)} := \ & \mathbb{E} \left[ \left. \left( \mathbf{A}^{(i)} - \mathbb{E} \left[ \left. \mathbf{A}^{(i)} \right| \mathbf{w} \right] \right)^2 \right| \mathbf{w} \right] \\
        = \ & \mathbb{E} \left[ \left( \mathbf{A}^{(i)} \right)^2 \left. \right| \mathbf{w} \right] - \left( \mathbb{E} \left[ \left. \mathbf{A}^{(i)} \right| \mathbf{w} \right] \right)^2
    \end{split}
\end{equation*}
for each $i \in \mathcal{S}$. Let us take a closer look at each entry of $\mathbf{M}^{(i)}$. Doing some straightforward algebra, one can observe that for every $(j, k) \in [n] \times [n]$,
\begin{equation}
    \label{eqn:proof_lemma:proof_lemma:exact_recovery_alg:worker_clustering_v2_v3}
    \begin{split}
        &\left[ \mathbb{E} \left[ \left( \mathbf{A}^{(i)} \right)^2 \left. \right| \mathbf{w} \right] \right]_{jk} \\
        = \ & \sum_{l \in [n] \setminus \left\{ j, k \right\}} \mathbb{E} \left[ \left. A_{jl}^{(i)} A_{lk}^{(i)} \right| \mathbf{w} \right] \\
        = \ & \sum_{l \in [n] \setminus \left\{ j, k \right\}} \mathbb{E} \left[ \left. M_{ij} M_{ik} \right| \mathbf{w} \right] \\
        = \ &
        \begin{cases}
            n-1 & \textnormal{if } j=k; \\
            (n-2) \Phi \left( \mathcal{Q}, \bm{\mu}, \bm{\nu} \right) \left( w_j, w_k \right) & \textnormal{otherwise}
        \end{cases}
    \end{split}
\end{equation}
and
\begin{equation}
    \label{eqn:proof_lemma:proof_lemma:exact_recovery_alg:worker_clustering_v2_v4}
    \begin{split}
        &\left[ \left( \mathbb{E} \left[ \left. \mathbf{A}^{(i)} \right| \mathbf{w} \right] \right)^2 \right]_{jk} \\
        = \ & \sum_{l \in [n] \setminus \left\{ j, k \right\}} \mathbb{E} \left[ \left. M_{ij} M_{il} \right| \mathbf{w} \right] \cdot \mathbb{E} \left[ \left. M_{il} M_{ik} \right| \mathbf{w} \right] \\
        = \ & \sum_{l \in [n] \setminus \left\{ j, k \right\}} \left( \sum_{t=1}^{d} \mu (t) \left\{ 2 \mathcal{Q} (t, w_j) - 1 \right\} \left\{ 2 \mathcal{Q} (t, w_l) - 1 \right\} \right) \\
        &\left( \sum_{t=1}^{d} \mu (t) \left\{ 2 \mathcal{Q} (t, w_l) - 1 \right\} \left\{ 2 \mathcal{Q} (t, w_k) - 1 \right\} \right) \\
        = \ & \sum_{l \in [n] \setminus \left\{ j, k \right\}} \Phi \left( \mathcal{Q}, \bm{\mu}, \bm{\nu} \right) \left( w_j, w_l \right) \Phi \left( \mathcal{Q}, \bm{\mu}, \bm{\nu} \right) \left( w_l, w_k \right).
    \end{split}
\end{equation}
By taking two pieces \eqref{eqn:proof_lemma:proof_lemma:exact_recovery_alg:worker_clustering_v2_v3} and \eqref{eqn:proof_lemma:proof_lemma:exact_recovery_alg:worker_clustering_v2_v4} collectively, we arrive at
\[
    \begin{split}
        M_{jk}^{(i)} = (n-1) - \sum_{l \in [n] \setminus \{ j \}} \left\{ \Phi \left( \mathcal{Q}, \bm{\mu}, \bm{\nu} \right) \left( w_j, w_l \right) \right\}^2
    \end{split}
\]
if $j = k$, and
\[
    \begin{split}
        &M_{jk}^{(i)}=\\
        & (n-2) \Phi \left( \mathcal{Q}, \bm{\mu}, \bm{\nu} \right) \left( w_j, w_k \right) \\
        &- \sum_{l \in [n] \setminus \{ j, k \}} \Phi \left( \mathcal{Q}, \bm{\mu}, \bm{\nu} \right) \left( w_j, w_l \right) \Phi \left( \mathcal{Q}, \bm{\mu}, \bm{\nu} \right) \left( w_l, w_k \right)
    \end{split}
\]
otherwise, and it can be observed that $- (n-1) \leq M_{jk}^{(i)} \leq n-1$ for every $(j, k) \in [n] \times [n]$. Consequently, we can conclude that $\left\| \mathbf{M}^{(i)} \right\|_{\infty} \leq n-1$ for every $i \in \mathcal{S}$, and this implies
\begin{equation}
    \label{eqn:proof_lemma:proof_lemma:exact_recovery_alg:worker_clustering_v2_v5}
    \sigma^2 = \left\| \sum_{i \in \mathcal{S}} \mathbf{M}^{(i)} \right\| \leq \sum_{i \in \mathcal{S}} \left\| \mathbf{M}^{(i)} \right\| \leq n \sum_{i \in \mathcal{S}} \left\| \mathbf{M}^{(i)} \right\|_{\infty} \leq r n^2.
\end{equation}
So from the matrix Bernstein inequality \cite{tropp2012user}, we have for any absolute constant $\gamma_2 > 0$ that
\begin{equation}
    \label{eqn:proof_lemma:proof_lemma:exact_recovery_alg:worker_clustering_v2_v6}
    \begin{split}
        &\mathbb{P} \left\{ \left. \left\| \mathbf{A} - \mathbb{E} \left[ \left. \mathbf{A} \right| \mathbf{w} \right] \right\| > \gamma_2 \cdot \sqrt{r} n \log n \right| \mathbf{w} \right\} \\
        \stackrel{\textnormal{(a)}}{\leq} \ & 2n \exp \left\{ - \frac{\gamma_{2}^2 \cdot rn^2 \left( \log n \right)^2}{2 \sigma^2 + \frac{4 \gamma_2}{3} \sqrt{r} n^2 \log n} \right\} \\
        \stackrel{\textnormal{(b)}}{\leq} \ & 2n \exp \left\{ - \frac{\gamma_{2}^2 \cdot rn^2 \left( \log n \right)^2}{2 r n^2 \log n + \frac{4 \gamma_2}{3} \cdot r n^2 \log n} \right\} \\
        = \ & 2n \exp \left( - \frac{\gamma_{2}^2}{2 + \frac{4 \gamma_2}{3}} \log n \right),
    \end{split}
\end{equation}
where the step (a) follows from \eqref{eqn:proof_lemma:proof_lemma:exact_recovery_alg:worker_clustering_v2_v2}, and the step (b) holds by plugging the bound \eqref{eqn:proof_lemma:proof_lemma:exact_recovery_alg:worker_clustering_v2_v5} of $\sigma^2$. By selecting the absolute constant $\gamma_2$ to be sufficiently large such that $\frac{\gamma_{2}^2}{2 + \frac{4 \gamma_2}{3}} \geq 12$, we may deduce that with probability at least $1 - 2n^{-11}$, 
\begin{equation*}
    \left\| \mathbf{A} - \mathbb{E} \left[ \left. \mathbf{A} \right| \mathbf{w} \right] \right\| \leq \gamma_2 \cdot \sqrt{r} n \log n,
\end{equation*}
and this finishes the proof of Lemma \ref{lemma:proof_lemma:exact_recovery_alg:worker_clustering_v2}.

\subsection{Proof of Lemma \ref{lemma:proof_prop:performance_SS_Q_in_M1_v1}}
\label{subsec:proof_lemma:proof_prop:performance_SS_Q_in_M1_v1}

\indent According to Definition \ref{defi:specialization_model}, we know that the random vectors $\left\{ \mathbf{M}_{i*} : i \in \mathcal{S} \right\}$ are conditionally independent given a \emph{pair of type vectors} $\left( \mathbf{t}, \mathbf{w} \right)$. Let $\mathbf{x}_{i*} := \left( x_{ij} : j \in [n] \right) \in \left\{ \pm 1 \right\}^n$ for $i \in \mathcal{S}$. Then, it's clear that
\begin{equation}
    \label{eqn:proof_lemma:proof_prop:performance_SS_Q_in_M1_v1_v1}
    \begin{split}
    &\mathbb{P} \left\{ \left. \left( \mathbf{M}_{i*} : i \in \mathcal{S} \right) = \left( \mathbf{x}_{i*} : i \in \mathcal{S} \right) \right| \mathbf{t}, \mathbf{w} \right\}\\
    & = \prod_{i \in \mathcal{S}} \mathbb{P} \left\{ \left. \mathbf{M}_{i*} = \mathbf{x}_{i*} \right| \mathbf{t}, \mathbf{w} \right\}.
    \end{split}
\end{equation}
So we reach
\begin{equation*}
    \begin{split}
        &\mathbb{P} \left\{ \left. \left( \mathbf{M}_{i*} : i \in \mathcal{S} \right) = \left( \mathbf{x}_{i*} : i \in \mathcal{S} \right) \right| \mathbf{w} \right\} \\ = \ & \mathbb{E}_{\mathbf{t} \sim \bm{\mu}^{\otimes m}} \left[ \mathbb{P} \left\{ \left. \left( \mathbf{M}_{i*} : i \in \mathcal{S} \right) = \left( \mathbf{x}_{i*} : i \in \mathcal{S} \right) \right| \mathbf{t}, \mathbf{w} \right\} \right] \\
        = \ & \mathbb{E}_{\mathbf{t} \sim \bm{\mu}^{\otimes m}} \left[ \prod_{i \in \mathcal{S}} \mathbb{P} \left\{ \left. \mathbf{M}_{i*} = \mathbf{x}_{i*} \right| \mathbf{t}, \mathbf{w} \right\} \right] \\
        \stackrel{\textnormal{(a)}}{=} \ & \mathbb{E}_{\mathbf{t} \sim \bm{\mu}^{\otimes m}} \left[ \prod_{i \in \mathcal{S}} \mathbb{P} \left\{ \left. \mathbf{M}_{i*} = \mathbf{x}_{i*} \right| t_i, \mathbf{w} \right\} \right] \\
        = \ & \prod_{i \in \mathcal{S}} \mathbb{E}_{t_i \sim \bm{\mu}} \left[ \mathbb{P} \left\{ \left. \mathbf{M}_{i*} = \mathbf{x}_{i*} \right| t_i, \mathbf{w} \right\} \right] \\
        = \ & \prod_{i \in \mathcal{S}} \mathbb{P} \left\{ \left. \mathbf{M}_{i*} = \mathbf{x}_{i*} \right| \mathbf{w} \right\},
    \end{split}
\end{equation*}
where the step (a) holds since
\begin{equation}
    \label{eqn:proof_lemma:proof_prop:performance_SS_Q_in_M1_v1_v2}
    \begin{split}
        &\mathbb{P} \left\{ \left. \mathbf{M}_{i*} \mathbf{x}_{i*} \right| \mathbf{t}, \mathbf{w} \right\} \\
        = \ & \prod_{j=1}^{n} \mathbb{P} \left\{ \left. M_{ij} = x_{ij} \right| \mathbf{t}, \mathbf{w} \right\} \\
        = \ & \prod_{j=1}^{n} \left[ F_{ij}^{\frac{1 + a_i x_{ij}}{2}} \left( 1 - F_{ij} \right)^{\frac{1 - a_i x_{ij}}{2}} \right] \\
        = \ & \prod_{w=1}^{d} \left[ \prod_{j \in \mathcal{W}_w} \mathcal{Q} \left( t_i, w \right)^{\frac{1 + a_i x_{ij}}{2}} \left( 1 - \mathcal{Q} \left( t_i, w \right) \right)^{\frac{1 - a_i x_{ij}}{2}} \right],
    \end{split}
\end{equation}
and one can observe that the last term of the equation \eqref{eqn:proof_lemma:proof_prop:performance_SS_Q_in_M1_v1_v2} depends only on $t_i$ among all the coordinates of the task type vector $\mathbf{t} \in [d]^m$. This completes the proof of Lemma \ref{lemma:proof_prop:performance_SS_Q_in_M1_v1}.

\bibliographystyle{IEEEtran}
\bibliography{IEEEabrv,main}

\begin{thebibliography}{10}
\providecommand{\url}[1]{#1}
\csname url@samestyle\endcsname
\providecommand{\newblock}{\relax}
\providecommand{\bibinfo}[2]{#2}
\providecommand{\BIBentrySTDinterwordspacing}{\spaceskip=0pt\relax}
\providecommand{\BIBentryALTinterwordstretchfactor}{4}
\providecommand{\BIBentryALTinterwordspacing}{\spaceskip=\fontdimen2\font plus
\BIBentryALTinterwordstretchfactor\fontdimen3\font minus
  \fontdimen4\font\relax}
\providecommand{\BIBforeignlanguage}[2]{{%
\expandafter\ifx\csname l@#1\endcsname\relax
\typeout{** WARNING: IEEEtran.bst: No hyphenation pattern has been}%
\typeout{** loaded for the language `#1'. Using the pattern for}%
\typeout{** the default language instead.}%
\else
\language=\csname l@#1\endcsname
\fi
#2}}
\providecommand{\BIBdecl}{\relax}
\BIBdecl

\bibitem{kim2021crowdsourced}
D.~Kim and H.~W. Chung, ``Crowdsourced labeling for worker-task specialization
  model,'' in \emph{{IEEE} International Symposium on Information Theory,
  {ISIT} 2021, Melbourne, Australia, July 12-20, 2021}.\hskip 1em plus 0.5em
  minus 0.4em\relax {IEEE}, 2021, pp. 3191--3195.

\bibitem{kim2022generalized}
D.~Kim, J.~Lee, and H.~W. Chung, ``A generalized worker-task specialization
  model for crowdsourcing: Optimal limits and algorithm,'' in \emph{2022 IEEE
  International Symposium on Information Theory (ISIT)}.\hskip 1em plus 0.5em
  minus 0.4em\relax IEEE, 2022, pp. 1483--1488.

\bibitem{dawid1979maximum}
A.~P. Dawid and A.~M. Skene, ``Maximum likelihood estimation of observer
  error-rates using the em algorithm,'' \emph{Journal of the Royal Statistical
  Society: Series C (Applied Statistics)}, vol.~28, no.~1, pp. 20--28, 1979.

\bibitem{jeong2023recovering}
H.~Jeong and H.~W. Chung, ``Recovering top-two answers and confusion
  probability in multi-choice crowdsourcing,'' in \emph{International
  Conference on Machine Learning}.\hskip 1em plus 0.5em minus 0.4em\relax PMLR,
  2023, pp. 14\,836--14\,868.

\bibitem{zhang2016spectral}
Y.~Zhang, X.~Chen, D.~Zhou, and M.~I. Jordan, ``Spectral methods meet em: A
  provably optimal algorithm for crowdsourcing,'' \emph{The Journal of Machine
  Learning Research}, vol.~17, no.~1, pp. 3537--3580, 2016.

\bibitem{dalvi2013aggregating}
N.~Dalvi, A.~Dasgupta, R.~Kumar, and V.~Rasgoti, ``Aggregating crowdsourced
  binary ratings,'' in \emph{International World Wide Web Conference}, 2013,
  pp. 185--294.

\bibitem{ghosh2011moderates}
A.~Ghosh, S.~Kale, and P.~McAfee, ``Who moderates the moderators?:
  crowdsourcing abuse detection in user-generated content,'' in
  \emph{Proceedings of the 12th ACM conference on Electronic commerce}.\hskip
  1em plus 0.5em minus 0.4em\relax ACM, 2011, pp. 167--176.

\bibitem{karger2013efficient}
D.~R. Karger, S.~Oh, and D.~Shah, ``Efficient crowdsourcing for multi-class
  labeling,'' in \emph{Proceedings of the ACM SIGMETRICS/international
  conference on Measurement and modeling of computer systems}, 2013, pp.
  81--92.

\bibitem{karger2014budget}
------, ``Budget-optimal task allocation for reliable crowdsourcing systems,''
  \emph{Operations Research}, vol.~62, no.~1, pp. 1--24, 2014.

\bibitem{karger2011iterative}
D.~Karger, S.~Oh, and D.~Shah, ``Iterative learning for reliable crowdsourcing
  systems,'' \emph{Advances in neural information processing systems}, vol.~24,
  2011.

\bibitem{li2014error}
H.~Li and B.~Yu, ``Error rate bounds and iterative weighted majority voting for
  crowdsourcing,'' \emph{arXiv preprint arXiv:1411.4086}, 2014.

\bibitem{liu2012variational}
Q.~Liu, J.~Peng, and A.~T. Ihler, ``Variational inference for crowdsourcing,''
  in \emph{Advances in neural information processing systems}, 2012, pp.
  692--700.

\bibitem{ok2016optimality}
J.~Ok, S.~Oh, J.~Shin, and Y.~Yi, ``Optimality of belief propagation for
  crowdsourced classification,'' in \emph{International Conference on Machine
  Learning}, 2016, pp. 535--544.

\bibitem{gao2013minimax}
C.~Gao and D.~Zhou, ``Minimax optimal convergence rates for estimating ground
  truth from crowdsourced labels,'' \emph{arXiv preprint arXiv:1310.5764},
  2013.

\bibitem{pmlr-v80-ma18b}
Y.~Ma, A.~Olshevsky, C.~Szepesvari, and V.~Saligrama, ``Gradient descent for
  sparse rank-one matrix completion for crowd-sourced aggregation of sparsely
  interacting workers,'' in \emph{Proceedings of the 35th International
  Conference on Machine Learning}, 2018, pp. 3335--3344.

\bibitem{ma2020adversarial}
Q.~Ma and A.~Olshevsky, ``Adversarial crowdsourcing through robust rank-one
  matrix completion,'' \emph{Advances in Neural Information Processing
  Systems}, vol.~33, pp. 21\,841--21\,852, 2020.

\bibitem{ibrahim2019crowdsourcing}
S.~Ibrahim, X.~Fu, N.~Kargas, and K.~Huang, ``Crowdsourcing via pairwise
  co-occurrences: Identifiability and algorithms,'' \emph{Advances in neural
  information processing systems}, vol.~32, 2019.

\bibitem{gao2016exact}
C.~Gao, Y.~Lu, and D.~Zhou, ``Exact exponent in optimal rates for
  crowdsourcing,'' in \emph{International Conference on Machine
  Learning}.\hskip 1em plus 0.5em minus 0.4em\relax PMLR, 2016, pp. 603--611.

\bibitem{raykar2010learning}
V.~C. Raykar, S.~Yu, L.~H. Zhao, G.~H. Valadez, C.~Florin, L.~Bogoni, and
  L.~Moy, ``Learning from crowds.'' \emph{Journal of machine learning
  research}, vol.~11, no.~4, 2010.

\bibitem{smyth1994inferring}
P.~Smyth, U.~Fayyad, M.~Burl, P.~Perona, and P.~Baldi, ``Inferring ground truth
  from subjective labelling of venus images,'' \emph{Advances in neural
  information processing systems}, vol.~7, 1994.

\bibitem{ipeirotis2010quality}
P.~G. Ipeirotis, F.~Provost, and J.~Wang, ``Quality management on amazon
  mechanical turk,'' in \emph{Proceedings of the ACM SIGKDD workshop on human
  computation}, 2010, pp. 64--67.

\bibitem{berend2014consistency}
D.~Berend and A.~Kontorovich, ``Consistency of weighted majority votes,''
  \emph{Advances in Neural Information Processing Systems}, vol.~27, 2014.

\bibitem{khetan2016achieving}
A.~Khetan and S.~Oh, ``Achieving budget-optimality with adaptive schemes in
  crowdsourcing,'' \emph{Advances in Neural Information Processing Systems},
  vol.~29, pp. 4844--4852, 2016.

\bibitem{zhou2014aggregating}
D.~Zhou, Q.~Liu, J.~Platt, and C.~Meek, ``Aggregating ordinal labels from
  crowds by minimax conditional entropy,'' in \emph{International conference on
  machine learning}.\hskip 1em plus 0.5em minus 0.4em\relax PMLR, 2014, pp.
  262--270.

\bibitem{shah2020permutation}
N.~B. Shah, S.~Balakrishnan, and M.~J. Wainwright, ``A permutation-based model
  for crowd labeling: Optimal estimation and robustness,'' \emph{IEEE
  Transactions on Information Theory}, vol.~67, no.~6, pp. 4162--4184, 2020.

\bibitem{shah2018reducing}
D.~Shah and C.~Lee, ``Reducing crowdsourcing to graphon estimation,
  statistically,'' in \emph{International Conference on Artificial Intelligence
  and Statistics}, 2018, pp. 1741--1750.

\bibitem{ho2013adaptive}
C.-J. Ho, S.~Jabbari, and J.~W. Vaughan, ``Adaptive task assignment for
  crowdsourced classification,'' in \emph{International Conference on Machine
  Learning}.\hskip 1em plus 0.5em minus 0.4em\relax PMLR, 2013, pp. 534--542.

\bibitem{ho2012online}
C.-J. Ho and J.~Vaughan, ``Online task assignment in crowdsourcing markets,''
  in \emph{Proceedings of the AAAI Conference on Artificial Intelligence},
  vol.~26, no.~1, 2012, pp. 45--51.

\bibitem{welinder2010multidimensional}
P.~Welinder, S.~Branson, P.~Perona, and S.~Belongie, ``The multidimensional
  wisdom of crowds,'' \emph{Advances in neural information processing systems},
  vol.~23, 2010.

\bibitem{zhou2012learning}
D.~Zhou, S.~Basu, Y.~Mao, and J.~Platt, ``Learning from the wisdom of crowds by
  minimax entropy,'' \emph{Advances in neural information processing systems},
  vol.~25, 2012.

\bibitem{zhou2015regularized}
D.~Zhou, Q.~Liu, J.~C. Platt, C.~Meek, and N.~B. Shah, ``Regularized minimax
  conditional entropy for crowdsourcing,'' \emph{arXiv preprint
  arXiv:1503.07240}, 2015.

\bibitem{khetan2018learning}
A.~Khetan, Z.~C. Lipton, and A.~Anandkumar, ``Learning from noisy
  singly-labeled data,'' in \emph{International Conference on Learning
  Representations}.

\bibitem{mazumdar2017clustering}
A.~Mazumdar and B.~Saha, ``Clustering with noisy queries,'' \emph{Advances in
  Neural Information Processing Systems}, vol.~30, 2017.

\bibitem{vinayak2016crowdsourced}
R.~K. Vinayak and B.~Hassibi, ``Crowdsourced clustering: Querying edges vs
  triangles,'' in \emph{Advances in Neural Information Processing Systems},
  2016, pp. 1316--1324.

\bibitem{abbe2015exact}
E.~Abbe, A.~S. Bandeira, and G.~Hall, ``Exact recovery in the stochastic block
  model,'' \emph{IEEE Transactions on information theory}, vol.~62, no.~1, pp.
  471--487, 2015.

\bibitem{kim2017community}
C.~Kim, A.~S. Bandeira, and M.~X. Goemans, ``Community detection in
  hypergraphs, spiked tensor models, and sum-of-squares,'' in
  \emph{International Conference on Sampling Theory and Applications (SampTA)},
  2017, pp. 124--128.

\bibitem{ahn2019community}
K.~Ahn, K.~Lee, and C.~Suh, ``Community recovery in hypergraphs,'' \emph{IEEE
  Transactions on Information Theory}, vol.~65, no.~10, pp. 6561--6579, 2019.

\bibitem{lee2020robust}
J.~Lee, D.~Kim, and H.~W. Chung, ``Robust hypergraph clustering via convex
  relaxation of truncated mle,'' \emph{IEEE Journal on Selected Areas in
  Information Theory}, vol.~1, no.~3, pp. 613--631, 2020.

\bibitem{dorfman1943detection}
R.~Dorfman, ``The detection of defective members of large populations,''
  \emph{The Annals of mathematical statistics}, vol.~14, no.~4, pp. 436--440,
  1943.

\bibitem{gallager1962low}
R.~Gallager, ``Low-density parity-check codes,'' \emph{IRE Transactions on
  information theory}, vol.~8, no.~1, pp. 21--28, 1962.

\bibitem{mackay2005fountain}
D.~J. MacKay, ``Fountain codes,'' \emph{IEEE Proceedings-Communications}, vol.
  152, no.~6, pp. 1062--1068, 2005.

\bibitem{arikan2009channel}
E.~Arikan, ``Channel polarization: A method for constructing capacity-achieving
  codes for symmetric binary-input memoryless channels,'' \emph{IEEE
  Transactions on Information Theory}, vol.~55, no.~7, pp. 3051--3073, 2009.

\bibitem{pittel2016satisfiability}
B.~Pittel and G.~B. Sorkin, ``The satisfiability threshold for k-xorsat,''
  \emph{Combinatorics, Probability and Computing}, vol.~25, no.~2, pp.
  236--268, 2016.

\bibitem{DBLP:journals/tit/KimC21}
D.~Kim and H.~W. Chung, ``Binary classification with {XOR} queries: Fundamental
  limits and an efficient algorithm,'' \emph{{IEEE} Trans. Inf. Theory},
  vol.~67, no.~7, pp. 4588--4612, 2021.

\bibitem{DBLP:conf/isit/KimC20}
------, ``Crowdsourced classification with {XOR} queries: An algorithm with
  optimal sample complexity,'' in \emph{{IEEE} International Symposium on
  Information Theory, {ISIT} 2020, Los Angeles, CA, USA, June 21-26,
  2020}.\hskip 1em plus 0.5em minus 0.4em\relax {IEEE}, 2020, pp. 2551--2555.

\bibitem{8437703}
H.~W. Chung, J.~O. Lee, and A.~O. Hero, ``Fundamental limits on data
  acquisition: Trade-offs between sample complexity and query difficulty,'' in
  \emph{2018 IEEE International Symposium on Information Theory (ISIT)}, 2018,
  pp. 681--685.

\bibitem{fei2018exponential}
Y.~Fei and Y.~Chen, ``Exponential error rates of sdp for block models: Beyond
  grothendieck’s inequality,'' \emph{IEEE Transactions on Information
  Theory}, vol.~65, no.~1, pp. 551--571, 2018.

\bibitem{ames2014guaranteed}
B.~P. Ames, ``Guaranteed clustering and biclustering via semidefinite
  programming,'' \emph{Mathematical Programming}, vol. 147, no. 1-2, pp.
  429--465, 2014.

\bibitem{vinayak2016similarity}
R.~K. Vinayak and B.~Hassibi, ``Similarity clustering in the presence of
  outliers: Exact recovery via convex program,'' in \emph{2016 IEEE
  International Symposium on Information Theory (ISIT)}.\hskip 1em plus 0.5em
  minus 0.4em\relax IEEE, 2016, pp. 91--95.

\bibitem{chen2014improved}
Y.~Chen, S.~Sanghavi, and H.~Xu, ``Improved graph clustering,'' \emph{IEEE
  Transactions on Information Theory}, vol.~60, no.~10, pp. 6440--6455, 2014.

\bibitem{chen2016statistical}
Y.~Chen and J.~Xu, ``Statistical-computational tradeoffs in planted problems
  and submatrix localization with a growing number of clusters and
  submatrices,'' \emph{The Journal of Machine Learning Research}, vol.~17,
  no.~1, pp. 882--938, 2016.

\bibitem{chen2018convexified}
Y.~Chen, X.~Li, and J.~Xu, ``Convexified modularity maximization for
  degree-corrected stochastic block models,'' \emph{The Annals of Statistics},
  vol.~46, no.~4, pp. 1573--1602, 2018.

\bibitem{huang2021solving}
B.~Huang, S.~Jiang, Z.~Song, and R.~Tao, ``Solving tall dense sdps in the
  current matrix multiplication time,'' \emph{arXiv preprint arXiv:2101.08208},
  2021.

\bibitem{jiang2020faster}
H.~Jiang, T.~Kathuria, Y.~T. Lee, S.~Padmanabhan, and Z.~Song, ``A faster
  interior point method for semidefinite programming,'' in \emph{2020 IEEE 61st
  annual symposium on foundations of computer science (FOCS)}.\hskip 1em plus
  0.5em minus 0.4em\relax IEEE, 2020, pp. 910--918.

\bibitem{wen2010alternating}
Z.~Wen, D.~Goldfarb, and W.~Yin, ``Alternating direction augmented lagrangian
  methods for semidefinite programming,'' \emph{Mathematical Programming
  Computation}, vol.~2, no.~3, pp. 203--230, 2010.

\bibitem{tropp2012user}
J.~A. Tropp, ``User-friendly tail bounds for sums of random matrices,''
  \emph{Foundations of computational mathematics}, vol.~12, no.~4, pp.
  389--434, 2012.

\bibitem{shi2000normalized}
J.~Shi and J.~Malik, ``Normalized cuts and image segmentation,'' \emph{IEEE
  Transactions on pattern analysis and machine intelligence}, vol.~22, no.~8,
  pp. 888--905, 2000.

\bibitem{von2007tutorial}
U.~Von~Luxburg, ``A tutorial on spectral clustering,'' \emph{Statistics and
  computing}, vol.~17, no.~4, pp. 395--416, 2007.

\bibitem{van2000asymptotic}
A.~W. Van~der Vaart, \emph{Asymptotic statistics}.\hskip 1em plus 0.5em minus
  0.4em\relax Cambridge university press, 2000, vol.~3.

\bibitem{watson1992characterization}
G.~A. Watson, ``Characterization of the subdifferential of some matrix norms,''
  \emph{Linear algebra and its applications}, vol. 170, no.~0, pp. 33--45,
  1992.

\bibitem{csiszar1998method}
I.~Csisz{\'a}r, ``The method of types [information theory],'' \emph{IEEE
  Transactions on Information Theory}, vol.~44, no.~6, pp. 2505--2523, 1998.

\bibitem{csiszar2011information}
I.~Csisz{\'a}r and J.~K{\"o}rner, \emph{Information theory: coding theorems for
  discrete memoryless systems}.\hskip 1em plus 0.5em minus 0.4em\relax
  Cambridge University Press, 2011.

\bibitem{sanov1958probability}
I.~N. Sanov, \emph{On the probability of large deviations of random
  variables}.\hskip 1em plus 0.5em minus 0.4em\relax United States Air Force,
  Office of Scientific Research, 1958.

\bibitem{klenke2013probability}
A.~Klenke, \emph{Probability theory: a comprehensive course}.\hskip 1em plus
  0.5em minus 0.4em\relax Springer Science \& Business Media, 2013.

\end{thebibliography}

\begin{IEEEbiographynophoto}{Doyeon Kim} received the B.S. degree in Electrical and Computer Engineering from University of Seoul in 2008, and M.S. and Ph.D. degrees in Electrical Engineering from Korea Advanced Institute of Science and Technology (KAIST) in 2018 and 2023, respectively. Since March 2023, he has been working at Samsung Electronics as a staff engineer. His research interests include statistical inference and machine learning.

\end{IEEEbiographynophoto}

\begin{IEEEbiographynophoto}{Jeonghwan Lee} received the B.S. degree in Mathematical Sciences with the highest honor at the College of Natural Sciences (summa cum laude) and the Presidential Award from Korea Advanced Institute of Science and Technology (KAIST) in 2022. He is currently a Ph.D. student in Statistics at the University of Chicago. He is a recipient of Doctoral Overseas Scholarship supported by the Kwanjeong Educational Foundation from 2022. His research interest includes high-dimensional statistics, theoretical foundations of machine learning (with emphasis on reinforcement learning and sequential decision-making, online learning, and learning in games), and optimization.
\end{IEEEbiographynophoto}

\begin{IEEEbiographynophoto}{Hye Won Chung}
(S'08--M'15) received the B.S. degree (with summa cum laude) from Korea Advanced Institute of Science and Technology (KAIST) in 2007, and the M.S. and Ph.D. degrees from Massachusetts Institute of Technology (MIT) in 2009 and 2014, respectively, all in Electrical Engineering and Computer Science. From August 2014 to May 2017, she worked as a Research Fellow in the Department of Electrical Engineering and Computer Science at the University of Michigan. Since June 2017, she has been working at School of Electrical Engineering at KAIST, where she is now an Associate Professor. Her research interests include information theory, data science, statistical inference and machine learning. 
\end{IEEEbiographynophoto}

\end{document}